# A Manifestly Causal Approach to Quantum Field Theory

A thesis submitted to the University of Manchester for the degree of

Doctor of Philosophy

in the Faculty of Science and Engineering

2025

Ross D. Jenkinson

School of Natural Sciences

Department of Physics and Astronomy

# Contents

**Word count**: 33300

# List of figures

# List of publications

[1] R. Dickinson, J. Forshaw, R. Jenkinson, and P. Millington, *Towards a Manifestly Causal Approach to Particle Scattering*, submitted to Physical Review D, arXiv: 2502.18551 [hep-ph], Feb. 2025.

[2] R. Dickinson, J. Forshaw, R. Jenkinson, and P. Millington, *A new study of the Unruh effect*, Classical and Quantum Gravity, vol. 42, no. 2, p. 025014, 2025. DOI: 10.1088/1361-6382/ad9c12. arXiv: 2409.12697 [hep-th].

# Acronyms

**BCH** Baker-Campbell-Hausdorff. 82, 84, 176

**BN** Bloch-Nordsieck. 12, 43, 94, 119, 176

**IR** infrared. 12, 19, 42–44, 46, 47, 49–53, 55, 58, 59, 72, 74, 75, 77, 80, 81, 94, 112, 119, 176, 177

**KLN** Kinoshita-Lee-Nauenberg. 12, 43, 58, 81, 94, 119, 176

**KMS** Kubo-Martin-Schwinger. 166

**LSZ** Lehmann–Symanzik–Zimmermann. 36–38, 45, 46, 55, 57, 58, 70, 73, 80, 81

**QCD** quantum chromodynamics. 28, 30, 31, 34, 39, 41–43, 56, 58, 59, 67, 94, 112, 119, 176

**QED** quantum electrodynamics. 28–31, 34, 35, 39, 42, 43, 94, 112, 119

**QFT** quantum field theory. 12, 13, 18, 19, 21, 25, 28, 31, 35, 39, 42, 44, 80–82, 90, 95, 121, 149, 176, 178

**RQI** relativistic quantum information. 85, 121, 148, 178

**UdW** Unruh-DeWitt. 12, 20, 121, 125, 126, 129, 147, 149, 150, 163, 167, 169–172, 175, 177, 178

**UV** ultraviolet. 12, 19, 42–44, 46, 47, 49, 50, 52, 53, 55, 59, 72, 73, 80, 215

# Abstract

We develop a probability-level, manifestly causal formalism for calculations in quantum field theory (QFT). The approach involves an implicit summation over final states, which makes causality manifest since retarded propagators emerge naturally. This inclusive summation over final states may also offer insights into the cancellation of infrared (IR) divergences in physical observables within gauge theories, in accordance with the Bloch-Nordsieck (BN) and Kinoshita-Lee-Nauenberg (KLN) theorems. To study this, we first conduct particle scattering calculations using conventional methods, determining the quark–antiquark production cross section at first-order in gluon corrections, with careful tracking and cancellation of both IR and ultraviolet (UV) divergences. We then apply the causal formalism to analogous processes in scalar field theory, introducing novel diagrams that represent algebraic terms at the probability level, akin to Feynman diagrams at the amplitude level. We present a list of rules that generate all probability-level diagrams for particle scattering processes in which one is fully inclusive over final states that contain no initial-state particles.

We also investigate the Unruh effect through the lens of the causal formalism. We calculate the transition rate of a uniformly accelerating Unruh-DeWitt (UdW) monopole detector coupled to a massive scalar field, from both the perspective of an inertial (Minkowski) observer and an accelerating (Rindler) observer. We confirm that the two perspectives give the same transition rate, despite the Rindler observer describing the Minkowski vacuum state as a thermal bath of particles. Numerical results for the transition rate are presented and explained, highlighting the transient effects caused by forcing the field to initially be in the Minkowski vacuum state. Finally, we review the literature regarding the response of an UdW detector on various trajectories in the spacetime of a $(3 + 1)$-dimensional Schwarzschild black hole, with a view to extending the analysis in the future using our causal formalism.

# Lay abstract

Our modern understanding of reality is that it is constructed from various *fields*, which pervade throughout the entire universe. Small excitations (bumps of energy) in these fields can only occur in discrete (*quantised*) amounts. These quantum excitations are what we call particles. This picture of reality is mathematically described by *quantum field theory (QFT)*.

Despite their incredible accuracy and predictivity, our current formulations of QFT are not perfect. Firstly, the law of cause and effect is at the heart of modern physics: nothing can travel faster than light, so any events which would require superluminal travel to affect one another must be causally disconnected. However, this is not always apparent in QFT calculations. Secondly, the mathematics we use in QFT often involves infinite terms, which does not make sense until we construct a sensible observable quantity for which these infinities cancel. The existence of these infinities in intermediate steps can make calculations difficult, both by hand and by computer.

In this thesis, we introduce a new approach to QFT in which the law of cause and effect is always obviously obeyed. We also investigate whether this new formalism may help avoid infinities in intermediate steps of calculations. We apply our approach to particle scattering and the Unruh effect. The Unruh effect is an effect baked into QFT that says that an accelerating observer perceives the emptiness of space not as a vacuum, but as a thermal bath of particles. There is no inconsistency between the perspectives of a stationary observer, who sees the space as empty, and the accelerating observer; both perspectives are true. The key is that they are defining particles differently. This leads to the conclusion that the concept of a 'particle' is observer-dependent. This is also true in gravitational fields, leading us to investigate how a detector responds near a black hole—an object which exhibits strong gravitational and quantum effects. Understanding how gravity and QFT work together is the biggest unknown question in theoretical physics.

# Declaration of originality

I hereby confirm that no portion of the work referred to in the thesis has been submitted in support of an application for another degree or qualification of this or any other university or other institute of learning.

# Copyright statement

# Acknowledgements

I would first like to express my deepest gratitude to my supervisors, Professor Jeff Forshaw and Professor Brian Cox. Without them, this project—and my career as a physicist so far—would not exist. Jeff has been a consistent source of clarity, opportunities, and inspiration, always able to explain the most complex subtleties in the most engaging way. Outside of physics, his friendship has made my Manchester experience richer. Brian has provided many insightful conversations, and I am repeatedly impressed by both his depth and breadth of scientific knowledge. He has created wonderful opportunities for me, and offered invaluable life and career advice.

My appreciation extends to the University of Manchester as a whole for curating a vibrant research environment and providing travel grants that enabled me to attend interesting workshops and conferences, such as the Quantum Field Theory in Curved Spacetimes 2024 workshop in Lisbon, for which I thank the organisers. I also thank the organisers of the Vitae 3-Minute Thesis competition and the Manchester particle physics group's annual Christmas meetings, which allowed me to hone my creative presentation skills.

I owe a great debt of gratitude to Zia Hayat and Callsign Ltd for funding this project and maintaining an interest in my research. I am especially grateful for the invitation to their offices in Abu Dhabi, where I was honoured to attend prestigious events and deliver presentations about my work. They were excellent hosts and provided an experience I will never forget. I also thank Dame Nancy Rothwell for her interest in my research and her enthusiasm to create further opportunities for me.

Collaboration with Dr Peter Millington and Robert Dickinson has been crucial to the success of our publications, and to them I am thankful for many enlightening conversations. Furthermore, if it weren't for the guidance and passion of previous tutors, I may never have pursued physics at all. Namely, I would like to thank Professor Andy Bunker, Professor Richard Berry, and Professor Chris Hays at the University of Oxford, and Nicholas Davies at Wilmslow High School.

I thank my wonderful friends, whom I have met throughout my life in Wilmslow, Oxford, and Manchester, who are responsible for my fondest memories and most eagerly awaited events. The greatest compliment I can give myself is the quality of the people who call me their friend.

My parents, Lorraine and Mark, deserve all of the credit (or none of the blame) for the person I am today and everything I have achieved. Their limitless love has always encouraged me to aim high and pursue what I am most interested in—a privilege not everyone has. I am eternally grateful for their work to provide the life they have given me. This sentiment is extended to my brothers, Ryan and Adam, and my extended family, especially my grandparents.

# Chapter 1

# Introduction

Quantum field theory (QFT) is the framework that unites quantum mechanics with the principles of special relativity, offering an elegant yet intricate language to describe Nature at a fundamental level. In traditional quantum mechanics, particles are treated as point-like entities whose dynamics are governed by the Schrödinger equation. QFT, by contrast, treats fields as the primary objects, with particles emerging as quantised excitations of the fields. Extending the quantum mechanical model to fields with an infinite number of degrees of freedom is essential for accommodating the relativistic processes that defy a simple single-particle interpretation, such as particle-antiparticle pair production and the interactions of virtual particles.

At its core, QFT provides a unified description of fundamental interactions. The computational methods of QFT yield theoretical predictions that agree with experiments to unprecedented precision [3–5], offering profound insights into the structure of our universe. The Standard Model [6–8], formulated within the framework of QFT, successfully describes the electromagnetic, weak, and strong forces. Moreover, in condensed matter systems [9, 10], statistical mechanics [11, 12], and even quantum gravity [13, 14], QFT has proven to be an important tool, offering deep insights into emergent phenomena.

Historically, some of the most significant advancements in QFT have emerged from studies of gravitational systems. For instance, the analysis of quantum fields in the curved spacetime near black holes led to the discovery of Hawking radiation [15–17], which revealed deep connections between QFT, thermodynamics, and general relativity. Likewise, the Unruh effect [18–22], which predicts that an accelerated observer perceives the vacuum as a thermal bath, reinforces the subtle interplay between acceleration, quantum fluctuations, and field theory. These breakthroughs not only underscore QFT's broad applicability but also motivate further investigation into the response of quantum fields in both flat and curved spacetimes.

A key technical aspect of QFT is the appearance of divergences when performing calculations. Ultraviolet (UV) divergences arise from contributions at arbitrarily high energies and are handled through renormalisation techniques that redefine parameters and lead to effective field theories valid at low energy scales [23–26]. In contrast, infrared (IR) divergences occur due to contributions at very low energies, requiring the careful definition of observables to yield finite predictions [27–29]. These IR divergences pose significant technical challenges, especially in more complex, higher-order calculations in perturbation theory.

A central pillar of QFT is the principle of causality. In a relativistic context, causality ensures that effects do not precede their causes. This is mathematically embedded in QFT through conditions such as micro-causality, where field operators are required to commute or anticommute at spacelike separations. Despite being built into the axioms of QFT, causality is often not manifest in perturbative calculations, leading to conceptual subtleties that continue to challenge our intuitive understanding. This hidden nature of causality exemplifies the dual character of QFT; it is both extraordinarily predictive and, in many respects, conceptually incomplete [25, 30, 31].

This thesis is motivated by a desire to elucidate the role of causality and the cancellation of divergences in QFT. To this end, we present a new formalism of QFT, originally established in Refs. [32–34], in which causality is made manifest through the appearance of field commutators and anticommutators. The investigation of causality in QFT forms an active area of research, with the causal formalism developed in this thesis seeing use in the field of causal set theory [35], and other studies also attempting to devise new formalisms of QFT in which causality is manifest [36–39].

Chapter 2 begins with an overview of the relevant background theory required to understand the content of this thesis. We specifically focus on the details of divergences in QFT, which sets the stage for the rigorous calculation of a particle-scattering cross section in Chapter 3. This calculation requires the cancellation of divergences, and comprehending how these cancellations yield finite, physically measurable observables is essential for accurate predictions in QFT. The ubiquity of the Feynman propagator, which is not manifestly causal, and the requirement of an inclusive observable then motivates the development of a new, manifestly causal formalism. This formalism is introduced in Chapter 4, where the Fermi two-atom problem is used to high-light the advantage of manifest causality.

Motivated by our scattering calculation, Chapter 5 employs the causal formalism to compute inclusive transition probabilities in scalar field theory. We present a new set of rules that generate diagrams at the probability level, akin to Feynman diagrams at the amplitude level. These diagrams explicitly include the retarded propagator, meaning that causality is evident. Chapter 6 then uses the causal formalism to explore the Unruh effect, rederiving known results and presenting new numerical results. Chapter 7 serves as a literature review and pedagogical introduction to the recent progress regarding the response of an Unruh-DeWitt (UdW) detector in the Schwarzschild spacetime of a black hole. Finally, Chapter 8 presents our conclusions.

The contents of Chapters 5 and 6 are published in Refs. [1] and [2]. Throughout this thesis, we adopt natural units $c = \hbar = k_{\mathrm{B}} = G = 1$ and the 'mostly-minus' metric signature $(+---)$. Hamiltonians and Lagrangians are denoted by $H$ and $L$, and Hamiltonian and Lagrangian *densities* are denoted by $\mathcal{H}$ and $\mathcal{L}$, such that $H = \int \mathrm{d}^3\mathbf{x}\, \mathcal{H}$ and $L = \int \mathrm{d}^3\mathbf{x}\, \mathcal{L}$.

# Chapter 2

# Background

## 2.1 Scalar Field Theory

Quantum field theory (QFT) arises as a natural framework for unifying quantum mechanics with special relativity, providing a consistent description of relativistic quantum particles. Scalar field theory serves as the simplest quantum field theory, describing spin-$0$ particles and forming the basis for more complex theories. The goal of scalar field theory is to construct a quantum description of a classical field and analyse its dynamics.

### 2.1.1 Quantising the Classical Theory

Consider a real-valued, classical scalar field $\phi \equiv \phi(x) \equiv \phi(\mathbf{x}, t)$ described by the Lagrangian density [23, 40]

$$\mathcal{L}_0 = \frac{1}{2}\left(\partial^\mu\phi\partial_\mu\phi - m^2\phi^2\right) = \frac{1}{2}\left(\dot{\phi}^2 - (\nabla\phi)^2 - m^2\phi^2\right) , \tag{2.1.1}$$

where $\partial^\mu = \partial/\partial x_\mu$ is the spacetime derivative, $\dot{\phi}(x) = \partial\phi/\partial t$, and $m$ is a parameter of the classical field theory (which will become the mass of the field quanta after quantisation). In natural units, the Lagrangian density has dimension $= 4$, and thus the scalar field, $\phi$, has dimension $= 1$.

The Euler-Lagrange equation [41–44],

$$\frac{\partial\mathcal{L}_0}{\partial\phi} = \partial_\mu\left(\frac{\partial\mathcal{L}_0}{\partial(\partial_\mu\phi)}\right) , \tag{2.1.2}$$

associated with Eq. (2.1.1) immediately yields the Klein-Gordon equation [45, 46],

$$(\partial^\mu \partial_\mu + m^2)\phi = 0. \tag{2.1.3}$$

This equation governs the free evolution of the field. Using Hamiltonian field theory [23, 47], the Hamiltonian density for this system is given by

$$\mathcal{H}_0 = \pi(x)\dot{\phi}(x) - \mathcal{L}_0 = \frac{1}{2}\left(\pi^2 + (\nabla\phi)^2 + m^2\phi^2\right)\,, \tag{2.1.4}$$

where $\pi \equiv \pi(x)$ is the momentum density conjugate to $\phi(x)$, given by

$$\pi(x) = \frac{\partial \mathcal{L}_0}{\partial \dot{\phi}} = \dot{\phi}\,. \tag{2.1.5}$$

If we Fourier expand the classical scalar field,

$$\phi(\mathbf{x}, t) = \int \frac{\mathrm{d}^3\mathbf{p}}{(2\pi)^3} e^{i\mathbf{p}\cdot\mathbf{x}} \phi(\mathbf{p}, t)\,, \tag{2.1.6}$$

the Klein-Gordon equation (Eq. (2.1.3)) becomes

$$\frac{\partial^2 \phi(\mathbf{p}, t)}{\partial t^2} = -\left(\mathbf{p}^2 + m^2\right)\phi(\mathbf{p}, t)\,. \tag{2.1.7}$$

This is exactly the equation of motion for a simple harmonic oscillator with frequency

$$\omega^2 = \mathbf{p}^2 + m^2. \tag{2.1.8}$$

Eq. (2.1.7) then tells us that that the scalar field in position space is an infinite set of simple harmonic oscillators, one for each wavenumber, $\mathbf{p}$. The quantisation of the simple harmonic oscillator is well-known [48–51], and thus we can similarly quantise the scalar field and its conjugate momentum as [52]

$$\phi(\mathbf{x}) = \int \frac{\mathrm{d}^3\mathbf{p}}{(2\pi)^3} \frac{1}{\sqrt{2\omega}} \left(a_\mathbf{p} e^{-i\mathbf{p}\cdot\mathbf{x}} + a_\mathbf{p}^\dagger e^{i\mathbf{p}\cdot\mathbf{x}}\right)\,, \tag{2.1.9}$$

$$\pi(\mathbf{x}) = \int \frac{\mathrm{d}^3\mathbf{p}}{(2\pi)^3} (-i)\sqrt{\frac{\omega}{2}} \left(a_\mathbf{p} e^{-i\mathbf{p}\cdot\mathbf{x}} + a_\mathbf{p}^\dagger e^{i\mathbf{p}\cdot\mathbf{x}}\right)\,, \tag{2.1.10}$$

where $a_\mathbf{p}$ and $a_\mathbf{p}^\dagger$ are the annihilation and creation operators, respectively, for quanta (particles) of the scalar field, analogous to those of a quantum harmonic oscillator. The scalar field, $\phi(\mathbf{x})$, has now been promoted to a quantum operator. The canonical quantisation relation relevant to the simple harmonic oscillator, $\left[a, a^\dagger\right] = 1$, has an analogous commutation relation for the scalar-field ladder operators,

$$[a_{\mathbf{p}}, a^\dagger_{\mathbf{p}'}] = (2\pi)^3 \delta^3(\mathbf{p} - \mathbf{p}'), \quad [a_{\mathbf{p}}, a_{\mathbf{p}'}] = [a^\dagger_{\mathbf{p}}, a^\dagger_{\mathbf{p}'}] = 0\,. \tag{2.1.11}$$

The Hamiltonian density of the system, Eq. (2.1.4), is written in terms of $\phi(\mathbf{x})$ and $\pi(\mathbf{x})$, and is hence also an operator. Using Eqs. (2.1.9) and (2.1.10), the Hamiltonian is given by

$$H_0 = \int \mathrm{d}^3\mathbf{x}\, \mathcal{H}_0 = \int \frac{\mathrm{d}^3\mathbf{p}}{(2\pi)^3}\, \omega_{\mathbf{p}} \left( a^\dagger_{\mathbf{p}} a_{\mathbf{p}} + \frac{1}{2}\left[a_{\mathbf{p}}, a^\dagger_{\mathbf{p}}\right] \right)\,. \tag{2.1.12}$$

The second term is proportional to the infinite constant term $\delta(0)$. This contribution comes from the sum over all modes of zero-point energies, $\omega_{\mathbf{p}}$. Since only energy differences are experimentally measurable, this term does not contribute to most observables, and is usually ignored [23, 24, 53].

Before considering interactions in scalar field theory, it is useful to write an expression for the scalar field as a function of $x = (t, \mathbf{x})$, i.e., in the Heisenberg picture instead of the Schrödinger picture. It can be shown that [23]

$$\phi(x) = \int \frac{\mathrm{d}^3\mathbf{p}}{(2\pi)^3} \frac{1}{\sqrt{2\omega_{\mathbf{p}}}} \left( a_{\mathbf{p}} e^{-ip\cdot x} + a^\dagger_{\mathbf{p}} e^{ip\cdot x} \right) \big|_{p^0=\omega_{\mathbf{p}}}\,. \tag{2.1.13}$$

From Eq. (2.1.11), it follows that

$$[\phi(\mathbf{x}, t), \pi(\mathbf{y}, t)] = i\delta^3(\mathbf{x} - \mathbf{y}), \tag{2.1.14}$$

$$[\phi(\mathbf{x}, t), \phi(\mathbf{y}, t)] = [\pi(\mathbf{x}, t), \pi(\mathbf{y}, t)] = 0. \tag{2.1.15}$$

These are the *equal-time commutation relations*.

### 2.1.2 Interactions

So far, free-particle states have been eigenstates of the Hamiltonian, which we call the free Hamiltonian, $H_0$ (or free Lagrangian, $L_0$). We now consider a Lagrangian density that includes interactions between particles, $\mathcal{L}_{\text{int}}$, which is treated as a pertur-bation on the free Lagrangian density, with the free theory serving as a basis,

$$\mathcal{L} = \mathcal{L}_0 + \mathcal{L}_{\text{int}} \quad \text{or} \quad \mathcal{H} = \mathcal{H}_0 + \mathcal{H}_{\text{int}} \,. \tag{2.1.16}$$

For scalar fields, the theory we consider in this thesis is 'phi-cubed' theory [54, 55],

$$\mathcal{L}_{\text{int}} = -\frac{\lambda}{3!}\phi^3 \Rightarrow \mathcal{H}_{\text{int}} = \frac{\lambda}{3!}\phi^3 \,, \tag{2.1.17}$$

where $\lambda$ is the coupling constant (of dimension $= 1$), which must be small for perturbation theory to be valid. The factor of $1/3!$ is a symmetry factor which corrects for overcounting, since there are three identical field operators, $\phi$.

The relationship between the Schrödinger picture field operator, $\phi(\mathbf{x}, t_0)$, at fixed time, $t_0$, and the Heisenberg picture field operator, $\phi(\mathbf{x}, t)$, is

$$\phi(\mathbf{x}, t) = e^{iH(t-t_0)}\phi(\mathbf{x}, t_0)e^{-iH(t-t_0)} \,. \tag{2.1.18}$$

We define the *interaction picture* field operator as

$$\phi_I(\mathbf{x}, t) = e^{iH_0(t-t_0)}\phi(\mathbf{x}, t_0)e^{-iH_0(t-t_0)} \,. \tag{2.1.19}$$

Similarly, we define the interaction picture Hamiltonian as

$$H_I(t) = e^{iH_0(t-t_0)}\, H_{\text{int}}\, e^{-iH_0(t-t_0)} = \int \mathrm{d}^3\mathbf{x}\, \frac{\lambda}{3!}\phi_I^3 \,. \tag{2.1.20}$$

The relationship between the full Heisenberg field, $\phi$, and the interaction picture field, $\phi_I$, is thus

$$\phi(\mathbf{x}, t) = e^{iH_{\text{int}}(t-t_0)}\phi_I(\mathbf{x}, t)e^{-iH_{\text{int}}(t-t_0)} \equiv U^\dagger(t, t_0)\, \phi_I(\mathbf{x}, t)U(t, t_0) \,, \tag{2.1.21}$$

where we have defined the *unitary time-evolution operator* as

$$U(t, t_0) \equiv e^{-iH_{\text{int}}(t-t_0)} \,. \tag{2.1.22}$$

The time-evolution operator can be written in terms of $H_I$ (and hence $\phi_I$) as

$$U(t, t_0) = 1 + (-i)\int_{t_0}^{t} \mathrm{d}t_1 H_I(t_1) + (-i)^2 \int_{t_0}^{t} \mathrm{d}t_1 \int_{t_0}^{t_1} \mathrm{d}t_2 H_I(t_1)H_I(t_2) + \cdots$$

$$\equiv \mathrm{T}\left\{\exp\left\{-i\int_{t_0}^{t} \mathrm{d}t' H_I(t')\right\}\right\} , \tag{2.1.23}$$

where T indicates time-ordering, such that operators are written from left to right ordered by the time they are evaluated at, latest to earliest. For the remainder of this thesis, we work in the interaction picture unless otherwise stated.

### 2.1.3 Wick's Theorem

Eq. (2.1.23) involves the time-ordered product of an arbitrary number of field operators, $\phi(x_i) = \phi_i$. This is calculated using Wick's theorem [56], which expresses the time-ordered product of field operators in terms of normal-ordered products, $N(\cdots)$, and Feynman propagators,

$$F^{\phi}_{xy} = \left\langle 0^{\phi}\middle|\mathrm{T}\{\phi_x\phi_y\}\middle|0^{\phi}\right\rangle . \tag{2.1.24}$$

The Feynman propagator is an important object in QFT, and it will be investigated further in Section 2.3. The normal-ordering operator moves all creation operators to the left of annihilation operators [23], e.g.,

$$N(a_{\mathbf{p}}a^{\dagger}_{\mathbf{k}}a_{\mathbf{q}}) = a^{\dagger}_{\mathbf{k}}a_{\mathbf{p}}a_{\mathbf{q}} . \tag{2.1.25}$$

The order of $a_{\mathbf{p}}$ and $a_{\mathbf{q}}$ is irrelevant, since they commute (Eq. (2.1.11)). Wick's theorem is stated as follows[1]:

$$\mathrm{T}\left\{\phi_1\phi_2\cdots\phi_n\right\} = N\left(\prod_{i=1}^{n}\phi_i + \sum_{x=1}^{n}\sum_{y=x+1}^{n}\left(F^{\phi}_{xy}\prod_{\substack{i=1\\ i\neq x,y}}^{n}\phi_i\right) + \cdots\right) . \tag{2.1.26}$$

In words, this means write all possible combinations of the product of Feynman propagators and remaining field operators, where all remaining field operators are normal ordered. For example, for four fields,

$$\begin{aligned}\mathrm{T}\left\{\phi_1\phi_2\phi_3\phi_4\right\} = N\big(&\phi_1\phi_2\phi_3\phi_4 + F^{\phi}_{12}\phi_3\phi_4 + F^{\phi}_{13}\phi_2\phi_4 + F^{\phi}_{14}\phi_2\phi_3\\ &+F^{\phi}_{23}\phi_1\phi_4 + F^{\phi}_{24}\phi_1\phi_3 + F^{\phi}_{34}\phi_1\phi_2\\ &+F^{\phi}_{12}F^{\phi}_{34} + F^{\phi}_{13}F^{\phi}_{24} + F^{\phi}_{14}F^{\phi}_{23}\big).\end{aligned} \tag{2.1.27}$$

[1] Wick's theorem is often stated in terms of *contractions* of two field operators, but Ref. [23] states that a contraction 'is exactly the Feynman propagator'.

Here, the normal ordering $N(\cdots)$ ensures that uncontracted field operators vanish in vacuum expectation values, since $a_{\mathbf{p}}\left|0\right\rangle = 0$, leading to

$$\left\langle 0\right| \mathrm{T}\left\{\phi_1\phi_2\phi_3\phi_4\right\}\left|0\right\rangle = F^{\phi}_{12}F^{\phi}_{34} + F^{\phi}_{13}F^{\phi}_{24} + F^{\phi}_{14}F^{\phi}_{23}. \tag{2.1.28}$$

This result generalises to any number of fields, allowing systematic computation of correlation functions in perturbation theory.

Wick's Theorem also holds for higher-spin theories, which are considered in the next section, so long as the time-ordering and normal-ordering operators are generalised to included minus signs for interchanges of fermion operators, $\psi$, since fermions obey Fermi-Dirac statistics [23].

## 2.2 Higher-Spin Theories

### 2.2.1 The Dirac Equation for Fermions

The Klein-Gordon equation in the previous section only describes spin-0 particles (i.e., scalar bosons), which means it does not describe the vast majority of particles in the Standard Model. Spin-$\frac{1}{2}$ fermions (particles of half-integer spin), such as leptons and quarks, are described by the Dirac Lagrangian density [57],

$$\mathcal{L}_{\text{Dirac}} = \bar{\psi}(i\gamma^\mu\partial_\mu - m)\psi\,, \tag{2.2.1}$$

where $\psi$ is the fermion field and $m$ is a free parameter (representing the mass of the quanta of the fermion field, after quantisation). The Dirac gamma matrices [23], $\gamma^\mu$, satisfy the Clifford algebra [58, 59],

$$\{\gamma^\mu, \gamma^\nu\} = 2g^{\mu\nu}\mathbb{I}\,, \tag{2.2.2}$$

where $g^{\mu\nu}$ is the metric tensor of Minkowski spacetime and $\mathbb{I}$ is the identity operator. The Dirac adjoint spinor is defined as $\bar{\psi} = \psi^\dagger\gamma^0$, since $\psi^\dagger\psi$ is not a Lorentz scalar (but $\bar{\psi}\psi$ is) and $\psi^\dagger\gamma^\mu\psi$ is not a Lorentz vector (but $\bar{\psi}\gamma^\mu\psi$ is). This Lagrangian density describes the dynamics of free fermions and serves as the foundation for introducing their interactions with gauge fields, such as photons and gluons.

The equation of motion which follows from the Euler-Lagrange equations is then

$$(i\gamma^\mu \partial_\mu - m)\psi = 0\,. \tag{2.2.3}$$

This is the free Dirac equation. The solutions to the Dirac equation describe spin-$\frac{1}{2}$ particles with positive and negative energy states, corresponding to particles and antiparticles.

The quantisation of the Dirac field follows from imposing equal-time *anticommutation* relations,

$$\begin{aligned} \{\psi_a(\mathbf{x},t), \psi_b^\dagger(\mathbf{y},t)\} &= \delta^3(\mathbf{p}-\mathbf{q})\,\delta_{ab}\,, \\ \{\psi_a(\mathbf{x},t), \psi_b(\mathbf{y},t)\} = \{\psi_a^\dagger(\mathbf{x},t), &\psi_b^\dagger(\mathbf{y},t)\} = 0\,, \end{aligned} \tag{2.2.4}$$

where $a$ and $b$ denote the spinor components of $\psi$, and $\delta_{ab}$ is the Kronecker delta function. Similarly to the scalar field, the mode expansion of the Dirac field operator (and its Dirac adjoint) is given by

$$\psi(x) = \int \frac{\mathrm{d}^3\mathbf{p}}{(2\pi)^3} \frac{1}{\sqrt{2\omega_\mathbf{p}}} \sum_s \left( b_\mathbf{p}^s \, u_p^s \, e^{-ip\cdot x} + d_\mathbf{p}^{s\dagger} \, v_p^s \, e^{ip\cdot x} \right) \,, \tag{2.2.5}$$

$$\bar{\psi}(x) = \int \frac{\mathrm{d}^3\mathbf{p}}{(2\pi)^3} \frac{1}{\sqrt{2\omega_\mathbf{p}}} \sum_s \left( d_\mathbf{p}^s \, \bar{v}_p^s \, e^{-ip\cdot x} + b_\mathbf{p}^{s\dagger} \, \bar{u}_p^s \, e^{ip\cdot x} \right) , \tag{2.2.6}$$

where the operators $b_\mathbf{p}^s$ and $b_\mathbf{p}^{s\dagger}$ are the annihilation and creation operators for particles of momentum $p$ and spin $s$, while $d_\mathbf{p}^s$ and $d_\mathbf{p}^{s\dagger}$ are the annihilation and creation operators for antiparticles with the same quantum numbers. The four-component Dirac spinors $u_p^s \equiv u^s(p)$ and $v_p^s \equiv v^s(p)$ are solutions to the Dirac equation associated with momentum $p$ and spin index $s$, with their adjoints given by $\bar{u}_p^s = u_p^{s\dagger}\gamma^0$ and $\bar{v}_p^s = v_p^{s\dagger}\gamma^0$. The energy $\omega_\mathbf{p} = \sqrt{\mathbf{p}^2 + m^2}$ ensures that the field satisfies the relativistic dispersion relation. It follows that the creation and annihilation operators have the anticommutation relations

$$\begin{aligned} \{b_\mathbf{p}^s, b_{\mathbf{p}'}^{s'\dagger}\} = \{d_\mathbf{p}^s, d_{\mathbf{p}'}^{s'\dagger}\} &= (2\pi)^3\delta^3(\mathbf{p}-\mathbf{p}')\delta_{ss'}\,, \\ \{b_\mathbf{p}^s, b_{\mathbf{p}'}^{s'}\} = \{d_\mathbf{p}^s, d_{\mathbf{p}'}^{s'}\} = \{b_\mathbf{p}^{s\dagger}, b_{\mathbf{p}'}^{s'\dagger}\} &= \{d_\mathbf{p}^{s\dagger}, d_{\mathbf{p}'}^{s'\dagger}\} = 0\,. \end{aligned} \tag{2.2.7}$$

These anticommutation relations mean that $(b_\mathbf{p}^{s\dagger})^2 = (d_\mathbf{p}^{s\dagger})^2 = 0$, such that a state cannot be filled twice. This is the Pauli exclusion principle [51, 60]. More generally, any multiparticle state is antisymmetric under the interchange of two particles, e.g., $b_\mathbf{p}^{s\dagger}\, b_{\mathbf{p}'}^{s'\dagger}\,|0\rangle = -b_{\mathbf{p}'}^{s'\dagger}\, b_\mathbf{p}^{s\dagger}\,|0\rangle$. Consequently, these particles obey Fermi-Dirac statis-

tics [61–64], not the Bose-Einstein statistics of integer-spin particles [63–66]. If we instead use commutation relations to quantise the Dirac field, we obtain solutions that allow violations of the spin-statistics theorem [67], leading to unphysical consequences such as negative probabilities. Thus, the use of anticommutators is a fundamental requirement for the consistency of fermionic QFT.

Furthermore, since the fermion field, $\psi$, is quantised using anticommutation relations, causality is encoded in the anticommutator of fields, rather than the commutator of fields as in scalar field theory. In other words, the Dirac field operator and its Dirac adjoint anticommute at spacelike separations.

### 2.2.2 Gauge Fields: Photons and Gluons

The Dirac Lagrangian is invariant under global gauge transformations, i.e., rotating the phase of the Dirac field, $\psi$. However, something interesting happens if we allow gauge transformations to arbitrarily change at different spacetime points (in other words, we make the gauge transformation spacetime dependent). These types of transformations are called *local* gauge transformations, and physics should not depend on these arbitrary choices.

In order to impose that our theory is locally gauge invariant, we must introduce new spin-1 fields to the Lagrangian. These fields are called *gauge fields*, and their corresponding spin-1 particles are called *gauge bosons*. This structure is a key feature of the Standard Model, where the electromagnetic, weak, and strong forces are all governed by gauge symmetry [7]. In this section, we introduce two gauge bosons: the photon and the gluon. The introduction of the photon field leads to quantum electrodynamics (QED), and the introduction of the gluon field leads to quantum chromodynamics (QCD).

#### Electromagnetic Interactions and the Photon Field

The Dirac equation is invariant under global $U(1)$ phase transformations of the form $\psi \to e^{i\alpha}\psi$. However, the derivative $\partial_\mu \psi$ is not invariant under local phase transformations $\alpha = \alpha(x)$. Requiring invariance under local gauge transformations leads to the definition of the covariant derivative,

$$D_\mu \psi = (\partial_\mu + ieA_\mu)\psi, \tag{2.2.8}$$

where $e = -|e|$ is the fundamental electric charge and $A_\mu$ is a gauge field (the photon field) that transforms as

$$A_\mu(x) \to A_\mu(x) - \frac{1}{e}\partial_\mu \alpha(x)\,. \tag{2.2.9}$$

This ensures gauge invariance under local $U(1)$ transformations. Substituting $D_\mu$ in place of $\partial_\mu$ in the Dirac Lagrangian introduces the interaction term $e\bar{\psi}\gamma^\mu A_\mu \psi$, describing the coupling between fermions and the photon.

The kinetic term for the photon field arises from the field strength tensor [23, 25],

$$F_{\mu\nu} = \partial_\mu A_\nu - \partial_\nu A_\mu, \tag{2.2.10}$$

leading to the free Lagrangian density for the photon,

$$\mathcal{L}_{\text{photon}} = -\frac{1}{4}F_{\mu\nu}F^{\mu\nu}. \tag{2.2.11}$$

This term describes the dynamics of the electromagnetic field and ensures that Maxwell's equations emerge from the Euler-Lagrange equations.

Combining the Dirac Lagrangian density (with the covariant derivative) and the free Lagrangian density of the photon, we get the Lagrangian density for QED[2],

$$\mathcal{L}_{\text{QED}} = \bar{\psi}(i\gamma^\mu \partial_\mu - m)\psi - \frac{1}{4}F_{\mu\nu}F^{\mu\nu} - e\bar{\psi}\gamma^\mu A_\mu \psi\,. \tag{2.2.12}$$

A fundamental result in QED, which follows from gauge invariance, is the *Ward-Takahashi identity* [68, 69]. This identity expresses the constraints imposed on Green's functions due to gauge symmetry and plays a crucial role in proving renormalisability (see Section 2.6). In Chapter 3, we explicitly show how ultraviolet divergences cancel in a scattering calculation, which is ultimately a consequence of this identity. At its core, the Ward-Takahashi identity arises from the Noether current associated with gauge symmetry.

[2]Strictly, we still have to *gauge fix* to remove the photon field's redundant degrees of freedom [23, 54]. For example, a gauge-fixing term of the form $-(\partial^\mu A_\mu)^2/2\xi$ can be added to the Lagrangian, with $\xi = 1$ defining the Feynman gauge. When calculating cross sections in Chapter 3, we will adopt the Feynman gauge.

The Ward-Takahashi identity is a generalisation of the *Ward identity*, which is a special case that applies specifically to on-shell amplitudes. The Ward identity is written as

$$k_\mu \mathcal{M}^\mu(k) = 0, \tag{2.2.13}$$

where $\mathcal{M}^\mu$ is the matrix element for a given QED process involving an external photon with momentum $k$. This equation encapsulates gauge invariance at the level of scattering amplitudes, ensuring that unphysical longitudinal photon polarisations do not contribute to observable quantities.

### Strong Interactions and the Gluon Field

In QCD, quarks interact through the exchange of gluons, which correspond to an $SU(3)$ gauge symmetry. The local $SU(3)$ transformation of a quark field, $\psi_i$, is given by

$$\psi_i \to U_{ij}(x)\psi_j, \quad U_{ij}(x) \in SU(3)\,, \tag{2.2.14}$$

where $i, j$ are the quark colour indices which run from 1 to $N_c$ (the number of colours), and $U_{ij}(x)$ is a spacetime-dependent transformation in the fundamental representation of $SU(3)$. To preserve gauge invariance, the derivative must be replaced by the covariant derivative

$$D_\mu \psi_i = (\partial_\mu \delta_{ij} - igG^a_\mu (T^a)_{ij})\psi_j\,. \tag{2.2.15}$$

Here, $G^a_\mu$ are the eight gluon fields, and $(T^a)_{ij}$ are the generators of $SU(3)$ satisfying the algebra

$$[T^a, T^b] = if^{abc}T^c\,, \tag{2.2.16}$$

where $f^{abc}$ are the structure constants of the $SU(3)$ algebra, and $a, b, c$ are the gluon colour indices. The gluon field strength tensor is given by [23, 25]

$$G^a_{\mu\nu} = \partial_\mu G^a_\nu - \partial_\nu G^a_\mu + gf^{abc}G^b_\mu G^c_\nu\,, \tag{2.2.17}$$

where the final term encapsulates the self-interaction of gluons due to the *non-Abelian* structure of the $SU(3)$ group, which is a characteristic feature of QCD. This leads to the free gluon Lagrangian density,

$$\mathcal{L}_{\text{gluon}} = -\frac{1}{4} G^a_{\mu\nu} G^{\mu\nu}_a \,, \qquad (2.2.18)$$

which governs the dynamics of the gluon fields.

Thus, in analogy to QED, we arrive at the Lagrangian density for QCD[3],

$$\mathcal{L}_{\text{QCD}} = \bar{\psi}_i (i\gamma^\mu \partial_\mu \delta_{ij} - m\, \delta_{ij}) \psi_j - \frac{1}{4} G^a_{\mu\nu} G^{\mu\nu}_a + e \bar{\psi}_i \gamma^\mu G^a_\mu (T^a)_{ij} \psi_j \,. \qquad (2.2.19)$$

The gauge invariance of QCD also leads to identities analogous to the Ward-Takahashi identity in QED, known as the Slavnov-Taylor identities [71–73], which impose constraints on Green's functions and ensure renormalisability.

In Chapter 3, we will use the results in this section to describe particle scattering processes. We will combine QED and QCD, since quarks couple to both photons and gluons.

## 2.3 Propagators and Functions

In this section, we investigate multiple invariant commutation and propagation functions in QFT. These definitions will be used throughout this thesis, unless otherwise stated.

### 2.3.1 Scalar Field Propagators and Functions

For scalar field theory, the invariant commutation functions are solutions to the *homogenous* Klein-Gordon equation with special boundary conditions, where 'homogeneous' means that there is no source term (the right-hand side of Eq. (2.1.3) remains zero). The invariant propagation functions are *Green's functions* of the Klein-Gordon equation, which means they are solutions to the *inhomogeneous* Klein-Gordon equation with a Dirac delta function as a source term.

[3]Again, strictly we should include a term in the Lagrangian to gauge fix the theory, and also introduce Faddeev-Popov ghost fields [70] to ensure unphysical states do not contribute to observable quantities. For the calculations in this thesis, we do need to consider ghost fields.

The notable solutions to the homogeneous Klein-Gordon equation are as follows [74]:

Positive Wightman Function:

$$\begin{aligned} \Delta_{xy}^{\phi(>)} \equiv \Delta^{\phi(>)}(x-y) &= \left\langle 0^{\phi} \middle| \phi(x)\phi(y) \middle| 0^{\phi} \right\rangle \\ &= \int \frac{\mathrm{d}^3\mathbf{k}}{(2\pi)^3} \frac{1}{2\omega_{\mathbf{k}}} e^{-ik\cdot(x-y)} \Big|_{k_0=\omega_{\mathbf{k}}} ; \end{aligned} \tag{2.3.1}$$

Negative Wightman Function:

$$\begin{aligned} \Delta_{xy}^{\phi(<)} \equiv \Delta^{\phi(<)}(x-y) &= \left\langle 0^{\phi} \middle| \phi(y)\phi(x) \middle| 0^{\phi} \right\rangle \\ &= \int \frac{\mathrm{d}^3\mathbf{k}}{(2\pi)^3} \frac{1}{2\omega_{\mathbf{k}}} e^{+ik\cdot(x-y)} \Big|_{k_0=\omega_{\mathbf{k}}} ; \end{aligned} \tag{2.3.2}$$

Pauli-Jordan Function:

$$\begin{aligned} \Delta_{xy}^{\phi} \equiv \Delta^{\phi}(x-y) &= \big[\phi(x), \phi(y)\big] \\ &= \int \frac{\mathrm{d}^3\mathbf{k}}{(2\pi)^3} \frac{1}{2\omega_{\mathbf{k}}} \big(e^{-ik\cdot(x-y)} - e^{ik\cdot(x-y)}\big) \Big|_{k_0=\omega_{\mathbf{k}}} ; \end{aligned} \tag{2.3.3}$$

Hadamard Function:

$$\begin{aligned} \Delta_{xy}^{\phi(H)} \equiv \Delta^{\phi(H)}(x-y) &= \langle 0^{\phi}| \big\{\phi(x), \phi(y)\big\} |0^{\phi}\rangle \\ &= \int \frac{\mathrm{d}^3\mathbf{k}}{(2\pi)^3} \frac{1}{2\omega_{\mathbf{k}}} \big(e^{-ik\cdot(x-y)} + e^{ik\cdot(x-y)}\big) \Big|_{k_0=\omega_{\mathbf{k}}} \\ &= 2\,\langle 0^{\phi}|\, \phi(x)\phi(y)\, |0^{\phi}\rangle - \Delta_{xy}^{\phi} ; \end{aligned} \tag{2.3.4}$$

where $\omega_{\mathbf{k}} = \sqrt{\mathbf{k}^2 + m^2}$ and $m$ is the mass of the field quanta. The notable Green's functions of the Klein-Gordon equation are as follows [74]:

Feynman Propagator:

$$\begin{aligned} F_{xy}^{\phi} \equiv F^{\phi}(x-y) &= \left\langle 0^{\phi} \middle| \mathrm{T}\big\{\phi(x)\phi(y)\big\} \middle| 0^{\phi} \right\rangle \\ &= \int \frac{\mathrm{d}^3\mathbf{k}}{(2\pi)^3} \frac{1}{2\omega_{\mathbf{k}}} \big(\Theta(x_0 - y_0) e^{-ik\cdot(x-y)} \\ &\qquad + \Theta(y_0 - x_0) e^{ik\cdot(x-y)}\big) \Big|_{k_0=\omega_{\mathbf{k}}} \\ &= i \int_{\mathcal{C}_{\mathrm{F}}} \frac{\mathrm{d}^4 k}{(2\pi)^4} \frac{e^{-ik\cdot(x-y)}}{k^2 - m^2} \\ &= i \int \frac{\mathrm{d}^4 k}{(2\pi)^4} \frac{e^{-ik\cdot(x-y)}}{k^2 - m^2 + i\epsilon} ; \end{aligned} \tag{2.3.5}$$

Retarded Propagator:

$$\begin{aligned} R^{\phi}_{xy} \equiv R^{\phi}(x-y) \equiv \Delta^{R}_{xy} &= \Theta(x_0 - y_0)\Delta(x-y) \\ &= \Theta(x_0 - y_0)\left[\phi(x), \phi(y)\right] \\ &= i \int_{\mathcal{C}_{\mathrm{R}}} \frac{\mathrm{d}^4 k}{(2\pi)^4} \frac{e^{-ik\cdot(x-y)}}{k^2 - m^2} \\ &= i \int \frac{\mathrm{d}^4 k}{(2\pi)^4} \frac{e^{-ik\cdot(x-y)}}{k^2 - m^2 + k_0 i\epsilon}\, ; \end{aligned} \tag{2.3.6}$$

where the contours $\mathcal{C}_{\mathrm{F}}$ and $\mathcal{C}_{\mathrm{R}}$ are shown in Figs. 2.1 and 2.2, respectively, and $\epsilon$ is a small positive parameter.

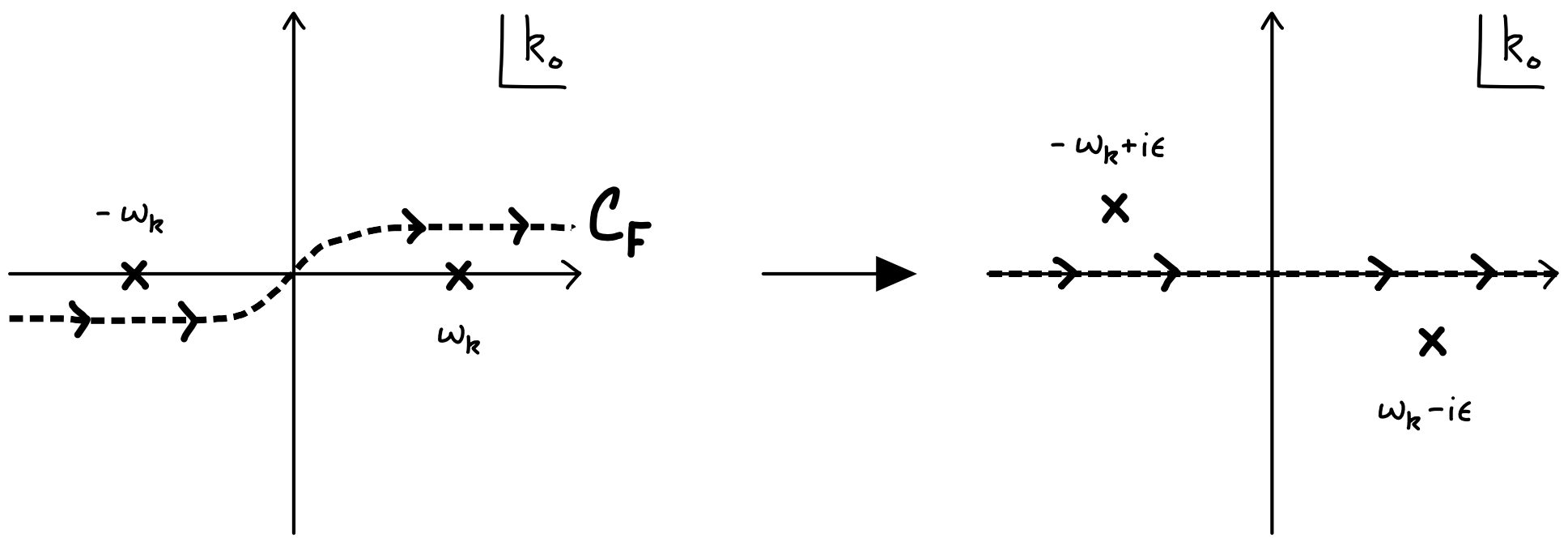

**Fig. 2.1.** The Feynman propagator is equivalently defined by integration along the contour $\mathcal{C}_{\mathrm{F}}$ or the $+i\epsilon$ pole prescription (with integration along the real axis), as in Eq. (2.3.5). With the $+i\epsilon$ pole prescription, the poles are at $\omega_{\mathbf{k}} - i\epsilon$ and $-\omega_{\mathbf{k}} + i\epsilon$.

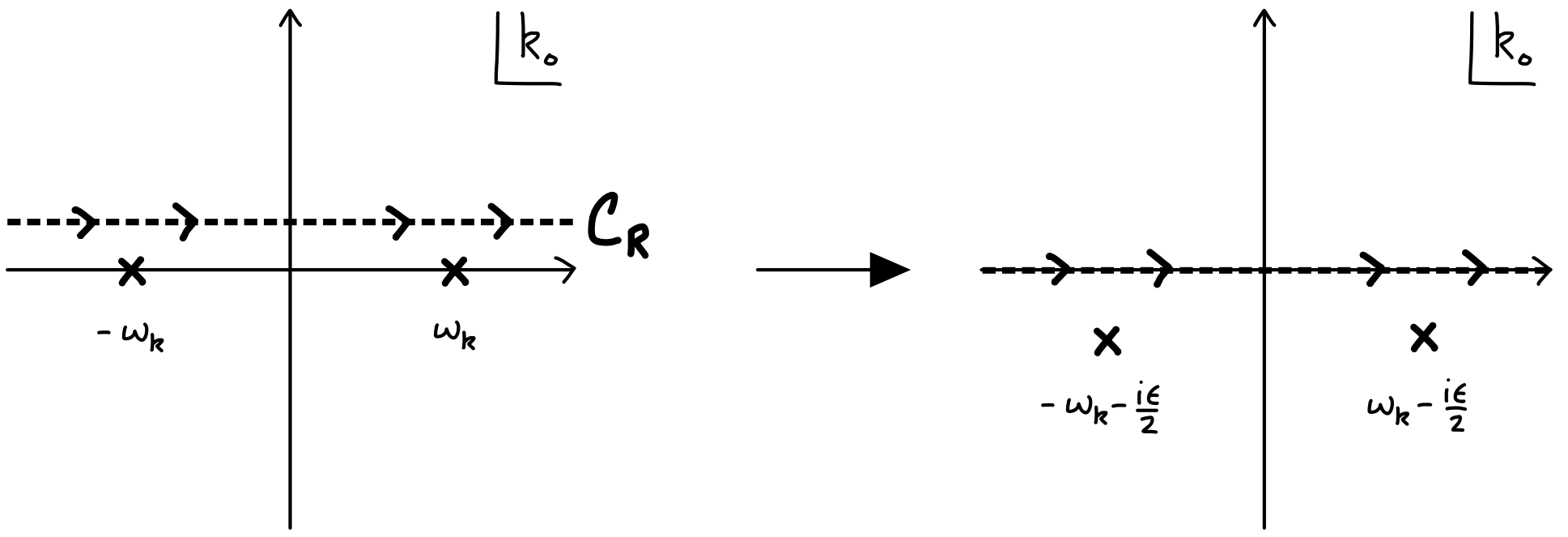

**Fig. 2.2.** The retarded propagator is equivalently defined by integration along the contour $\mathcal{C}_{\mathrm{R}}$ or the $+k_0 i\epsilon$ pole prescription (with integration along the real axis), as in Eq. (2.3.6). With the $+k_0 i\epsilon$ pole prescription, the poles are at $\omega_{\mathbf{k}} - \frac{i\epsilon}{2}$ and $-\omega_{\mathbf{k}} - \frac{i\epsilon}{2}$.

Note that the Feynman and retarded propagators are sometimes defined with an additional factor of $(i)^{-1}$, which is how they are defined in Section 4.4 and Chapter 6. Also, the retarded propagator is sometimes defined with $\mathrm{sgn}(k_0)i\epsilon$ in the denominator instead of $k_0 i\epsilon$, so that $\epsilon$ has the same dimensions in the Feynman and retarded propagators.

The Feynman propagator represents the propagation of a virtual particle between two points, and appears in the Feynman rules for constructing scattering matrix elements (see Section 2.5). However, the Feynman propagator does not vanish for spacelike separations [23], so does this mean that particles can propagate at superluminal speeds?

The key to preserving causality is to ask the question whether a *measurement* performed at one point can affect a measurement at a spacelike-separated second point. This is encoded by the commutator $\Delta^{\phi}_{xy} \equiv [\phi_x, \phi_y]$, since if this commutator vanishes then one measurement cannot affect the other. The equal-time commutation relations for scalar field theory (Eqs. (2.1.15)) show that the commutator of field operators vanishes at equal times, but what about general times? Using Eq. (2.3.1),

$$\begin{aligned}[\phi_x, \phi_y] &= \langle 0^{\phi}| \, [\phi_x, \phi_y] \, |0^{\phi}\rangle = \langle 0^{\phi}| \, \phi_x \phi_y \, |0^{\phi}\rangle - \langle 0^{\phi}| \, \phi_y \phi_x \, |0^{\phi}\rangle \\ &= \Delta^{\phi(>)}(x-y) - \Delta^{\phi(>)}(y-x) \,, \end{aligned} \tag{2.3.7}$$

where we have used the fact that the commutator of fields is a c-number (classical number, as opposed to quantum operator) so we are free to write $[\phi_x, \phi_y] = \big\langle 0^{\phi} \big| [\phi_x, \phi_y] \big| 0^{\phi} \big\rangle$. Each of the Wightman functions are separately invariant under continuous Lorentz transformations [23]. For spacelike separations, there exists a continuous Lorentz transformation $(x - y) \to (y - x)$. Consequently, Eq. (2.3.7)—and hence the retarded propagator, Eq. (2.3.6)—vanishes for spacelike separations.

### 2.3.2 Higher-Spin Field Propagators

In addition to scalar fields, propagators play a crucial role in the study of spinor, vector, and gauge fields. Specifically, the Feynman propagators in this section appear in the Feynman rules for QED and QCD.

For a Dirac field, the Feynman propagator is given by [23]

$$\begin{aligned} S_{xy} \equiv S_F(x-y) &= \big\langle 0^{\psi} \big| \mathrm{T}\big\{\psi(x)\bar{\psi}(y)\big\} \big| 0^{\psi} \big\rangle \\ &= i \int \frac{\mathrm{d}^4 k}{(2\pi)^4} \frac{(\gamma^{\mu} k_{\mu} + m) e^{-ik\cdot(x-y)}}{k^2 - m^2 + i\epsilon} \,. \end{aligned} \tag{2.3.8}$$

The retarded propagator for the Dirac field is defined as [23]

$$S^R_{xy} = \Theta(x_0 - y_0) \langle 0^\psi | \{\psi(x), \bar{\psi}(y)\} |0^\psi\rangle$$
$$= i \int \frac{\mathrm{d}^4 k}{(2\pi)^4} \frac{(\gamma^\mu k_\mu + m) e^{-ik\cdot(x-y)}}{k^2 - m^2 + k_0 i\epsilon} , \tag{2.3.9}$$

where we note the appearance of the anticommutator instead of the commutator, due to the Fermi-Dirac statistics of the Dirac field.

In QED, the photon field satisfies gauge constraints, leading to different choices of propagator. In the *Feynman gauge*, the Feynman propagator is [23]

$$D^{xy}_{\mu\nu} \equiv D^F_{\mu\nu}(x-y) = \big\langle 0 \big| \mathrm{T}\big\{A_\mu(x) A_\nu(y)\big\} \big| 0 \big\rangle \tag{2.3.10}$$
$$= -i \int \frac{\mathrm{d}^4 k}{(2\pi)^4} \frac{g_{\mu\nu} e^{-ik\cdot(x-y)}}{k^2 + i\epsilon} . \tag{2.3.11}$$

In quantum chromodynamics (QCD), the gluon propagator also depends on the choice of gauge. In the Feynman gauge, the Feynman propagator for a gluon field is [23]

$$D^{ab}_{\mu\nu}(x-y) = \big\langle 0 \big| \mathrm{T}\big\{A^a_\mu(x) A^b_\nu(y)\big\} \big| 0 \big\rangle \tag{2.3.12}$$
$$= -i\delta^{ab} \int \frac{\mathrm{d}^4 k}{(2\pi)^4} \frac{g_{\mu\nu} e^{-ik\cdot(x-y)}}{k^2 + i\epsilon} . \tag{2.3.13}$$

## 2.4 The S-Matrix and the LSZ Reduction Formula

The $S$-matrix, or scattering matrix, is a fundamental object in QFT that encodes the transition amplitudes between initial and final asymptotic states in a scattering process. It provides the theoretical framework for calculating observable quantities such as cross-sections and decay rates.

The $S$-matrix is formally defined as the operator that relates the in-state $|\mathrm{in}\rangle$ (before the interaction) to the out-state $|\mathrm{out}\rangle$ (after the interaction) [23, 25, 54, 75, 76],

$$S\,|\mathrm{in}\rangle = |\mathrm{out}\rangle \ . \tag{2.4.1}$$

In the interaction picture, the $S$-matrix is the long-time limit of the unitary time-evolution operator[4], $U(t, t_0)$, but with the assumption that states evolve freely in the distant past and future (i.e., the interaction Hamiltonian vanishes asymptotically). Thus, it is defined as

$$S \equiv \lim_{t\to\infty} U(t, -t) = \mathrm{T}\left\{\exp\left(-i\int_{-\infty}^{\infty} \mathrm{d}^4x H_{\mathrm{int}}(x)\right)\right\}, \tag{2.4.2}$$

where T denotes time-ordering. In this way, the $S$-matrix does not describe time evolution at intermediate times but rather the overall scattering process.

The transition amplitude between an initial state $|i\rangle$ and a final state $|f\rangle$ is given by the $S$-matrix element

$$S_{fi} = \langle f|S|i\rangle\,. \tag{2.4.3}$$

Even in an interacting theory, there is a non-zero probability that the particles do not interact (e.g., they completely miss each other). To isolate the part of the $S$-matrix that is due to interactions, it is conventional to define the $T$-matrix [23, 25, 54, 77],

$$S = \mathbb{I} + iT, \tag{2.4.4}$$

where $\mathbb{I}$ is the identity operator. The physically relevant quantity for scattering processes is therefore the $T$-matrix element,

$$T_{fi} = \langle f|T|i\rangle, \tag{2.4.5}$$

which is directly related to the *invariant matrix element*, $\mathcal{M}_{fi}$, by factoring out momentum-conserving $\delta$-functions:

$$T_{fi} = (2\pi)^4\delta^4\left(\sum p_i - \sum p_f\right)\mathcal{M}_{fi}. \tag{2.4.6}$$

The invariant matrix element contains all the dynamical information about the interaction and is computed using Feynman rules derived from the Lagrangian (see Section 2.5).

In order to calculate $S$-matrix elements, we compute time-ordered correlation functions of fields. One must then use the *Lehmann–Symanzik–Zimmermann (LSZ) reduc-*

[4]or, equivalently, a sequence of time-evolution operators, since $U(t_a, t_c) = U(t_a, t_b)U(t_b, t_c)$.

*tion formula* [23, 78, 79]:

$$\langle \mathbf{q}_1 \dots \mathbf{q}_n | S | \mathbf{p}_1 \dots \mathbf{p}_N \rangle = \left( \prod_{k=1}^{n} \int \mathrm{d}^4 x_k \, e^{-iq_k \cdot x_k} \frac{(q_k^2 - m_k^2 + i\epsilon)}{i\sqrt{Z_k}} \right) \times \left( \prod_{j=1}^{N} \int \mathrm{d}^4 y_j \, e^{ip_j \cdot y_j} \frac{(p_j^2 - m_j^2 + i\epsilon)}{i\sqrt{Z_j}} \right) \times \langle 0 | \mathrm{T}\{\phi(x_1) \dots \phi(x_n) \phi(y_1) \dots \phi(y_N)\} | 0 \rangle \,, \qquad (2.4.7)$$

for $N$ initial-state particles of momenta $\{p_1, \dots, p_N\}$, $n$ final-state particles of momenta $\{q_1, \dots, q_n\}$, and $m$ is the mass of each particle. The quantity $Z$ is the field-strength renormalisation constant, defined as the residue of the single-particle pole in the two-point function of fields [23, 25] (we calculate $Z$ for a specific process in Section 3.4.2). In the case where there are different types of external particle (e.g. $q$ and $e^-$), each external particle contributes a unique factor of $\sqrt{Z_i}$ (e.g. $\sqrt{Z_q}$ and $\sqrt{Z_e}$), obtained from each of their two-point functions. These renormalisation factors appear because we are working in bare (as opposed to renormalised) perturbation theory (see Section 2.6.2).

In words, the LSZ reduction formula says that in order to calculate an $S$-matrix element, one must compute the Fourier-transformed time-ordered correlation function (using the momentum-space Feynman rules in Section 2.5). Then, due to the factors of $p^2 - m^2$, remove all terms in the time-ordered product except those with poles of the form $(p^2 - m^2)^{-1}$. These correspond to propagators of on-shell particles, and the $S$-matrix is given by the residue of these poles. Thus, the LSZ reduction formula projects one-particle asymptotic states out from time-ordered products of fields, defining the $S$-matrix.

It will prove useful to express the LSZ reduction formula directly in terms of Feynman diagrams. To do this, consider the four-point function, shown in Fig. 2.3. The Feynman diagram has been separated into all possible *amputated* diagrams (labelled 'Amp.') and isolated *self-energy* corrections to the external legs (patterned circles). To 'amputate' a diagram, start at the tip of the external leg and follow it into the diagram, then 'cut' the line at the final possible point in which such a cut separates the external leg (and any self-contained loop corrections) from the rest of the diagram.

Since $Z$ is defined as the residue of the single-particle pole in the two-point function, we know that the contribution of the four exact two-point propagators in Fig. 2.3 con-

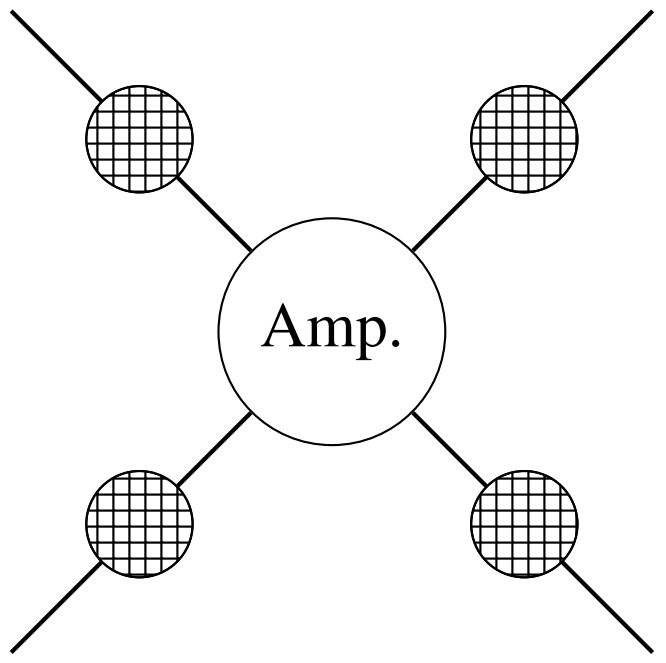

**Fig. 2.3.** Feynman diagram representing the four-point function in scalar field theory. The central circle labelled 'Amp.' contains all amputated diagrams. The patterned circles represent the exact scalar propagator, including corrections to all orders.

tains

$$\frac{iZ}{p_1^2 - m^2 + i\epsilon} \frac{iZ}{p_2^2 - m^2 + i\epsilon} \frac{iZ}{q_1^2 - m^2 + i\epsilon} \frac{iZ}{q_2^2 - m^2 + i\epsilon}, \tag{2.4.8}$$

where $\{p_1, p_2, q_1, q_2\}$ are the momenta of the four external particles and we have assumed $Z$ and $m$ are equivalent for all particles (i.e., they are all the same type of particle). These propagator factors exactly cancel with those in Eq. (2.4.7), except for an overall factor of $\sqrt{Z}$ for each external particle. In this way, the LSZ reduction formula 'picks out' the two-point propagator poles in the time-ordered correlation function and forces the external particles on-shell. All other contributions from the time-ordered correlation function are then multiplied by zero (the on-shell propagator factors in Eq. (2.4.7)).

Thus, by explicitly accounting for the contributions of the two-point propagators, the LSZ formula can be written in a form which relates the $S$-matrix elements directly to amputated Feynman diagrams,

$$\langle \mathbf{q}_1 \cdots \mathbf{q}_n | \, S \, | \mathbf{p}_1 \mathbf{p}_2 \rangle = \left(\sqrt{Z}\right)^{n+2}$$

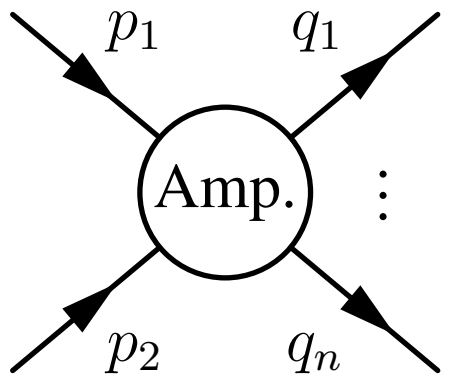

, (2.4.9)

where we have specialised to the case of two initial-state particles, which will be the case in Chapter 3. Note that any spinor or polarisation factors associated with external legs are still to be included.

## 2.5 Feynman Rules

In this section, we specify the momentum-space Feynman rules for scalar $\phi^3$ theory, QED, and QCD. Feynman rules provide a systematic method for computing scattering amplitudes in quantum field theory [76, 80] . They translate the graphical elements of Feynman diagrams into mathematical factors, which can be derived from terms in the interaction Lagrangian. Further details on Feynman diagrams can be found in most textbooks on QFT [23–25, 54].

### 2.5.1 Feynman Rules for Scalar Field Theory

For a real scalar field with interaction term $\frac{\lambda}{3!}\phi^3$, the Feynman rules are as follows:

- Internal propagator:

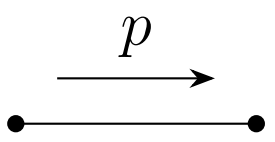

$$= \frac{i}{p^2 - m^2 + i\epsilon}\,; \tag{2.5.1}$$

- Vertex:

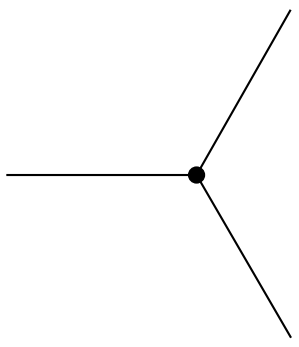

$$= -i\lambda\,; \tag{2.5.2}$$

- External leg:

$$= 1\,; \tag{2.5.3}$$

- Impose momentum conservation at each vertex;
- Integrate over each undetermined loop momentum with measure $\int \mathrm{d}^D k/(2\pi)^D$, in $D$ dimensions;
- Divide by symmetry factors, $S$, where necessary.

### 2.5.2 Feynman Rules for QED

QED describes interactions between electrons, positrons, and photons. Electrons and positrons are denoted by solid lines with an arrow pointing in the direction of fermion flow (such that the arrow is parallel to momentum flow for electrons and antiparallel to momentum flow for positrons). Photons are denoted by a wiggly line. The Feynman rules are as follows:

- Internal fermion propagator:

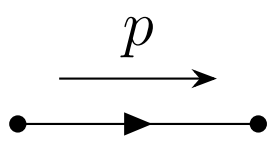

$$= \frac{i(\gamma^\mu p_\mu + m)}{p^2 - m^2 + i\epsilon}\,; \tag{2.5.4}$$

- Internal photon propagator:

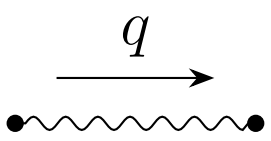

$$= \frac{-i\left(g^{\mu\nu} + (\xi - 1)\frac{q^\mu q^\nu}{q^2}\right)}{q^2 + i\epsilon}\,; \tag{2.5.5}$$

- Fermion-photon vertex:

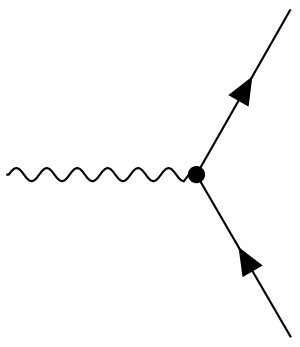

$$= -ie\gamma^\mu\,; \tag{2.5.6}$$

- External incoming electron:

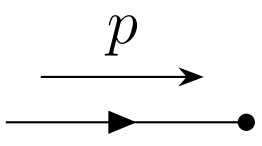

$$= u_s(p)\,; \tag{2.5.7}$$

- External outgoing electron:

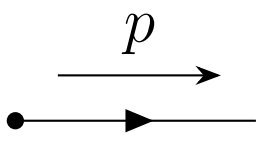

$$= \bar{u}_s(p)\,; \tag{2.5.8}$$

- External incoming positron:

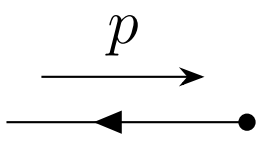

$$= \bar{v}_s(p)\,; \tag{2.5.9}$$

- External outgoing positron:

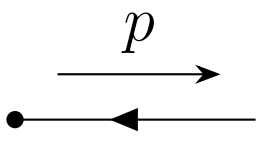

$$= v_s(p)\,; \tag{2.5.10}$$

- External incoming photon:

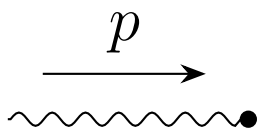

$$= \epsilon^{\mu}(p)\,; \tag{2.5.11}$$

- External outgoing photon:

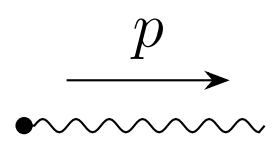

$$= \epsilon^{*\mu}(p)\,; \tag{2.5.12}$$

- Impose momentum conservation at each vertex;
- Integrate over each undetermined loop momentum with measure $\int \mathrm{d}^D k/(2\pi)^D$, in $D$ dimensions;

where $\xi$ is the gauge parameter used to gauge fix the theory (in this thesis, we use the Feynman gauge, $\xi = 1$) and $\epsilon^{\mu}(p)$ is the polarisation vector of the external photon. For internal photon propagators, the uncontracted Lorentz indices label the Lorentz indices associated with the vertex factors on either side of the propagator. For internal antifermion propagators, the direction of momentum is antiparallel to the direction of the fermion flow, and so $p_\mu \to -p_\mu$ in the propagator factor. When constructing a matrix element, it is important to follow the arrows denoting the fermion flow.

### 2.5.3 Feynman Rules for QCD

QCD describes the interactions between quarks (solid lines) and gluons (curly lines). The Feynman rules relevant to the calculations in Chapter 3 are as follows:

- Quark propagator:

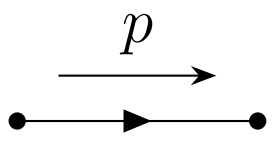

$$= \frac{i\delta^{ij}(\gamma^{\mu}p_{\mu} + m)}{p^2 - m^2 + i\epsilon}\,; \tag{2.5.13}$$

- Gluon propagator:

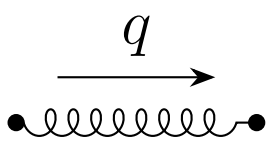

$$= \frac{-i\delta^{ab}}{q^2 + i\epsilon}\left(g^{\mu\nu} + (\xi - 1)\frac{q^{\mu}q^{\nu}}{q^2 + i\epsilon}\right)\,; \tag{2.5.14}$$

- Quark-gluon vertex:

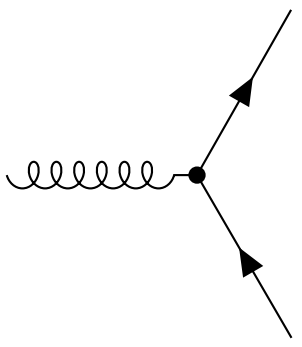

$$= ig\gamma^\mu (T^a)_{ij} \, ; \tag{2.5.15}$$

- External leg factors are identical to those in QED, except quarks and gluons carry colour indices;
- Impose momentum conservation at each vertex;
- Integrate over each undetermined loop momentum with measure $\int \mathrm{d}^D k/(2\pi)^D$, in $D$ dimensions;

where $i, j$ are the quark colour indices and $a, b, c$ are the gluon colour indices. We have omitted gluon self-interactions and ghost fields, since they will not be considered in this thesis. Matrix elements are constructed similarly to those of QED. In Chapter 3, we consider a process which involves both QED and QCD. In this case, there is a quark-photon vertex similar to Eq. (2.5.6):

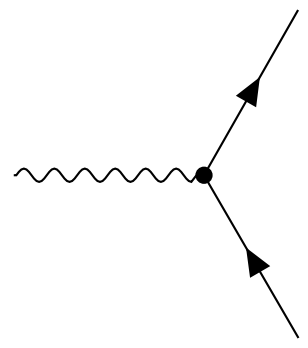

$$= -iee_q \gamma^\mu \delta^{ij} \, ; \tag{2.5.16}$$

where the $\delta^{ij}$ ensures that the two fermions have matching colour and $e_q$ is the dimensionless ratio of the quark's electric charge to $e = -|e|$ (e.g., $e_q = -\frac{2}{3}$ for an up quark and $e_q = \frac{1}{3}$ for a down quark).

## 2.6 Ultraviolet and Infrared Divergences

Calculations in QFT often involve terms which tend to infinity in the high-energy or low-energy limits. These divergences are classified as *infrared (IR) divergences* (for divergences in the low-energy limit) and *ultraviolet (UV) divergences* (for divergences in the high-energy limit). Not only do these divergences have different physical origins, but they are treated (*regularised*) in different ways. Despite this, many calculations ignore the distinction between the two types of divergence.

In this section, we explain the difference between these divergences and examine some of the divergent integrals that are ubiquitous in QFT. We derive a general solution for a divergent integral using *dimensional regularisation*, which is a technique used to parameterise the divergences in the intermediate steps of a calculation. In doing so, we are able to keep track of divergent terms. This is crucial for our calculation in Chapter 3, where we calculate the first-order gluon corrections to quark-antiquark production, and show that IR and UV divergences cancel independently.

### 2.6.1 Infrared Divergences

IR divergences appear in loop integrals with massless virtual particles and in phase-space integrals with massless real emissions. IR divergences manifest in two primary ways:

- Soft divergences: These occur when the energy of an internal loop particle or real emission approaches zero.
- Collinear divergences: These arise when particles are emitted collinearly with an external leg.

In both of these cases, the process is degenerate with a process in which the particle does not appear at all. In other words, experimental detectors cannot distinguish between states differing by arbitrarily soft/collinear particle emissions or by virtual corrections involving negligible energy transfer. The detectors therefore naturally sum over these degenerate states [23]. To treat them separately in the mathematics is therefore unphysical.

However, physical observables, such as cross-sections and decay rates, must be IR finite, even if individual Feynman diagrams contain divergences. This is ensured by the Bloch-Nordsieck (BN) theorem [27] (for Abelian gauge theories such as QED) and the Kinoshita-Lee-Nauenberg (KLN) theorem [28, 29] (for non-Abelian gauge theories such as QCD). These theorems state that when summing over all degenerate final states, IR divergences cancel. This cancellation ensures that measurable quantities remain well-defined.

This cancellation can be formally expressed, at the level of cross sections, as

$$\sigma_{\text{physical}} = \sigma_{\text{virtual}} + \sigma_{\text{real}}, \tag{2.6.1}$$

where the IR divergences present in $\sigma_{\text{virtual}}$ are precisely cancelled by those in $\sigma_{\text{real}}$. We will see this explicit cancellation of IR divergences in Chapter 3.

### 2.6.2 Ultraviolet Divergences and Renormalisation

UV divergences arise from momentum integrals extending to arbitrarily high energy scales, since we do not know the true, fundamental theory of Nature at these high energies. If left untreated, these divergences would render predictions meaningless. The procedure by which such infinities are systematically removed is called *renormalisation*, which leads to finite, physically meaningful predictions.

There are enough details of renormalisation to write a textbook (e.g., Ref. [26]), but all of the subtleties and systematics of renormalisation are not required for this thesis. Instead, this section serves as a brief introduction to renormalisation (as in Ref. [77]) and an explanation of how it will be used in Chapter 3.

So far, quantities such as the mass and coupling constant have appeared as free parameters of a Lagrangian, able to take on any value. In interacting field theories, these initial (*bare*) parameters receive corrections from increasing orders of particle interactions. In order to align these theories with reality, we take measurements to determine the physical values of the parameters[5]. We can then directly use the physical (*renormalised*) parameters in our theory instead of the radiatively corrected bare parameters. This is the basic idea of renormalisation, and it is required even if there are no UV divergences [24, 77]. If the corrections to the bare parameters involve UV divergences, then changing variables to the renormalised parameters offers a set of finite parameters, often making calculations more tractable.

But how can we use QFT to make predictions at all if we do not know the fundamental theory at all energies? By the Heisenberg uncertainty principle, large violations of energy conservation can only occur for very short time scales, $\Delta t \sim 1/\Delta E$, meaning that high-energy effects must appear local (confined to short distances) for experiments at accessible energy levels. Local interactions are described by interaction terms of the Lagrangian, so these high-energy effects will result in shifts to the parameters of the interaction Lagrangian (e.g., mass and coupling constants). This justifies the procedure of including these high-energy corrections in the parameters of the interaction Lagrangian. We only observe the physical, renormalised parameters, which combine the unphysical, bare parameters with the high-energy corrections (renormalisation factors).

[5]These parameters are not observables, and thus cannot be directly measured. Instead, they are inferred from observables.

For example, consider a Lagrangian density containing bare fields and parameters, denoted by a subscript $B$,

$$\mathcal{L} = \frac{1}{2}\partial_\mu\phi_B\,\partial^\mu\phi_B - \frac{1}{2}m_B^2\phi_B^2 - \frac{\lambda_B}{3!}\phi_B^3\,. \tag{2.6.2}$$

To express the theory in terms of renormalised quantities, we introduce renormalisation factors,

$$\mathcal{L} = \frac{1}{2}Z_2\partial_\mu\phi_R\,\partial^\mu\phi_R - \frac{1}{2}Z_0 m_R^2\phi_R^2 - \frac{Z_1\lambda_R}{3!}\phi_R^3\,, \tag{2.6.3}$$

where $\phi_R, m_R, \lambda_R$ are the renormalised field, mass, and coupling, which are finite and measurable. The renormalisation factors $Z_i$ are chosen to absorb divergences appearing in loop corrections, ensuring that physical observables remain finite.

Rewriting this, we separate the Lagrangian density into the renormalised Lagrangian density, $\mathcal{L}_R$, and a counterterm Lagrangian density, $\mathcal{L}_{\text{ct}}$,

$$\mathcal{L} = \mathcal{L}_R + \mathcal{L}_{\text{ct}}\,, \tag{2.6.4}$$

where

$$\mathcal{L}_R = \frac{1}{2}\partial_\mu\phi_R\,\partial^\mu\phi_R - \frac{1}{2}m_R^2\phi_R^2 - \frac{\lambda_R}{3!}\phi_R^3\,, \tag{2.6.5}$$

and

$$\mathcal{L}_{\text{ct}} = \frac{1}{2}(Z_2-1)\partial_\mu\phi_R\,\partial^\mu\phi_R - \frac{1}{2}(Z_0-1)m_R^2\phi_R^2 - \frac{1}{3!}(Z_1-1)\lambda_R\phi_R^3\,. \tag{2.6.6}$$

The renormalisation constants are always chosen to explicitly cancel the divergences introduced by loop corrections, ensuring finite results order by order in perturbation theory.

The factor $Z_2$ is exactly the same field-strength renormalisation, $Z$, that appears in the LSZ reduction formula (Eq. (2.4.9)) [23]. It ensures that physical states are properly normalised, preventing divergences from propagating into $S$-matrix elements. Consequently, we can either work with renormalised parameters and explicit counterterms (*renormalised perturbation theory*) or bare parameters, with renormalisation factors appearing in the LSZ reduction formula (*bare perturbation theory*). The two approaches are exactly equivalent. In Chapter 3, we work with bare parameters and explicitly see how the UV divergences from loop corrections cancel when the LSZ formula is applied.

### 2.6.3 Feynman Parameterisation

Throughout Section 2.6.4 and Chapter 3, momentum integrals will involve a product of propagators. Feynman parametrisation is a technique used to combine denominators of propagators into a single quadratic polynomial in the momentum variable [23, 25, 80]. The remaining momentum integral is then spherically symmetric, allowing analytic integration without difficulty. The general identity is given by

$$\prod_{i=1}^{n} \frac{1}{D_i} = \int_0^1 \mathrm{d}x_1 \cdots \int_0^1 \mathrm{d}x_n \, \delta\Big(1 - \sum_{i=1}^{n} x_i\Big) \frac{(n-1)!}{(\sum_{i=1}^{n} x_i D_i)^n}\,, \tag{2.6.7}$$

where $x_1, \cdots, x_n$ are auxiliary parameters to be integrated over the range $0$ to $1$. For example, for two propagator denominators,

$$\frac{1}{AB} = \int_0^1 \mathrm{d}x \int_0^1 \mathrm{d}y \, \delta(1-x-y) \frac{1}{(xA+yB)^2} = \int_0^1 \mathrm{d}x \, \frac{1}{[xA+(1-x)B]^2}\,. \tag{2.6.8}$$

### 2.6.4 Divergent Integrals

Before we consider a calculation which involves divergences (Chapter 3), we should examine the divergent integrals which will arise in such calculations. We will evaluate integrals of the form,

$$J(D, \alpha, \beta, a^2) \equiv \int \frac{\mathrm{d}^D k}{(2\pi)^D} \frac{(k^2)^\alpha}{(k^2 - a^2 + i\epsilon)^\beta}, \tag{2.6.9}$$

in $D = 4 + 2\varepsilon$ spacetime dimensions, with $|\varepsilon| \ll 1$. The reason we evaluate these integrals in $D$ dimensions is to regulate UV ($k \to \infty$) and IR ($k \to 0$) divergences. These divergences will occur when $D = 4$, so we can temporarily parameterise the divergences with $\epsilon$, before taking $\epsilon \to 0$ at the end of the calculation.

UV divergences can be regulated by slightly decreasing the dimensionality of the in-tegral (i.e. $\varepsilon_{\text{UV}} < 0$), and IR divergences can be regulated by slightly increasing the dimensionality of the integral (i.e. $\varepsilon_{\text{IR}} > 0$). This is *dimensional regularisation*. Typically, the calculation for $R_q = \sigma(e^+e^- \to q\bar{q})/\sigma(e^+e^- \to \mu^+\mu^-)$ is done with a single dimensional regularisation parameter, $\varepsilon$, which regulates both IR and UV divergences. Whilst this can be made to give the correct result, it is disingenuous to how dimensional regularisation truly works. In Chapter 3, we treat the two types of divergences separately, and find that the distinct types of divergence cancel independently of each other.

### Evaluating the Integral

It is worth performing the integral in Eq. (2.6.9). First, we transform from the usual, Lorentzian 4-momentum, $k^\mu$, to the Euclidean 4-momentum, $k_E^\mu$, with a *Wick rotation* [81],

$$k^\mu = (k^0, \mathbf{k}) = (ik_E^0, \mathbf{k}_E) \tag{2.6.10}$$

$$\Rightarrow k^2 = (k^0)^2 - \mathbf{k}^2 = -(k_E^0)^2 - \mathbf{k}_E^2 \equiv -k_E^2\,, \tag{2.6.11}$$

$$\mathrm{d}k^0 = i\,\mathrm{d}k_E^0 \Rightarrow \mathrm{d}^D k = i\,\mathrm{d}^D k_E\ . \tag{2.6.12}$$

This Wick rotation is shown in Fig. 2.4. To understand why we can Wick rotate, consider the closed contour on the left of Fig. 2.4. The integration along this contour must vanish due to the residue theorem, since there are no enclosed poles. Since the integral along the curved parts of the contour must also be zero (the integrand in Eq. (2.6.9) vanishes for $|k_0^2| \to \infty$), the integral along the real and imaginary axes must be equal and opposite. This highlights the importance of the pole prescription of the Feynman propagator (as defined in Eq. (2.3.5) and shown in Fig. 2.1).

We can now write Eq. (2.6.9) as

$$\begin{aligned} J(D,\alpha,\beta,a^2) &= i\int \frac{\mathrm{d}^D k_E}{(2\pi)^D}\frac{(-1)^\alpha (k_E^2)^\alpha}{(-k_E^2-a^2)^\beta} \\ &= i(-1)^{\alpha-\beta}\int \frac{\mathrm{d}^D k_E}{(2\pi)^D}\frac{(k_E^2)^\alpha}{(k_E^2+a^2)^\beta}, \end{aligned} \tag{2.6.13}$$

where we have safely taken $\epsilon \to 0$ since the denominator is always positive and does not lead to pole ambiguities (i.e., taking $\epsilon \to 0$ no longer moves the poles onto the contour of integration). In these coordinates, the integral is spherically symmetric

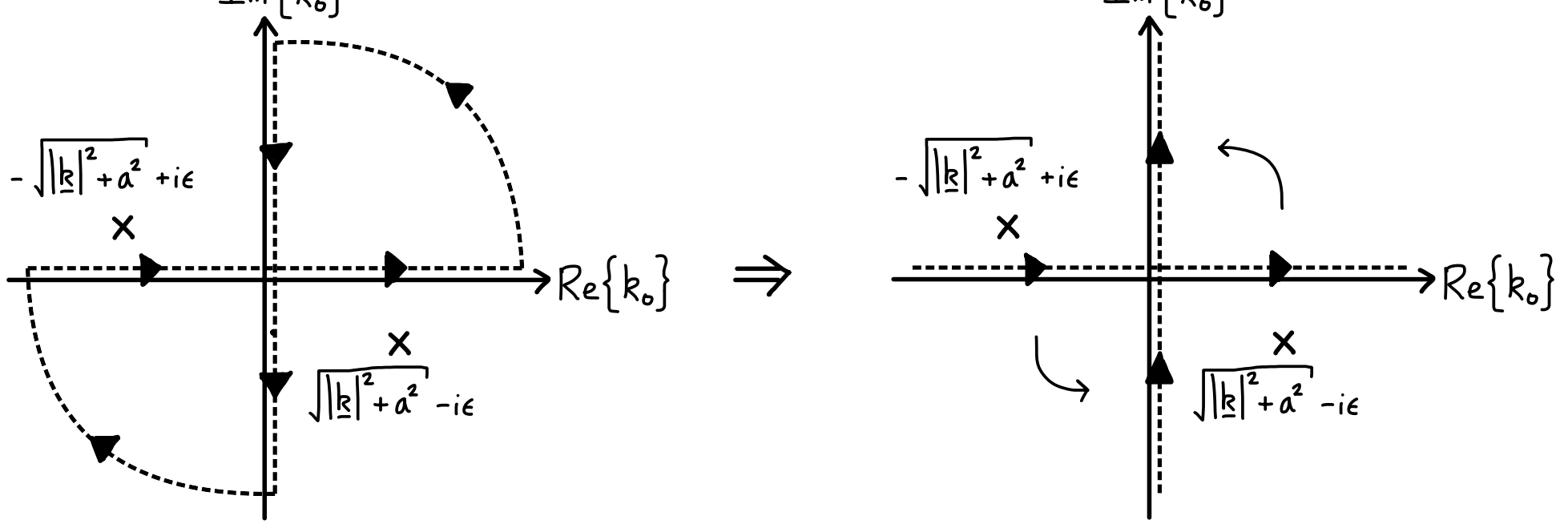

**Fig. 2.4. Left:** The integral along this contour vanishes due to the residue theorem, since there are no enclosed poles. Since the integral along the curved parts of the contour must also be zero, the integral along the real and imaginary axes must be equal and opposite. **Right:** The Wick rotation of the integration contour between Lorentzian 4-momentum coordinates, $k^\mu$, to Euclidean 4-momentum coordinates, $k_E^\mu$. The rotation is 90° anticlockwise in the complex plane, which is permitted due to the position of the poles.

and thus

$$\begin{aligned} J(D,\alpha,\beta,a^2) &= \frac{i(-1)^{\alpha-\beta}}{(2\pi)^D}\int \mathrm{d}\Omega_D \int_0^\infty \mathrm{d}|k_E|\,|k_E|^{D-1}\frac{|k_E|^{2\alpha}}{(|k_E^2|+a^2)^\beta} \\ &= \frac{2i(-1)^{\alpha-\beta}}{(4\pi)^{D/2}\Gamma(D/2)}\int_0^\infty \mathrm{d}|k_E|\,\frac{|k_E|^{2\alpha+D-1}}{(|k_E^2|+a^2)^\beta}, \end{aligned} \tag{2.6.14}$$

where $\int \mathrm{d}\Omega_D = 2\pi^{D/2}/\Gamma(D/2)$ is the ‘surface area’ of a $D$-dimensional unit sphere, and the Gamma function, $\Gamma(z)$, is defined as [82–85]

$$\Gamma(z) = \int_0^\infty t^{z-1}e^{-t}\,\mathrm{d}t\ , \quad \mathrm{Re}\{z\} > 0. \tag{2.6.15}$$

We can evaluate the remaining integral with a change of variable,

$$u \equiv \frac{a^2}{(|k_E|^2+a^2)}, \tag{2.6.16}$$

such that

$$\lim_{|k_E|\to\infty} u = 0 \tag{2.6.17}$$

$$\lim_{|k_E|\to 0} u = 1 \tag{2.6.18}$$

$$|k_E| = \left(\frac{a^2}{u}(1-u)\right)^{1/2} \tag{2.6.19}$$

$$\mathrm{d}u = \frac{-2a^2|k_E|}{(|k_E|^2+a^2)^2}\,\mathrm{d}|k_E| \Rightarrow \mathrm{d}|k_E| = -\frac{1}{2}\,\mathrm{d}u\,(1-u)^{-1/2}(a^{-2})\left(\frac{a^2}{u}\right)^{3/2}. \tag{2.6.20}$$

Therefore, Eq. (2.6.14) can be written as

$$J(D,\alpha,\beta,a^2) = \frac{i(-1)^{\alpha-\beta}(a^2)^{\alpha-\beta+D/2}}{(4\pi)^{D/2}\Gamma(D/2)} \int_0^1 \mathrm{d}u\, u^{\beta-\alpha-D/2-1}(1-u)^{\alpha+D/2-1}, \quad (2.6.21)$$

where the limits of integration have been swapped, picking up a minus sign. Finally, we apply the Euler Beta function, defined as [82, 84, 85]

$$\mathcal{B}(m,n) = \int_0^1 \mathrm{d}u\, u^{m-1}(1-u)^{n-1} = \frac{\Gamma(m)\Gamma(n)}{\Gamma(n+m)}. \quad (2.6.22)$$

The final expression for Eq. (2.6.9) is

$$J(D,\alpha,\beta,a^2) = \frac{i}{(4\pi)^{D/2}}(a^2)^{D/2}(-a^2)^{\alpha-\beta}\frac{\Gamma(\beta-\alpha-D/2)\Gamma(\alpha+D/2)}{\Gamma(\beta)\Gamma(D/2)}. \quad (2.6.23)$$

The Gamma function, $\Gamma(z)$, has simple poles at $z = 0, -1, -2, \ldots$, which will lead to divergences. It can be Laurent expanded [86, 87] as

$$\Gamma(z) = \frac{1}{z} - \gamma_E + \mathcal{O}(z), \quad (2.6.24)$$

where $\gamma_E$ is the Euler-Mascheroni constant [84, 85, 88], and the expansion is only valid in the region $z > 0$ with $z \in \mathbb{R}$ and $|z| \ll 1$. Since $\varepsilon_{\text{UV}} < 0$ and $\varepsilon_{\text{IR}} > 0$, the only valid expansions of the Gamma function in $\varepsilon$ are

$$\Gamma(-\varepsilon_{\text{UV}}) = -\frac{1}{\varepsilon_{\text{UV}}} - \gamma_E + \mathcal{O}(\varepsilon_{\text{UV}}) \quad \text{and} \quad (2.6.25)$$

$$\Gamma(\varepsilon_{\text{IR}}) = \frac{1}{\varepsilon_{\text{IR}}} - \gamma_E + \mathcal{O}(\varepsilon_{\text{IR}}). \quad (2.6.26)$$

Another source of divergences in Eq. (2.6.23) are indeterminations of the type $0^0$, which may arise from the $(a^2)^{\alpha-\beta+D/2}$ term. This will result in both UV and IR divergences. We will treat this type of divergence in Section 3.4.2.

It should be noted that we haven't considered tensor integrals (Eq. (2.6.9) is a scalar integral). Tensor integrals can be written in terms of the scalar integral as [89],

$$\int \frac{\mathrm{d}^D k}{(2\pi)^D}\frac{(k^2)^\alpha k^\mu k^\nu}{(k^2-a^2+i\epsilon)^\beta} = \frac{g^{\mu\nu}}{D}J(D,\alpha+1,\beta,a^2)\,. \quad (2.6.27)$$

Any integral with an odd power of $k^\mu$ in the numerator will be zero by symmetry.

**Example: UV Divergence**

Consider the integral,

$$I^{\mu}_{\text{UV}} \equiv \int \frac{\mathrm{d}^D k}{(2\pi)^D} \frac{k^{\mu}}{(k^2 + i\epsilon)((k+p)^2 - m^2 + i\epsilon)}, \tag{2.6.28}$$

then Feynman parameterise and shift the integration variable,

$$\begin{aligned} I^{\mu}_{\text{UV}} &= \int_0^1 \mathrm{d}x \int \frac{\mathrm{d}^D k}{(2\pi)^D} \frac{k^{\mu}}{[(k+px)^2 - a^2 + i\epsilon]^2} \\ &= \int_0^1 \mathrm{d}x \int \frac{\mathrm{d}^D k}{(2\pi)^D} \frac{k^{\mu} - xp^{\mu}}{[k^2 - a^2 + i\epsilon]^2}. \end{aligned} \tag{2.6.29}$$

Now we note that any integral of this form with an odd power of $k^{\mu}$ in the numerator is zero by symmetry, i.e.

$$\int \frac{\mathrm{d}^D k}{(2\pi)^D} \frac{(k^2)^n k^{\mu}}{[k^2 - a^2 + i\epsilon]^m} = 0, \tag{2.6.30}$$

and thus,

$$\begin{aligned} I^{\mu}_{\text{UV}} &= \int_0^1 \mathrm{d}x \int \frac{\mathrm{d}^D k}{(2\pi)^D} \frac{-xp^{\mu}}{[k^2 - a^2 + i\epsilon]^2} = -p^{\mu} \int_0^1 \mathrm{d}x\, x J(D, 0, 2, a^2) \\ &= \frac{-ip^{\mu}}{(4\pi)^{D/2}} \Gamma(2 - D/2) \int_0^1 \mathrm{d}x\, x \Big( -p^2 x(1-x) + m^2 x \Big)^{D/2-2}, \end{aligned} \tag{2.6.31}$$

where we have used Eq. (2.6.23) to write the integral in terms of Gamma functions. Taking $D = 4$ at this point would cause a divergence due to $\Gamma(0)$. One may also think there will be a divergence of the form $0^0$ in the integrand when $x = 0$, but the extra factor of $x$ in the integrand makes sure this converges (note that if $p^2 = m^2 = 0$ then this will indeed result in an indetermination and hence further divergences—both UV and IR—as we will see in Section 3.4.2).

So the only divergence in Eq. (2.6.31) is due to $\Gamma(\beta - \alpha - D/2) = \Gamma(0)$. We can understand the nature of this divergence by examining Eq. (2.6.21). For $D = 4$, the only possible divergence would come from the $u^{\beta-\alpha-D/2-1}$ term when $\beta - \alpha - D/2 = 0$ (as in this example) and $u = 0$. From Eq. (2.6.17) we know that $u = 0$ corresponds to $|k_E| \to \infty \Rightarrow |k| \to \infty$. Thus, a divergence due to $\beta - \alpha - D/2 = 0$ is an UV divergence.

**Example: IR Divergence**

Consider the integral,

$$I_{\text{IR}} = \int \frac{\text{d}^D k}{(2\pi)^D} \frac{1}{(k^2 + i\epsilon)\,((k+p_1)^2 + i\epsilon)\,((k+p_2)^2 + i\epsilon)}\,, \tag{2.6.32}$$

where $p_1^2 = p_2^2 = 0$. We can guess intuitively that there may be some problems as $k \to 0$ because the integrand becomes (for $\epsilon \to 0$),

$$\frac{1}{k^2(2k\cdot p_1)(2k\cdot p_2)}\,, \tag{2.6.33}$$

which exhibits two types of IR divergences:

1. $k^2 = 0$ "soft divergence"
2. $k\cdot p_1 = 0$ or $k \cdot p_2 = 0$ "collinear divergence"

This is not enough to prove that there are IR divergences ($k^2 \to 0$ also seems problematic for Eq. (2.6.28), but we have shown that it is not), so let us evaluate Eq. (2.6.32) and see what happens. First, we apply Feynman parameterisation,

$$I_{\text{IR}} = \int_0^1 \text{d}x \int_0^1 \text{d}y \int \frac{\text{d}^D k}{(2\pi)^D} \frac{2x}{[(k + p_1 xy - p_3 x(1-y))^2 - a^2 + i\epsilon]^3}\,, \tag{2.6.34}$$

where $a^2 = -2(p_1\cdot p_2)x^2 y(1-y)$. Then we shift variables $k \to k - p_1 xy + p_2 x(1-y)$ and recognise that we have an integral in the form of Eq. (2.6.9),

$$I_{\text{IR}} = \int_0^1 \text{d}x \int_0^1 \text{d}y \int \frac{\text{d}^D k}{(2\pi)^D} \frac{2x}{[k^2 - a^2 + i\epsilon]^3} = 2\int_0^1 \text{d}x \int_0^1 \text{d}y\, x J(D,0,3,a^2)\,. \tag{2.6.35}$$

Using Eq. (2.6.23) to write this in terms of Gamma functions, we get

$$I_{\text{IR}} = \frac{-i}{(4\pi)^{D/2}}\Gamma(3 - D/2)(-2p_1\cdot p_2)^{D/2-3} \int_0^1 \text{d}x\, x^{D-5} \int_0^1 \text{d}y\, y^{D/2-3}(1-y)^{D/2-3}\,. \tag{2.6.36}$$

If one were to take $D = 4$ at this point, there would be divergences from the $x$ and $y$ integrals in three regions of the integral space:

$$1)\quad x = 0\,; \qquad 2)\quad y = 0\,; \qquad 3)\quad y = 1\,.$$

Examining Eq. (2.6.34), we can see what each of these regions physically correspond to. The first region ($x = 0$) results in the denominator of the integrand becoming $k^6$. Clearly, the integral will diverge for $k = 0$. The second region ($y = 0$) gives a denominator $k^4(k^2 + 6k \cdot p_1)$, which diverges for $k = 0$ or $k \cdot p_1 = 0 \wedge k \to 0$. The third region ($y = 1$) gives a denominator $k^4(k^2 - 6k \cdot p_2)$, which diverges for $k = 0$ or $k \cdot p_2 = 0 \wedge k \to 0$. These are exactly the soft and collinear divergences we expected. These are the only divergences present, so $I_{\text{IR}}$ is IR divergent and not UV divergent.

Thus, to successfully evaluate $I_{\text{IR}}$, we must work in $D = 4 + 2\varepsilon_{\text{IR}}$ dimensions, with $\varepsilon_{\text{IR}} > 0$ in order to regulate these IR divergences. Using the Euler Beta function (Eq. (2.6.22)), it is then straightforward to arrive at the result,

$$I_{\text{IR}} = \frac{i}{(4\pi)^2} \frac{1}{2p_2 \cdot p_3} \left( \frac{-2p_1 \cdot p_2}{4\pi} \right)^{\varepsilon_{\text{IR}}} \frac{\Gamma(1 - \varepsilon_{\text{IR}})}{\Gamma(1 + 2\varepsilon_{\text{IR}})} \Gamma^2(\varepsilon_{\text{IR}}). \tag{2.6.37}$$

The structure of the IR poles can then be seen when one expands the Gamma functions using

$$\Gamma(\varepsilon_{\text{IR}}) = \frac{1}{\varepsilon_{\text{IR}}} - \gamma_E + \frac{1}{12}(\pi^2 + 6\gamma_E^2)\varepsilon_{\text{IR}} + \mathcal{O}\big(\varepsilon_{\text{IR}}{}^2\big), \tag{2.6.38}$$

$$\Gamma(1 \mp \varepsilon_{\text{IR}}) = 1 \pm \gamma_E \varepsilon_{\text{IR}} + \frac{1}{12}(\pi^2 + 6\gamma_e^2)\varepsilon_{\text{IR}}{}^2 + \mathcal{O}\big(\varepsilon_{\text{IR}}{}^3\big), \tag{2.6.39}$$

$$(f)^{\varepsilon_{\text{IR}}} = 1 + \varepsilon_{\text{IR}} \ln f + \frac{\varepsilon_{\text{IR}}{}^2}{2!}(\ln f)^2 + \ldots, \tag{2.6.40}$$

$$\ln(-1) = \ln\big(e^{i\pi}\big) = -i\pi, \tag{2.6.41}$$

being careful to include the $\mathcal{O}(\varepsilon_{\text{IR}}{}^2)$ terms when necessary as they will multiply with the $\varepsilon_{\text{IR}}{}^{-2}$ terms from $\Gamma^2(\varepsilon_{\text{IR}})$ to give constants. In Eq. (2.6.41), we have taken $\ln(-1) = -i\pi$ as a matter of convention in line with the Sokhotski-Plemelj theorem [90, 91], since $\ln(-1) = -i\pi + 2in\pi \,, \forall\, n \in \mathbb{Z}$.

### Summary of UV and IR Divergences

Integrals in the form of Eq. (2.6.9) can exhibit UV and IR divergences.

- When $\beta - \alpha - D/2 = 0$, this corresponds to $k \to \infty$ and hence these are UV divergences, and must be regulated with a small negative dimension $\varepsilon_{\text{UV}}$ such that $D = 4 + 2\varepsilon_{\text{UV}}$.

- When the divergences arise from the integration of the Feynman parameters, this corresponds to $k \to 0$ and hence are IR divergences (both soft and collinear), and must be regulated with a small positive dimension $\varepsilon_{\text{IR}}$ such that $D = 4+2\varepsilon_{\text{IR}}$.
- In the massless limit, $a^2 = 0$ and there will be an indetermination of the form $0^0$. This will result in both UV and IR divergences and will be treated in analysis of the self-energy diagram in Section 3.4.2.

### Loop Integrals

We now collect and evaluate the integrals which will arise in Chapter 3, so that we can simply refer back to them when needed. The full derivation of the integrals can be found in Ref. [89].

$$\begin{aligned} C_1 &\equiv \int \frac{\mathrm{d}^D k}{(2\pi)^D} \frac{1}{(k^2+i\epsilon)\left((k+q_1)^2+i\epsilon\right)\left((k-q_2)^2+i\epsilon\right)} \\ &= \frac{i}{(4\pi)^2}\frac{1}{2q_1\cdot q_2}\left(\frac{-2q_1\cdot q_2}{4\pi}\right)^{\varepsilon_{\text{IR}}}\frac{\Gamma(1-\varepsilon_{\text{IR}})}{\Gamma(1+2\varepsilon_{\text{IR}})}\Gamma^2(\varepsilon_{\text{IR}}) \end{aligned} \tag{2.6.42a}$$

$$\begin{aligned} C_2 &\equiv \int \frac{\mathrm{d}^D k}{(2\pi)^D} \frac{k^\mu}{(k^2+i\epsilon)\left((k+q_1)^2+i\epsilon\right)\left((k-q_2)^2+i\epsilon\right)} \\ &= (q_2^\mu - q_1^\mu)\frac{i}{(4\pi)^2}\frac{1}{2q_1\cdot q_2}\left(\frac{-2q_1\cdot q_2}{4\pi}\right)^{\varepsilon_{\text{IR}}}\frac{\Gamma(1+\varepsilon_{\text{IR}})\Gamma(1-\varepsilon_{\text{IR}})}{\Gamma(2+2\varepsilon_{\text{IR}})}\Gamma(\varepsilon_{\text{IR}}) \end{aligned} \tag{2.6.42b}$$

$$\begin{aligned} C_3 &\equiv \int \frac{\mathrm{d}^D k}{(2\pi)^D} \frac{k^2}{(k^2+i\epsilon)\left((k+q_1)^2+i\epsilon\right)\left((k-q_2)^2+i\epsilon\right)} \\ &= \frac{-i}{(4\pi)^2}(2q_1\cdot q_2)^{\varepsilon_{\text{UV}}}\left[\frac{1}{\hat{\varepsilon}_{\text{UV}}} - 2 - i\pi\right] \end{aligned} \tag{2.6.42c}$$

$$\begin{aligned} C_4 &\equiv \int \frac{\mathrm{d}^D k}{(2\pi)^D} \frac{k^\mu k^\nu}{(k^2+i\epsilon)\left((k+q_1)^2+i\epsilon\right)\left((k-q_2)^2+i\epsilon\right)} \\ &= \frac{g^{\mu\nu}}{4}\frac{-i}{(4\pi)^2}(2q_1\cdot q_2)^{\varepsilon_{\text{UV}}}\left[\frac{1}{\hat{\varepsilon}_{\text{UV}}} - 3 - i\pi\right] \\ &+ \frac{i}{(4\pi)^2}\frac{1}{2q_1\cdot q_2}\left(\frac{-2q_1\cdot q_2}{4\pi}\right)^{\varepsilon_{\text{IR}}}\frac{\Gamma(1-\varepsilon_{\text{IR}})}{\Gamma(3+2\varepsilon_{\text{IR}})} \\ &\times \left[\left(q_1^\mu q_1^\nu + q_2^\mu q_2^\nu\right)\Gamma(2+\varepsilon_{\text{IR}})\Gamma(\varepsilon_{\text{IR}}) - \left(q_1^\mu q_2^\nu + q_1^\nu q_2^\mu\right)\Gamma^2(1+\varepsilon_{\text{IR}})\right] \end{aligned} \tag{2.6.42d}$$

where

$$\frac{1}{\hat{\varepsilon}} \equiv \frac{1}{\varepsilon} + \gamma_E - \ln(4\pi)\,. \tag{2.6.43}$$

and we have used

$$(2q_1 \cdot q_2)^{\varepsilon_{\rm UV}} \approx 1 + \varepsilon_{\rm UV} \ln(2q_1 \cdot q_2)\,. \tag{2.6.44}$$

# Chapter 3

# Traditional Particle Scattering

In this chapter, we will calculate cross sections for electron-positron annihilation, given by [23, 25, 54],

$$\sigma(e^+e^- \to \gamma^* \to X) = \frac{1}{2s}\int \mathrm{dPS}\,\left|\mathcal{M}_{fi}^{\mathrm{LSZ}}\right|^2, \tag{3.0.1}$$

where $\gamma^*$ denotes a virtual photon, $X$ indicates any final-state particles, $s \equiv (p_1+p_2)^2$ is the Mandelstam variable [92] for the square of the centre-of-mass energy (where $p_1$ and $p_2$ are the 4-momenta of the electron and positron), dPS is the measure for the final-state phase space, and $\mathcal{M}_{fi}^{\mathrm{LSZ}}$ is the sum of invariant matrix elements associated with all relevant Feynman diagrams, with corrections from the Lehmann–Symanzik–Zimmermann (LSZ) reduction formula (see Section 2.4).

The main goal of this chapter is to calculate the cross section for quark-antiquark production to first order in gluon corrections, i.e., up to $\mathcal{O}(e^4g^2)$, where $e$ is the electromagnetic coupling constant and $g$ is the strong coupling constant. The calculation is fully inclusive over the final state, except demanding that it contains one quark and one antiquark. This is denoted $e^-e^+ \to \gamma^* \to q\bar{q}\,X$, where $X$ either represents nothing or the real emission of a gluon (up to $\mathcal{O}(e^4g^2)$).

This calculation demonstrates the cancellation of infrared (IR) and ultraviolet (UV) divergences, but is often performed without full rigour, with IR and UV divergences either artificially cancelled against each other or treated together in renormalisation. In this chapter, we keep these divergences separate: UV divergences cancel among themselves due to the equality of the field-strength and vertex renormalisation constants, while IR divergences cancel due to the degeneracy of the process $e^-e^+ \to \gamma^* \to q\bar{q}g$ with the tree-level process $e^-e^+ \to \gamma^* \to q\bar{q}$, in the soft (low-energy) or collinear limits.

## 3.1 Lagrangian and Feynman Diagrams

The total Lagrangian density for a system consisting of electrons, muons, quarks, photons, and gluons is given by (see Section 2.2)

$$\begin{aligned}\mathcal{L} = &\underbrace{\sum_{l=e,\mathrm{m}} \Big(\bar{\psi}_l(i\gamma^\mu\partial_\mu - m_l)\psi_l\Big) + \bar{\psi}^i_q(i\gamma^\mu\partial_\mu - m_q)\psi^i_q - \frac{1}{4}F_{\mu\nu}F^{\mu\nu} - \frac{1}{4}G^a_{\mu\nu}G^{\mu\nu}_a}_{\mathcal{L}_0} \\ &\underbrace{+ \sum_{l=e,\mathrm{m}} \Big(-e\bar{\psi}_l\gamma^\mu A_\mu\psi_l\Big) - e_q e\bar{\psi}^i_q\gamma^\mu A_\mu\psi^i_q + g\bar{\psi}^i_q\gamma^\mu G^a_\mu(T^a)^{ij}\psi^j_q}_{\mathcal{L}_{\text{int}}}\end{aligned} \tag{3.1.1}$$

where $\mathcal{L}_0$ is the free Lagrangian density and $\mathcal{L}_{\text{int}}$ includes the interaction terms. We only consider one flavour of quark then sum over quark flavours at the end of our calculations.

The free Lagrangian density, $\mathcal{L}_0$, describes the kinetic and mass terms for the electron, muon, quark, photon, and gluon fields. The terms $\bar{\psi}_e(i\gamma^\mu\partial_\mu - m_e)\psi_e$ and $\bar{\psi}_\mathrm{m}(i\gamma^\mu\partial_\mu - m_\mathrm{m})\psi_\mathrm{m}$ describe the free Dirac fields for the electron and muon, with their respective masses $m_e$ and $m_\mathrm{m}$. The term $\bar{\psi}^i_q(i\gamma^\mu\partial_\mu - m_q)\psi^i_q$ describes the free Dirac field for a single quark flavour with mass $m_q$ and colour index $i$. The terms $-\frac{1}{4}F_{\mu\nu}F^{\mu\nu}$ and $-\frac{1}{4}G^a_{\mu\nu}G^{\mu\nu}_a$ are the kinetic terms for the photon field, $A_\mu$, and the gluon field, $G^a_\mu$.

The interaction terms, $\mathcal{L}_{\text{int}}$, describe how the fermions couple to the gauge fields. They are: The electron-photon interaction term $-e\bar{\psi}_e\gamma^\mu A_\mu\psi_e$, where $e = -|e|$ is the fundamental electric charge; the muon-photon interaction term $-e\bar{\psi}_\mathrm{m}\gamma^\mu A_\mu\psi_\mathrm{m}$, identical in form to the electron's interaction; the quark-photon interaction term $-e_q e\bar{\psi}^i_q\gamma^\mu A_\mu\psi^i_q$, where $e_q$ is the dimensionless ratio of the quark's electric charge to $e = -|e|$; the quark-gluon interaction term $g\bar{\psi}^i_q\gamma^\mu G^a_\mu(T^a)^{ij}\psi^j_q$, where $g$ is the strong coupling constant, and $T^a$ are the $SU(3)$ generators in the fundamental representation.

In Section 3.2, we calculate the cross section for muon-antimuon production, $\sigma(e^-e^+ \to \gamma^* \to \mu^-\mu^+)$. This process involves no quarks and hence no corrections from quantum chromodynamics (QCD). In Section 3.3, we adjust the muon-antimuon cross section to give the tree-level result for the quark-antiquark production cross section. Since the coupling of the quarks to gluons is stronger than that of quarks to pho-tons (i.e., $|g| \gg |e|$) [93], the largest first-order corrections to the tree-level quark-antiquark production cross section are gluon corrections. Accordingly, Section 3.4 calculates $\sigma(e^-e^+ \to \gamma^* \to q\bar{q}\,X)$ to $\mathcal{O}(e^4g^2)$.

Using the LSZ reduction formula from Section 2.4, the $S$-matrix for the process $e^-e^+ \to \gamma^* \to q\bar{q}$ is given by

$$\langle \mathbf{q}_1\mathbf{q}_2|\, S\, |\mathbf{p}_1\mathbf{p}_2\rangle = Z_e\, Z_q \sum (\text{amputated diagrams}) \tag{3.1.2}$$

where $\{p_1, p_2, q_1, q_2\}$ are the 4-momenta of the electron, positron, quark and antiquark respectively, $Z_e$ is the electron (or positron) field-strength renormalisation, and $Z_q$ is the quark (or antiquark) field-strength renormalisation constant. Thus, the total, LSZ-corrected invariant matrix element is given by

$$\mathcal{M}_{fi}^{\text{LSZ}} = Z_e\, Z_q \sum_{\text{amputated}} \mathcal{M}_{fi} \tag{3.1.3}$$

where $\mathcal{M}_{fi}$ are the amputated correlation amplitudes (defined in Section 2.4) connecting an electron-positron pair to a quark-antiquark pair, calculated directly from Feynman diagrams using the Feynman rules in Section 2.5.

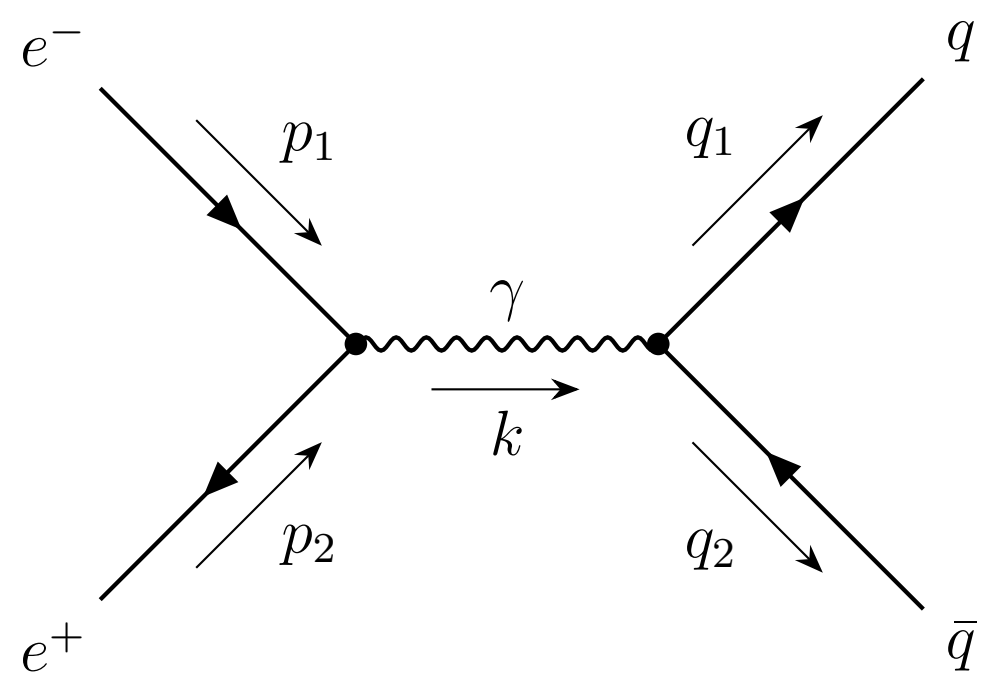

**Fig. 3.1.** Tree-level Feynman diagram for quark-antiquark production from an electron-positron pair.

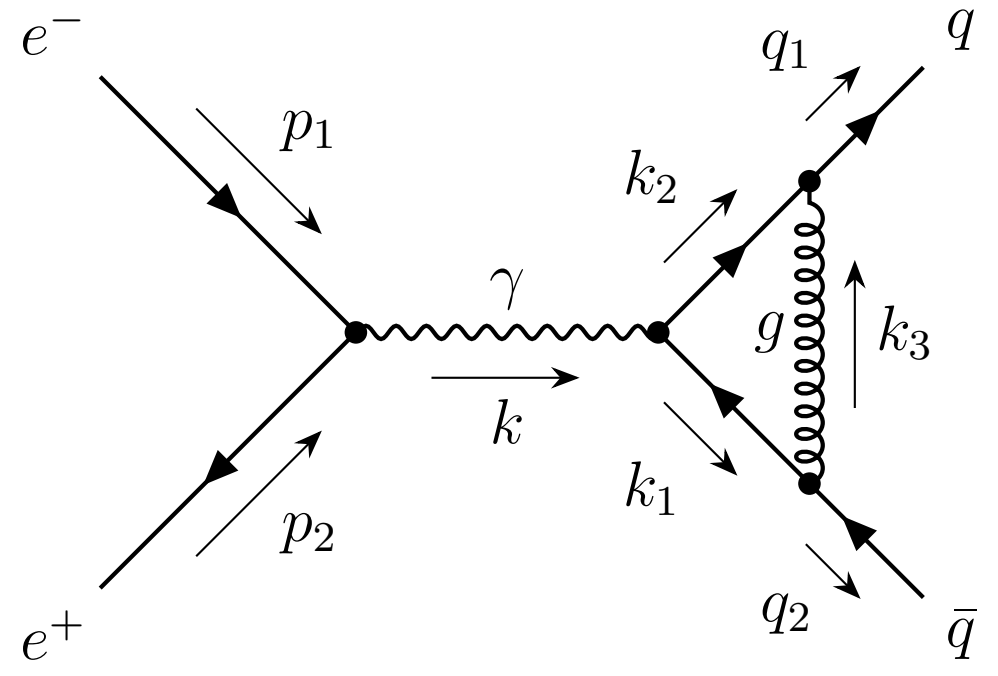

**Fig. 3.2.** Feynman diagram for the gluon vertex correction to quark-antiquark production.

The relevant amputated Feynman diagrams are the tree-level diagram, $\mathcal{M}_{fi}^{\text{tree}}$, shown in Fig. 3.1, and the vertex correction, $\mathcal{M}_{fi}^{\text{vertex}}$, shown in Fig. 3.2. The tree-level diagram is $\mathcal{O}(e^2)$ and the vertex correction is $\mathcal{O}(e^2 g^2)$. The electron field-strength renormalisation constant, $Z_e$, equals 1 to first order in the electromagnetic coupling, $e$, and since we are not considering any higher order corrections in $e$, we can ignore this constant. The quark field-strength renormalisation constant, $Z_q \equiv Z$, can be written as $Z = 1 + \delta Z$, where $\delta Z$ includes the $\mathcal{O}(g^2)$ corrections. Higher orders will not be considered in this chapter. Thus, up to order $\mathcal{O}(e^2 g^2)$,

$$\begin{aligned} \mathcal{M}_{fi}^{\text{LSZ}} &= Z\left(\mathcal{M}_{fi}^{\text{tree}} + \mathcal{M}_{fi}^{\text{vertex}}\right) + \cdots \\ &= \mathcal{M}_{fi}^{\text{tree}} + \delta Z\, \mathcal{M}_{fi}^{\text{tree}} + \mathcal{M}_{fi}^{\text{vertex}} + \cdots , \end{aligned} \tag{3.1.4}$$

where $\cdots$ indicates higher order terms. Taking the modulus squared,

$$\begin{aligned} |\mathcal{M}_{fi}^{\text{LSZ}}|^2 &= |\mathcal{M}_{fi}^{\text{tree}}|^2 + 2\,\delta Z\, |\mathcal{M}_{fi}^{\text{tree}}|^2 + \mathcal{M}_{fi}^{\text{vertex}} \mathcal{M}_{fi}^{\text{tree}\dagger} + \mathcal{M}_{fi}^{\text{tree}} \mathcal{M}_{fi}^{\text{vertex}\dagger} + \cdots \\ &= |\mathcal{M}_{fi}^{\text{tree}}|^2 + 2\,\delta Z\, |\mathcal{M}_{fi}^{\text{tree}}|^2 + 2\,\text{Re}\Big\{\mathcal{M}_{fi}^{\text{vertex}} \mathcal{M}_{fi}^{\text{tree}\dagger}\Big\} + \cdots \end{aligned} \tag{3.1.5}$$

Since the only relevant correction from the LSZ formula is the term $2\,\delta Z\, |\mathcal{M}_{fi}^{\text{tree}}|^2$, this term will be referred to as the LSZ correction term, and we drop the superscript 'LSZ' for the total invariant matrix element.

We will find that Eq. (3.1.5) is IR divergent. This is because we have not considered all relevant contributions. To calculate the inclusive cross section $\sigma(e^+e^- \to \gamma^* \to q\bar{q}\,X)$ to $\mathcal{O}(e^4 g^2)$, we must consider the lowest-order 2-to-3 cross section involving the real emission of a gluon, $\sigma_{\text{r.e.}} \equiv \sigma(e^+e^- \to \gamma^* \to q\bar{q}g)$. In the soft (low-energy) or collinear limit, this process is degenerate with the 2-to-2 process, since neither the 2-to-2 cross section nor the soft/collinear real emission can be measured individually by a real detector; only their sum is physically observable. This is in line with the Kinoshita-Lee-Nauenberg (KLN) theorem [28, 29], which ensures the cancellation of infrared divergences in physical observables in QCD. The cross section we calculate is therefore that which is associated with the probability of either of these processes occurring.

The lowest-order contributions from the real-emission process, $\mathcal{M}_{fi}^{\text{r.e.}}$, are shown in Fig. 3.3. Since $|\mathcal{M}_{fi}^{\text{r.e.}}|^2$ is already $\mathcal{O}(e^4 g^2)$, we do not consider corrections from field-strength renormalisation constants in the LSZ formula, which contribute at higher orders. Due to the three-particle final state, we sum this contribution as a distinguish-able process (i.e. we add at the cross section level, not the amplitude level). Thus, the total, first-order cross section we wish to calculate is

$$\sigma_1 = \sigma_{q\bar{q}} + \sigma_{\text{r.e.}}\,, \tag{3.1.6}$$

where $\sigma_{q\bar{q}} \equiv \sigma(e^- e^+ \to \gamma^* \to q\bar{q})$ is calculated using the matrix element in Eq. (3.1.5). This first-order cross section is both UV-finite and IR-finite.

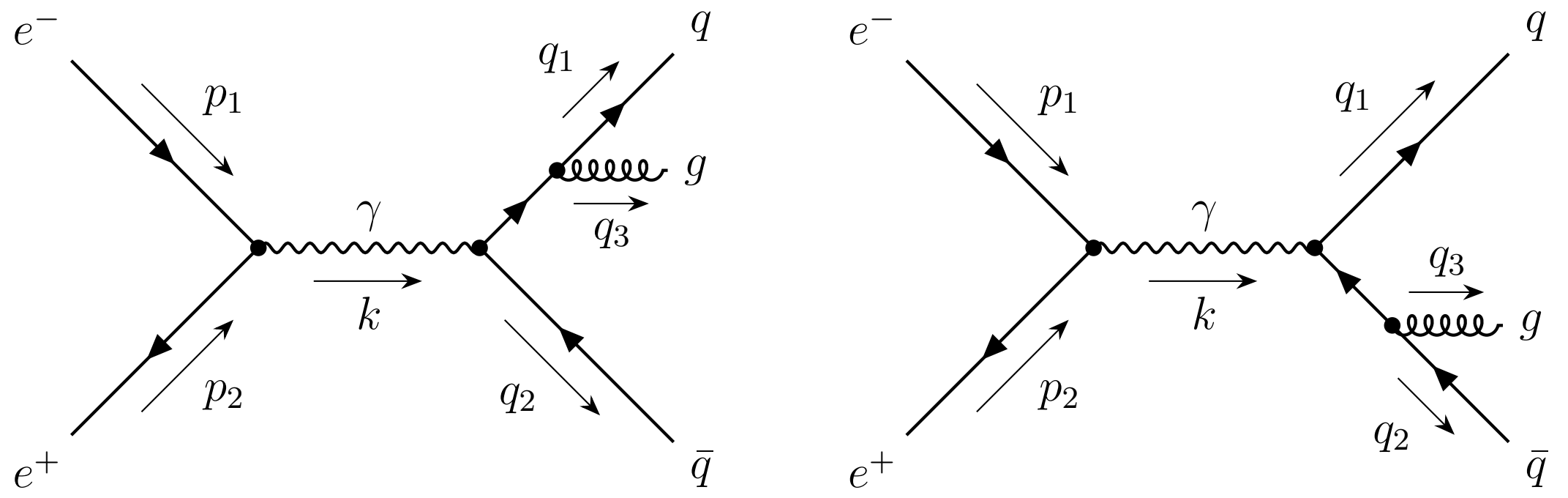

**Fig. 3.3.** Feynman diagrams for the real emission of a gluon in quark-antiquark production.

## 3.2 Tree level $e^+e^- \to \mu^+\mu^-$ cross section

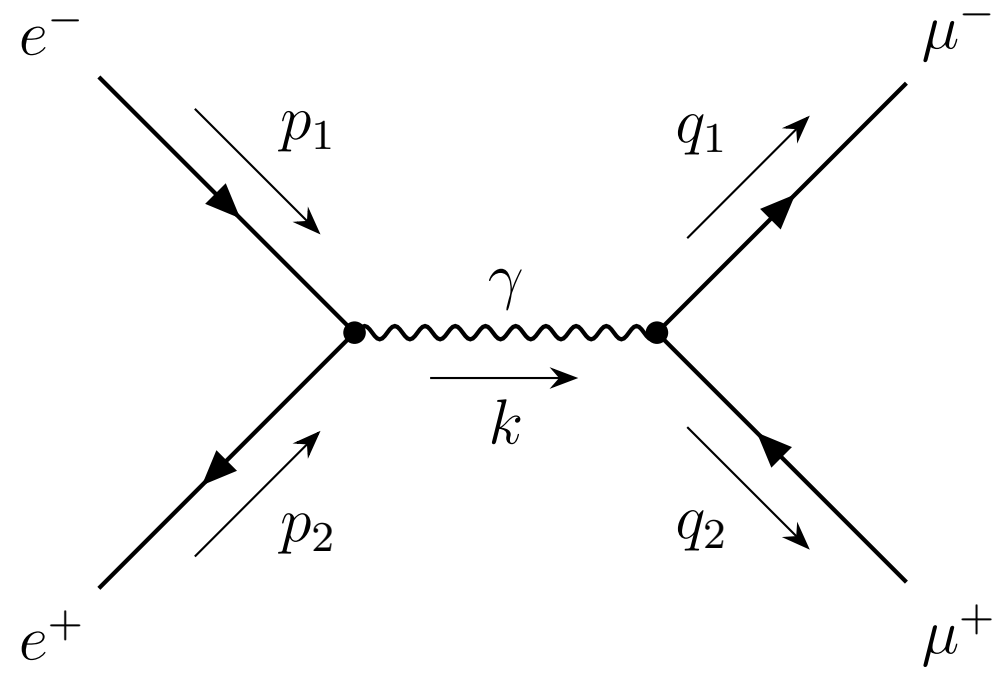

**Fig. 3.4.** Tree-level Feynman diagram for muon-antimuon production from an electron-positron pair.

### 3.2.1 Calculating the Matrix Element

Before considering the production of a quark-antiquark pair, we first consider a slightly more simple process: muon-antimuon production. The tree-level cross section for muon-antimuon production, $\sigma_0(e^+e^- \to \gamma^* \to \mu^+\mu^-)$, is described by the Feynman diagram in Fig. 3.4. We consider this process since there are no hadrons or quarks, so we do not have to consider QCD corrections, and there are no $t$-channel or $u$-channel contributions, unlike electron-positron production. In the full Standard Model, there is a similar process mediated by a $Z$ boson instead of a photon, but we can make this contribution small by assuming we are far from the $Z$ boson's rest energy. This means that this calculation will give useful results for the tree-level, Standard Model contribution.

We calculate the matrix element in $D$ dimensions so that the results can be used when we employ dimensional regularisation later in this chapter. In $D$ dimensions, it is conventional [25] to take

$$e \to \mu^{\frac{4-D}{2}} e \,, \tag{3.2.1}$$

where $\mu$ is an arbitrary parameter of mass dimension 1. This ensures that the coupling, $e$, remains dimensionless.

Using the Feynman rules for QED (Section 2.5.2), the matrix element, $\mathcal{M}_{fi}$, associated with Fig. 3.4 is expressed as

$$i\mathcal{M}_{fi} = \left[\bar{v}^{s'}(p_2)(-i\mu^{\frac{4-D}{2}} e\gamma^{\mu})u^{s}(p_1)\right]\left[\frac{-ig_{\mu\nu}}{k^2 + i\epsilon}\right]\left[\bar{u}^{r}(q_1)(-i\mu^{\frac{4-D}{2}} e\gamma^{\nu})v^{r'}(q_2)\right], \tag{3.2.2}$$

where we have chosen the Feynman gauge for the internal photon propagator. Taking the modulus squared of this expression, we get

$$\begin{aligned} \left|\mathcal{M}_{fi}\right|^2 = \left[\frac{g_{\mu\nu}g_{\rho\sigma}}{k^4}\right] & \underbrace{\left[\mu^{(4-D)}e^2\bar{v}^{s'}(p_2)(\gamma^{\mu})u^{s}(p_1)\bar{u}^{s}(p_1)(\gamma^{\rho})v^{s'}(p_2)\right]}_{\text{Initial-state particles, } \tilde{L}^{\mu\rho}} \\ \times & \underbrace{\left[\mu^{(4-D)}e^2\bar{u}^{r}(q_1)(\gamma^{\nu})v^{r'}(q_2)\bar{v}^{r'}(q_2)(\gamma^{\sigma})u^{r}(q_1)\right]}_{\text{Final-state particles, } \tilde{H}^{\nu\sigma}} , \end{aligned} \tag{3.2.3}$$

where $\epsilon \to 0$ is trivial due to the lack of loops. It is useful to label the initial-state term as $\tilde{L}^{\mu\rho}$ and the final-state term as $\tilde{H}^{\nu\sigma}$ ($L$ stands for leptons, as the initial-state particles will always be an electron and a positron, the $H$ stands of hadrons, as the final-state particles will later be a quark-antiquark pair, and the tilde is to indicate that the particle spins are yet to be summed over).

If incoming electron and positron beams are unpolarised, we *average* over the spins states $s$ and $s'$. Conversely, muon detectors are usually blind to polarisation, so we *sum* over spin states $r$ and $r'$. Together, this leads to

$$\frac{1}{2}\sum_{s}\frac{1}{2}\sum_{s'}\sum_{r}\sum_{r'}\left|\mathcal{M}_{fi}\right|^2 = \frac{1}{4}\sum_{\text{spins}}\left|\mathcal{M}_{fi}\right|^2 \equiv \left|\mathcal{M}_{fi}\right|^2_{\Sigma}\,. \tag{3.2.4}$$

Summing over spin states of the particle spinors gives [23],

$$\sum_{s} u^s(p)\bar{u}^s(p) = \not{p} + m \quad , \quad \sum_{s} v^s(p)\bar{v}^s(p) = \not{p} - m\,, \tag{3.2.5}$$

where the slashed notation denotes a contraction between the Dirac gamma matrices, $\gamma^\mu$, and the 4-momentum, such that $\not{p} = \gamma^\mu p_\mu$.

Let's focus on $\tilde{L}^{\mu\rho}$ for now. If we write the components of $\tilde{L}^{\mu\rho}$ using spinor index notation, we can rearrange the $v$ spinor at the end to the beginning,

$$\begin{aligned} L^{\mu\rho} = \frac{1}{4}\sum_{s,s'}\tilde{L}^{\mu\rho} &= \frac{\mu^{(4-D)}e^2}{4}\sum_{s,s'} v_d^{s'}(p_2)\bar{v}_a^{s'}(p_2)\gamma^\mu_{ab}u^s_b(p_1)\bar{u}^s_c(p_1)\gamma^\rho_{cd} \\ &= \frac{\mu^{(4-D)}e^2}{4}(\not{p}_2 - m_e)_{da}\,\gamma^\mu_{ab}\,(\not{p}_1 + m_e)_{bc}\,\gamma^\rho_{cd} \\ &= \frac{\mu^{(4-D)}e^2}{4}\text{tr}\big[(\not{p}_2 - m_e)\gamma^\mu(\not{p}_1 + m_e)\gamma^\rho\big]\,, \end{aligned} \tag{3.2.6}$$

where the definition of the trace, $\text{tr}[\cdots]$, has been applied in the final line [23, 47, 94]. This trace of $D$-dimensional gamma matrices can be calculated using gamma matrix identities [23, 25, 47] or computationally. The result is

$$L^{\mu\rho} = \mu^{(4-D)}e^2\left(p_2^\mu\,p_1^\rho + p_2^\rho\,p_1^\mu - g^{\mu\rho}(p_1\cdot p_2 + m_e^2)\right)\,. \tag{3.2.7}$$

As $\tilde{H}^{\nu\sigma}$ is of the same form for this diagram, the result is similar, but for a factor of $4$ since the spins are summed over, not averaged,

$$H^{\nu\sigma} = \sum_{r,r'}\tilde{H}^{\nu\sigma} = 4\mu^{(4-D)}e^2\left(q_2^\nu\,q_1^\sigma + q_2^\sigma\,q_1^\nu - g^{\nu\sigma}(q_1\cdot q_2 + m_\mu^2)\right)\,. \tag{3.2.8}$$

Writing Eq. (3.2.3) as

$$\left|\mathcal{M}_{fi}\right|^2_{\Sigma} = \frac{1}{k^4}g_{\mu\nu}g_{\rho\sigma}L^{\mu\rho}H^{\nu\sigma} \tag{3.2.9}$$

and substituting in Eqs. (3.2.7) and (3.2.8),

$$|\mathcal{M}_{fi}|^2_\Sigma = \frac{4\mu^{2(4-D)}e^4}{k^4}\Bigg[2(p_1\cdot q_1)(p_2\cdot q_2)+2(p_1\cdot q_2)(p_2\cdot q_1)+(D-4)(p_1\cdot p_2)(q_1\cdot q_2)$$
$$+(D-2)(q_1\cdot q_2)m_e^2+(D-2)(p_1\cdot p_2)m_\mu^2+Dm_\mu^2 m_e^2\Bigg]\,, \tag{3.2.10}$$

where we have used $g_{\mu\nu}g^{\mu\nu} = D$. Taking the high-energy limit such that the masses of the electron and muon are negligible compared to their energies, we find,

$$|\mathcal{M}_{fi}|^2_\Sigma = \frac{4\mu^{2(4-D)}e^4}{k^4}\Bigg[2(p_1\cdot q_1)(p_2\cdot q_2)+2(p_1\cdot q_2)(p_2\cdot q_1)+(D-4)(p_1\cdot p_2)(q_1\cdot q_2)\Bigg]\,. \tag{3.2.11}$$

We can evaluate this expression in the centre-of-mass frame and then integrate over final-state momenta. However, this would involve explicit angular dependence, making the integration complicated in $D$ dimensions. So, before we continue, we shall introduce a simplification that will be used throughout this chapter.

### 3.2.2 Simplifying the Interactions to Decays

The calculation of a general cross section $\sigma(e^+e^- \to \gamma^* \to X)$ can be simplified to the calculation of a decay rate $\Gamma(\gamma^* \to X)$ multiplied by a factor common to all interactions with the same initial state. This will be shown in $D$ dimensions. This is useful because the expressions will be less algebraically complex, less repetitive, and not involve angular dependence (making the phase-space integral easier) [23, 25].

A cross section is given by Eq. (3.0.1),

$$\sigma(e^+e^- \to \gamma^* \to X) = \frac{1}{2s}\int \mathrm{dPS}\,|\mathcal{M}_{fi}|^2\,. \tag{3.2.12}$$

From Eq. (3.2.9), we see that

$$|\mathcal{M}_{fi}|^2_\Sigma = \frac{1}{k^4}g_{\mu\nu}g_{\rho\sigma}L^{\mu\rho}H^{\nu\sigma} = \frac{1}{k^4}L_{\nu\sigma}H^{\nu\sigma}\,. \tag{3.2.13}$$

The phase-space integral involves only the final-state trace, giving

$$\int \mathrm{dPS}\, H^{\nu\sigma}. \tag{3.2.14}$$

This will integrate over all parameters of the final state except for Lorentz invariant scalars and the ($D$-dimensional) 4-vector $k^\mu$, which hence must characterise the final state. Since[1]

$$k_\nu H^{\nu\sigma} = H^{\nu\sigma} k_\sigma = 0\,, \tag{3.2.15}$$

Eq. (3.2.14) must have the form,

$$\int \mathrm{dPS}\, H^{\nu\sigma} = \Big(g^{\nu\sigma} - \frac{k^\nu k^\sigma}{k^2}\Big) \cdot H\,, \tag{3.2.16}$$

where $H$ is a scalar. Combining Eq. (3.2.13) and Eq. (3.2.16), we find,

$$\begin{aligned} \int \mathrm{dPS}\, \left|\mathcal{M}_{fi}\right|^2_\Sigma &= \frac{1}{k^4} L_{\nu\sigma} \Big(g^{\nu\sigma} - \frac{k^\nu k^\sigma}{k^2}\Big) \cdot H \\ &= \frac{1}{k^4} L_{\nu\sigma} g^{\nu\sigma} \cdot H\,, \end{aligned} \tag{3.2.17}$$

where we have used $L_{\nu\sigma} k^\nu = 0$. Now consider separately contracting each of the traces $L_{\mu\nu}$ and $H_{\rho\sigma}$,

$$\begin{aligned} \big(g^{\mu\nu} L_{\mu\nu}\big) \cdot \int \mathrm{dPS}\, \big(g^{\rho\sigma} H_{\rho\sigma}\big) &= \big(g^{\mu\nu} L_{\mu\nu}\big) \cdot g^{\rho\sigma} \Big(g_{\rho\sigma} - \frac{k_\rho k_\sigma}{k^2}\Big) \cdot H \\ &= \big(g^{\mu\nu} L_{\mu\nu}\big) \cdot \Big(D - 1\Big) \cdot H \\ &= (D-1) \big(g^{\mu\nu} L_{\mu\nu}\big) \cdot H\,. \end{aligned} \tag{3.2.18}$$

Combining Eqs. (3.2.17) and (3.2.18) to eliminate $H$, we find,

$$\int \mathrm{dPS}\, \left|\mathcal{M}_{fi}\right|^2_\Sigma = \frac{1}{(D-1)k^4} \big(g^{\mu\nu} L_{\mu\nu}\big) \cdot \int \mathrm{dPS}\, \big(g^{\rho\sigma} H_{\rho\sigma}\big)\,. \tag{3.2.19}$$

This result states that rather than needing to calculate the total matrix element squared, one can calculate and contract the initial-state and final-state traces separately. This is very useful as the factor of $g^{\mu\nu} L_{\mu\nu}$ will be the same for all processes we consider

[1] Ref. [23] claims this is true based on general principles. This is related to fermion current conservation, *not* the Ward identity (since the photon is not an external particle).

(since they all involve the same initial-state particles),

$$\begin{aligned} g^{\mu\nu} L_{\mu\nu} &= \mu^{(4-D)} e^2 g^{\mu\nu} \left( p_{2\mu} p_{1\nu} + p_{2\nu} p_{1\mu} - g_{\mu\nu}(p_1 \cdot p_2 + m_e^2) \right) \\ &= \frac{(2-D)}{2} \mu^{(4-D)} e^2 s \,, \end{aligned} \tag{3.2.20}$$

where $s = 2p_1 \cdot p_2$ and the masses have been taken to zero. This means that we can factor $g^{\mu\nu} L_{\mu\nu}$ out of all diagrams, and it will cancel when we calculate the ratio of cross sections with the same initial particles. Thus, we need only compute the trace of the final state particles. This is equivalent to considering a decay from a virtual photon, $\gamma^*$, to the same final particles (although this interaction is unphysical). This

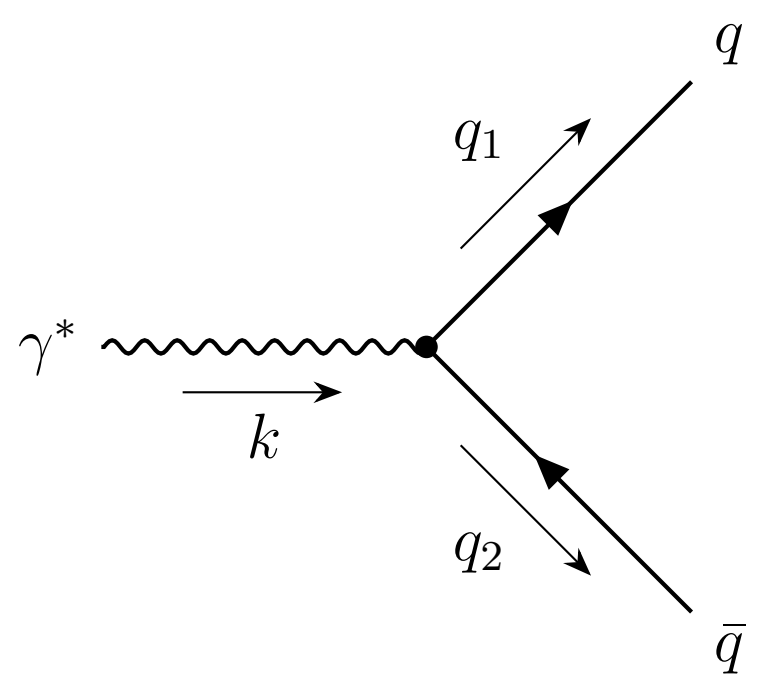

**Fig. 3.5.** The "decay" of a virtual photon to a quark-antiquark pair.

interaction (pictured in Fig. 3.5) has the decay rate [23, 25, 54]

$$\Gamma(\gamma^* \to X) = \frac{1}{2\sqrt{s}} \int \mathrm{dPS} \left| \mathcal{M}_{fi,\gamma^*} \right|^2 , \tag{3.2.21}$$

with the squared matrix element given by (using the Feynman rules in Section 2.5)

$$\left| \mathcal{M}_{fi,\gamma^*} \right|^2 = \epsilon_\rho \epsilon_\sigma^* \left[ \mu^{(4-D)} e^2 \bar{u}^r(q_1)(\gamma^\rho) v^{r'}(q_2) \bar{v}^{r'}(q_2)(\gamma^\sigma) u^r(q_1) \right] = \epsilon_\rho \epsilon_\sigma^* \tilde{H}^{\rho\sigma} \,. \tag{3.2.22}$$

To relate this to $g^{\rho\sigma} H_{\rho\sigma}$, we can use the relation for summing over photon polarisations [23],

$$\sum_\lambda \epsilon_\rho(\lambda) \epsilon_\sigma^*(\lambda) = - \left( g_{\rho\sigma} + (\eta - 1) \frac{k_\rho k_\sigma}{k^2} \right) \,, \tag{3.2.23}$$

where $\lambda = \pm 1$ is the helicity of the photon, $k_\rho$ is the 4-momentum of the incoming virtual photon, and $\eta$ is an arbitrary gauge parameter ($\eta = 1$ is the Feynman gauge, $\eta = 0$ is the Landau gauge [95]). We choose the Feynman gauge for consistency with Eq. (3.2.2), but any choice of gauge will give the following result due to Eq. (3.2.15). Once we sum over photon helicities (as well as the spins of the final-state particles), we get,

$$\left|\mathcal{M}_{fi,\gamma^*}\right|^2_{\Sigma} = -g_{\rho\sigma}H^{\rho\sigma} = -g^{\rho\sigma}H_{\rho\sigma}\,, \tag{3.2.24}$$

and hence, from Eq. (3.2.21),

$$\Gamma(\gamma^* \to X) = -\frac{1}{2\sqrt{s}}\int \mathrm{dPS}\, g^{\rho\sigma}H_{\rho\sigma}\,. \tag{3.2.25}$$

Combining Eqs. (3.0.1), (3.2.19), (3.2.20), and (3.2.25),

$$\sigma(e^+e^- \to \gamma^* \to X) = \frac{\mu^{(4-D)}e^2}{4s^2}\frac{(2-D)}{(D-1)}\int \mathrm{dPS}\, g^{\rho\sigma}H_{\rho\sigma} \tag{3.2.26}$$

$$= \frac{\mu^{(4-D)}e^2}{2s^{3/2}}\frac{(D-2)}{(D-1)}\Gamma(\gamma^* \to X)\,. \tag{3.2.27}$$

These equations allow us to calculate the relevant cross sections in this chapter via simpler decay rates.

### 3.2.3 Integrating over the Two-Particle Phase Space

From Eq. (3.2.26) we see that, for muon-antimuon or quark-antiquark production, we need to calculate

$$\int \mathrm{dPS}_2\, g^{\rho\sigma}H_{\rho\sigma(\mathrm{tree})}\,, \tag{3.2.28}$$

where $\mathrm{dPS}_2$ indicates the two-particle phase space and $H_{\rho\sigma(\mathrm{tree})}$ now denotes the tree-level final-state trace. Using Eq. (3.2.8),

$$g^{\rho\sigma}H_{\rho\sigma(\mathrm{tree})} = 4\mu^{(4-D)}e^2(2-D)\,(q_1\cdot q_2) = 2(2-D)\mu^{(4-D)}e^2 s. \tag{3.2.29}$$

Returning to Eq. (3.2.26), we find,

$$\sigma_0(e^+e^- \to \gamma^* \to \mu^+\mu^-) = \frac{\mu^{2(4-D)}e^4}{2s}\frac{(2-D)^2}{(D-1)}\int \mathrm{dPS}_2\,. \tag{3.2.30}$$

With no angular dependence, the phase space integral is no longer too complicated. For two massless particles, the integrated two-particle phase space in $D$ dimensions is commonly expressed in one of two equivalent ways [25, 89],

$$\int \mathrm{dPS}_2 = \left(\frac{s}{4\pi}\right)^{\frac{D-4}{2}}\frac{2^{-D}}{\sqrt{\pi}\,\Gamma((D-1)/2)} = \frac{1}{8\pi}\left(\frac{s}{4\pi}\right)^{\frac{D-4}{2}}\frac{\Gamma(D/2-1)}{\Gamma(D-2)}\,, \tag{3.2.31}$$

where $\Gamma(z)$ is the Gamma function defined by Eq. (2.6.15). Thus,

$$\sigma_0(e^+e^- \to \mu^+\mu^-) = \frac{\mu^{2(4-D)}e^4}{16\pi s}\left(\frac{s}{4\pi}\right)^{\frac{D-4}{2}}\frac{(D-2)^2}{(D-1)}\frac{\Gamma(D/2-1)}{\Gamma(D-2)}\,. \qquad (3.2.32)$$

This is a $D$-dimensional expression for the tree-level contribution to this interaction. Although this is finite for $D = 4$, the $D$-dimensional corrections will be required when the tree-level diagram interferes with one-loop diagrams. Taking $D = 4$,

$$\sigma_0(e^+e^- \to \mu^+\mu^-) = \frac{e^4}{12\pi s} = \frac{4\pi\alpha_{\text{em}}^2}{3s}\,, \qquad (3.2.33)$$

where the fine structure constant, $\alpha_{\text{em}}$, is defined as,

$$\alpha_{\text{em}} = \frac{e^2}{4\pi}\,. \qquad (3.2.34)$$

From here on out, it is understood that the mediating particle in all processes in this chapter is a virtual photon, so it is not written explicitly.

## 3.3 Tree level $e^+e^- \to q\bar{q}$ cross section

The cross section for quark-antiquark production, $\sigma(e^+e^- \to q\bar{q})$, has a tree-level contribution, $\sigma_0(e^+e^- \to q\bar{q})$, where the corresponding Feynman diagram is shown in Fig. 3.1. We can calculate $\sigma_0(e^+e^- \to q\bar{q})$ in a very similar way to $\sigma_0(e^+e^- \to \mu^+\mu^-)$. One will arrive at exactly the same answer as Eq. (3.2.33), except with an additional factor of $e_q^2$, the squared ratio between quark charge and muon (electron) charge. There is also a colour factor, $N_c$, which arises from summing over the free colour indices in the vertex factors, i.e., $\delta_{ij}\delta_{ji} = N_c$ (see Eq. (2.5.16)),

$$\sigma_{\text{tree-level}} \equiv \sigma_0 = \left(N_c e_q^2\right)\frac{4\pi\alpha^2}{3E_{cm}^2} \qquad (3.3.1)$$

where $N_c$ is the number of colours. Calculating the ratio of Eq. (3.3.1) and Eq. (3.2.33),

$$R_{q(0)} \equiv \frac{\sigma_0(e^+e^- \to q\bar{q})}{\sigma_0(e^+e^- \to \mu^+\mu^-)} = N_c e_q^2. \qquad (3.3.2)$$

Summing over all quarks with mass satisfying $4m_q^2 < s$ and assuming all quarks hadronise with unit probability, one arrives at the well-known tree-level result,

$$R_0 \equiv \frac{\sigma_0(e^+e^- \to \text{hadrons})}{\sigma_0(e^+e^- \to \mu^+\mu^-)} = N_c \sum_q e_q^2. \tag{3.3.3}$$

## 3.4 First Order QCD Corrections to $e^+e^- \to q\bar{q}\,X$

In this section, we calculate the ratio of the cross sections of $e^+e^- \to q\bar{q}\,X$ and $e^+e^- \to \mu^+\mu^-$ to first order in QCD corrections, which is denoted,

$$R_q \equiv \frac{\sigma(e^+e^- \to q\bar{q}\,X)}{\sigma(e^+e^- \to \mu^+\mu^-)}. \tag{3.4.1}$$

These first-order QCD corrections involve gluons coupling to the quark and anti-quark. The interactions involve no weak bosons, and we take the incoming and outgoing particles to be massless. We initially ignore colour factors and indices, but consider their effects at the end of the calculation.

### 3.4.1 Vertex Correction

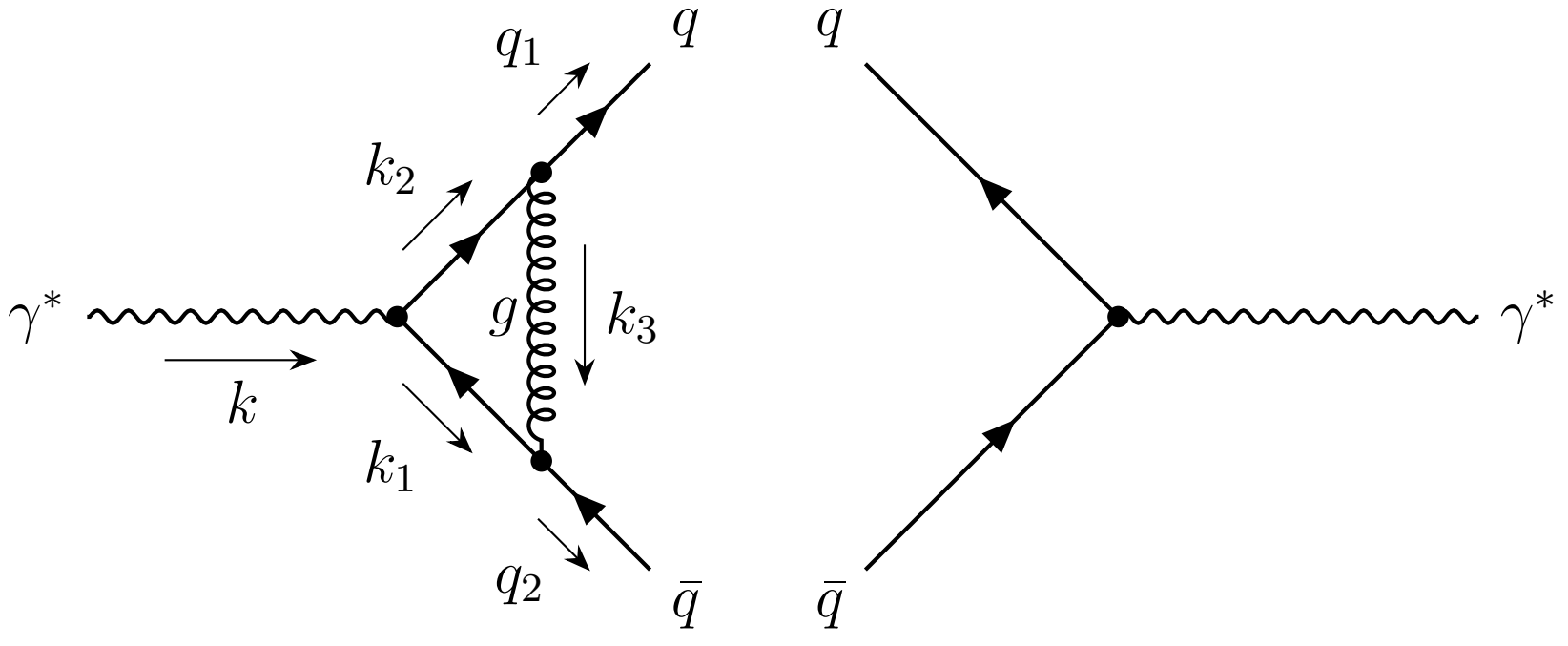

**Fig. 3.6.** Feynman diagrams representing the product of the virtual vertex correction decay term with the conjugate of the tree-level diagram. This contribution is $\mathcal{O}\left(e^2g^2\right)$. The trace follows the fermion lines around the loop formed by combining the amplitude with the complex amplitude.

As in Eq. (3.1.5), the $\mathcal{O}(e^4g^2)$ contribution involving the vertex correction comes from the cross term of the vertex correction amplitude interfering with the tree-level amplitude, shown diagrammatically in Fig. 3.6. Taking the trace around the fermion loop and using Eq. (3.2.19), we find

$$\int \text{dPS}\left(\mathcal{M}_{fi}^{\text{vertex}}\mathcal{M}_{fi}^{\text{tree}\,\dagger}\right)_\Sigma = \frac{1}{(D-1)k^4}\left(g^{\mu\nu}L_{\mu\nu}\right)\int \text{dPS}\left(g^{\rho\sigma}H_{\rho\sigma}^{(\text{vertex})}\right). \tag{3.4.2}$$

where $H^{\rho\sigma}_{\text{(vertex)}}$, which is the only thing we need to calculate, is given by,

$$\begin{aligned} H^{\rho\sigma}_{\text{(vertex)}} =& -2ie_q^2e^2g^2\mu^{2(4-D)}\int\frac{\mathrm{d}^Dk_3}{(2\pi)^D}\left[\frac{1}{k_3^2+i\epsilon}\left(g_{\alpha\beta}+(\xi-1)\frac{k_{3_\alpha}k_{3_\beta}}{k_3^2+i\epsilon}\right)\right] \\ &\times\left[\frac{1}{\left(k_2^2-m_q^2+i\epsilon\right)\left(k_1^2-m_q^2+i\epsilon\right)}\right] \\ &\times\,\text{tr}\left[(\not{q}_1+m_q)\gamma^\alpha(\not{k}_2+m_q)\gamma^\sigma(\not{k}_1+m_q)\gamma^\beta(\not{q}_2-m_q)\gamma^\rho\right] \end{aligned} \tag{3.4.3}$$

where $g$ is the strong coupling constant, $e_qe$ is the coupling for the photon to the quark/antiquark, $\xi$ is the gluon gauge parameter, and a factor of 2 is included due to an identical contribution from the conjugate diagram (as in Eq. (3.1.5)). Since we are evaluating the integral in $D$ dimensions, the strong coupling constants have been rescaled by explicitly writing an extra factor of $\mu^{\frac{4-D}{2}}$ per coupling, in order to maintain their usual dimensions. We can use conservation of momentum to make the following substitutions,

$$k_2^\mu = q_1^\mu + k_3^\mu \quad , \quad k_1^\mu = q_2^\mu - k_3^\mu \quad , \quad k^\mu = q_1^\mu + q_2^\mu . \tag{3.4.4}$$

For simplicity, we will work in the massless limit ($m_e = m_q = 0$), hence,

$$p_1^2 = p_2^2 = q_1^2 = q_2^2 = 0\,, \tag{3.4.5}$$

$$k^2 = 2q_1\cdot q_2 \equiv s\,. \tag{3.4.6}$$

Calculating the trace computationally, contracting with $g_{\rho\sigma}$, and adopting the Feynman gauge ($\xi = 1$), we get

$$\begin{aligned} g_{\rho\sigma}H^{\rho\sigma}_{\text{(vertex)}} =& -2ie_q^2e^2g^2\mu^{2(4-D)}\int\frac{\mathrm{d}^Dk_3}{(2\pi)^D}(D-2) \\ &\times\left[\frac{1}{(k_3^2+i\epsilon)\left((q_1+k_3)^2+i\epsilon\right)\left((q_2-k_3)^2+i\epsilon\right)}\right] \\ &\times\Big[16(q_1\cdot q_2)(q_1\cdot k_3)-16(q_1\cdot q_2)(q_2\cdot k_3)-16(q_1\cdot q_2)^2 \\ &\qquad +16(q_1\cdot k_3)(q_2\cdot k_3)+4(D-4)(q_1\cdot q_2)k_3^2\Big] \\ =& -2ie_q^2e^2g^2\mu^{2(4-D)}\int\frac{\mathrm{d}^Dk_3}{(2\pi)^D}(D-2)\times \\ &\left[\frac{8s(q_1\cdot k_3)-8s(q_2\cdot k_3)-4s^2+16(q_1\cdot k_3)(q_2\cdot k_3)+2(D-4)sk_3^2}{(k_3^2+i\epsilon)\left((q_1+k_3)^2+i\epsilon\right)\left((q_2-k_3)^2+i\epsilon\right)}\right]. \end{aligned} \tag{3.4.7}$$

We can write this using Eqs. (2.6.42),

$$g_{\rho\sigma}H^{\rho\sigma}_{\text{(vertex)}} = -\,2ie_q^2e^2g^2\mu^{2(4-D)}(D-2)\times \\ \left[8s(q_{1\mu}-q_{2\mu})C^{\mu}_{2,\text{IR}} - 4s^2C_{1,\text{IR}} + 16q_{1\mu}q_{2\mu}C^{\mu\nu}_{4,\text{IR}} + 2s(D-4)C_{3,\text{IR}}\right] . \tag{3.4.8}$$

Now we substitute in the final expressions for the $C$ integrals,

$$\begin{aligned} g_{\rho\sigma}H^{\rho\sigma}_{\text{(vertex)}} =& -\,2ie_q^2e^2g^2\mu^{2(4-D)}(D-2)\times \\ &\Bigg[8s(2q_1\cdot q_2)\frac{i}{(4\pi)^2}\frac{1}{2q_1\cdot q_2}\left(\frac{-2q_1\cdot q_2}{4\pi}\right)^{\varepsilon_{\text{IR}}}\frac{\Gamma(1-\varepsilon_{\text{IR}})\Gamma(1+\varepsilon_{\text{IR}})}{\Gamma(2+2\varepsilon_{\text{IR}})}\Gamma(\varepsilon_{\text{IR}}) \\ &- 4s^2\frac{i}{(4\pi)^2}\frac{1}{2q_1\cdot q_2}\left(\frac{-2q_1\cdot q_2}{4\pi}\right)^{\varepsilon_{\text{IR}}}\frac{\Gamma(1-\varepsilon_{\text{IR}})}{\Gamma(1+2\varepsilon_{\text{IR}})}\Gamma^2(\varepsilon_{\text{IR}}) \\ &+ 4(q_1\cdot q_2)\frac{-i}{(4\pi)^2}(2q_1\cdot q_2)^{\varepsilon_{\text{UV}}}\left[\frac{1}{\hat{\varepsilon}_{\text{UV}}} - 3 - i\pi\right] \\ &- 16\frac{i}{(4\pi)^2}\frac{1}{2q_1\cdot q_2}\left(\frac{-2q_1\cdot q_2}{4\pi}\right)^{\varepsilon_{\text{IR}}}\frac{\Gamma(1-\varepsilon_{\text{IR}})}{\Gamma(3+2\varepsilon_{\text{IR}})}(q_1\cdot q_2)^2\Gamma^2(1+\varepsilon_{\text{IR}}) \\ &+ 4s\varepsilon_{\text{UV}}\frac{-i}{(4\pi)^2}(2q_1\cdot q_2)^{\varepsilon_{\text{UV}}}\left[\frac{1}{\hat{\varepsilon}_{\text{UV}}} - 2 - i\pi\right]\Bigg] \\ =& -\,2ie_q^2e^2g^2\mu^{2(4-D)}(D-2)\Bigg[8s\frac{i}{(4\pi)^2}\left(\frac{-s}{4\pi}\right)^{\varepsilon_{\text{IR}}}\frac{\Gamma(1-\varepsilon_{\text{IR}})\Gamma(1+\varepsilon_{\text{IR}})}{\Gamma(2+2\varepsilon_{\text{IR}})}\Gamma(\varepsilon_{\text{IR}}) \\ &- 4s\frac{i}{(4\pi)^2}\left(\frac{-s}{4\pi}\right)^{\varepsilon_{\text{IR}}}\frac{\Gamma(1-\varepsilon_{\text{IR}})}{\Gamma(1+2\varepsilon_{\text{IR}})}\Gamma^2(\varepsilon_{\text{IR}}) \\ &+ 2s\frac{-i}{(4\pi)^2}s^{\varepsilon_{\text{UV}}}\left[\frac{1}{\hat{\varepsilon}_{\text{UV}}} - 3 - i\pi\right] \\ &- 4s\frac{i}{(4\pi)^2}\left(\frac{-s}{4\pi}\right)^{\varepsilon_{\text{IR}}}\frac{\Gamma(1-\varepsilon_{\text{IR}})}{\Gamma(3+2\varepsilon_{\text{IR}})}\Gamma^2(1+\varepsilon_{\text{IR}}) \\ &+ 4s\varepsilon_{\text{UV}}\frac{-i}{(4\pi)^2}s^{\varepsilon_{\text{UV}}}\left[\frac{1}{\hat{\varepsilon}_{\text{UV}}} - 2 - i\pi\right]\Bigg] \\ =& -\,8ise_q^2e^2g^2\Bigg[2\frac{i}{(4\pi)^2}(-1)^{\varepsilon_{\text{IR}}}\left(\frac{s}{\mu^4}\right)^{\varepsilon_{\text{IR}}}\left(\frac{1}{4\pi}\right)^{\varepsilon_{\text{IR}}}(1+\varepsilon_{\text{IR}})\Gamma(1-\varepsilon_{\text{IR}}) \\ &\left(2\frac{\Gamma(1+\varepsilon_{\text{IR}})\Gamma(\varepsilon_{\text{IR}})}{\Gamma(2+2\varepsilon_{\text{IR}})} - \frac{\Gamma^2(\varepsilon_{\text{IR}})}{\Gamma(1+2\varepsilon_{\text{IR}})} - \frac{\Gamma^2(1+\varepsilon_{\text{IR}})}{\Gamma(3+2\varepsilon_{\text{IR}})}\right) \\ &+ \frac{-i}{(4\pi)^2}\left(\frac{s}{\mu^4}\right)^{\varepsilon_{\text{UV}}}\left(\frac{1}{4\pi}\right)^{\varepsilon_{\text{UV}}}\left[\frac{1}{\varepsilon_{\text{UV}}} + \gamma_E - i\pi\right] + \mathcal{O}(\varepsilon_{\text{UV}})\Bigg] , \end{aligned} \tag{3.4.9}$$

where, in the final equation, $D$ has been taken to $4 + 2\varepsilon_{\text{IR}}$ or $4 + 2\varepsilon_{\text{UV}}$ for the IR-divergent or UV-divergent terms, respectively. Since Eq. (3.1.5) involves the real part of the vertex-tree cross term, we keep only the real parts of $(-1)^{\varepsilon_{\text{IR}}}$ and exclude the $i\pi$ term (in other words, the $i\pi$ will cancel between $\mathcal{M}_{fi}^{\text{vertex}}\mathcal{M}_{fi}^{\text{tree}\dagger}$ and $\mathcal{M}_{fi}^{\text{tree}}\mathcal{M}_{fi}^{\text{vertex}\dagger}$). Thus,

$$
\begin{aligned}
g_{\rho\sigma}H^{\rho\sigma}_{\text{(vertex)}} = &-\frac{16se_q^2e^2g^2}{(4\pi)^2}\left(1-\frac{\pi^2{\varepsilon_{\text{IR}}}^2}{2}+\mathcal{O}\left({\varepsilon_{\text{IR}}}^4\right)\right)\left(\frac{s}{\mu^4}\right)^{\varepsilon_{\text{IR}}}\left(\frac{1}{4\pi}\right)^{\varepsilon_{\text{IR}}}(1+\varepsilon_{\text{IR}})\Gamma(1-\varepsilon_{\text{IR}}) \\
&\left(2\frac{\Gamma(1+\varepsilon_{\text{IR}})\Gamma(\varepsilon_{\text{IR}})}{\Gamma(2+2\varepsilon_{\text{IR}})}-\frac{\Gamma^2(\varepsilon_{\text{IR}})}{\Gamma(1+2\varepsilon_{\text{IR}})}-\frac{\Gamma^2(1+\varepsilon_{\text{IR}})}{\Gamma(3+2\varepsilon_{\text{IR}})}\right) \\
&+\frac{8se_q^2e^2g^2}{(4\pi)^2}\left(\frac{s}{\mu^4}\right)^{\varepsilon_{\text{UV}}}\left(\frac{1}{4\pi}\right)^{\varepsilon_{\text{UV}}}\left[\frac{1}{\varepsilon_{\text{UV}}}+\gamma_E\right].
\end{aligned}
\tag{3.4.10}
$$

Before integrating over the two-particle phase space, we must account for the renormalisation factors arising from the residue of the fermion two-point function, as the LSZ reduction formula demands.

## 3.4.2 LSZ Correction

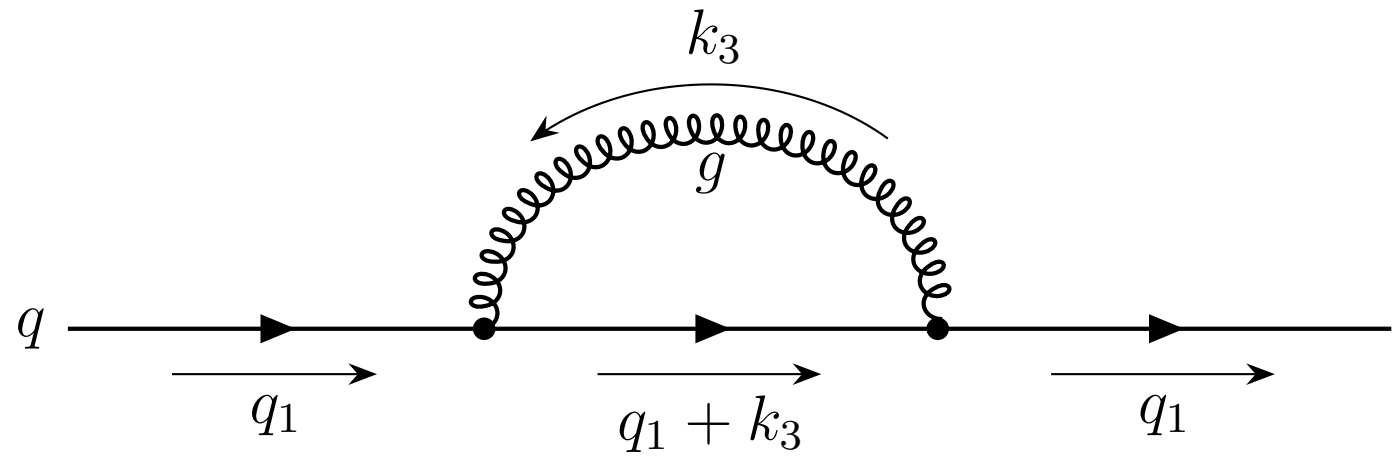

**Fig. 3.7.** Feynman diagram representing the self-energy correction of a fermion. This is the $\mathcal{O}(g^2)$ contribution to the two-point correlation function, which we denote as $\Sigma_g$.

We now consider the 'correction term' arising from the LSZ reduction formula, which is the second term in Eq. (3.1.5). This is just $2\,\delta Z$ multiplied by the tree-level contribution. The field strength renormalisation constant, $Z = 1 + \delta Z$, is defined as the $\mathcal{O}(g^2)$ part of the residue of the single-particle pole in the two-point function of fields [23, 25, 89] (see Section 2.4). Therefore, to calculate $Z$, we must calculate the first-order (in $g^2$) contribution to the two-point correlation function, $\Sigma_g$. This contribution is shown in Fig. 3.7 and, using Feynman rules, is given by

$$
\begin{aligned}
i\Sigma_g(\not{q}_1) = -g^2\mu^{4-D}\int\frac{\mathrm{d}^Dk_3}{(2\pi)^D}\gamma^\alpha\frac{\not{q}_1+\not{k}_3+m_q}{(q_1+k_3)^2-m_q^2+i\epsilon}\gamma^\beta \\
\times\left[\frac{1}{k_3^2+i\epsilon}\left(g_{\alpha\beta}+(\xi-1)\frac{k_{3_\alpha}k_{3_\beta}}{k_3^2+i\epsilon}\right)\right].
\end{aligned}
\tag{3.4.11}
$$

Taking the massless limit and adopting the Feynman gauge ($\xi = 1$), this expression

simplifies to

$$i\Sigma_g(\not{q}_1) = -g^2\mu^{4-D}\int \frac{\mathrm{d}^D k_3}{(2\pi)^D}\gamma^\alpha \frac{\not{q}_1 + \not{k}_3}{(q_1+k_3)^2 + i\epsilon}\gamma_\alpha \left[\frac{1}{k_3^2 + i\epsilon}\right] . \tag{3.4.12}$$

There is no trace to evaluate, since we are not multiplying by a conjugate amplitude. Instead, we use the $D$ dimensional gamma matrix identity [25, 89], $\gamma^\mu\gamma^\nu\gamma_\mu = (2-D)\gamma^\nu$, such that

$$i\Sigma_g(\not{q}_1) = -g^2\mu^{4-D}\int \frac{\mathrm{d}^D k_3}{(2\pi)^D}\frac{(2-D)(\not{q}_1 + \not{k}_3)}{\big((q_1+k_3)^2 + i\epsilon\big)\big(k_3^2 + i\epsilon\big)} . \tag{3.4.13}$$

With Feynman parameterisation (Section 2.6.3), we can write this as

$$i\Sigma_g(\not{q}_1) = -g^2\mu^{4-D}\int \frac{\mathrm{d}^D k_3}{(2\pi)^D}\int_0^1 \mathrm{d}x\, \frac{(2-D)(\not{q}_1 + \not{k}_3)}{[(k_3 + xq_1)^2 - a^2 + i\epsilon]^2} , \tag{3.4.14}$$

where $a^2 \equiv -q_1^2 x(1-x)$. Shifting the integration variable $k_3 \to k_3 - xq_1$, and remembering that the term with an odd power of $k_3$ in the numerator will vanish [89], we find

$$\begin{aligned} i\Sigma_g(\not{q}_1) &= -g^2\mu^{4-D}(2-D)\not{q}_1\int_0^1 \mathrm{d}x\,(1-x)J(D,0,2,a^2) \\ &= -g^2\mu^{4-D}(2-D)\not{q}_1\frac{i}{(4\pi)^{D/2}}\Gamma(2-D/2)(-q_1^2)^{D/2-2}\frac{\Gamma(D/2)\Gamma(D/2-1)}{\Gamma(D-1)} , \end{aligned} \tag{3.4.15}$$

where we have used Eq. (2.6.23) then evaluated the $x$ integral using the Euler Beta function (Eq. (2.6.22)). If one were to take $q_1^2 = 0$ and $D = 4$ at this point, there would be an indetermination of the form $0^0$. In order to regularise this, we use

$$\begin{aligned} \int_{-q_1^2}^{\infty}\frac{\mathrm{d}x}{x}x^n &= (-q_1^2)^n\left(-\frac{1}{n}\right) \quad \text{for} \quad \mathrm{Re}\{n\} < 0 , \\ \int_{-q_1^2}^{\infty}\frac{\mathrm{d}x}{x}x^{-n} &= (-q_1^2)^{-n}\left(\frac{1}{n}\right) \quad \text{for} \quad \mathrm{Re}\{n\} > 0 , \end{aligned} \tag{3.4.16}$$

and therefore

$$\begin{aligned} (-q_1^2)^{D/2-2}(2-D/2)^{-1} &= \int_{-q_1^2}^{\infty}\frac{\mathrm{d}x}{x}x^{D/2-2} \\ &= \left[\int_{-q_1^2}^{s}\frac{\mathrm{d}x}{x}x^{D/2-2} + \int_s^\infty \frac{\mathrm{d}x}{x}x^{D/2-2}\right] , \end{aligned} \tag{3.4.17}$$

where we have cut the total integration region at $s$, the characteristic energy scale of the process, in order to to separately integrate in the IR and UV regions of momentum. When $q_1^2 \to 0$, the first integral is only convergent if $D/2 - 2 > 0$, so we take $D = 4 + 2\varepsilon_{\text{IR}}$ since $\varepsilon_{\text{IR}} > 0$. Similarly, the second integral is only convergent when $D/2 - 2 < 0$, so we take $D = 4 + 2\varepsilon_{\text{UV}}$ since $\varepsilon_{\text{UV}} < 0$. The result is then

$$(-q_1^2)^{D/2-2}(2 - D/2)^{-1} = \frac{s^{\varepsilon_{\text{IR}}}}{\varepsilon_{\text{IR}}} - \frac{s^{\varepsilon_{\text{UV}}}}{\varepsilon_{\text{UV}}}\,. \tag{3.4.18}$$

Returning to Eq. (3.4.15) and taking $D = 4 + 2\varepsilon_{\text{IR}}$ or $D = 4 + 2\varepsilon_{\text{UV}}$ as required, we get,

$$\begin{aligned}
i\Sigma_g(\not{q}_1) &= -g^2\mu^{4-D}(2-D)\not{q}_1\frac{i}{(4\pi)^{D/2}}\Gamma(3-D/2)\frac{\Gamma(D/2)\Gamma(D/2-1)}{\Gamma(D-1)} \\
&\qquad\qquad \times\left[\left.\frac{2s^{(D-4)/2}}{D-4}\right|_{D=4+2\varepsilon_{\text{IR}}} - \left.\frac{2s^{(D-4)/2}}{D-4}\right|_{D=4+2\varepsilon_{\text{UV}}}\right] \\
&= \frac{2ig^2\not{q}_1}{(4\pi)^2} \\
&\quad\times\Bigg((1+\varepsilon_{\text{IR}})\left(\frac{s}{\mu^2}\right)^{\varepsilon_{\text{IR}}}(4\pi)^{-\varepsilon_{\text{IR}}}\Gamma(1-\varepsilon_{\text{IR}})\frac{\Gamma(2+\varepsilon_{\text{IR}})\Gamma(1+\varepsilon_{\text{IR}})}{\Gamma(3+2\varepsilon_{\text{IR}})}\left[\frac{1}{\varepsilon_{\text{IR}}}\right] \\
&\quad-(1+\varepsilon_{\text{UV}})\left(\frac{s}{\mu^2}\right)^{\varepsilon_{\text{UV}}}(4\pi)^{-\varepsilon_{\text{UV}}}\Gamma(1-\varepsilon_{\text{UV}})\frac{\Gamma(2+\varepsilon_{\text{UV}})\Gamma(1+\varepsilon_{\text{UV}})}{\Gamma(3+2\varepsilon_{\text{UV}})}\left[\frac{1}{\varepsilon_{\text{UV}}}\right]\Bigg) \\
&= \frac{ig^2\not{q}_1}{(4\pi)^2}\Bigg(\left(\frac{s}{\mu^2}\right)^{\varepsilon_{\text{IR}}}\left(\frac{1}{4\pi}\right)^{\varepsilon_{\text{IR}}}(1+(\gamma_E-1)\varepsilon_{\text{IR}}+\ldots)\left[\frac{1}{\varepsilon_{\text{IR}}}\right] \\
&\quad-\left(\frac{s}{\mu^2}\right)^{\varepsilon_{\text{UV}}}\left(\frac{1}{4\pi}\right)^{\varepsilon_{\text{UV}}}(1+(\gamma_E-1)\varepsilon_{\text{UV}}+\ldots)\left[\frac{1}{\varepsilon_{\text{UV}}}\right]\Bigg) \\
&= \frac{ig^2\not{q}_1}{(4\pi)^2}\Bigg(\left(\frac{s}{4\pi\mu^2}\right)^{\varepsilon_{\text{IR}}}\left[\frac{1}{\varepsilon_{\text{IR}}}+\gamma_E-1+\ldots\right]-\left(\frac{s}{4\pi\mu^2}\right)^{\varepsilon_{\text{UV}}}\left[\frac{1}{\varepsilon_{\text{UV}}}+\gamma_E-1+\ldots\right]\Bigg)\,,
\end{aligned} \tag{3.4.19}$$

where the iterative property of the Gamma function has been used to write $(2 - D/2)\Gamma(2 - D/2) = \Gamma(3 - D/2)$. This expression exhibits both UV and IR divergences.

It can be shown that the residue of this self-energy term is given by [23, 25]

$$Z = \left(1 + \left.\frac{\mathrm{d}\Sigma_g}{\mathrm{d}\not{k}_1}\right|_{\not{k}_1=0}\right)^{-1} \approx 1 - \left.\frac{\mathrm{d}\Sigma_g}{\mathrm{d}\not{k}_1}\right|_{\not{k}_1=0}\,. \tag{3.4.20}$$

Thus,

$$\begin{aligned}\delta Z \equiv Z - 1 &= -\frac{\mathrm{d}\Sigma_g}{\mathrm{d}\not{q}_1}\\ &= -\frac{g^2}{(4\pi)^2}\left(\left(\frac{s}{4\pi\mu^2}\right)^{\varepsilon_{\text{IR}}}\left[\frac{1}{\varepsilon_{\text{IR}}} + \gamma_E - 1 + \ldots\right] - \left(\frac{s}{4\pi\mu^2}\right)^{\varepsilon_{\text{UV}}}\left[\frac{1}{\varepsilon_{\text{UV}}} + \gamma_E - 1 + \ldots\right]\right).\end{aligned} \tag{3.4.21}$$

We can now calculate the full LSZ correction term using the tree-level hadronic trace (Eq. (3.2.29) multiplied by $e_q^2$),

$$\begin{aligned}2\,\delta Z\, g^{\rho\sigma} H^{(\text{tree})}_{\rho\sigma} &= -\frac{4(2-D)e_q^2 e^2 g^2 \mu^{4-D} s}{(4\pi)^2}\left(\left(\frac{s}{4\pi\mu^2}\right)^{\varepsilon_{\text{IR}}}\left[\frac{1}{\varepsilon_{\text{IR}}} + \gamma_E - 1 + \ldots\right]\right.\\ &\qquad\left. - \left(\frac{s}{4\pi\mu^2}\right)^{\varepsilon_{\text{UV}}}\left[\frac{1}{\varepsilon_{\text{UV}}} + \gamma_E - 1 + \ldots\right]\right)\\ &= \frac{8e^2 g^2 s}{(4\pi)^2}\left(\left(\frac{s}{4\pi\mu^4}\right)^{\varepsilon_{\text{IR}}}(1+\varepsilon_{\text{IR}})\left[\frac{1}{\varepsilon_{\text{IR}}} + \gamma_E - 1 + \ldots\right]\right.\\ &\qquad\left. - \left(\frac{s}{4\pi\mu^4}\right)^{\varepsilon_{\text{UV}}}(1+\varepsilon_{\text{UV}})\left[\frac{1}{\varepsilon_{\text{UV}}} + \gamma_E - 1 + \ldots\right]\right)\\ &= \frac{8e^2 g^2 s}{(4\pi)^2}\left(\left(\frac{s}{4\pi\mu^4}\right)^{\varepsilon_{\text{IR}}}\left[\frac{1}{\varepsilon_{\text{IR}}} + \gamma_E + \ldots\right] - \left(\frac{s}{4\pi\mu^4}\right)^{\varepsilon_{\text{UV}}}\left[\frac{1}{\varepsilon_{\text{UV}}} + \gamma_E + \ldots\right]\right).\end{aligned} \tag{3.4.22}$$

We can define the hadronic trace first-order in virtual corrections as the sum of this LSZ term and the vertex correction in Eq. (3.4.10),

$$\begin{aligned}g^{\rho\sigma} H^{(\text{virtual})}_{\rho\sigma} &\equiv g^{\rho\sigma} H^{(\text{vertex})}_{\rho\sigma} + 2\,\delta R\, g^{\rho\sigma} H^{(\text{tree})}_{\rho\sigma}\\ &= -\frac{16 s e_q^2 e^2 g^2}{(4\pi)^2}\left(1 - \frac{\pi^2 {\varepsilon_{\text{IR}}}^2}{2} + \mathcal{O}\big({\varepsilon_{\text{IR}}}^4\big)\right)\left(\frac{s}{4\pi\mu^4}\right)^{\varepsilon_{\text{IR}}}(1+\varepsilon_{\text{IR}})\Gamma(1-\varepsilon_{\text{IR}})\\ &\left(2\frac{\Gamma(1+\varepsilon_{\text{IR}})\Gamma(\varepsilon_{\text{IR}})}{\Gamma(2+2\varepsilon_{\text{IR}})} - \frac{\Gamma^2(\varepsilon_{\text{IR}})}{\Gamma(1+2\varepsilon_{\text{IR}})} - \frac{\Gamma^2(1+\varepsilon_{\text{IR}})}{\Gamma(3+2\varepsilon_{\text{IR}})}\right)\\ &+ \frac{8 s e_q^2 e^2 g^2}{(4\pi)^2}\left(\frac{s}{4\pi\mu^4}\right)^{\varepsilon_{\text{IR}}}\left[\frac{1}{\varepsilon_{\text{IR}}} + \gamma_E + \ldots\right],\end{aligned} \tag{3.4.23}$$

where all UV divergent terms have cancelled.

This cancellation of UV-divergent terms might initially seem like a fortunate coincidence. However, it can be shown that, to all orders of perturbation theory, the field-strength renormalisation factor, $Z$ (often denoted $Z_2$), exactly cancels the vertex correction, which is captured by another renormalisation constant, $Z_1$. More succinctly, $Z_1 = Z_2$ [23, 25, 54]. This is a consequence of the Slavnov-Taylor identities [72, 73, 96], which are the non-Abelian generalisations of the Ward-Takahashi identity [68, 69].

### 3.4.3 Integrating over the Two-Particle Phase Space

The next step is to integrate over the two-particle phase-space, which (for an integrand which is not dependent on the centre-of-mass scattering angle and the particles are massless) is given by Eq. (3.2.31), which can be written in terms of $\varepsilon_{\text{IR}}$ as

$$\int \mathrm{dPS}_2 = \frac{1}{8\pi}\left(\frac{s}{4\pi}\right)^{\varepsilon_{\text{IR}}} \frac{\Gamma(1+\varepsilon_{\text{IR}})}{\Gamma(2+2\varepsilon_{\text{IR}})}\,. \tag{3.4.24}$$

Since Eq. (3.4.10) does not depend on the centre-of-mass scattering angle, we can combine it with Eq. (3.4.24) to give

$$\begin{aligned}\int \mathrm{dPS}_2\, g_{\rho\sigma} H^{\rho\sigma}_{\text{virtual}} = &-\frac{2se_q^2e^2g^2}{\pi(4\pi)^2}\left(\frac{s^2}{(4\pi)^2\mu^4}\right)^{\varepsilon_{\text{IR}}}(1+\varepsilon_{\text{IR}})\left(1-\frac{\pi^2{\varepsilon_{\text{IR}}}^2}{2}\right)\frac{\Gamma(1+\varepsilon_{\text{IR}})}{\Gamma(2+2\varepsilon_{\text{IR}})}\Gamma(1-\varepsilon_{\text{IR}})\\ &\left(2\frac{\Gamma(1+\varepsilon_{\text{IR}})\Gamma(\varepsilon_{\text{IR}})}{\Gamma(2+2\varepsilon_{\text{IR}})}-\frac{\Gamma^2(\varepsilon_{\text{IR}})}{\Gamma(1+2\varepsilon_{\text{IR}})}-\frac{\Gamma^2(1+\varepsilon_{\text{IR}})}{\Gamma(3+2\varepsilon_{\text{IR}})}\right)\\ &+\frac{se_q^2e^2g^2}{\pi(4\pi)^2}\left(\frac{s^2}{(4\pi)^2\mu^4}\right)^{\varepsilon_{\text{IR}}}\frac{\Gamma(1+\varepsilon_{\text{IR}})}{\Gamma(2+2\varepsilon_{\text{IR}})}\left[\frac{1}{\varepsilon_{\text{IR}}}+\gamma_E\right]\\ = &-\frac{se_q^2e^2g^2}{\pi(4\pi)^2}\left(\frac{s^2}{(4\pi)^2\mu^4}\right)^{\varepsilon_{\text{IR}}}\frac{(1+\varepsilon_{\text{IR}})}{\Gamma(2+2\varepsilon_{\text{IR}})}\left(-\frac{2}{{\varepsilon_{\text{IR}}}^2}+\frac{3}{\varepsilon_{\text{IR}}}-8+\pi^2+\mathcal{O}(\varepsilon_{\text{IR}})\right).\end{aligned} \tag{3.4.25}$$

From Eq. (3.2.26), the cross section contribution from the virtual correction terms is thus

$$\begin{aligned}\sigma_{\text{virtual}}(e^+e^-\to q\bar{q}) = &\frac{e_q^2e^4g^2}{2\pi(4\pi)^2 s}\left(\frac{s^2}{(4\pi)^2\mu^6}\right)^{\varepsilon_{\text{IR}}}\frac{(1+\varepsilon_{\text{IR}})^2}{(3+2\varepsilon_{\text{IR}})\Gamma(2+2\varepsilon_{\text{IR}})}\\ &\left(-\frac{2}{{\varepsilon_{\text{IR}}}^2}+\frac{3}{\varepsilon_{\text{IR}}}-8+\pi^2+\mathcal{O}(\varepsilon_{\text{IR}})\right).\end{aligned} \tag{3.4.26}$$

This cross section has IR divergences. In order to arrive at a result which is IR finite, we must consider the effect of the real emission of gluons.

### 3.4.4 Real Emission

As previously explained, in order to calculate the inclusive cross section for quark-antiquark production to $\mathcal{O}(e^4g^2)$, we must also consider the process that involves the real emission of a gluon. The term involving the real emission of a gluon will need to be treated differently when integrating over the phase space, as there are three final-state particles. It is the IR divergences that arise during this process which cancel with the IR divergences in Eq. (3.4.26).

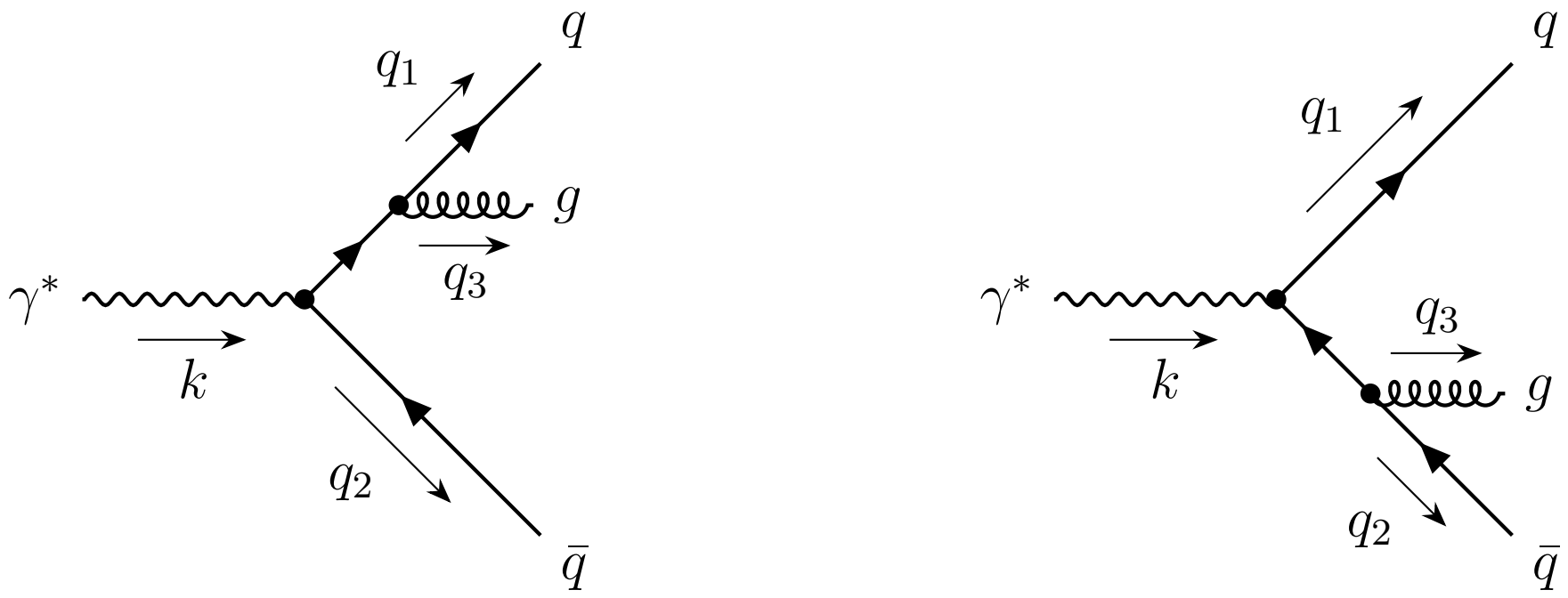

**Fig. 3.8.** Feynman diagrams representing the real-emission decay process, with the gluon emitted from either final-state particle. These contributions are summed and then squared, such that they are $\mathcal{O}(e^2g^2)$. The trace follows the fermion lines around the loop formed by combining the amplitude with the complex amplitude.

First, let us calculate the hadronic trace associated with the Feynman diagrams in Fig. 3.8. After summing then squaring these diagrams, one can write the result as the sum of four traces (the squares of each diagram plus the two cross terms).

$$
\begin{aligned}
H^{\rho\sigma}_{\text{r.e.}} = &- e_q^2 e^2 g^2 \mu^{2(4-D)} \left( g_{\alpha\beta} + (\xi - 1)\frac{q_{3\alpha} q_{3\beta}}{q_3^2 + i\epsilon} \right) \\
&\left[ \frac{\text{tr}\left[(\not{q}_1 + \not{q}_3 + m_q)\gamma^\alpha(\not{q}_1 + m_q)\gamma^\beta(\not{q}_1 + \not{q}_3 + m_q)\gamma^\sigma(\not{q}_2 - m_q)\gamma^\rho\right]}{\left((q_1+q_3)^2 - m_q^2 + i\epsilon\right)\left((q_1+q_3)^2 - m_q^2 - i\epsilon\right)} + \right. \\
&\frac{\text{tr}\left[(\not{q}_1 + m_q)\gamma^\sigma(\not{q}_2 + \not{q}_3 + m_q)\gamma^\alpha(\not{q}_2 - m_q)\gamma^\beta(\not{q}_2 + \not{q}_3 + m_q)\gamma^\rho\right]}{\left((q_2+q_3)^2 - m_q^2 + i\epsilon\right)\left((q_2+q_3)^2 - m_q^2 - i\epsilon\right)} - \\
&\frac{\text{tr}\left[(\not{q}_1 + m_q)\gamma^\alpha(\not{q}_1 + \not{q}_3 + m_q)\gamma^\sigma(\not{q}_2 - m_q)\gamma^\beta(\not{q}_2 + \not{q}_3 + m_q)\gamma^\rho\right]}{\left((q_1+q_3)^2 - m_q^2 + i\epsilon\right)\left((q_2+q_3)^2 - m_q^2 - i\epsilon\right)} - \\
&\left. \frac{\text{tr}\left[(\not{q}_1 + \not{q}_3 + m_q)\gamma^\alpha(\not{q}_1 + m_q)\gamma^\sigma(\not{q}_2 + \not{q}_3 + m_q)\gamma^\beta(\not{q}_2 - m_q)\gamma^\rho\right]}{\left((q_1+q_3)^2 - m_q^2 - i\epsilon\right)\left((q_2+q_3)^2 - m_q^2 + i\epsilon\right)} \right]
\end{aligned}
\tag{3.4.27}
$$

where we have used the same polarisation summation for the gluon helicities as we did with the photon helicities in Eq. (3.2.23). Notice the minus sign in front of the final two terms (the cross terms). This is due to one internal fermion propagator having momentum in the opposite direction to the fermion flow, so this internal propagator factor has a negative sign. We can evaluate these traces computationally in $D$ dimensions, contract with $g_{\sigma\rho}$, take the massless limit, and adopt the Feynman gauge, result-

ing in

$$
\begin{aligned}
g_{\sigma\rho} H^{\rho\sigma}_{\text{r.e.}} &= -e_q^2 e^2 g^2 \mu^{2(4-D)} \\
&\times \left[ \frac{8(D-2)^2 (q_1 \cdot q_3)(q_2 \cdot q_3)}{((q_1+q_3)^2 + i\epsilon)\,((q_1+q_3)^2 - i\epsilon)} + \frac{8(D-2)^2 (q_1 \cdot q_3)(q_2 \cdot q_3)}{((q_2+q_3)^2 + i\epsilon)\,((q_2+q_3)^2 - i\epsilon)} \right. \\
&+ \frac{8(D-2)(2(q_1 \cdot q_2)^2 + 2(q_1 \cdot q_2)(q_1 \cdot q_3 + q_2 \cdot q_3) + (D-4)(q_1 \cdot q_3)(q_2 \cdot q_3)}{((q_1+q_3)^2 + i\epsilon)\,((q_2+q_3)^2 - i\epsilon)} \\
&+ \left. \frac{8(D-2)(2(q_1 \cdot q_2)^2 + 2(q_1 \cdot q_2)(q_1 \cdot q_3 + q_2 \cdot q_3) + (D-4)(q_1 \cdot q_3)(q_2 \cdot q_3)}{((q_1+q_3)^2 - i\epsilon)\,((q_2+q_3)^2 + i\epsilon)} \right],
\end{aligned}
\tag{3.4.28}
$$

where we have taken $q_3^2 = 0$. We will now rewrite this expression in terms of dimensionless energy fractions,

$$
x_i = \frac{2 q_i \cdot k}{s}\,, \quad i = 1, 2, 3\,. \tag{3.4.29}
$$

Conservation of momentum, $q_1 + q_2 + q_3 = k$, then implies that

$$
x_1 + x_2 + x_3 = 2. \tag{3.4.30}
$$

Thus, we can write Lorentz invariants of the form

$$
\begin{aligned}
(q_1 + q_3)^2 &= 2 q_1 \cdot q_3 = (k - q_2)^2 = s(1 - x_2), \\
(q_2 + q_3)^2 &= 2 q_2 \cdot q_3 = (k - q_1)^2 = s(1 - x_1), \\
(q_1 + q_2)^2 &= 2 q_1 \cdot q_2 = (k - q_3)^2 = s(1 - x_3).
\end{aligned}
\tag{3.4.31}
$$

This change of basis will not only simplify our expression in Eq. (3.4.28), but prove very useful when we come to examine the phase-space integral of these real-emission processes. Substituting Eqs. (3.4.31) into Eq. (3.4.28), we find,

$$
\begin{aligned}
& g_{\sigma\rho} H^{\rho\sigma}_{\text{r.e.}} = \\
& - e_q^2 e^2 g^2 \mu^{2(4-D)} \left[ \frac{2(D-2)^2 (1-x_1)}{(1-x_2)} + \frac{2(D-2)^2 (1-x_2)}{(1-x_1)} \right. \\
& + \left. \frac{4(D-2)\Big(2(1-x_3)^2 + 2(1-x_3)\big((1-x_2) + (1-x_1)\big) + (D-4)(1-x_2)(1-x_1)\Big)}{(1-x_2)\,(1-x_1)} \right],
\end{aligned}
\tag{3.4.32}
$$

where we have taken $\epsilon \to 0$, which is permitted since the energy of the real emission is fixed and we will only integrate over its 3-momentum (i.e. there is no loop integral, and hence no $q^0$ integral). We use the conservation of momentum, Eq. (3.4.30), to eliminate $x_3$ from Eq. (3.4.32),

$$
\begin{aligned}
g_{\sigma\rho} H^{\rho\sigma}_{\text{r.e.}} =& - e_q^2 e^2 g^2 \mu^{2(4-D)} \Bigg[ \frac{2(D-2)^2(1-x_1)}{(1-x_2)} + \frac{2(D-2)^2(1-x_2)}{(1-x_1)} \\
&+ \frac{4(D-2)\big(2(x_1+x_2-1) + (D-4)(1-x_1)(1-x_2)\big)}{(1-x_2)\,(1-x_1)} \Bigg] \\
=& - \frac{2e_q^2 e^2 g^2 \mu^{2(4-D)}(D-2)}{(1-x_1)(1-x_2)} \Big( (D-2)(1-2x_1+x_1^2) + (D-2)(1-2x_2+x_2^2) \\
&+ 4x_1 + 4x_2 - 4 + 2(D-4)(1-x_1-x_2+x_1x_2) \Big) \\
=& - \frac{2e_q^2 e^2 g^2 \mu^{2(4-D)}(D-2)}{(1-x_1)(1-x_2)} \Big( D(x_1^2+x_2^2+2x_1x_2-4x_1-4x_2+4) \\
&- 2(x_1^2+x_2^2+4x_1x_2-8x_1-8x_2+8) \Big) \\
=& - 8e_q^2 e^2 g^2 \mu^{2(4-D)}(1+\varepsilon_{\text{IR}}) \frac{x_1^2+x_2^2+\varepsilon_{\text{IR}}(x_1+x_2-2)^2}{(1-x_1)(1-x_2)},
\end{aligned}
\tag{3.4.33}
$$

where we have set $D = 4 + 2\varepsilon_{\text{IR}}$ with $\varepsilon_{\text{IR}} > 0$. We choose to use the positive, IR dimension, $\varepsilon_{\text{IR}}$, because we can see that the above expression diverges at $x_1 = 1$ or $x_2 = 1$ (and therefore so will the cross section after integrating over the phase space). Using Eqs. (3.4.31), these two limits correspond to $q_1 \cdot q_3 = 0$ and $q_2 \cdot q_3 = 0$, respectively. In the massless limit,

$$
q_i \cdot q_3 = \omega_i \omega_g (1 - \cos\theta_{i3}) \,, \quad i = 1, 2 \tag{3.4.34}
$$

where $\omega_i$ is the energy of the quark ($i = 1$) or antiquark ($i = 2$), $\omega_g$ is the energy of the gluon, and $\cos\theta_{i3}$ is the relative angle between the quark/antiquark and the gluon. Thus, equating this quantity to zero in the massless limit requires the gluon to have no energy or for it to be collinear with the quark/antiquark. These two cases are exactly the soft IR divergence and collinear IR divergence described in Section 2.6, hence we must regulate the phase-space integral with a small positive dimension, $\varepsilon_{\text{IR}}$.

Now that we have calculated the hadronic trace, we need to integrate over the three-particle phase space. Here, the expression required to integrate the three-particle phase space in $D = 4 + 2\varepsilon_{\text{IR}}$ dimensions for the real emission is [95]

$$\begin{aligned}\int \mathrm{dPS}_{\text{r.e.}} =& \frac{s}{16(2\pi)^3}\Big(\frac{4\pi}{s}\Big)^{-2\varepsilon_{\text{IR}}}\frac{1}{\Gamma(2+2\varepsilon_{\text{IR}})} \\ &\times \int_0^1 \mathrm{d}x_1 \int_{1-x_1}^1 \mathrm{d}x_2 \Big[(1-x_1)(1-x_2)(x_1+x_2-1)\Big]^{\varepsilon_{\text{IR}}} .\end{aligned} \tag{3.4.35}$$

Combining Eqs. (3.4.33) and (3.4.35),

$$\begin{aligned}\int \mathrm{dPS}_{\text{r.e.}}\, g_{\sigma\rho}H^{\rho\sigma}_{\text{r.e.}} =& -\frac{e_q^2 e^2 g^2 s \mu^{2(4-D)}}{2(2\pi)^3}\Big(\frac{4\pi}{s}\Big)^{-2\varepsilon_{\text{IR}}}\frac{(1+\varepsilon_{\text{IR}})}{\Gamma(2+2\varepsilon_{\text{IR}})} \\ &\times \int_0^1 \mathrm{d}x_1 \int_{1-x_1}^1 \mathrm{d}x_2 \Big[(1-x_1)(1-x_2)(x_1+x_2-1)\Big]^{\varepsilon_{\text{IR}}} \\ &\times \frac{x_1^2+x_2^2+\varepsilon_{\text{IR}}(x_1+x_2-2)^2}{(1-x_1)(1-x_2)} .\end{aligned} \tag{3.4.36}$$

To evaluate this integral, we perform a change of variables $x_1 = x$ and $x_2 = 1 - vx$,

$$\begin{aligned}\int \mathrm{dPS}_{\text{r.e.}}\, g_{\sigma\rho}H^{\rho\sigma}_{\text{r.e.}} =& -\frac{e_q^2 e^2 g^2 s \mu^{2(4-D)}}{2(2\pi)^3}\Big(\frac{s^2}{(4\pi)^2}\Big)^{\varepsilon_{\text{IR}}}\frac{(1+\varepsilon_{\text{IR}})}{\Gamma(2+2\varepsilon_{\text{IR}})} \\ &\times \int_0^1 \mathrm{d}x \int_0^1 \mathrm{d}v\, x\Big[(1-x)vx^2(1-v)\Big]^{\varepsilon_{\text{IR}}} \\ &\times \frac{x^2+1-2vx+v^2x^2+\varepsilon_{\text{IR}}(x(1-v)-1)^2}{vx(1-x)} \\ =& -\frac{e_q^2 e^2 g^2 s \mu^{2(4-D)}}{2(2\pi)^3}\Big(\frac{s^2}{(4\pi)^2}\Big)^{\varepsilon_{\text{IR}}}\frac{(1+\varepsilon_{\text{IR}})}{\Gamma(2+2\varepsilon_{\text{IR}})} \\ &\times \int_0^1 \mathrm{d}x \int_0^1 \mathrm{d}v\, (1-x)^{\varepsilon_{\text{IR}}} v^{\varepsilon_{\text{IR}}} x^{2\varepsilon_{\text{IR}}+1}(1-v)^{\varepsilon_{\text{IR}}} \\ &\times \Big[(1+\varepsilon_{\text{IR}})\Big(x^{-1}(1-x)v^{-1}+vx(1-x)^{-1}\Big) \\ &\quad + 2v^{-1}(1-v)(1-x)^{-1} + 2\varepsilon_{\text{IR}}\Big] .\end{aligned} \tag{3.4.37}$$

Writing the integral as factors of $x^n$, $v^n$, $(1-x)^n$, and $(1-v)^n$, we can use the Euler Beta function (Eq. (2.6.22)) to find,

$$\begin{aligned}\int \mathrm{dPS}_{\text{r.e.}}\, g_{\sigma\rho}H^{\rho\sigma}_{\text{r.e.}} =& -\frac{e_q^2 e^2 g^2 s \mu^{2(4-D)}}{2(2\pi)^3}\Big(\frac{s^2}{(4\pi)^2}\Big)^{\varepsilon_{\text{IR}}}\frac{(1+\varepsilon_{\text{IR}})}{\Gamma(2+2\varepsilon_{\text{IR}})} \\ &\times \Bigg[\frac{2(1+\varepsilon_{\text{IR}})\Gamma(\varepsilon_{\text{IR}})\Gamma(\varepsilon_{\text{IR}}+1)\Gamma(\varepsilon_{\text{IR}}+2)}{\Gamma(3\varepsilon_{\text{IR}}+3)} \\ &\quad + \frac{2\Gamma^2(\varepsilon_{\text{IR}})\Gamma(\varepsilon_{\text{IR}}+2)}{\Gamma(3\varepsilon_{\text{IR}}+2)} + \frac{2\varepsilon_{\text{IR}}\Gamma^3(1+\varepsilon_{\text{IR}})}{\Gamma(3\varepsilon_{\text{IR}}+3)}\Bigg] .\end{aligned} \tag{3.4.38}$$

Carefully expanding the Gamma functions leads to the result,

$$\int \mathrm{dPS}_{\text{r.e.}}\, g_{\sigma\rho} H^{\rho\sigma}_{\text{r.e.}} = -\frac{e_q^2 e^2 g^2 s \mu^{2(4-D)}}{2(2\pi)^3} \left(\frac{s^2}{(4\pi)^2}\right)^{\varepsilon_{\text{IR}}} \frac{(1+\varepsilon_{\text{IR}})}{\Gamma(2+2\varepsilon_{\text{IR}})} \left[\frac{2}{{\varepsilon_{\text{IR}}}^2} - \frac{3}{\varepsilon_{\text{IR}}} + \frac{19}{2} - \pi^2\right]. \tag{3.4.39}$$

From Eq. (3.2.26), we arrive at the cross section for quark-antiquark production with the real emission of a gluon,

$$\begin{aligned}\sigma_{\text{r.e.}}(e^+e^- \to q\bar{q}\gamma) = & \frac{e_q^2 e^4 g^2 \mu^{3(4-D)}}{4(2\pi)^3 s} \left(\frac{s^2}{(4\pi)^2}\right)^{\varepsilon_{\text{IR}}} \frac{(1+\varepsilon_{\text{IR}})^2}{(3+2\varepsilon_{\text{IR}})\Gamma(2+2\varepsilon_{\text{IR}})} \\ & \left[\frac{2}{{\varepsilon_{\text{IR}}}^2} - \frac{3}{\varepsilon_{\text{IR}}} + \frac{19}{2} - \pi^2 + \mathcal{O}(\varepsilon_{\text{IR}})\right].\end{aligned} \tag{3.4.40}$$

### 3.4.5 Total Cross Section

Finally, we can combine Eqs. (3.4.26) and (3.4.40) to get,

$$\begin{aligned}\sigma_1(e^+e^- \to q\bar{q}\,X) &= \sigma_{\text{virtual}}(e^+e^- \to q\bar{q}) + \sigma_{\text{r.e.}}(e^+e^- \to q\bar{q}g) \\ &= \lim_{\varepsilon_{\text{IR}}\to 0} \frac{e_q^2 e^4 g^2}{2\pi(4\pi)^2 s} \left(\frac{s^2}{(4\pi)^2\mu^6}\right)^{\varepsilon_{\text{IR}}} \frac{(1+\varepsilon_{\text{IR}})^2}{(3+2\varepsilon_{\text{IR}})\Gamma(2+2\varepsilon_{\text{IR}})} \\ &\quad \left[\left(-\frac{2}{{\varepsilon_{\text{IR}}}^2} + \frac{3}{\varepsilon_{\text{IR}}} - 8 + \pi^2\right) + \left(\frac{2}{{\varepsilon_{\text{IR}}}^2} - \frac{3}{\varepsilon_{\text{IR}}} + \frac{19}{2} - \pi^2\right) + \mathcal{O}(\varepsilon_{\text{IR}})\right] \\ &= \lim_{\varepsilon_{\text{IR}}\to 0} \frac{e_q^2 e^4 g^2}{2\pi(4\pi)^2 s} \left(\frac{s^2}{(4\pi)^2\mu^6}\right)^{\varepsilon_{\text{IR}}} \frac{(1+\varepsilon_{\text{IR}})^2}{(3+2\varepsilon_{\text{IR}})\Gamma(2+2\varepsilon_{\text{IR}})} \left[\frac{3}{2} + \mathcal{O}(\varepsilon_{\text{IR}})\right] \\ &= \lim_{\varepsilon_{\text{IR}}\to 0} \frac{e_q^2 e^4 g^2}{2\pi(4\pi)^2 s} \left(\frac{s^2}{(4\pi)^2\mu^6}\right)^{\varepsilon_{\text{IR}}} \left[\frac{1}{2} + \mathcal{O}(\varepsilon_{\text{IR}})\right] \\ &= \frac{e_q^2 e^4 g^2}{(4\pi)^3 s} \\ &= \frac{e_q^2 \alpha_{\text{em}}^2 \alpha_{\text{s}}}{s},\end{aligned} \tag{3.4.41}$$

where

$$\alpha_s = \frac{g^2}{4\pi}. \tag{3.4.42}$$

Thus, the first order contribution to the cross section is

$$\sigma_1 = \frac{e_q^2 \alpha_{\text{em}}^2 \alpha_s}{s} = \frac{3\sigma_0 \alpha_s}{4\pi N_c}, \tag{3.4.43}$$

where $\sigma_0$ is the tree-level cross section, given in Eq. (3.3.1). The final step is to account for the colour factor for this process, which up until now has been entirely ignored. The colour factor is the number of colours, $N_c$, multiplied by $SU(3)$ invariant, $C_F = 4/3$ [23]. Thus, $\sigma_1$ becomes

$$\sigma_1 = \frac{\sigma_0 \alpha_s}{\pi}\,, \tag{3.4.44}$$

such that, up to first order,

$$\sigma(e^+e^- \to q\bar{q}X) = \sigma_0\Big(1 + \frac{\alpha_s}{\pi}\Big)\,. \tag{3.4.45}$$

Taking the ratio of this cross section and the tree-level muon-antimuon production cross section (Eq. (3.2.33)), we arrive at the result:

$$R_q = \frac{\sigma(e^+e^- \to q\bar{q}X)}{\sigma(e^+e^- \to \mu^+\mu^-)} = N_c e_q^2\Big(1 + \frac{\alpha_s}{\pi}\Big)\,. \tag{3.4.46}$$

Summing over quark flavours, we get

$$R \equiv \frac{\sigma(e^+e^- \to \text{hadrons})}{\sigma(e^+e^- \to \mu^+\mu^-)} = N_c \sum_q e_q^2\Big(1 + \frac{\alpha_s}{\pi}\Big)\,. \tag{3.4.47}$$

## 3.5 Summary

In this chapter, we calculated the ratio of the cross sections for quark-antiquark production and muon-antimuon production, up to first order in gluon corrections (Eq. (3.4.46)). This calculation exhibits both UV and IR divergences, making it a good example to showcase how both types of divergences are typically dealt with in traditional quantum field theory (QFT).

We defined a cross section first-order in virtual corrections as the vertex correction cross section, corrected using the LSZ reduction formula. This cross section was no longer UV divergent, since the UV divergences in the vertex and LSZ terms exactly cancelled—a consequence of the Slavnov-Taylor identities.

The virtual corrections cross section was still IR divergent, since the process $e^-e^+ \to q\bar{q}g$ was not accounted for. This process is degenerate with the 2-to-2 cross section in the limit of the gluon becoming soft or collinear. To this end, we summed the cross section for this process and the first-order 2-to-2 cross section and the IR divergences cancelled, as expected from the KLN theorem.

Similar calculations become more complicated for higher-order corrections or different scattering processes. For example, even the first-order photon corrections to the muon-antimuon production cross section from Section 3.2 involve further technical considerations, since one must also account for vacuum polarisation (the quark-loop correction to the virtual photon, $\gamma^*$).

With this in mind, it would be useful to develop a formulation of QFT which intrinsically sums over final states of certain Hilbert spaces (i.e., certain particles). This inclusive summation may inherently account for degenerate final states, such as the soft/collinear real emission of a gluon, helping to cancel IR divergences. In Chapters 4–6, we present and utilise a new, manifestly causal formalism for calculations in QFT, which allows for the immediate summation over final states. Chapter 5 applies this formalism to particle scattering in scalar field theory, and we see that the real-emission terms are not separable from the vertex and self-energy (analogous to LSZ corrections) diagrams. Optimistically, this may imply the cancellation of IR divergences when applied to gauge theories.

# Chapter 4

# Manifestly Causal QFT

In relativistic quantum field theories (QFTs), causality is encoded in the vanishing of the commutator (or anticommutator) of field operators. However, as we saw in Chapter 3, calculations are usually performed in a manner in which causality is not manifest, due to the ubiquity of the Feynman propagator as opposed to the retarded propagator (which is manifestly causal). Furthermore, Bogoliubov's *Condition of Causality* [97] shows that causality is manifest and meaningful only at the probability level. For example, in scattering calculations, the $S$-matrix does not directly demonstrate causality because it represents only the transition amplitude, not the full probabilistic outcome of a process. This distinction underscores the importance of observable probabilities over amplitudes when discussing physical principles like causality.

With this in mind, a manifestly causal, probability-level formalism for calculations in QFT was developed [32–34] by applying a generalisation [98] of the Baker-Campbell-Hausdorff (BCH) lemma [74, 99–104] to the transition probability, naturally expressed in terms of commutators of fields. This formalism has proven interesting in the studies of particle scattering [1] (Chapter 5), the Unruh effect [2] (Chapter 6), semi-inclusive observables [105], and causal set theory [35].

In this chapter, we introduce and develop this manifestly causal formalism, building a general toolbox for use on more specific problems later (Chapters 5 and 6). Section 4.4 describes the Fermi two-atom problem, highlighting the incorrect, acausal conclusion one might reach if naïvely applying traditional QFT techniques, and subsequently applies the manifestly causal formalism, following Refs. [33, 106].

## 4.1 Defining the Probability

Suppose that a system is initially ($t = t_{\text{in}}$) described by a density operator $\rho_0$ and that a measurement outcome is described by an effect operator $E$. In general, $E$ is an element of a Positive Operator-Valued Measure, which is a generalisation of projection operators [107, 108]. The probability of the measurement outcome, $\mathbb{P}$, is then given by

$$\mathbb{P} \;=\; \text{tr}(E\rho_t)\,, \tag{4.1.1}$$

where

$$\rho_t \;\equiv\; U_{t,t_{\text{in}}}\,\rho_0\,U^{\dagger}_{t,t_{\text{in}}} \tag{4.1.2}$$

is the density operator at time $t$ and

$$U_{t,t_{\text{in}}} = \text{T}\left\{\exp\left(\frac{1}{i}\int_{t_{\text{in}}}^{t}\text{d}t'\,H_{\text{int}}(t')\right)\right\} \tag{4.1.3}$$

is the unitary evolution operator (T indicates time ordering) for a given interaction Hamiltonian, $H_{\text{int}}(t')$. From Eq. (4.1.1), the probability of the measurement outcome for a pure initial state, $\rho_0 = |i\rangle\langle i|$, can be written as

$$\mathbb{P} \;=\; \left\langle i \,\middle|\, U^{\dagger}_{t,t_{\text{in}}}\,E\,U_{t,t_{\text{in}}} \,\middle|\, i\right\rangle . \tag{4.1.4}$$

In the case of an exclusive final state, $|f\rangle$, the effect operator, $E$, becomes a projection operator,

$$E = |f\rangle\langle f|\,, \tag{4.1.5}$$

meaning the probability is given by

$$\mathbb{P} \;=\; \langle i|\,U^{\dagger}_{t,t_{\text{in}}}\,|f\rangle\langle f|\,U_{t,t_{\text{in}}}\,|i\rangle \;=\; \left|\,\langle f|\,U_{t,t_{\text{in}}}\,|i\rangle\,\right|^2 . \tag{4.1.6}$$

This is the usual expression for the probability of an initial state, $|i\rangle$, evolving into a final state, $|f\rangle$.

Rather than calculating the amplitude and squaring, we can instead work directly with the expression in Eq. (4.1.4). We do so using the BCH lemma [74, 99–104],

$$e^{A}Oe^{-A} = O + [A, O] + \frac{1}{2!}[A, [A, O]] + \frac{1}{3!}\Big[A, \big[A, [A, O]\big]\Big] + \cdots \tag{4.1.7}$$

where $O$ and $A$ are two operators, which leads to [98]

$$\begin{aligned}
U^{\dagger}_{t,t_{\text{in}}}\, O\, U_{t,t_{\text{in}}} = O& \\
&+ \left(\frac{1}{i}\right)\int_{t_{\text{in}}}^{t} \mathrm{d}t_1\, [O, H_{\text{int}}(t_1)] \\
&+ \left(\frac{1}{i}\right)^{2}\int_{t_{\text{in}}}^{t} \mathrm{d}t_1 \int_{t_{\text{in}}}^{t_1} \mathrm{d}t_2\, \big[[O, H_{\text{int}}(t_1)], H_{\text{int}}(t_2)\big] + \\
&\vdots \\
&+ \left(\frac{1}{i}\right)^{j}\int_{t_{\text{in}}}^{t} \mathrm{d}t_1 \int_{t_{\text{in}}}^{t_1} \mathrm{d}t_2 \cdots \int_{t_{\text{in}}}^{t_{j-1}} \mathrm{d}t_j \\
&\qquad \Big[\cdots\big[[O, H_{\text{int}}(t_1)], H_{\text{int}}(t_2)\big]\cdots, H_{\text{int}}(t_j)\Big] \\
&\vdots
\end{aligned} \tag{4.1.8}$$

where we are working in the interaction picture. Thus, the probability can be written as

$$\mathbb{P} = \sum_{j=0}^{\infty}\int_{t_{\text{in}}}^{t} \mathrm{d}t'_1 \mathrm{d}t'_2 \ldots \mathrm{d}t'_j \Theta_{12\ldots j} \langle i \,|\, \mathcal{F}_j \,|\, i\rangle\,, \tag{4.1.9}$$

$$\text{where} \quad \Theta_{ijk\ldots} \equiv \begin{cases} 1, & \text{if } t'_i > t'_j > t'_k \ldots \\ 0, & \text{otherwise} \end{cases}, \tag{4.1.10}$$

$$\mathcal{F}_0 = E, \tag{4.1.11}$$

$$\text{and} \quad \mathcal{F}_j = \frac{1}{i}\left[\mathcal{F}_{j-1}, H_{\text{int}}(t'_j)\right]. \tag{4.1.12}$$

In general, $E$ may be written as the sum of the effect operators of each final state, which themselves are products of Hermitian operators in different Hilbert spaces, i.e.,

$$E = \sum_{\kappa}\prod_{i} E^{\mathscr{H}_i}_{(\kappa)}, \tag{4.1.13}$$

where the superscripts indicate the Hilbert spaces of each operator, and $\{\kappa\}$ is the set of final states. If all final states of a particular Hilbert space are summed over, then the effect operator in that Hilbert space is the identity operator, $E^{\mathscr{H}_i} = \mathbb{I}^{\mathscr{H}_i}$. If this is done for all Hilbert spaces, then the calculation is completely inclusive and calculates the probability of 'anything at all' happening. It is trivial to see from Eq. (4.1.4) that this probability is 1, as expected.

In this thesis, we apply this formalism exclusively to Hamiltonians involving up to three distinct Hilbert spaces, and therefore the effect operator will include only a single product of, at most, three Hermitian operators,

$$E = E^X E^Y E^Z, \tag{4.1.14}$$

where $X$, $Y$, and $Z$ denote the three Hilbert spaces.

It should be noted that Eq. (4.1.1) applies for a mixed initial state as well, and so this formalism does not rely on a pure initial state. This generalisation may prove useful for studies in which mixed initial states are common, such as in the field of relativistic quantum information (RQI) [109, 110]. In this thesis, we only consider pure initial states.

## 4.2 Calculating the $\mathcal{F}_j$ Operators

The operator $\mathcal{F}_j$ corresponds to an interaction of order $j$, that is to say there are $j$ vertices in the interaction at probability level. One can see from Eq. (4.1.12) that $\mathcal{F}_j$ will contain a series of nested commutators of products of Hermitian operators in different Hilbert spaces (originating from the Hamiltonian). The first task is to separate the Hilbert spaces, writing $\mathcal{F}_j$ in terms of nested commutators/anticommutators that each contain Hermitian operators from the same Hilbert space.

We shall consider an interaction Hamiltonian that is the sum of two products of two operators, selected from three Hilbert spaces, such that one of the Hilbert spaces in shared between the terms,

$$H_{\text{int}}(t'_j) = \int \mathrm{d}^3\mathbf{x}_j \, \left( g_Y A^X_j B^Y_j + g_Z \mathcal{A}^X_j C^Z_j \right) \, , \tag{4.2.1}$$

where $A^X$, $\mathcal{A}^X$ are operators in Hilbert space $X$, $B^Y$ is an operator in Hilbert space $Y$, $C^Z$ is an operator in Hilbert space $Z$, and $g_Y$ and $g_Z$ are two different constants.

We choose this Hamiltonian since it is a generalisation of all of the Hamiltonians considered later in this thesis. For the Fermi two-atom problem considered in Section 4.4, $A^X$ and $\mathcal{A}^X$ are the same operator but evaluated at different points in space, and the constants are the same ($g_Y = g_Z = 1$). Chapter 5 considers particle scattering interactions, where the constants are different (as they represent couplings between different fields) but $A^X = \mathcal{A}^X$. Chapter 6 considers the Unruh effect using a Hamiltonian of this form with only one term, i.e., $g_Z = 0$.

We can now calculate $\mathcal{F}_1$ using Eq. (4.1.12) and substituting in $\mathcal{F}_0$ (Eq. (4.1.11)) and the interaction Hamiltonian (4.2.1), giving

$$\begin{aligned}\mathcal{F}_1 &= \int \mathrm{d}^3\mathbf{x}_1\ \frac{1}{i}\Big[E, g_Y A_1^X B_1^Y + g_Z \mathcal{A}_1^X C_1^Z\Big] \\ &= \int \mathrm{d}^3\mathbf{x}_1\ \frac{1}{i}\Big[E^X E^Y E^Z, g_Y A_1^X B_1^Y + g_Z \mathcal{A}_1^X C_1^Z\Big],\end{aligned} \tag{4.2.2}$$

where we have used Eq. (4.1.14) for our effect operator, $E$. We can use the following commutation relation to expand this commutator:

$$\left[A^X B^Y, \alpha^X \beta^Y\right] = \frac{1}{2}\left[A^X, \alpha^X\right]\left\{B^Y, \beta^Y\right\} + \frac{1}{2}\left\{A^X, \alpha^X\right\}\left[B^Y, \beta^Y\right], \tag{4.2.3}$$

where $A^X$ and $\alpha^X$ ($B^Y$ and $\beta^Y$) are different operators in the same Hilbert space, $X$ ($Y$). Eq. (4.2.2) now becomes

$$\begin{aligned}\mathcal{F}_1 = \int \mathrm{d}^3\mathbf{x}_1\ \Big(&\frac{1}{2i}\left[E^X, g_Y A_1^X\right]\left\{E^Y, B_1^Y\right\}E^Z + \frac{1}{2i}\left\{E^X, g_Y A_1^X\right\}\left[E^Y, B_1^Y\right]E^Z \\ &+ \frac{1}{2i}\left[E^X, g_Z \mathcal{A}_1^X\right]E^Y\left\{E^Z, C_1^Z\right\} + \frac{1}{2i}\left\{E^X, g_Z \mathcal{A}_1^X\right\}E^Y\left[E^Z, C_1^Z\right]\Big).\end{aligned} \tag{4.2.4}$$

We now define the useful notation:

$$\begin{aligned}&\text{Hilbert space } X: && \mathcal{E}^{\cdots Y}_{\ldots k} := \frac{1}{i}\left[\mathcal{E}^{\cdots}_{\ldots}, g_Y A_k^X\right], && \mathcal{E}^{\cdots Y}_{\ldots \underline{k}} := \left\{\mathcal{E}^{\cdots}_{\ldots}, g_Y A_k^X\right\}, \\ & && \mathcal{E}^{\cdots Z}_{\ldots k} := \frac{1}{i}\left[\mathcal{E}^{\cdots}_{\ldots}, g_Z \mathcal{A}_k^X\right], && \mathcal{E}^{\cdots Z}_{\ldots \underline{k}} := \left\{\mathcal{E}^{\cdots}_{\ldots}, g_Z \mathcal{A}_k^X\right\},\end{aligned} \tag{4.2.5}$$

$$\begin{aligned}&\text{Hilbert space } Y: && E^Y_{\ldots k} := \frac{1}{i}\left[E^Y_{\ldots}, B_k^Y\right], && E^Y_{\ldots \underline{k}} := \left\{E^Y_{\ldots}, B_k^Y\right\}, \\ &\text{Hilbert space } Z: && E^Z_{\ldots k} := \frac{1}{i}\left[E^Z_{\ldots}, C_k^Z\right], && E^Z_{\ldots \underline{k}} := \left\{E^Z_{\ldots}, C_k^Z\right\},\end{aligned} \tag{4.2.6}$$

where $\mathcal{E} \equiv E^X$. The use of $\mathcal{E}$ instead of $E$ for Hilbert space X is to highlight the difference in its structure. The presence of two different operators, $A^X$ and $\mathcal{A}^X$, from Hilbert space $X$ in each term of the interaction Hamiltonian (Eq. (4.2.1)) means that the $\mathcal{E}^{\cdots}_{\cdots}$ operators will be nested (anti)commutators of both operators, and we should keep track of this using the superscript indices. Furthermore, should the constants $g_Y$ and $g_Z$ be different, the superscript indices of $\mathcal{E}^{\cdots}_{\cdots}$ also tell us the number of each constant the overall term is multiplied by.

Using the above notation, we can write Eq. (4.2.4) as

$$\mathcal{F}_1 = \int \mathrm{d}^3\mathbf{x}_1 \; \frac{1}{2} \left( \mathcal{E}^Y_1 E^Y_{\underline{1}} E^Z + \mathcal{E}^Y_{\underline{1}} E^Y_1 E^Z + \mathcal{E}^Z_1 E^Y E^Z_{\underline{1}} + \mathcal{E}^Z_{\underline{1}} E^Y E^Z_1 \right) . \tag{4.2.7}$$

One can condense this expression by introducing underdot notation [33], where, e.g.,

$$\begin{aligned}
E^Y_{\underset{\circ}{k}} \mathcal{E}^Y_{\underset{\bullet}{k}} &:= E^Y_k \mathcal{E}^Y_{\underline{k}} + E^Y_{\underline{k}} \mathcal{E}^Y_k , \\
E^Y_{\underset{\circ\circ}{kl}} \mathcal{E}^{YY}_{\underset{\bullet\bullet}{kl}} &:= E^Y_{kl} \mathcal{E}^{YY}_{\underline{k}\underline{l}} + E^Y_{k\underline{l}} \mathcal{E}^{YY}_{\underline{k}l} + E^Y_{\underline{k}l} \mathcal{E}^{YY}_{k\underline{l}} + E^Y_{\underline{k}\underline{l}} \mathcal{E}^{YY}_{kl} , \\
E^Y_{\underset{\circ}{k}} E^Z_{\underset{\circ}{l}} \mathcal{E}^{YZ}_{\underset{\bullet\bullet}{kl}} &:= E^Y_k E^Z_l \mathcal{E}^{YZ}_{\underline{k}\underline{l}} + E^Y_{\underline{k}} E^Z_l \mathcal{E}^{YZ}_{k\underline{l}} + E^Y_k E^Z_{\underline{l}} \mathcal{E}^{YZ}_{\underline{k}l} + E^Y_{\underline{k}} E^Z_{\underline{l}} \mathcal{E}^{YZ}_{kl} ,
\end{aligned} \tag{4.2.8}$$

which encodes a sum over all possible permutations of commutation/anticommutation pairs. An index with underdots must be underlined once, and only once, across the operators from different Hilbert spaces. This reflects how Eq. (4.2.3) involves one term with a commutator and anticommutator, and then another term with the types of commutators flipped. We then sum over the possibilities.

Through repeated use of Eq. (4.1.12), one can find expressions for higher order $\mathcal{F}_j$ operators. In summary,

$$\mathcal{F}_0 = E^Y E^Z \mathcal{E} , \tag{4.2.9}$$

$$\mathcal{F}_1 = \int \mathrm{d}^3\mathbf{x}_1 \; \frac{1}{2} \big( E^Y_{\underset{\circ}{1}} E^Z \mathcal{E}^Y_{\underset{\bullet}{1}} + E^Y E^Z_{\underset{\circ}{1}} \mathcal{E}^Z_{\underset{\bullet}{1}} \big) , \tag{4.2.10}$$

$$\mathcal{F}_2 = \int \mathrm{d}^3\mathbf{x}_1 \, \mathrm{d}^3\mathbf{x}_2 \; \frac{1}{4} \big( E^Y_{\underset{\circ\circ}{12}} E^Z \mathcal{E}^{YY}_{\underset{\bullet\bullet}{12}} + E^Y_{\underset{\circ}{1}} E^Z_{\underset{\circ}{2}} \mathcal{E}^{YZ}_{\underset{\bullet\bullet}{12}} + E^Y_{\underset{\circ}{2}} E^Z_{\underset{\circ}{1}} \mathcal{E}^{ZY}_{\underset{\bullet\bullet}{12}} + E^Y E^Z_{\underset{\circ\circ}{12}} \mathcal{E}^{ZZ}_{\underset{\bullet\bullet}{12}} \big) , \tag{4.2.11}$$

$$\mathcal{F}_3 = \int \prod_{\kappa=1}^{3} \left( \mathrm{d}^3\mathbf{x}_\kappa \right) \frac{1}{8} \big( E^Y_{\underset{\circ\circ\circ}{123}} E^Z \mathcal{E}^{YYY}_{\underset{\bullet\bullet\bullet}{123}} + E^Y_{(\underset{\circ\circ}{12}} E^Z_{\underset{\circ}{3})} \mathcal{E}^{(YYZ)}_{(\underset{\bullet\bullet\bullet}{123})} + E^Y_{(\underset{\circ}{1}} E^Z_{\underset{\circ\circ}{23})} \mathcal{E}^{(YZZ)}_{(\underset{\bullet\bullet\bullet}{123})} + E^Y E^Z_{\underset{\circ\circ\circ}{123}} \mathcal{E}^{ZZZ}_{\underset{\bullet\bullet\bullet}{123}} \big) . \tag{4.2.12}$$

etc.

We have, once again, reduced the expressions, this time by using parentheses around some of the indices. The parentheses indicate a summation over all permutations of those indices which result in unique terms (remembering that the time indices are time-ordered within each operator). For example, we need not consider $E^{Y}_{\underset{\circ\circ}{31}} E^{Z}_{\underset{\circ}{2}} \mathcal{E}^{YZY}_{\underset{\bullet\bullet\bullet}{123}}$, as this is not unique to $E^{Y}_{\underset{\circ\circ}{13}} E^{Z}_{\underset{\circ}{2}} \mathcal{E}^{YZY}_{\underset{\bullet\bullet\bullet}{123}}$. The superscript letter indices on the $\mathcal{E}$ operators will shuffle such that they will always remain aligned with the subscript number index corresponding to the $E$ operator which also carries that number index. An example of expanding the parentheses is given below, using the second term in Eq. (4.2.12):

$$E^{Y}_{(\underset{\circ\circ}{12}} E^{Z}_{\underset{\circ}{3})} \mathcal{E}^{(YYZ)}_{(\underset{\bullet\bullet\bullet}{123})} = E^{Y}_{\underset{\circ\circ}{12}} E^{Z}_{\underset{\circ}{3}} \mathcal{E}^{YYZ}_{\underset{\bullet\bullet\bullet}{123}} + E^{Y}_{\underset{\circ\circ}{13}} E^{Z}_{\underset{\circ}{2}} \mathcal{E}^{YZY}_{\underset{\bullet\bullet\bullet}{123}} + E^{Y}_{\underset{\circ\circ}{23}} E^{Z}_{\underset{\circ}{1}} \mathcal{E}^{ZYY}_{\underset{\bullet\bullet\bullet}{123}} . \tag{4.2.13}$$

We are now suitably equipped to write the general formula for $\mathcal{F}_n$,

$$\mathcal{F}_n = 2^{-n} \int \prod_{\kappa=1}^{n} \left(\mathrm{d}^3\mathbf{x}_\kappa\right) \sum_{a=0}^{n} E^{Y}_{(\underset{\circ}{1}\underset{\cdots}{\ldots}\underset{\circ}{a}} E^{Z}_{\underset{\circ}{a+1}\underset{\cdots}{\ldots}\underset{\circ}{n})} \mathcal{E}^{(Y \ldots Y\ \ Z \ldots Z)}_{(\underset{\bullet}{1}\underset{\cdots}{\ldots}\underset{\bullet}{a}\,\underset{\bullet}{a+1}\underset{\cdots}{\ldots}\underset{\bullet}{n})} . \tag{4.2.14}$$

Despite its appearance, the above equation is fairly straightforward. In words, it is simply the sum of all distinct products of operators of the form $2^{-n} E^{S}_{\ldots} E^{D}_{\ldots} \mathcal{E}^{\cdots}_{\ldots}$, with every index $\{1, \ldots, n\}$ appearing once on an $E$ operator and once on an $\mathcal{E}$ operator. For the subscript indices $i \ldots j$, if $i > j$ then this set is the empty set, resulting in a factor of $E$ with no indices (the original effect operator for that Hilbert space). Likewise, if $n = 0$, then there is no integral over space.

## 4.3 Calculating Expectation Values

Returning to Eq. (4.1.9), the next step after calculating $\mathcal{F}_j$ is to calculate the expectation value of this operator, given the initial state of the system, $|i\rangle$. Of course, one could expand the nested commutators in $\mathcal{F}_j$, but this would result in similar expressions as if one were to use the traditional QFT methods. For example, if one were to consider the probability of an electron-positron pair annihilating to a muon-antimuon or quark-antiquark pair, the expressions would return to exactly the same expressions as in Chapter 3.

The method for evaluating expectation values depends on the specific process under consideration, including the choice of initial state and the Hermitian operators in the effect operator and Hamiltonian. However, in any case, one should endeavour to utilise this new commutator form of the probability, in an attempt to gain further insight. As we shall see, working directly with commutators can result in a manifestly causal result, something which the traditional QFT method cannot claim.

## 4.4 Fermi Two-Atom Problem

Armed with the arsenal of general equations from the previous sections, we can turn our attention to a specific quantum system. To start, we will consider two atoms (a 'source' atom, $S$, and a 'detector' atom, $D$) in a scalar field, $\phi$. The full analysis of this system using the causal formalism of this chapter is performed in Ref. [33], with different types of measurements. Rather than repeating this, we shall describe the system and highlight the main results for a local measurement. This will allow us to quickly jump to the main conclusion; we shall see that this new method obtains manifestly causal results.

There are three Hilbert spaces for this closed system: $\mathscr{H} = \mathscr{H}^S \times \mathscr{H}^D \times \mathscr{H}^\phi$. We shall only consider interactions between the atoms and the fields, not field self-interactions. The full Hamiltonian is then

$$H = H_0 + H_{\text{int}} = H_0^S + H_0^D + H_0^\phi + H^{S\phi} + H^{D\phi} \tag{4.4.1}$$

where the superscripts indicate the Hilbert spaces (or product of Hilbert spaces) where the operators act. The atoms $S$ and $D$ are static and interact with the field at the fixed, spatial points $\mathbf{x}^S$ and $\mathbf{x}^D$, so the distance between the atoms is $R \equiv |\mathbf{x}^D - \mathbf{x}^S|$. Under the free Hamiltonian, $H_0$, each atom $X \in \{S, D\}$ has a complete set of bound states, $\{|n^X\rangle\}$, with eigenvalues given by $H_0^X |n^X\rangle = \Omega_n^X |n^X\rangle$. The atoms interact with the field via transition moments, $\mu_{mn}^X$, which are taken to be monopole moments in this toy model. Thus, the full interaction-picture Hamiltonian is

$$\begin{aligned} H = &\underbrace{\sum_n \Omega_n^S |n^S\rangle\langle n^S| + \sum_n \Omega_n^D |n^D\rangle\langle n^D| + \int \mathrm{d}^3\mathbf{x} \left(\tfrac{1}{2}(\partial_t\phi)^2 + \tfrac{1}{2}(\nabla\phi)^2 + \tfrac{1}{2}m^2\phi^2\right)}_{H_0} \\ &\underbrace{+ \sum_{m,n} \mu_{mn}^S \, e^{i\Omega_{mn}^S t_j'} \, |m^S\rangle\langle n^S| \, \phi(\mathbf{x}^S, t_j') + \sum_{m,n} \mu_{mn}^D \, e^{i\Omega_{mn}^D t_j'} \, |m^D\rangle\langle n^D| \, \phi(\mathbf{x}^D, t_j')}_{H_{\text{int}}(t_j')} \, , \end{aligned} \tag{4.4.2}$$

where $\Omega^X_{mn} = \Omega^X_m - \Omega^X_n$.

Suppose we wanted to calculate the probability that the detector atom, $D$, transitions from its ground state to an excited state due to the absorption of a field quantum, given that the source atom, $S$, transitions from an excited state to its ground state via the emission of a field quantum. Using the traditional QFT approach, this probability, $\mathbb{P}_{\text{Fermi}}$, is given by

$$\mathbb{P}_{\text{Fermi}} = |\langle f| U_{t,0} |i\rangle|^2 = \left|\langle g^S q^D 0^\phi| U_{t,0} |p^S g^D 0^\phi\rangle\right|^2 , \qquad (4.4.3)$$

where $|g^X\rangle$ denotes the ground state of each atom, $|p^S\rangle$ denotes the source atom being in an excited state, $|q^D\rangle$ denotes the detector atom being in an excited state, $|0^\phi\rangle$ denotes the vacuum state of the $\phi$ field, and the unitary time evolution operator, $U_{t,0}$, is given by Eq. (4.1.3). Perturbatively expanding $U_{t,0}$ and using the interaction Hamiltonian in Eq. (4.4.2), the lowest order contribution is

$$\mathbb{P}_{\text{Fermi}} = \left|\int_0^t \mathrm{d}t_1 \int_0^t \mathrm{d}t_2\, \mu^S_{gp}\, \mu^D_{qg}\, e^{i\Omega^S_{gp} t_1}\, e^{i\Omega^D_{qg} t_2}\, \Delta^{SD(F)}_{12}\right|^2 + \cdots , \qquad (4.4.4)$$

where the Feynman propagator, $\Delta^{SD(F)}_{12}$, is defined by

$$\Delta^{SD(F)}_{12} \equiv \langle 0^\phi| \mathrm{T}\left\{\phi^S_1 \phi^D_2\right\} |0^\phi\rangle , \qquad (4.4.5)$$

where $\phi^X_j \equiv \phi(\mathbf{x}^X, t')$. For $t < R$, Eq. (4.4.4) is non-zero, since the Feynman propagator does not vanish for spacelike separations. This is known as the Fermi two-atom problem[1], since it seems to imply that source atom can affect the detector atom despite being causally disconnected.

This is not that case. The key to resolving this issue is to realise that the measurement on the detector atom should be a local measurement. The measurement only determines the state of atom $D$; it does not measure atom $S$ or the state of the field. It is no surprise that such an impossible simultaneous 'measurement' would be acausal. This means that $\mathbb{P}_{\text{Fermi}}$ is the probability of just one of several processes that should be considered. What should actually vanish for spacelike separations is the sum of the probabilities for all processes that depend on the state of $S$.

[1] Fermi actually claimed that the result was zero for spacelike separations [111], since he erroneously approximated an integral over positive frequencies by one over both positive and negative frequencies [112].

This underscores the importance of developing a formalism in which causality is explicit. It also highlights the advantage of our formalism in its ability to inclusively sum over all final states in certain Hilbert spaces. We now tackle the Fermi problem using our manifestly causal framework.

If we define

$$M^X(t'_j) \coloneqq \sum_{mn} \mu^X_{mn}\, e^{i\Omega^X_{mn} t'_j}\, |m^X\rangle \langle n^X| , \quad \text{for } X \in \{S, D\} , \tag{4.4.6}$$

then one can write the interaction Hamiltonian in the form

$$H_{\text{int}}(t'_j) = M^S(t'_j)\, \phi^S_j + M^D(t'_j)\, \phi^D_j . \tag{4.4.7}$$

As $M^X(t'_j)$ and $\phi^X_j$ are both Hermitian operators, the interaction Hamiltonian is of the same form as Eq. (4.2.1), with: the constants $g_Y = g_Z = 1$; $\mathscr{H}^\phi$ operators $A^X = \phi^S_j$ and $\mathcal{A}^X = \phi^D_j$; $\mathscr{H}^S$ operators $M^S(t'_j)$; and $\mathscr{H}^D$ operators $M^D(t'_j)$. The field operators are evaluated at a specific point, so there is no need for an integral over space. We can use Eq. (4.2.14), once we define the initial and final states.

We would like to determine the sensitivity of the detector atom to changes in the initial state of the source atom, so it makes sense to consider the initial density matrix,

$$\rho_0 = \gamma\, |i_p\rangle \langle i_p| + (1-\gamma)\, |i_g\rangle \langle i_g| \ , \tag{4.4.8}$$

where

$$|i_p\rangle = |p^S\rangle \otimes |g^D\rangle \otimes |0^\phi\rangle \equiv |p^S\, g^D\, 0^\phi\rangle \ , \tag{4.4.9}$$

$$|i_g\rangle = |g^S\rangle \otimes |g^D\rangle \otimes |0^\phi\rangle \equiv |g^S\, g^D\, 0^\phi\rangle \ . \tag{4.4.10}$$

Accordingly, the state $|i_p\rangle$ corresponds to an excited source atom, whilst $|i_g\rangle$ corresponds to a ground state source atom; both $|i_p\rangle$ and $|i_g\rangle$ specify that the detector atom is in its ground state and the field has no excitations. Eq. (4.4.8) therefore means that the system is in a mixed state between atom $S$ being in an excited state or ground state. The amount of mixing between either possibility is defined by the constant parameter, $\gamma$.

As stated previously, we are considering a local measurement. This means that our measurement only determines the state of atom $D$, and we must leave the states of atom $S$ and the field unspecified. This information is encoded entirely in an effect operator of the form

$$E = \sum_{n,\alpha} |n^S\, q^D\, \alpha^\phi\rangle \langle n^S\, q^D\, \alpha^\phi| = \mathbb{I}^S\, |q^D\rangle \langle q^D|\, \mathbb{I}^\phi\ , \tag{4.4.11}$$

where all possible states of atom $S$ and the field have been summed over to give the identity operator, $\mathbb{I}$, in their respective Hilbert spaces.

We can now use Eqs. (4.1.9) and (4.1.12) to compute the probabilities that we measure this final state, given that atom $S$ is either excited or in its ground state. We expect signals between atom $S$ and atom $D$ to travel slower than the speed of light, and so it is variations to the initial state (Eq. (4.4.8)), controlled by $\gamma$, which we expect to be causal. Consequently, we define the sensitivity of the detector as

$$\sigma_{pg} := \frac{\mathrm{d}\mathbb{P}}{\mathrm{d}\gamma} = \mathbb{P}_p - \mathbb{P}_g\ , \tag{4.4.12}$$

where, using Eq. (4.4.8),

$$\mathbb{P} = \gamma\, \mathbb{P}_p + (1-\gamma)\, \mathbb{P}_g\, , \tag{4.4.13}$$

and, using Eq. (4.1.9),

$$\mathbb{P}_{p,g} = \sum_{j=0}^{\infty} \int_0^t \mathrm{d}t'_1 \mathrm{d}t'_2 \ldots \mathrm{d}t'_j \Theta_{12\ldots j}\, \langle i_{p,g}\,|\, \mathcal{F}_j\,|\, i_{p,g}\rangle\ , \tag{4.4.14}$$

for initial time $t_{\text{in}} = 0$. Defining the detector sensitivity in this way means that any contributions which are independent of changes to the initial state (such as those due to vacuum fluctuations) will cancel. The first non-zero contribution to $\sigma_{pg}$ arises at fourth order [33]:

$$\begin{aligned}
\langle i_p|\, \mathcal{F}_4\, |i_p\rangle \;\supset\; & \langle p^S g^D 0^\phi|\ \tfrac{1}{16}\Big( E^D_{12} E^S_{\underline{3}\underset{\circ}{\underline{4}}} \mathcal{E}^{DDSS}_{\underline{1}\underline{2}3\underset{\bullet}{\underline{4}}} + E^D_{13} E^S_{\underline{2}\underset{\circ}{\underline{4}}} \mathcal{E}^{DSDS}_{\underline{1}2\underline{3}\underset{\bullet}{\underline{4}}} + E^D_{1\underset{\circ}{\underline{4}}} E^S_{\underline{2}3} \mathcal{E}^{DSSD}_{\underline{1}2\underline{3}\underset{\bullet}{\underline{4}}} \Big)\ |p^S g^D 0^\phi\rangle \\
=\; & \tfrac{1}{16} \langle E^D_{12}\rangle \Big( \langle E^S_{\underline{3}4}\rangle \langle \mathcal{E}^{DDSS}_{\underline{1}\underline{2}3\underline{4}}\rangle + \langle E^S_{\underline{3}\underline{4}}\rangle \langle \mathcal{E}^{DDSS}_{\underline{1}\underline{2}34}\rangle \Big) \\
& + \tfrac{1}{16} \langle E^D_{13}\rangle \Big( \langle E^S_{\underline{2}4}\rangle \langle \mathcal{E}^{DSDS}_{\underline{1}2\underline{3}\underline{4}}\rangle + \langle E^S_{\underline{2}\underline{4}}\rangle \langle \mathcal{E}^{DSDS}_{\underline{1}2\underline{3}4}\rangle \Big) \\
& + \tfrac{1}{16} \langle E^D_{14}\rangle\ \langle E^S_{\underline{2}3}\rangle\ \langle \mathcal{E}^{DSSD}_{\underline{1}2\underline{3}\underline{4}}\rangle + \tfrac{1}{16} \langle E^D_{1\underline{4}}\rangle\ \langle E^S_{\underline{2}3}\rangle\ \langle \mathcal{E}^{DSSD}_{\underline{1}2\underline{3}4}\rangle \qquad (4.4.15) \\
=\; & 2 \sum_n |\mu^S_{pn}|^2\, |\mu^D_{qg}|^2\ \times \\
& \Big\{ \cos \Omega^D_{qg} t'_{12} \Big( \sin \Omega^S_{pn} t'_{34}\, \Delta^{DS(\mathrm{H})}_{24} + \cos \Omega^S_{pn} t'_{34}\, \Delta^{DS(\mathrm{R})}_{24} \Big) \Delta^{DS(\mathrm{R})}_{13}
\end{aligned}$$

$$+ \cos\Omega^D_{qg}t'_{12}\Big( \sin\Omega^S_{pn}t'_{34}\,\Delta^{DS(\mathrm{H})}_{14} + \cos\Omega^S_{pn}t'_{34}\,\Delta^{DS(\mathrm{R})}_{14}\Big)\Delta^{DS(\mathrm{R})}_{23}$$

$$+ \cos\Omega^D_{qg}t'_{13}\Big( \sin\Omega^S_{pn}t'_{24}\,\Delta^{DS(\mathrm{H})}_{34} + \cos\Omega^S_{pn}t'_{24}\,\Delta^{DS(\mathrm{R})}_{34}\Big)\Delta^{DS(\mathrm{R})}_{12}$$

$$+ \sin\Omega^S_{pn}t'_{23}\Big( \cos\Omega^D_{qg}t'_{14}\,\Delta^{SD(\mathrm{H})}_{34} + \sin\Omega^D_{qg}t'_{14}\,\Delta^{SD(\mathrm{R})}_{34}\Big)\Delta^{DS(\mathrm{R})}_{12}\Big\} \ . \tag{4.4.16}$$

where $t'_{ij} = t'_i - t'_j$ and the symbol $\supset$ indicates we have only kept the terms which will be non-zero upon calculating $\sigma_{pg} = \mathbb{P}_p - \mathbb{P}_g$. The retarded propagator ($\Delta^{XY(\mathrm{R})}_{ij}$) and the Hadamard function ($\Delta^{XY(\mathrm{H})}_{ij}$) are defined as follows:

$$\Delta^{XY(\mathrm{R})}_{ij} \equiv \Theta_{ij}\,\langle 0|\ \frac{1}{i}\big[\phi^X_i, \phi^Y_j\big]\ |0\rangle \ , \tag{4.4.17}$$

$$\Delta^{XY(\mathrm{H})}_{ij} \equiv \langle 0|\ \big\{\phi^X_i, \phi^Y_j\big\}\ |0\rangle \tag{4.4.18}$$

Between Eqs. (4.4.15) and (4.4.16), the expectation values have been evaluated; details on how these have been computed can be found in Appendices A and B of Ref. [33].

The revelation of this method is that every term in Eq. (4.4.16) contains a retarded propagator, $\Delta^{DS(\mathrm{R})}_{ij}$, with $0 < t'_j < t'_i < t$. This implies that every term in $\sigma_{pg}$ vanishes when $t < R$. Whilst the above expressions are fixed order, this statement holds for all orders [33]. This is entirely consistent with the demands of causality: measurements of the state of the detector atom are completely insensitive to changes in the initial state of the source atom, for times less than the time it would take for light (information) to travel between the two atoms. Thus, this new method of QFT calculations gives manifestly causal results for local measurements.

# Chapter 5

# Manifestly Causal Particle Scattering

## 5.1 Introduction

In this chapter, we apply the manifestly causal formalism explained in Chapter 4 to particle scattering processes. We consider inclusive final states where we demand only that there are no initial-state particles in the final state. Since all of the expressions are at the probability level, they are algebraically more complicated than in the usual amplitude-level approach. Therefore, we work with simpler scalar field theories, which offer an instructive analogy to gauge theories such as quantum electrodynamics (QED) and quantum chromodynamics (QCD). The result is a new probability-level, diagrammatic method for calculating scattering probabilities in which the retarded propagator plays a key role, making causality explicit. The work in this chapter was published in Ref. [1].

The appearance of retarded propagators may offer a novel window on infrared (IR) divergences, which arise in part due to low-energy (soft) massless particles having arbitrarily large wavelengths that have effects over infinite separations. As explained in Section 2.3, retarded propagators are zero for spacelike separations and carry a distinct analytic structure compared to the Feynman propagator, and this approach may help quell these infinite-distance contributions. The Bloch-Nordsieck (BN) [27] and the Kinoshita-Lee-Nauenberg (KLN) [28, 29] theorems ensure that, while individual amplitudes may diverge, physical observables (like cross sections and decay rates) remain finite when real and virtual emissions are properly summed, as we saw in Chapter 3. Our formalism allows one to sum implicitly over all relevant final states from the beginning of the calculation and thereby explore the IR behaviour from a new angle.

An example of a typical gauge theory process of interest is electron-positron ($e^+e^-$)

pair annihilation to a quark-antiquark ($q\bar{q}$) pair, mediated by a photon ($\gamma$), with gluon ($g$) corrections to the final state. The Feynman diagram for this process is pictured on the right of Fig. 5.1. This process is studied in detail, using the traditional quantum field theory (QFT) approach, in Chapter 3. A scalar analogue of this process is $\psi\psi \to \chi \to \phi\phi$ with $h$ corrections to the final state, where $\psi, \chi, \phi$, and $h$ are all real scalar fields, as pictured on the left of Fig. 5.1. In this chapter, we first consider the decay process $\chi \to X$ (Sections 5.3 and 5.4), where $X$ represents a general final state, before turning to the annihilation process $\psi\psi \to X$ (Section 5.5). For both processes, we are fully inclusive over final states, except for demanding that the final state contains no initial-state particles. The work in this chapter allows us to demonstrate how our new formalism plays out in a simple example, without the technical complications of spinor and gauge structure.

Section 5.2 defines a new diagrammatic method and outlines the general rules for generating the set of diagrams relevant to any scalar-field scattering process for which the final state contains anything except initial-state particles. Sections 5.3-5.5 contain algebraic derivations of the complete sets of relevant diagrams without using the rules. We observe that distinct final states are intrinsically summed over and are not separable in our results, and the retarded self-energy [113, 114] emerges naturally. The calculations look complicated in the intermediate steps, but they reduce considerably. This suggests the existence of a more fundamental set of rules. Section 5.6 concludes.

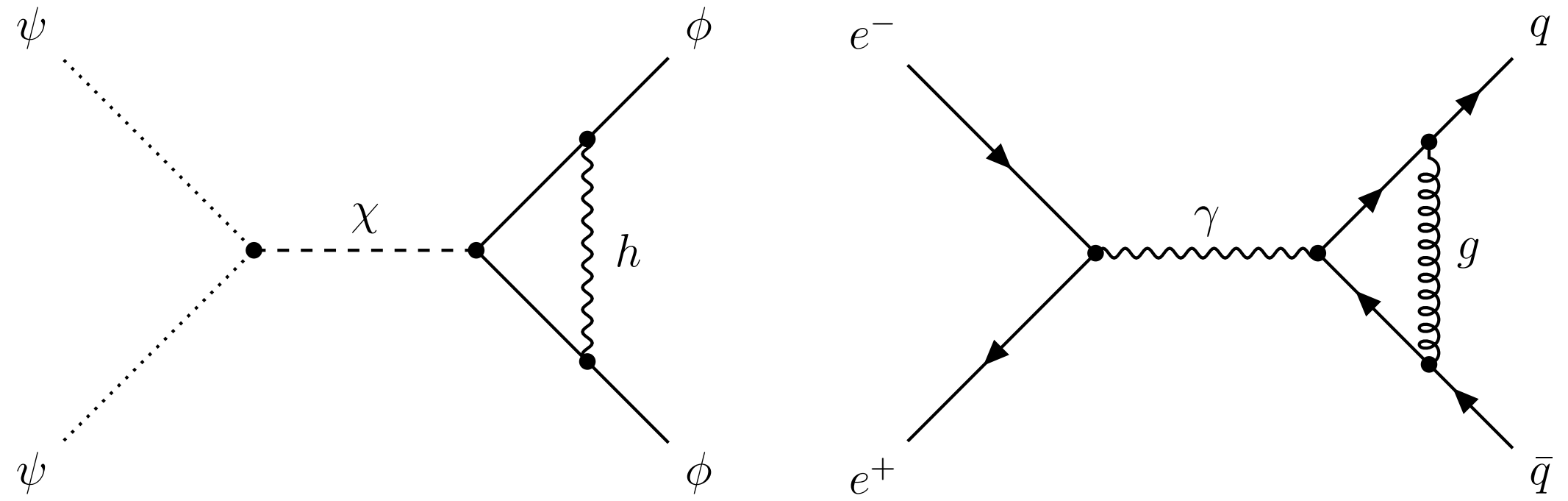

**Fig. 5.1.** Examples of traditional, amplitude-level diagrams for the processes of interest. **Right:** The gauge theory process $e^-e^+ \to \gamma \to q\bar{q}$, with gluon ($g$) corrections. **Left:** An analogous toy-model in which $\psi, \chi, \phi$ and $h$ are all real scalar fields.

## 5.2 Diagrams and Rules

Here, we present a set of rules akin to Feynman rules, but applicable at the probability level. Let us first explain the notation used in the diagrams. For a diagram which contributes to $\langle \mathcal{F}_n \rangle$, where $\langle \cdots \rangle \equiv \langle\, i \,|\cdots|\, i \,\rangle$:

- Times run from $t_n$ (earliest) on the left to $t_1$ (latest) on the right and are denoted by dots.
- Lines between two dots represent propagators.
- Lines only connected to one dot represent plane waves from the initial state at $t \to -\infty$ (these are drawn vertically).
- Feynman propagators are denoted by black lines in all Hilbert spaces.
- Retarded propagators are denoted by red lines in all Hilbert spaces, with an arrow pointing from the earlier time to the later time.
- Plane waves highlighted in yellow carry momentum $p$ out of the diagram (i.e., carry momentum $-p$ into the vertex), and unhighlighted plane waves carry momentum $p$ into the diagram.

For the scalar field theories considered, we indicate $\psi$-space contributions with dotted lines, $\chi$-space contributions with dashed lines, $\phi$-space contributions with solid lines, and $h$-space contributions with wiggly lines, as pictured in Fig. 5.1.

We can now describe the rules which can be used to construct a diagram that contributes to $\langle \mathcal{F}_n \rangle$. The following rules apply for a calculation which demands that there are no initial-state particles in the final state and which is fully inclusive over all other Hilbert spaces (i.e., $E^{\mathscr{H}_i} = \mathbb{I}$ in all Hilbert spaces, except for the Hilbert space of the initial-state fields, in which $E^{\mathscr{H}_i} = |0\rangle\langle 0|$). The rules are conjectured based on explicit calculations in Sections 5.3–5.5. The rules are:

1. Draw $n$ time points, which will later become vertices.
2. Connect one initial state (regardless of the number of field quanta) to $t_1$ (the latest time)[1].

[1]This assumes that the Hilbert space associated with the initial state *only* has particles in the initial state, i.e., the calculation is to $\mathcal{O}(g_i^2)$ in the coupling constant associated with the initial state Hilbert space, $g_i$.

3. Connect the other initial state to any other time.

4. Ensure each vertex which is not connected to an initial state is connected to a later vertex by a retarded propagator of any field.

5. Connect the remaining vertices with retarded or Feynman propagators, using the interaction vertices from the interaction Hamiltonian.

6. If the two initial states are connected by a chain of retarded propagators (where each point on the chain connects to a later time), then momentum $p$ must flow into the $t_1$ initial state and out of the other initial state. Otherwise, choose either momentum flow.

We highlight that, as a consequence of Rule 4, every vertex is connected to at least one of the initial states by a chain of retarded propagators.

The pre-factor for a diagram which contributes to $\mathbb{P}$ through $\langle \mathcal{F}_n \rangle$ can be calculated as follows:

- $\times 1/2\omega_i$ for each initial particle of energy $\omega_i$ in the initial state $|i\rangle$
- $\times S g_i$ for each vertex, where $g_i$ is the associated coupling constant and $S$ is the symmetry factor if particles are indistinguishable
- $\times(-1)(-i)^n$ (these factors come from the incoming plane waves and the total number of commutators in the effect operators)
- $\times(-1)$ for each retarded propagator
- $\times 1/2$ for each loop of retarded propagators between two (and only two) vertices
- $\times 1/2$ for each loop of Feynman propagators between two (and only two) vertices
- $\times 1/2$ for each instance of a Feynman propagator connected to the same vertex at each end (Feynman propagator tadpole)
- Integrate over time-ordered spacetime points, $\int \mathrm{d}^4x_1 \ldots \mathrm{d}^4x_n \Theta_{12\ldots n}$, where $\Theta_{12\ldots n}$ is defined by Eq. (4.1.10)

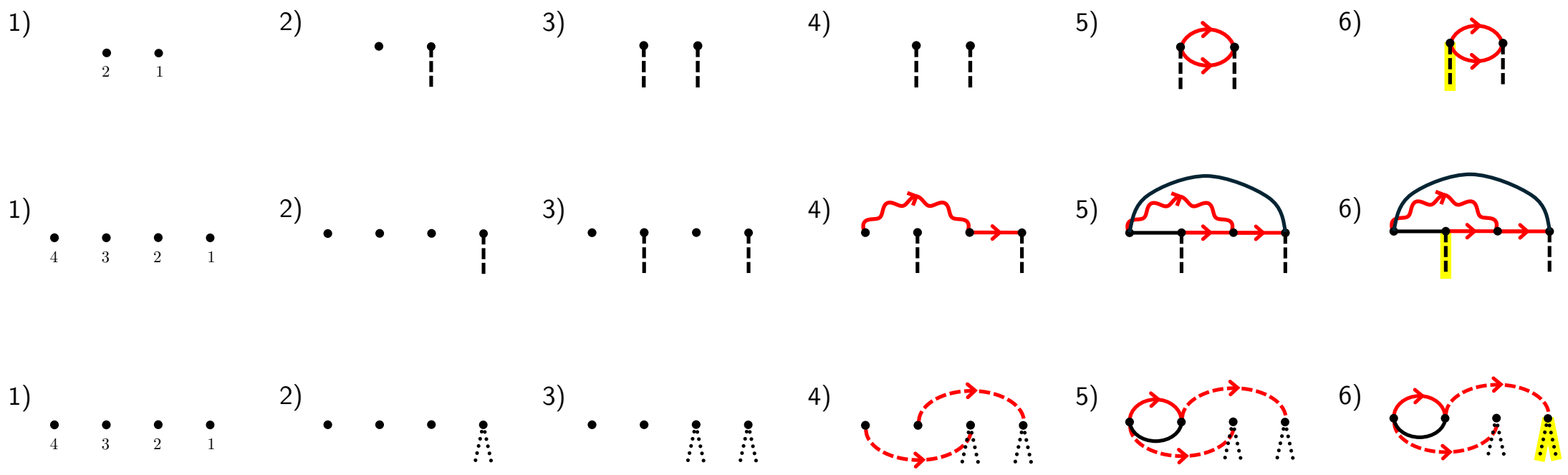

**Fig. 5.2.** Examples of generating a diagram using the rules. Each step corresponds to each rule number. Feynman propagators are denoted by black lines and retarded propagators are denoted by red lines with an arrow pointing from the earlier time to the later time. **Top:** A diagram for the tree-level contribution to $\chi \to \phi\phi$. Step 4 is redundant since there are no non-initial-state vertices. The pre-factor for this diagram is $2g_\chi^2/2\omega_{\mathbf{p}}$. This specific diagram is generated in Section 5.3, and can be seen in Fig. 5.4 (the final diagram). **Middle:** A diagram for a first-order $h$ correction to $\chi \to \phi\phi$. The pre-factor for this diagram is $16g_\chi^2 g_h^2/2\omega_{\mathbf{p}}$. This specific diagram is generated in Section 5.4, and can be seen in Fig. 5.9 (4th line from the top, 2nd diagram from the right). **Bottom:** A diagram for the annihilation process to $\psi\psi \to X$. The pre-factor for this diagram is $16g_\chi^2 g_h^2/4\omega_{\mathbf{p}_1}\omega_{\mathbf{p}_2}$. This specific diagram is generated in Section 5.5, and can be seen in Fig. 5.11 (7th line from the top, 1st diagram from the left).

The total probability, $\mathbb{P}$, is calculated by the summation of all possible diagrams. The total number of unique diagrams in Sections 5.3-5.5 matches the number determined using these rules, serving as a cross-check for their validity. To illustrate how to construct a diagram, three examples are given in Fig. 5.2.

## 5.3 Inclusive Decay: Lowest Order

To calculate the decay probability of a particle of a scalar field, $\chi$, consider the interaction Hamiltonian,

$$H_{\text{int}}(t_j) = \int \mathrm{d}^3\mathbf{x}_j \, \left(g_\chi \phi_j^2 \chi_j + g_h \phi_j^2 h_j\right) \, , \tag{5.3.1}$$

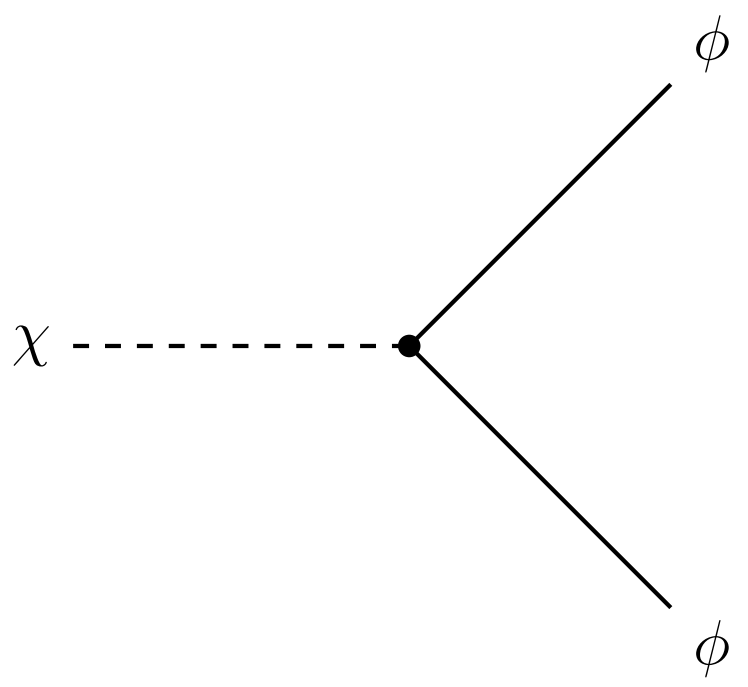

**Fig. 5.3.** The traditional tree-level Feynman diagram for the process $\chi \to \phi\phi$.

where $\phi_j \equiv \phi(x_j)$ and $h_j \equiv h(x_j)$ are scalar fields, and $g_\chi$ and $g_h$ are coupling constants. This interaction Hamiltonian, like all other interaction Hamiltonians in this chapter, is of the exact form of Eq. (4.2.1), and therefore we can use the causal formalism described in Chapter 4. Since $\chi$ only couples to $\phi^2$, the decay $\chi \to \phi\phi$ is the only lowest-order process which can occur. The scalar field, $h$, will be used in Section 5.4 to investigate higher-order corrections. The traditional, amplitude-level Feynman diagram for the lowest-order process is shown in Fig. 5.3.

The initial state is encoded in an initial density operator

$$\rho_0 = |0^h\, p^\chi\, 0^\phi\rangle\, \langle 0^h\, p^\chi\, 0^\phi|\,. \tag{5.3.2}$$

This means that the system initially has one $\chi$ particle, of momentum $p$, and no other field excitations. For the final state, we consider any number of excitations in the $\phi$ and $h$ fields and no excitations in the $\chi$ field. This is encoded in an effect operator

$$\begin{aligned} E &= \sum_{n,\alpha} |n^h\, 0^\chi\, \alpha^\phi\rangle\, \langle n^h\, 0^\chi\, \alpha^\phi| = \mathbb{I}^h\ |0^\chi\rangle\, \langle 0^\chi|\ \mathbb{I}^\phi\ , \\ \text{i.e.,}\quad & E^h = \mathbb{I}^h\ , \quad E^\chi = |0^\chi\rangle\, \langle 0^\chi|\ , \quad E^\phi = \mathbb{I}^\phi\ , \end{aligned} \tag{5.3.3}$$

where we have used the completeness of states to sum over all states in the $\phi$ and $h$ Hilbert spaces, resulting in identity operators. It is in this way that inclusive observables take a straightforward form at the probability level.

This set-up allows us to use the rules in Section 5.2, but we first calculate the transition probability directly. The result can then be used to verify the rules.

If we wanted to completely reproduce the expressions of the traditional scattering calculation, we would set $E^h = |0^h\rangle\, \langle 0^h|$ (at $\mathcal{O}(g_h^0)$), $E^\chi = |0^\chi\rangle\, \langle 0^\chi|$, and $E^\phi = |q_1^\phi, q_2^\phi\rangle\, \langle q_1^\phi, q_2^\phi|$. The probability would then factorise into amplitude and conjugate amplitude, as in Chapter 3.

Returning to the effect operator in Eq. (5.3.3), we now define the notation analogous to Eq. (4.2.5) and (4.2.6):

$$\begin{aligned} \text{Hilbert space } \phi\text{:}\qquad & \mathcal{E}^{\cdots h}_{\cdots k} := \frac{1}{i}\big[\mathcal{E}^{\cdots}_{\cdots}, g_h\phi_k^2\big], \qquad \mathcal{E}^{\cdots h}_{\cdots \underline{k}} := \big\{\mathcal{E}^{\cdots}_{\cdots}, g_h\phi_k^2\big\}, \\ & \mathcal{E}^{\cdots \chi}_{\cdots k} := \frac{1}{i}\big[\mathcal{E}^{\cdots}_{\cdots}, g_\chi\phi_k^2\big], \qquad \mathcal{E}^{\cdots \chi}_{\cdots \underline{k}} := \big\{\mathcal{E}^{\cdots}_{\cdots}, g_\chi\phi_k^2\big\}, \\ \text{Hilbert space } h\text{:}\qquad & E^h_{\ldots k} := \frac{1}{i}\big[E^h_{\ldots}, h_k\big], \qquad E^h_{\ldots \underline{k}} := \big\{E^h_{\ldots}, h_k\big\}, \end{aligned} \tag{5.3.4}$$

$$\text{Hilbert space } \chi: \qquad E^{\chi}_{\ldots k} := \frac{1}{i}[E^{\chi}_{\ldots}, \chi_k], \qquad E^{\chi}_{\ldots \underline{k}} := \{E^{\chi}_{\ldots}, \chi_k\}, \tag{5.3.5}$$

where $\mathcal{E} \equiv E^{\phi}$. The general formula for $\mathcal{F}_n$ is given by Eq. (4.2.14),

$$\mathcal{F}_n = 2^{-n} \int \prod_{\kappa=1}^{n} \left(\mathrm{d}^3\mathbf{x}_\kappa\right) \sum_{a=0}^{n} E^{h}_{(\underset{\circ}{1} \cdots \underset{\circ}{a}} \, E^{\chi}_{\underset{\circ}{a+1} \cdots \underset{\circ}{n})} \, \mathcal{E}^{(h \ldots h \;\; \chi \ldots \chi)}_{(\underset{\bullet}{1} \cdots \underset{\bullet}{a} \, \underset{\bullet}{a+1} \cdots \underset{\bullet}{n})} \, . \tag{5.3.6}$$

This expression has been written in a condensed form using the underdot notation defined in Eq. (4.2.8) and the parentheses notation defined in Eq. (4.2.13).

The lowest order non-zero contribution to $\mathbb{P}$ is from $\mathcal{F}_2$. Using Eq. (5.3.6),

$$\mathcal{F}_2 = \frac{1}{4} \int \mathrm{d}^3\mathbf{x}_1 \mathrm{d}^3\mathbf{x}_2 \left( E^{h}_{\underset{\circ}{1}\underset{\circ}{2}} \, E^{\chi} \, \mathcal{E}^{hh}_{\underset{\bullet}{1}\underset{\bullet}{2}} + E^{h}_{\underset{\circ}{1}} \, E^{\chi}_{\underset{\circ}{2}} \, \mathcal{E}^{h\chi}_{\underset{\bullet}{1}\underset{\bullet}{2}} + E^{h}_{\underset{\circ}{2}} \, E^{\chi}_{\underset{\circ}{1}} \, \mathcal{E}^{\chi h}_{\underset{\bullet}{1}\underset{\bullet}{2}} + E^{h} \, E^{\chi}_{\underset{\circ}{1}\underset{\circ}{2}} \, \mathcal{E}^{\chi\chi}_{\underset{\bullet}{1}\underset{\bullet}{2}} \right) \, . \tag{5.3.7}$$

Due to our choices of the initial density operator, $\rho_0$, and the effect operator, $E$, only the final term contributes once we take the expectation value of $\mathcal{F}_2$, as in Eq. (4.1.9). The first subscript index of $\mathcal{E}^{\cdots}_{\cdots}$ must be underlined, since a commutator with $\mathcal{E} = \mathbb{I}^{\phi}$ would vanish. Thus, the only contribution from Eq. (5.3.7) is

$$\begin{aligned} \mathcal{F}_2 &= \frac{1}{4} \int \mathrm{d}^3\mathbf{x}_1 \mathrm{d}^3\mathbf{x}_2 \, E^{\chi}_{1\underset{\circ}{2}} \, \mathcal{E}^{\chi\chi}_{\underline{1}\underset{\bullet}{2}} \\ &= \frac{1}{4} \int \mathrm{d}^3\mathbf{x}_1 \mathrm{d}^3\mathbf{x}_2 \, \left( E^{\chi}_{12} \, \mathcal{E}^{\chi\chi}_{\underline{1}\underline{2}} + E^{\chi}_{1\underline{2}} \, \mathcal{E}^{\chi\chi}_{\underline{1}2} \right) \, . \end{aligned} \tag{5.3.8}$$

Therefore,

$$\langle i | \, \mathcal{F}_2 \, | i \rangle = \frac{1}{4} \int \mathrm{d}^3\mathbf{x}_1 \mathrm{d}^3\mathbf{x}_2 \, \left( \langle p^{\chi} | \, E^{\chi}_{12} \, | p^{\chi} \rangle \, \langle 0^{\phi} | \, \mathcal{E}^{\chi\chi}_{\underline{1}\underline{2}} \, | 0^{\phi} \rangle + \langle p^{\chi} | \, E^{\chi}_{1\underline{2}} \, | p^{\chi} \rangle \, \langle 0^{\phi} | \, \mathcal{E}^{\chi\chi}_{\underline{1}2} \, | 0^{\phi} \rangle \right) \, . \tag{5.3.9}$$

Evaluating the $\chi$-space expectation value first gives

$$\begin{aligned} \langle p^{\chi} | \, E^{\chi}_{1\underset{\circ}{2}} \, | p^{\chi} \rangle &= \langle 0^{\chi} | \, a(p) \, E^{\chi}_{1\underset{\circ}{2}} \, a^{\dagger}(p) \, | 0^{\chi} \rangle \\ &= \frac{1}{i} \left( \frac{1}{i} \right)^{(1-\eta_2)/2} \langle 0^{\chi} | \, a(p) \left[ \left[ | 0^{\chi} \rangle \langle 0^{\chi} | , \chi_1 \right], \chi_2 \right]_{\eta_2} a^{\dagger}(p) \, | 0^{\chi} \rangle \\ &= - \left( \frac{1}{i} \right)^{(3-\eta_2)/2} \langle 0^{\chi} | \, a(p) \, \chi_1 \, | 0^{\chi} \rangle \langle 0^{\chi} | \, \chi_2 \, a^{\dagger}(p) \, | 0^{\chi} \rangle \\ &\qquad + \eta_2 \, \langle 0^{\chi} | \, a(p) \, \chi_2 \, | 0^{\chi} \rangle \langle 0^{\chi} | \, \chi_1 \, a^{\dagger}(p) \, | 0^{\chi} \rangle \; , \end{aligned} \tag{5.3.10}$$

where we have introduced the following compact notation for commutators and anti-commutators:

$$\begin{aligned} \big[A, B\big]_\eta &= AB + \eta BA = (1+\eta)\, AB - \eta\big[A, B\big] \\ &= \begin{cases} \big[A, B\big] & \text{if } \eta = -1\,, \\ \big\{A, B\big\} & \text{if } \eta = +1\,, \end{cases} \end{aligned} \tag{5.3.11}$$

such that $\eta_2 = 1$ for $E^{\chi}_{1\underline{2}}$ and $\eta_2 = -1$ for $E^{\chi}_{12}$. Using the mode expansion of the $\chi$ field,

$$\begin{aligned} \langle p^\chi |\, E^{\chi}_{1\underset{\circ}{2}} \,| p^\chi\rangle &= -\left(\frac{1}{i}\right)^{(3-\eta_2)/2} \int \frac{\mathrm{d}^3\mathbf{k}_1\, e^{ik_1\cdot x_1}}{(2\pi)^3\sqrt{2\omega_1}} \int \frac{\mathrm{d}^3\mathbf{k}_2\, e^{-ik_2\cdot x_2}}{(2\pi)^3\sqrt{2\omega_2}} \\ &\qquad\qquad \times \langle 0^\chi|\, a(p)\, a^\dagger(k_1)\, |0^\chi\rangle\, \langle 0^\chi|\, a(k_2)\, a^\dagger(p)\, |0^\chi\rangle \\ &\quad + \eta_2 \;\text{[same as above but with } 1 \leftrightarrow 2] \\ &= -\left(\frac{1}{i}\right)^{(3-\eta_2)/2} \int \frac{\mathrm{d}^3\mathbf{k}_1}{\sqrt{2\omega_1}}\, e^{ik_1\cdot x_1} \int \frac{\mathrm{d}^3\mathbf{k}_2}{\sqrt{2\omega_2}}\, e^{-ik_2\cdot x_2}\, \delta^3(\mathbf{p}-\mathbf{k}_1)\, \delta^3(\mathbf{k}_2-\mathbf{p}) \\ &\quad + \eta_2 \;\text{[same as above but with } 1 \leftrightarrow 2] \\ &= \left(\frac{1}{i}\right)^{(3-\eta_2)/2} \frac{1}{2\omega_{\mathbf{p}}} \big(-e^{ip\cdot x_1}\, e^{-ip\cdot x_2} + \eta_2\, e^{ip\cdot x_2}\, e^{-ip\cdot x_1}\big)\,, \end{aligned} \tag{5.3.12}$$

where $\omega_1 \equiv p_1^0$, $\omega_2 \equiv p_2^0$, and $\omega_1 = \omega_2 = \omega_{\mathbf{p}} \equiv \sqrt{\mathbf{p}^2 + m_\chi^2}$ is the energy of the incoming particle.

Now consider the $\phi$-space expectation value,

$$\begin{aligned} \langle 0^\phi|\, \mathcal{E}^{\chi\chi}_{\underline{1}\underset{\bullet}{2}} \,|0^\phi\rangle &= \left(\frac{1}{i}\right)^{(1-\epsilon_2)/2} g_\chi^2\, \langle 0^\phi|\, \Big[\Big\{\mathbb{I}^\phi, \phi_1^2\Big\}, \phi_2^2\Big]_{\epsilon_2} |0^\phi\rangle \\ &= 2\left(\frac{1}{i}\right)^{(1-\epsilon_2)/2} g_\chi^2\, \langle 0^\phi|\, \big[\phi_1^2, \phi_2^2\big]_{\epsilon_2} |0^\phi\rangle\,, \end{aligned} \tag{5.3.13}$$

where $\epsilon_2 = -\eta_2$. The effect operator being the identity operator has resulted in the commutator of products of fields. On this occasion, the commutator is simple, but Eq. (4.1.12) shows that this will result in nested commutators when we consider higher orders. To evaluate these, we use Eq. (23) from Ref. [115],

$$[f(\phi_1, \ldots, \phi_{n-1}), g(\phi_n)] = -\underbrace{\sum_{k_1} \cdots \sum_{k_{n-1}}}\left(\prod_{i=1}^{n-1} \frac{(-[\phi_i, \phi_n])^{k_i}}{k_i!}\right)\left(\partial^{k_1}_{\phi_1} \ldots \partial^{k_{n-1}}_{\phi_{n-1}} f \partial^{k}_{\phi_n} g\right)\,, \tag{5.3.14}$$

where

$$k = \sum_{i=1}^{n-1} k_i \tag{5.3.15}$$

and the indices within the underbrace ($\underbrace{\quad}$) are not all simultaneously zero.

Using Eqs. (5.3.11) and (5.3.14),

$$\begin{aligned}
\langle 0^\phi | \mathcal{E}^{\chi\chi}_{\underset{\bullet}{12}} |0^\phi\rangle &= \left(\frac{1}{i}\right)^{(1-\epsilon_2)/2} \Big[ 2\, g_\chi^2\, \langle 0^\phi | (1+\epsilon_2)\phi_1^2\phi_2^2 |0^\phi\rangle \\
&\qquad - 2\, g_\chi^2\, \epsilon_2 \langle 0^\phi | \Big( \Delta^\phi_{12} \partial_1 \left[\phi_1^2\right] \partial_2 \left[\phi_2^2\right] \\
&\qquad\qquad - \frac{1}{2} (\Delta^\phi_{12})^2 \partial_1^2 \left[\phi_1^2\right] \partial_2^2 \left[\phi_2^2\right] \Big) |0^\phi\rangle \Big] \\
&= \left(\frac{1}{i}\right)^{(1-\epsilon_2)/2} \Big[ 2\, g_\chi^2\, (1+\epsilon_2) \langle 0^\phi | \phi_1^2\phi_2^2 |0^\phi\rangle \\
&\qquad - 8\, g_\chi^2\, \epsilon_2\, \Delta^\phi_{12} \langle 0^\phi | \phi_1\phi_2 |0^\phi\rangle + 4\, g_\chi^2\, \epsilon_2\, (\Delta^\phi_{12})^2 \Big] \,,
\end{aligned} \tag{5.3.16}$$

where we have introduced the Pauli-Jordan function, $\Delta^\phi_{xy}$ (Eq. (2.3.3)). Substituting Eqs. (5.3.12) and (5.3.16) into Eq. (5.3.9), we have

$$\begin{aligned}
\langle i | \mathcal{F}_2 | i \rangle &= \frac{g_\chi^2}{4i^2} \int \mathrm{d}^3\mathbf{x}_1 \mathrm{d}^3\mathbf{x}_2 \Bigg( \frac{1}{2\omega_\mathbf{p}} \left( -e^{ip\cdot x_1}\, e^{-ip\cdot x_2} - \, e^{ip\cdot x_2}\, e^{-ip\cdot x_1} \right) \\
&\qquad \Big( 4\, \langle 0^\phi | \phi_1^2\phi_2^2 |0^\phi\rangle - 8\, \Delta_{12} \langle 0^\phi | \phi_1\phi_2 |0^\phi\rangle + 4\; \Delta_{12}^2 \Big) \\
&\qquad + \frac{1}{2\omega_\mathbf{p}} \left( -e^{ip\cdot x_1}\, e^{-ip\cdot x_2} + \, e^{ip\cdot x_2}\, e^{-ip\cdot x_1} \right) \\
&\qquad \Big( 8\, \Delta_{12} \langle 0^\phi | \phi_1\phi_2 |0^\phi\rangle - 4\, \Delta_{12}^2 \Big) \Bigg) \\
&= \frac{g_\chi^2}{2\omega_\mathbf{p}} \int \mathrm{d}^3\mathbf{x}_1 \mathrm{d}^3\mathbf{x}_2 \Bigg( e^{ip\cdot x_1}\, e^{-ip\cdot x_2} \langle 0^\phi | \phi_1^2\phi_2^2 |0^\phi\rangle \\
&\qquad + \, e^{ip\cdot x_2}\, e^{-ip\cdot x_1} \Big( \langle 0^\phi | \phi_1^2\phi_2^2 |0^\phi\rangle - 4\, \Delta^\phi_{12} \langle 0^\phi | \phi_1\phi_2 |0^\phi\rangle + 2\, (\Delta^\phi_{12})^2 \Big) \Bigg) \,.
\end{aligned} \tag{5.3.17}$$

Since Eq. (4.1.9) includes $\Theta_{12}$, we can write

$$\Theta_{12} \langle 0^\phi | \phi_1^n \phi_2^m |0^\phi\rangle = \Theta_{12} \langle 0^\phi | \mathrm{T}\{\phi_1^n\phi_2^m\} |0^\phi\rangle \,, \tag{5.3.18}$$

where $n$ and $m$ are positive integers, and use Wick's Theorem (see Section 2.1.3) to

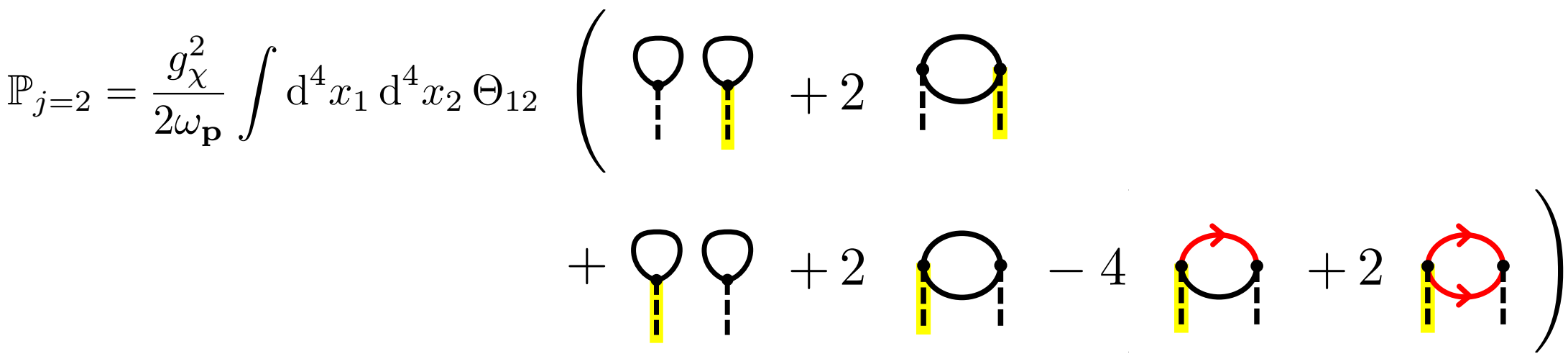

**Fig. 5.4.** A diagrammatic representation of the tree-level decay probability (Eq. (5.3.19)).

obtain

$$\mathbb{P}_{j=2} = \frac{g_\chi^2}{2\omega_{\mathbf{p}}} \int \mathrm{d}^4x_1 \mathrm{d}^4x_2\, \Theta_{12} \left( e^{ip\cdot x_1}\, e^{-ip\cdot x_2} \left( F^\phi_{11} F^\phi_{22} + 2\, F^\phi_{12} F^\phi_{12} \right) \right.$$
$$\left. + \; e^{ip\cdot x_2}\, e^{-ip\cdot x_1} \left( F^\phi_{11} F^\phi_{22} + 2\, F^\phi_{12} F^\phi_{12} - 4\, R^\phi_{12} F^\phi_{12} + 2\, (R^\phi_{12})^2 \right) \right) , \qquad (5.3.19)$$

where we have introduced the Feynman propagator (Eq. (2.3.5)), $F^\phi_{xy}$, and the retarded propagator (Eq. (2.3.6)), $R^\phi_{xy}$. Eq. (5.3.19) is expressed diagrammatically in Fig. 5.4. These are exactly the diagrams which would be generated using the rules in Section 5.2.

Since

$$\Theta_{12}\, F^\phi_{12} = \Theta_{12}\, \Delta^{\phi(>)}_{12} , \qquad (5.3.20)$$
$$\Theta_{12}\, R^\phi_{12} = \Theta_{12}\, \Delta^\phi_{12} = \Theta^\phi_{12} \left( \Delta^{\phi(>)}_{12} - \Delta^{\phi(<)}_{12} \right) , \qquad (5.3.21)$$

where the Wightman functions, $\Delta^{\phi(>)}_{xy}$ and $\Delta^{\phi(<)}_{xy}$, are defined in Eqs. (2.3.1) and (2.3.2), we can rewrite Eq. (5.3.19) as

$$\mathbb{P}_{j=2} = \frac{2g_\chi^2}{2\omega_{\mathbf{p}}} \int \mathrm{d}^4x_1 \mathrm{d}^4x_2\, \Theta_{12} \left( e^{ip\cdot x_1}\, e^{-ip\cdot x_2} \left( \Delta^{\phi(>)}_{12} \right)^2 + e^{ip\cdot x_2}\, e^{-ip\cdot x_1} \left( \Delta^{\phi(<)}_{12} \right)^2 \right) . \qquad (5.3.22)$$

We have ignored the $F^\phi_{11} F^\phi_{22}$ terms since these contributions are not allowed kinematically (alternatively, we could have chosen $E^\phi = \mathbb{I}^\phi - |0^\phi\rangle\, \langle 0^\phi|$ such that these contributions would cancel). The part of the integrand after the $\Theta$-function is symmetric under the exchange of $t_1 \leftrightarrow t_2$, so we can replace the $\Theta$-function with $(1/2!)$. By substituting in the momentum-integral representations of the Wightman functions, the decay probability can then be expressed as

$$\mathbb{P}_{j=2} = \frac{2g_\chi^2}{2\omega_{\mathbf{p}}} \int \frac{\mathrm{d}^3\mathbf{q}_1}{(2\pi)^3} \frac{1}{2\omega_{q_1}} \int \frac{\mathrm{d}^3\mathbf{q}_2}{(2\pi)^3} \frac{1}{2\omega_{q_2}} \left[ \delta^4(p - q_1 - q_2) \right]^2 (2\pi)^8 . \qquad (5.3.23)$$

This is the usual tree-level probability for scalar particle decay, integrated over all final state momenta [54].

## 5.4 Inclusive Decay: First Order

In this section, we will generate diagrams which represent the first-order $h$ corrections to the decay probability in Section 5.3. As such, the Hamiltonian, initial state, and effect operator remain the same as Section 5.3. Some examples of the traditional, amplitude-level Feynman diagrams for this process are shown in Fig. 5.5.

To calculate these corrections, we require

$$\mathcal{F}_4 = \frac{1}{16} \int \mathrm{d}^3\mathbf{x}_1\, \mathrm{d}^3\mathbf{x}_2\, \mathrm{d}^3\mathbf{x}_3\, \mathrm{d}^3\mathbf{x}_4 \sum_{a=0}^{4} E^{h}_{(\underset{\circ}{1}\cdots\underset{\circ}{a}}\, E^{\chi}_{\underset{\circ}{a+1}\cdots\underset{\circ}{4})}\, \mathcal{E}^{(h\ldots h\;\;\chi\ldots\chi)}_{(\underset{\bullet}{1}\cdots\underset{\bullet}{a}\,\underset{\bullet}{a+1}\cdots\underset{\bullet}{4})} \,. \tag{5.4.1}$$

The relevant term is of order $\mathcal{O}\big(g_h^2 g_\chi^2\big)$, which is the $a = 2$ term,

$$\mathcal{F}_4^{(a=2)} = \frac{1}{16} \int \mathrm{d}^3\mathbf{x}_1\, \mathrm{d}^3\mathbf{x}_2\, \mathrm{d}^3\mathbf{x}_3\, \mathrm{d}^3\mathbf{x}_4 E^{h}_{(\underset{\circ}{1}\underset{\circ}{2}}\, E^{\chi}_{\underset{\circ}{3}\underset{\circ}{4})}\, \mathcal{E}^{(hh\chi\chi)}_{(\underset{\bullet}{1}\underset{\bullet}{2}\underset{\bullet}{3}\underset{\bullet}{4})} \,. \tag{5.4.2}$$

For clarity and simplicity, we will write a term with a general permutation of indices as

$$E^{h}_{\underset{\circ}{i}\underset{\circ}{j}}\, E^{\chi}_{\underset{\circ}{k}\underset{\circ}{l}}\, \mathcal{E}^{(hh\chi\chi)}_{\underset{\bullet}{1}\underset{\bullet}{2}\underset{\bullet}{3}\underset{\bullet}{4}} \,. \tag{5.4.3}$$

Since $E^h = \mathbb{I}^h$ and $E^\phi = \mathbb{I}^\phi$, the first index of the $\phi$-space and $h$-space operators must be underlined (a commutator with the identity operator is zero):

$$E^{h}_{\underline{i}\underset{\circ}{j}}\, E^{\chi}_{\underset{\circ}{k}\underset{\circ}{l}}\, \mathcal{E}^{(hh\chi\chi)}_{\underline{1}\underset{\bullet}{2}\underset{\bullet}{3}\underset{\bullet}{4}} \,. \tag{5.4.4}$$

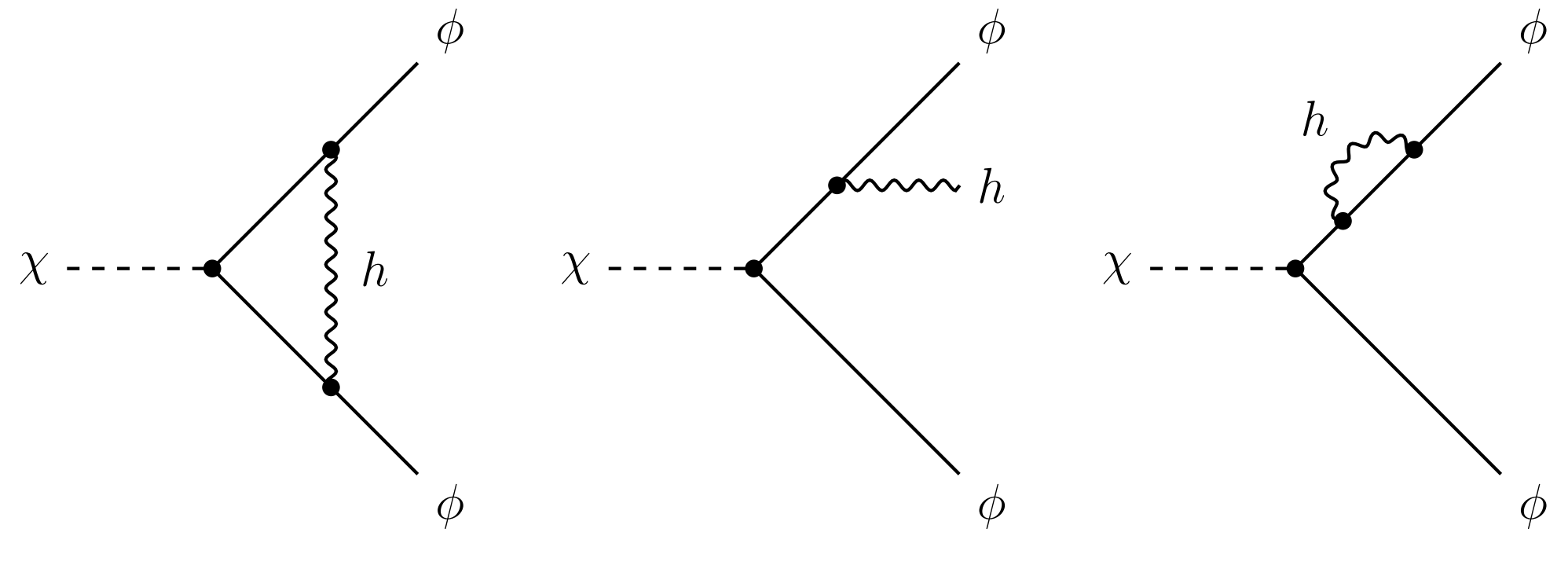

**Fig. 5.5.** The Feynman diagrams for the process $\chi \to \phi\phi$ with first-order $h$ corrections.

Now that the index '1' is underlined on the $\phi$-space operator, it cannot be underlined on either of the other operators. This means that $i \neq 1$. However, the index 1 must be the first index of an operator, so $k = 1$:

$$E^h_{\underset{\circ}{\underline{i}j}}\, E^\chi_{\underset{\circ}{1\underline{l}}}\, \mathcal{E}^{(hh\chi\chi)}_{\underset{\bullet\bullet\bullet}{\underline{1}234}} \;. \tag{5.4.5}$$

In total, there will always be four non-underlined indices. From Eqs. (5.3.4) and (5.3.5), we know that each non-underlined index results in a factor of $1/i$. Since $(1/i)^4 = 1$, we will simply ignore these factors of $i$ in this section. Expanding Eq. (5.4.5) gives

$$\begin{aligned} E^h_{\underset{\circ}{\underline{i}j}}\, E^\chi_{\underset{\circ}{1\underline{l}}}\, \mathcal{E}^{(hh\chi\chi)}_{\underset{\bullet\bullet\bullet}{\underline{1}234}} &= E^h_{\underline{2}3}\, E^\chi_{14}\, \mathcal{E}^{\chi hh\chi}_{\underline{1}2\underline{3}\underline{4}} + E^h_{\underline{2}3}\, E^\chi_{1\underline{4}}\, \mathcal{E}^{\chi hh\chi}_{\underline{1}2\underline{3}4} + E^h_{\underline{2}\underline{3}}\, E^\chi_{14}\, \mathcal{E}^{\chi hh\chi}_{\underline{1}23\underline{4}} + E^h_{\underline{2}\underline{3}}\, E^\chi_{1\underline{4}}\, \mathcal{E}^{\chi hh\chi}_{\underline{1}234} \\ &+ E^h_{\underline{2}4}\, E^\chi_{13}\, \mathcal{E}^{\chi h\chi h}_{\underline{1}2\underline{3}\underline{4}} + E^h_{\underline{2}4}\, E^\chi_{1\underline{3}}\, \mathcal{E}^{\chi h\chi h}_{\underline{1}23\underline{4}} + E^h_{\underline{2}\underline{4}}\, E^\chi_{13}\, \mathcal{E}^{\chi h\chi h}_{\underline{1}2\underline{3}4} + E^h_{\underline{2}\underline{4}}\, E^\chi_{1\underline{3}}\, \mathcal{E}^{\chi hh\chi}_{\underline{1}234} \\ &+ E^h_{\underline{3}4}\, E^\chi_{12}\, \mathcal{E}^{\chi\chi hh}_{\underline{1}\underline{2}3\underline{4}} + E^h_{\underline{3}4}\, E^\chi_{1\underline{2}}\, \mathcal{E}^{\chi\chi hh}_{\underline{1}23\underline{4}} + E^h_{\underline{3}\underline{4}}\, E^\chi_{12}\, \mathcal{E}^{\chi\chi hh}_{\underline{1}\underline{2}34} + E^h_{\underline{3}\underline{4}}\, E^\chi_{1\underline{2}}\, \mathcal{E}^{\chi\chi hh}_{\underline{1}234} \;. \end{aligned} \tag{5.4.6}$$

Taking the expectation value of this expression, we can evaluate term-by-term and one Hilbert space at a time. The expectation value of the $h$-space operator is straightforward:

$$\langle 0^h |\, E^h_{\underset{\circ}{\underline{i}j}}\, | 0^h \rangle = \langle 0^h |\, \Big[ \Big\{ \mathbb{I}^h, h_i \Big\}, h_j \Big]_{\lambda_j} | 0^h \rangle = 2\, \langle 0^h |\, \big[ h_i, h_j \big]_{\lambda_j} | 0^h \rangle \;. \tag{5.4.7}$$

This is either proportional to a Pauli-Jordan function ($\lambda_j = -1$) or a Hadamard function ($\lambda_j = +1$) (Eq. (2.3.4)). We can evaluate the expectation value of the $\chi$-space operator as we did in Section 5.3, giving

$$\langle p^\chi |\, E^\chi_{\underset{\circ}{1\underline{l}}}\, | p^\chi \rangle = \frac{1}{2\omega_{\mathbf{p}}} \big( -e^{ip\cdot x_1}\, e^{-ip\cdot x_l} + \eta_l\, e^{ip\cdot x_l}\, e^{-ip\cdot x_1} \big) \,. \tag{5.4.8}$$

The $\phi$-space operator in Eq. (5.4.5) is more complicated. It can be written as

$$\mathcal{E}_{\underset{\bullet\bullet\bullet}{\underline{1}234}} = \Big[ \mathcal{E}_{\underset{\bullet\bullet}{\underline{1}23}}, \phi_4^2 \Big]_{\epsilon_4} , \tag{5.4.9}$$

where the Hilbert spaces are no longer denoted, and we understand that there will be an overall factor of $g_h^2 g_\chi^2$. We shall therefore first consider

$$\mathcal{E}_{\underset{\bullet\bullet}{\underline{1}23}} = \Big[ \mathcal{E}_{\underset{\bullet}{\underline{1}2}}, \phi_3^2 \Big]_{\epsilon_3} = (1 + \epsilon_3)\, \mathcal{E}_{\underset{\bullet}{\underline{1}2}}\, \phi_3^2 - \epsilon_3 \Big[ \mathcal{E}_{\underset{\bullet}{\underline{1}2}}, \phi_3^2 \Big] \,, \tag{5.4.10}$$

where we have temporarily ignored coupling constants, and use Eqs. (5.3.14) and (5.3.16) to get

$$
\begin{aligned}
\mathcal{E}_{\underline{1}\underset{\bullet}{2}\underset{\bullet}{3}} = 2\,(1+\epsilon_3)\,\Big(\,(1+\epsilon_2)\,\phi_1^2\,\phi_2^2\,\phi_3^2 - 4\epsilon_2\Delta_{12}\phi_1\phi_2\phi_3^2 + 2\epsilon_2\,(\Delta_{12})^2\phi_3^2\Big) \\
+ 4\epsilon_3\Big(-2\,\Delta_{13}\Big((1+\epsilon_2)\phi_1\,\phi_2^2\,\phi_3 - 2\,\epsilon_2\Delta_{12}\phi_2\,\phi_3\Big) \\
- 2\,\Delta_{23}\Big((1+\epsilon_2)\phi_1^2\,\phi_2\,\phi_3 - 2\,\epsilon_2\Delta_{12}\phi_1\,\phi_3\Big) \\
+ 4\Delta_{13}\Delta_{23}\Big((1+\epsilon_2)\Delta_{13}^2\phi_2^2 - \epsilon_2\Delta_{12}\Big) \\
+ (1+\epsilon_2)(\Delta_{13})^2\,\phi_2^2 - (1+\epsilon_2)(\Delta_{23})^2\,\phi_1^2\Big)\,, \qquad (5.4.11)
\end{aligned}
$$

where $\Delta_{xy} \equiv \Delta_{xy}^{\phi}$. Substituting this into Eq. (5.4.9) and using Eq. (5.3.14),

$$
\begin{aligned}
\mathcal{E}_{\underline{1}\underset{\bullet}{2}\underset{\bullet}{3}\underset{\bullet}{4}} = g_h^2 g_\chi^2 \Bigg\{ 2\,(1+\epsilon_4)\Bigg(\,(1+\epsilon_3)\Big[(1+\epsilon_2)\,\phi_1^2\,\phi_2^2\,\phi_3^2\,\phi_4^2 - 4\,\epsilon_2\Delta_{12}\,\phi_1\,\phi_2\,\phi_3^2\,\phi_4^2 \\
+ 2\epsilon_2\,(\Delta_{12})^2\,\phi_3^2\,\phi_4^2\Big] \\
+ 2\epsilon_3\Big[-2\,\Delta_{13}\Big((1+\epsilon_2)\,\phi_1\,\phi_2^2\,\phi_3\,\phi_4^2 - 2\,\epsilon_2\,\Delta_{12}\,\phi_2\,\phi_3\,\phi_4^2\Big) \\
- 2\,\Delta_{23}\Big((1+\epsilon_2)\,\phi_1^2\,\phi_2\,\phi_3\,\phi_4^2 - 2\,\epsilon_2\,\Delta_{12}\,\phi_1\,\phi_3\,\phi_4^2\Big) \\
+ 4\,\Delta_{13}\,\Delta_{23}\Big((1+\epsilon_2)\,\phi_1\,\phi_2\,\phi_4^2 - \epsilon_2\,\Delta_{12}\,\phi_4^2\Big) \\
+ (1+\epsilon_2)(\Delta_{13})^2\,\phi_2^2\,\phi_4^2 + (1+\epsilon_2)\,(\Delta_{23})^2\phi_1^2\,\phi_4^2\Big]\Bigg) \\
+ 4\,\epsilon_4\Bigg(-2\,\Delta_{14}\Big[(1+\epsilon_3)\Big((1+\epsilon_2)\,\phi_1\,\phi_2^2\,\phi_3^2\,\phi_4 - 2\,\epsilon_2\,\Delta_{12}\,\phi_2\,\phi_3^2\,\phi_4\Big) \\
+ 2\,\epsilon_3\Big(-(1+\epsilon_2)\,\Delta_{13}\,\phi_2^2\,\phi_3\,\phi_4 \\
- 2\,\Delta_{23}\big((1+\epsilon_2)\,\phi_1\,\phi_2\,\phi_3\,\phi_4 - \epsilon_2\,\Delta_{12}\,\phi_3\,\phi_4\big) \\
+ 2\,(1+\epsilon_2)\,\Delta_{13}\,\Delta_{23}\,\phi_2\,\phi_4 + (1+\epsilon_2)\,(\Delta_{23})^2\,\phi_1\,\phi_4\Big)\Big] \\
- 2\,\Delta_{24}\Big[(1+\epsilon_3)\Big((1+\epsilon_2)\,\phi_1^2\,\phi_2\,\phi_3^2\,\phi_4 - 2\,\epsilon_2\,\Delta_{12}\,\phi_1\,\phi_3^2\,\phi_4\Big) \\
+ 2\,\epsilon_3\Big(-(1+\epsilon_2)\,\Delta_{23}\,\phi_1^2\,\phi_3\,\phi_4 \\
- 2\,\Delta_{13}\big((1+\epsilon_2)\,\phi_1\,\phi_2\,\phi_3\,\phi_4 - \epsilon_2\,\Delta_{12}\,\phi_3\,\phi_4\big) \\
+ 2\,(1+\epsilon_2)\,\Delta_{13}\,\Delta_{23}\,\phi_1\,\phi_4 + (1+\epsilon_2)\,(\Delta_{13})^2\,\phi_2\,\phi_4\Big)\Big] \\
- 2\,\Delta_{34}\Big[(1+\epsilon_3)\Big((1+\epsilon_2)\,\phi_1^2\,\phi_2^2\,\phi_3\,\phi_4 - 4\,\epsilon_2\,\Delta_{12}\,\phi_1\,\phi_2\,\phi_3\,\phi_4
\end{aligned}
$$

$$
\begin{aligned}
&\qquad + 2\,\epsilon_2\,(\Delta_{12})^2\,\phi_3\,\phi_4\Big) \\
&\quad + 2\,\epsilon_3\Big(-\Delta_{13}\big((1+\epsilon_2)\,\phi_1\,\phi_2^2\,\phi_4 - 2\,\epsilon_2\,\Delta_{12}\,\phi_2\,\phi_4\big) \\
&\qquad - \Delta_{23}\big((1+\epsilon_2)\,\phi_1^2\,\phi_2\,\phi_4 - 2\,\epsilon_2\,\Delta_{12}\,\phi_1\,\phi_4\big)\Big)\Bigg] \\
&+ 4\,\Delta_{14}\,\Delta_{24}\Bigg[(1+\epsilon_3)\Big((1+\epsilon_2)\,\phi_1\,\phi_2\,\phi_3^2 - \epsilon_2\,\Delta_{12}\,\phi_3^2\Big) \\
&\qquad + 2\,\epsilon_3\,(1+\epsilon_2)\Big(-\Delta_{13}\,\phi_2\,\phi_3 - \Delta_{23}\,\phi_1\,\phi_3 + \Delta_{13}\,\Delta_{23}\Big)\Bigg] \\
&+ 4\,\Delta_{14}\,\Delta_{34}\Bigg[(1+\epsilon_3)\Big((1+\epsilon_2)\,\phi_1\,\phi_2^2\,\phi_3 - 2\,\epsilon_2\,\Delta_{12}\,\phi_2\,\phi_3\Big) \\
&\qquad + \epsilon_3\Big(-(1+\epsilon_2)\,\Delta_{13}\,\phi_2^2 - 2\,\Delta_{23}\big((1+\epsilon_2)\,\phi_1\,\phi_2 - \epsilon_2\,\Delta_{12}\big)\Big)\Bigg] \\
&+ 4\,\Delta_{24}\,\Delta_{34}\Bigg[(1+\epsilon_3)\Big((1+\epsilon_2)\,\phi_1^2\,\phi_2\,\phi_3 - 2\,\epsilon_2\,\Delta_{12}\,\phi_1\,\phi_3\Big) \\
&\qquad + \epsilon_3\Big(-(1+\epsilon_2)\,\Delta_{23}\,\phi_1^2 - 2\,\Delta_{13}\big((1+\epsilon_2)\,\phi_1\,\phi_2 - \epsilon_2\,\Delta_{12}\big)\Big)\Bigg] \\
&+ (\Delta_{14})^2\,(1+\epsilon_2)\Bigg[(1+\epsilon_3)\,\phi_2^2\,\phi_3^2 + 2\,\epsilon_3\Big(-2\,\Delta_{23}\,\phi_2\,\phi_3 + (\Delta_{23})^2\Big)\Bigg] \\
&+ (\Delta_{24})^2\,(1+\epsilon_2)\Bigg[(1+\epsilon_3)\,\phi_1^2\,\phi_3^2 + 2\,\epsilon_3\Big(-2\,\Delta_{13}\,\phi_1\,\phi_3 + (\Delta_{13})^2\Big)\Bigg] \\
&+ (\Delta_{34})^2\,(1+\epsilon_3)\Bigg[(1+\epsilon_2)\,\phi_1^2\,\phi_2^2 + 2\,\epsilon_2\Big(-2\,\Delta_{12}\,\phi_1\,\phi_2 + (\Delta_{12})^2\Big)\Bigg]\Bigg)\Bigg\}\,.
\end{aligned}
\tag{5.4.12}
$$

Using Eqs. (5.4.7), (5.4.8), and (5.4.12), we can evaluate the expectation value of each of the terms in Eq. (5.4.6) (see Appendix A). Note that for any given expectation value, terms with an odd number of $\Delta$s have a relative minus sign to those with an even number, owing to the minus sign before the commutator in Eq. (5.3.14). This is reflected in the pre-factor rules in Section 5.2.

Since

$$
\Theta_{ij\ldots n}\,\langle 0^\phi|\,\phi_i\phi_j\ldots\phi_n\,|0^\phi\rangle = \Theta_{ij\ldots n}\,\langle 0^\phi|\,\mathrm{T}\{\phi_i\phi_j\ldots\phi_n\}\,|0^\phi\rangle\;, \tag{5.4.13}
$$

we can use Wick's theorem to express products of fields as Feynman propagators. Each expectation value in Eq. (5.4.5) is shown algebraically in Appendix A (unsimplified). To convert these expectation values to probabilities, we simply use Eq. (4.1.9).

The diagrams can be grouped into different categories based on the topology of the diagram. The diagrams within each category are the same up to the exchange of times.

The categories are:

- Disconnected — One of the initial state $\chi$s produces a $\phi$ loop. The rest of the diagram is entirely disconnected to this initial state.
- Tadpoles — The $h$ propagator produces a $\phi$ loop.
- Oscillations — The initial state $\chi$s convert into the $h$ propagator via a loop of two $\phi$ propagators.
- Vertex — The $h$ propagator connects across the $h\phi^2$ vertex, connecting two different $\phi$ 'legs' of the diagram.
- Self-Energies — The $h$ propagator connects the same $\phi$ 'leg' of the diagram with itself.

Disconnected and tadpole diagrams will vanish upon considering correct momentum conservation. Oscillations can be made to have arbitrarily small contributions by varying the mass of the $h$-field compared to the mass of the $\chi$-field. In the analogous gauge theory process, $\gamma \to q\bar{q} \to g \to q\bar{q}$, oscillation diagrams equal zero due to colour conservation.

Figs. 5.6–5.10 show the results diagrammatically. The $h$-field Hadamard function has been rewritten in terms of Feynman and retarded propagators using Eq. (2.3.4) and the time-ordering $\Theta$-function. Duplicate diagrams either sum or cancel, reducing the total number of terms. These are exactly the diagrams which would be generated had we used the rules in Section 5.2.

$$\mathbb{P} \supset \int \mathrm{d}^4x_1\, \mathrm{d}^4x_2\, \mathrm{d}^4x_3\, \mathrm{d}^4x_4\, \Theta_{1234} \frac{2g_h^2 g_\chi^2}{2\omega_{\mathbf{p}}} \times$$

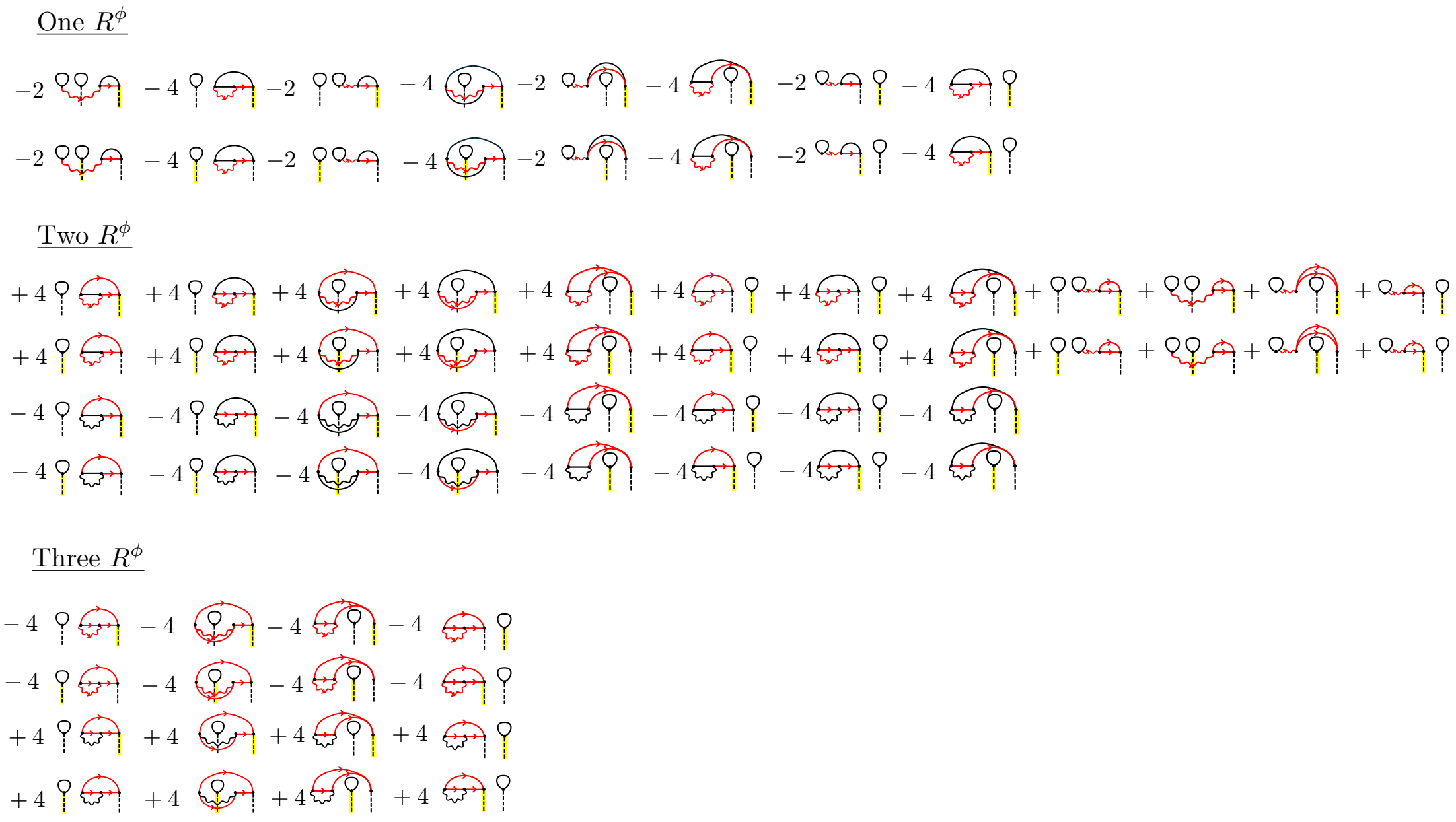

**Fig. 5.6.** Probability-level diagrammatic representation of **disconnected** diagrams.

$$\mathbb{P} \supset \int \mathrm{d}^4x_1\, \mathrm{d}^4x_2\, \mathrm{d}^4x_3\, \mathrm{d}^4x_4\, \Theta_{1234} \frac{2g_h^2 g_\chi^2}{2\omega_{\mathbf{p}}} \times$$

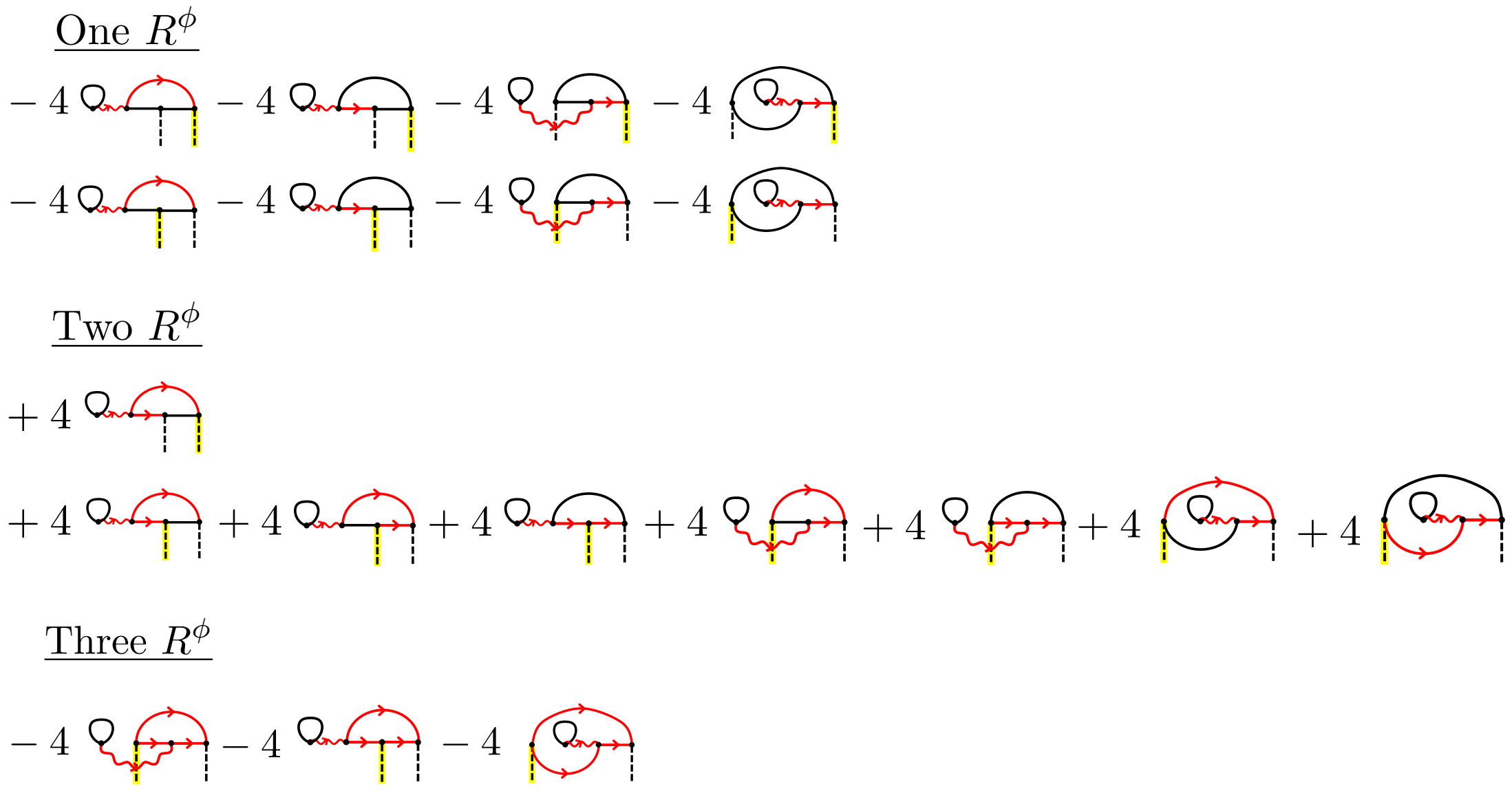

**Fig. 5.7.** Probability-level diagrammatic representation of **tadpole** diagrams.

$$\mathbb{P} \supset \int \mathrm{d}^4x_1\, \mathrm{d}^4x_2\, \mathrm{d}^4x_3\, \mathrm{d}^4x_4\, \Theta_{1234} \frac{2g_h^2 g_\chi^2}{2\omega_{\mathbf{p}}} \times$$

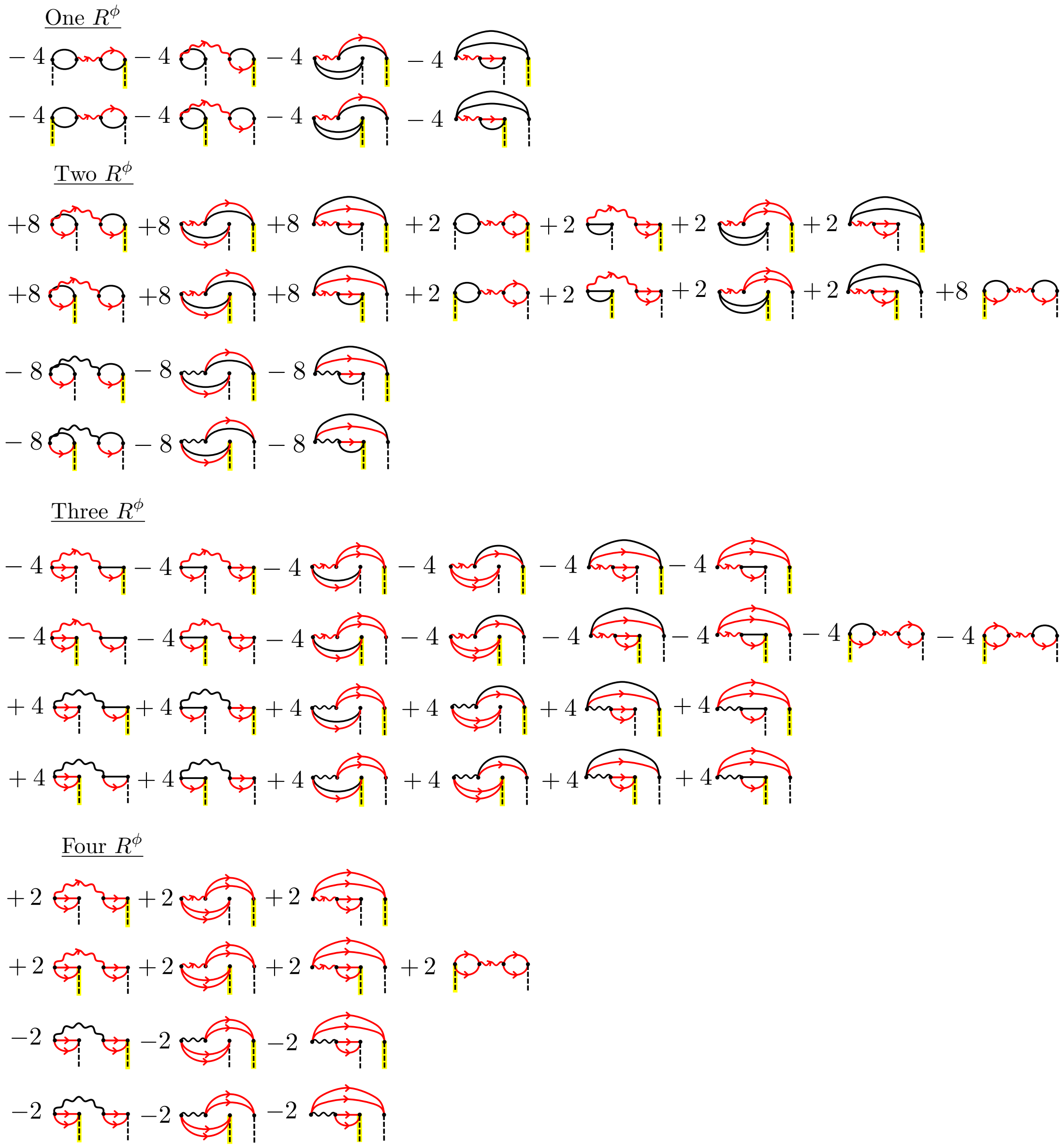

**Fig. 5.8.** Probability-level diagrammatic representation of **oscillation** diagrams.

$$\mathbb{P} \supset \int \mathrm{d}^4x_1\, \mathrm{d}^4x_2\, \mathrm{d}^4x_3\, \mathrm{d}^4x_4\, \Theta_{1234} \frac{2g_h^2 g_\chi^2}{2\omega_{\mathbf{p}}} \times$$

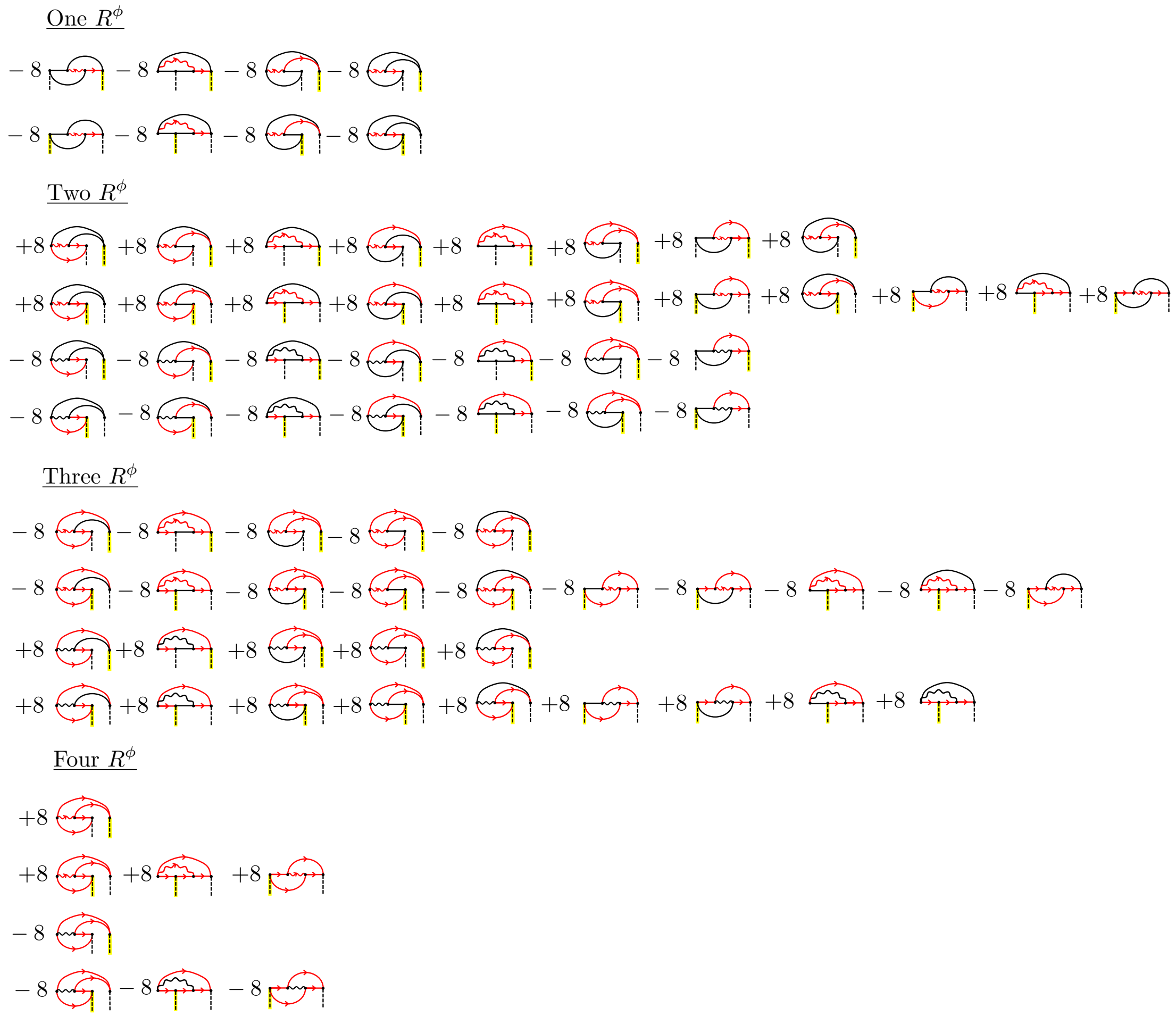

**Fig. 5.9.** Probability-level diagrammatic representation of **vertex** diagrams.

$$\mathbb{P} \supset \int \mathrm{d}^4x_1\, \mathrm{d}^4x_2\, \mathrm{d}^4x_3\, \mathrm{d}^4x_4\, \Theta_{1234} \frac{2g_h^2 g_\chi^2}{2\omega_{\mathbf{p}}} \times$$

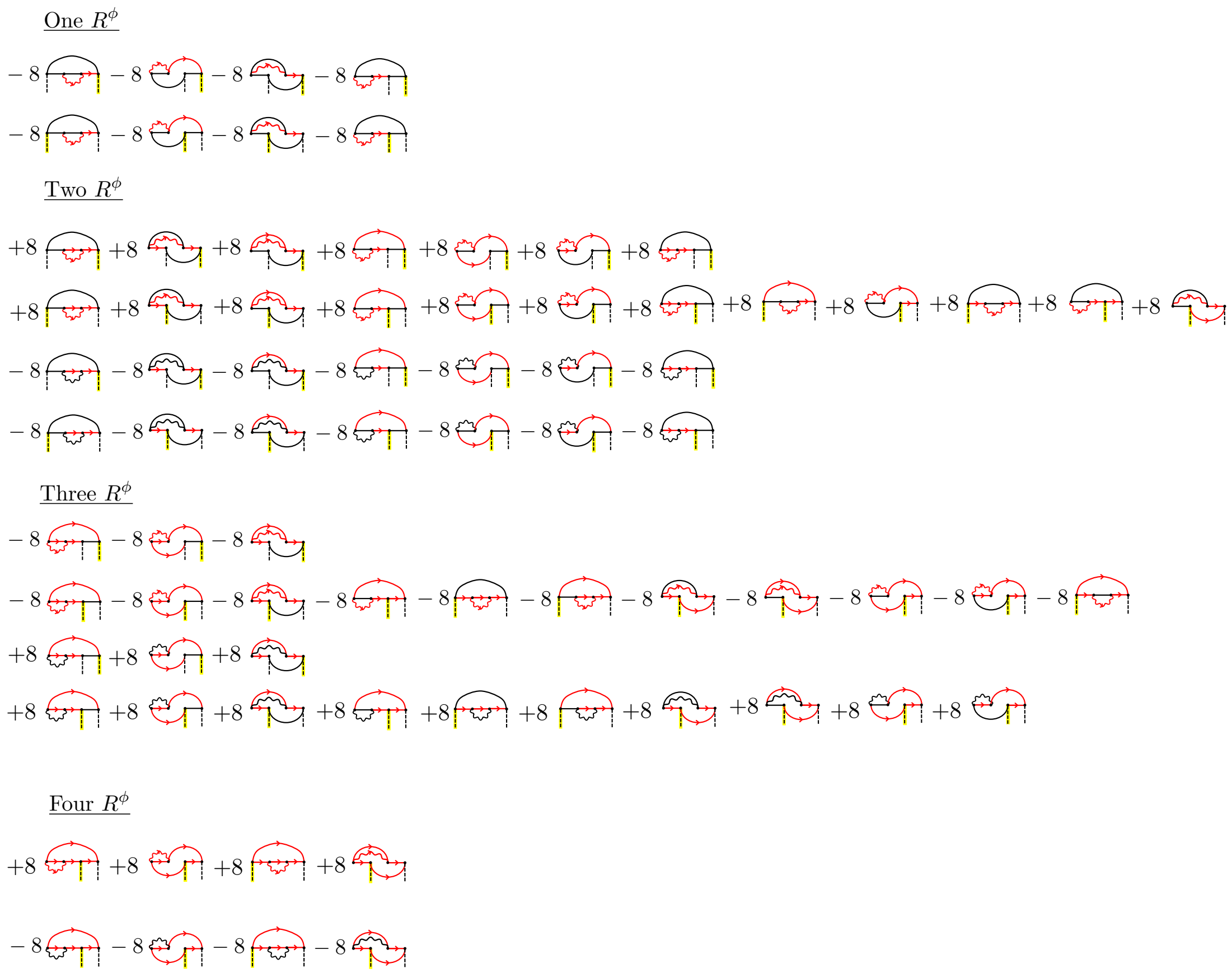

**Fig. 5.10.** Probability-level diagrammatic representation of **self-energy** diagrams.

Note that there are no distinct 'real emission' diagrams, pictured in the centre of Fig. 5.5, due to the inclusivity of the effect operator. Instead, these contributions are contained within the 'vertex' and 'self-energy' diagrams. Since IR divergences appear in the traditional approach as an artifact of separating vertex, self-energy, and real emission terms, one might be optimistic that the inclusivity of this approach may lead to inherent cancellation of IR divergences.

The total decay probability can be calculated as in Section 5.3. However, scalar $\phi^3$ theory is severely IR divergent in four dimensions [55], and attempting to evaluate the integrals will not serve a useful purpose. Instead, this section provides a proof-of-concept for future calculations in gauge theories (QED and QCD), in which we expect to find similar rules and diagrams, and the theory has a better-defined IR limit.

## 5.5 Inclusive Annihilation: First Order

We now consider an annihilation process between two particles described by the interaction Hamiltonian

$$H_{\text{int}}(t_j) = \int \mathrm{d}^3\mathbf{x}_j \, \left( g_\psi \psi_j^2 \chi_j + g_\phi \phi_j^2 \chi_j \right) \, , \tag{5.5.1}$$

where $\psi_j \equiv \psi(x_j)$ is another real scalar field. The relevant initial density matrix is

$$\rho_0 = |0^\chi \, p_1^\psi p_2^\psi \, 0^\phi\rangle \, \langle 0^\chi \, p_1^\psi p_2^\psi \, 0^\phi| \, , \tag{5.5.2}$$

i.e., the system initially has two $\psi$ particles, of momentum $p_1$ and $p_2$, and no other field excitations. We choose the effect operator

$$\begin{aligned} E &= \sum_{n,\alpha} |n^\chi \, 0^\psi \, \alpha^\phi\rangle \, \langle n^\chi \, 0^\psi \, \alpha^\phi| = \mathbb{I}^\chi \; |0^\psi\rangle \, \langle 0^\psi| \; \mathbb{I}^\phi \, , \\ \text{i.e.,} \quad E^\chi &= \mathbb{I}^\chi \, , \quad E^\psi = |0^\psi\rangle \, \langle 0^\psi| \, , \quad E^\phi = \mathbb{I}^\phi \, , \end{aligned} \tag{5.5.3}$$

such that we consider all possible final states with zero quanta of the $\psi$-field.

Since the structure of the Hamiltonian differs slightly to the decay studied in Sections 5.3 and 5.4, our notation is now

$$\begin{aligned} &\text{Hilbert space } \chi: && \mathcal{E}^{\cdots\psi}_{\ldots k} := \frac{1}{i}[\mathcal{E}^{\cdots}_{\ldots}, g_\psi \chi_k], && \mathcal{E}^{\cdots\psi}_{\ldots \underline{k}} := \{\mathcal{E}^{\cdots}_{\ldots}, g_\psi \chi_k\}, \\ & && \mathcal{E}^{\cdots\phi}_{\ldots k} := \frac{1}{i}[\mathcal{E}^{\cdots}_{\ldots}, g_\phi \chi_k], && \mathcal{E}^{\cdots\phi}_{\ldots \underline{k}} := \{\mathcal{E}^{\cdots}_{\ldots}, g_\phi \chi_k\}, \end{aligned} \tag{5.5.4}$$

$$\begin{aligned} &\text{Hilbert space } \psi: && E^{\psi}_{\ldots k} := \frac{1}{i}\big[E^{\psi}_{\ldots}, \psi_k^2\big], && E^{\psi}_{\ldots \underline{k}} := \big\{E^{\psi}_{\ldots}, \psi_k^2\big\}, \\ &\text{Hilbert space } \phi: && E^{\phi}_{\ldots k} := \frac{1}{i}\big[E^{\chi}_{\ldots}, \phi_k^2\big], && E^{\phi}_{\ldots \underline{k}} := \big\{E^{\phi}_{\ldots}, \phi_k^2\big\}, \end{aligned} \tag{5.5.5}$$

and $\mathcal{E} = E^\chi$.

The lowest-order contributions come from the process $\psi\psi \to \chi$, contained within $\mathcal{F}_2$, which are trivial and uninteresting. Instead, we consider the next order of contributions, which come from $\mathcal{F}_4$. Specifically, we isolate and examine the $\mathcal{O}\big(g_\psi^2 g_\phi^2\big)$ contributions, which correspond to the $a = 2$ term in $\mathcal{F}_4$,

$$\mathcal{F}_4^{(a=2)} = \frac{1}{16} \int \mathrm{d}^3\mathbf{x}_1 \, \mathrm{d}^3\mathbf{x}_2 \, \mathrm{d}^3\mathbf{x}_3 \, \mathrm{d}^3\mathbf{x}_4 E^{\psi}_{(\underset{\circ}{1}\underset{\circ}{2}} \, E^{\phi}_{\underset{\circ}{3}\underset{\circ}{4})} \, \mathcal{E}^{(\psi\psi\phi\phi)}_{(\underset{\bullet}{1}\underset{\bullet}{2}\underset{\bullet}{3}\underset{\bullet}{4})} \, . \tag{5.5.6}$$

Similar to the first-order $h$ corrections to the inclusive decay in Section 5.4, the inclusive effect operators $\mathcal{E} = \mathbb{I}^{\chi}$ and $E^{\phi} = \mathbb{I}^{\phi}$ result in the only non-zero terms in Eq. (5.5.6) having the general form

$$E^{\psi}_{\underset{\circ}{1}j}\, E^{\phi}_{\underset{\circ}{\underline{k}}l}\, \mathcal{E}^{\psi(\psi\phi\phi)}_{\underline{1}\underset{\bullet}{2}\underset{\bullet}{3}\underset{\bullet}{4}}\ . \tag{5.5.7}$$

As in Section 5.4, there will always be four non-underlined indices. Due to Eqs. (5.5.4) and (5.5.5), this will always result in a factor of $(1/i)^4 = 1$, so these factors of $i$ will be ignored for the rest of this section.

The expectation value of the $\psi$-space operator is

$$\begin{aligned}\langle p_1^{\psi} p_2^{\psi}|\, E^{\psi}_{\underset{\circ}{1}j}\, |p_1^{\psi} p_2^{\psi}\rangle &= \langle p_1^{\psi} p_2^{\psi}| \left[\left[|0^{\psi}\rangle\,\langle 0^{\psi}|\,, \psi_1^2\right], \psi_j^2\right]_{\eta_j} |p_1^{\psi} p_2^{\psi}\rangle \\ &= \frac{4}{(2\omega_{\mathbf{p}_1})(2\omega_{\mathbf{p}_2})}\big(-e^{ip_1\cdot x_1} e^{ip_2\cdot x_1} e^{-ip_1\cdot x_j} e^{-ip_2\cdot x_j} \\ &\qquad + \eta_j e^{ip_1\cdot x_j} e^{ip_2\cdot x_j} e^{-ip_1\cdot x_1} e^{-ip_2\cdot x_1}\big)\,.\end{aligned} \tag{5.5.8}$$

The expectation value of the $\phi$-space operator is

$$\begin{aligned}\langle 0^{\phi}|\, E^{\phi}_{\underset{\circ}{\underline{k}}l}\, |0^{\phi}\rangle &= \langle 0^{\phi}| \left[\left\{\mathbb{I}^{\phi}, \phi_k^2\right\}, \phi_l^2\right]_{\lambda_l} |0^{\phi}\rangle = 2\,\langle 0^{\phi}|\,\left[\phi_k^2, \phi_l^2\right]_{\lambda_l} |0^{\phi}\rangle \\ &= 2\,(1+\lambda_j)\,\langle 0^{\phi}|\,\phi_k^2\phi_l^2\,|0^{\phi}\rangle - 8\,\lambda_l\,\Delta^{\phi}_{kl}\,\langle 0^{\phi}|\,\phi_k\phi_l\,|0^{\phi}\rangle + 4\,\lambda_l\,(\Delta^{\phi}_{kl})^2\,.\end{aligned} \tag{5.5.9}$$

To find the $\chi$-space effect operator, we start with a lower order and build up using Eq. (5.3.14) (temporarily ignoring the coupling constants $g_{\psi}$ and $g_{\phi}$ in $\mathcal{E}_{\underline{1}\underset{\bullet}{2}}$ and $\mathcal{E}_{\underline{1}\underset{\bullet}{2}\underset{\bullet}{3}}$, and reinstating them in $\mathcal{E}_{\underline{1}\underset{\bullet}{2}\underset{\bullet}{3}\underset{\bullet}{4}}$),

$$\mathcal{E}_{\underline{1}\underset{\bullet}{2}} = \left[\{\mathbb{I}^{\chi}, \chi_1\}\,, \chi_2\right]_{\epsilon_2} = 2\big[\chi_1, \chi_2\big]_{\epsilon_2} = 2\,(1+\epsilon_2)\,\chi_1\,\chi_2 - 2\,\epsilon_2\,\Delta^{\chi}_{12} \tag{5.5.10}$$

$$\begin{aligned}\mathcal{E}_{\underline{1}\underset{\bullet}{2}\underset{\bullet}{3}} &= \left[\mathcal{E}_{\underline{1}\underset{\bullet}{2}}, \chi_3\right]_{\epsilon_3} = (1+\epsilon_3)\mathcal{E}_{\underline{1}\underset{\bullet}{2}}\,\chi_3 - \epsilon_3\left[\mathcal{E}_{\underline{1}\underset{\bullet}{2}}, \chi_3\right] \\ &= (1+\epsilon_3)\big(2\,(1+\epsilon_2)\chi_1\chi_2\chi_3 - 2\,\epsilon_2\Delta^{\chi}_{12}\chi_3\big) - 2\,\epsilon_3\,(1+\epsilon_2)\big(\Delta^{\chi}_{13}\chi_2 + \Delta^{\chi}_{23}\chi_1\big)\end{aligned} \tag{5.5.11}$$

$$\mathcal{E}_{\underline{1}\underset{\bullet}{2}\underset{\bullet}{3}\underset{\bullet}{4}} = g_{\psi}^2 g_{\phi}^2\left[\mathcal{E}_{\underline{1}\underset{\bullet}{2}\underset{\bullet}{3}}, \chi_4\right]_{\epsilon_4} = g_{\psi}^2 g_{\phi}^2\,(1+\epsilon_4)\mathcal{E}_{\underline{1}\underset{\bullet}{2}\underset{\bullet}{3}}\,\chi_4 - g_{\psi}^2 g_{\phi}^2\,\epsilon_4\left[\mathcal{E}_{\underline{1}\underset{\bullet}{2}\underset{\bullet}{3}}, \chi_4\right]$$

$$
\begin{aligned}
&= g_\psi^2 g_\phi^2\,(1+\epsilon_4)\bigg(2\,(1+\epsilon_3)\Big((1+\epsilon_2)\chi_1\chi_2\chi_3\chi_4 - \epsilon_2\Delta^\chi_{12}\chi_3\chi_4\Big) \\
&\qquad\qquad - 2\,(1+\epsilon_2)\,\epsilon_3\Big(\Delta^\chi_{13}\chi_2\chi_4 + \Delta^\chi_{23}\chi_1\chi_4\Big)\bigg) \\
&\quad - 2\,g_\psi^2 g_\phi^2\,\epsilon_4\bigg((1+\epsilon_2)\Delta^\chi_{14}\Big((1+\epsilon_3)\chi_2\chi_3 - \epsilon_3\Delta^\chi_{23}\Big) \\
&\qquad\qquad + (1+\epsilon_2)\Delta^\chi_{24}\Big((1+\epsilon_3)\chi_1\chi_3 - \epsilon_3\Delta^\chi_{13}\Big) \\
&\qquad\qquad + (1+\epsilon_3)\Delta^\chi_{34}\Big((1+\epsilon_2)\chi_1\chi_2 - \epsilon_2\Delta^\chi_{12}\Big)\bigg)\,.
\end{aligned} \tag{5.5.12}
$$

The sum of operators over all non-zero permutations of time indices is

$$
\begin{aligned}
E^\psi_{1(\underset{\circ}{2}}\,E^\phi_{\underline{3}\underset{\circ}{4})}\,\mathcal{E}^{\psi(\psi\phi\phi)}_{\underline{1}\underset{\bullet}{2}\underset{\bullet}{3}\underset{\bullet}{4}} &= E^\psi_{12}\,E^\phi_{\underline{3}4}\,\mathcal{E}^{\psi\psi\phi\phi}_{\underline{1}\underline{2}3\underline{4}} + E^\psi_{1\underline{2}}\,E^\phi_{\underline{3}4}\,\mathcal{E}^{\psi\psi\phi\phi}_{\underline{1}23\underline{4}} + E^\psi_{12}\,E^\phi_{\underline{3}\underline{4}}\,\mathcal{E}^{\psi\psi\phi\phi}_{\underline{1}\underline{2}34} + E^\psi_{1\underline{2}}\,E^\phi_{\underline{3}\underline{4}}\,\mathcal{E}^{\psi\psi\phi\phi}_{\underline{1}234} \\
&+ E^\psi_{13}\,E^\phi_{\underline{2}4}\,\mathcal{E}^{\psi\phi\psi\phi}_{\underline{1}2\underline{3}\underline{4}} + E^\psi_{1\underline{3}}\,E^\phi_{\underline{2}4}\,\mathcal{E}^{\psi\phi\psi\phi}_{\underline{1}23\underline{4}} + E^\psi_{13}\,E^\phi_{\underline{2}\underline{4}}\,\mathcal{E}^{\psi\phi\psi\phi}_{\underline{1}2\underline{3}4} + E^\psi_{1\underline{3}}\,E^\phi_{\underline{2}\underline{4}}\,\mathcal{E}^{\psi\phi\psi\phi}_{\underline{1}234} \\
&+ E^\psi_{14}\,E^\phi_{\underline{2}3}\,\mathcal{E}^{\psi\phi\phi\psi}_{\underline{1}2\underline{3}\underline{4}} + E^\psi_{1\underline{4}}\,E^\phi_{\underline{2}3}\,\mathcal{E}^{\psi\phi\phi\psi}_{\underline{1}2\underline{3}4} + E^\psi_{14}\,E^\phi_{\underline{2}\underline{3}}\,\mathcal{E}^{\psi\phi\phi\psi}_{\underline{1}23\underline{4}} + E^\psi_{1\underline{4}}\,E^\phi_{\underline{2}\underline{3}}\,\mathcal{E}^{\psi\phi\phi\psi}_{\underline{1}234}\,.
\end{aligned} \tag{5.5.13}
$$

From Eq. (5.5.12), we have

$$
\mathcal{E}^{(\psi\psi\phi\phi)}_{\underline{1}23\underline{4}} = \mathcal{E}^{(\psi\psi\phi\phi)}_{\underline{1}234} = 0\,, \tag{5.5.14}
$$

so Eq. (5.5.13) becomes

$$
\begin{aligned}
E^\psi_{1(\underset{\circ}{2}}\,E^\phi_{\underline{3}\underset{\circ}{4})}\,\mathcal{E}^{\psi(\psi\phi\phi)}_{\underline{1}\underset{\bullet}{2}\underset{\bullet}{3}\underset{\bullet}{4}} &= E^\psi_{12}\,E^\phi_{\underline{3}4}\,\mathcal{E}^{\psi\psi\phi\phi}_{\underline{1}\underline{2}3\underline{4}} + E^\psi_{12}\,E^\phi_{\underline{3}\underline{4}}\,\mathcal{E}^{\psi\psi\phi\phi}_{\underline{1}\underline{2}34} + E^\psi_{13}\,E^\phi_{\underline{2}4}\,\mathcal{E}^{\psi\phi\psi\phi}_{\underline{1}2\underline{3}\underline{4}} + E^\psi_{13}\,E^\phi_{\underline{2}\underline{4}}\,\mathcal{E}^{\psi\phi\psi\phi}_{\underline{1}2\underline{3}4} \\
&+ E^\psi_{14}\,E^\phi_{\underline{2}3}\,\mathcal{E}^{\psi\phi\phi\psi}_{\underline{1}2\underline{3}\underline{4}} + E^\psi_{1\underline{4}}\,E^\phi_{\underline{2}3}\,\mathcal{E}^{\psi\phi\phi\psi}_{\underline{1}2\underline{3}4}\,.
\end{aligned} \tag{5.5.15}
$$

Using our equations for each Hilbert space and substituting into Eq. (4.1.9), we find

$$
\begin{aligned}
\mathbb{P} = &- \frac{4\,g_\psi^2 g_\phi^2}{(2\omega_{\mathbf{p}_1})(2\omega_{\mathbf{p}_2})}\int \mathrm{d}^4x_1\,\mathrm{d}^4x_2\,\mathrm{d}^4x_3\,\mathrm{d}^4x_4\,\Theta_{1234} \\
&\Bigg(\Big(e^{ip_1\cdot x_1}e^{ip_2\cdot x_1}e^{-ip_1\cdot x_2}e^{-ip_2\cdot x_2} + e^{ip_1\cdot x_2}e^{ip_2\cdot x_2}e^{-ip_1\cdot x_1}e^{-ip_2\cdot x_1}\Big) \\
&\Big(2\Delta^\phi_{34}\,\langle\phi_3\phi_4\rangle - (\Delta^\phi_{34})^2\Big) \\
&\Big(2\,\Delta^\chi_{13}\,\langle\chi_2\,\chi_4\rangle + 2\,\Delta^\chi_{23}\,\langle\chi_1\,\chi_4\rangle - \Delta^\chi_{14}\Delta^\chi_{23} - \Delta^\chi_{13}\Delta^\chi_{24}\Big) \\
&+ \Big(e^{ip_1\cdot x_1}e^{ip_2\cdot x_1}e^{-ip_1\cdot x_2}e^{-ip_2\cdot x_2} + e^{ip_1\cdot x_2}e^{ip_2\cdot x_2}e^{-ip_1\cdot x_1}e^{-ip_2\cdot x_1}\Big) \\
&\Big(\langle\phi_3^2\phi_4^2\rangle - 2\Delta^\phi_{34}\,\langle\phi_3\phi_4\rangle + (\Delta^\phi_{34})^2\Big)\Big(\Delta^\chi_{14}\Delta^\chi_{23} + \Delta^\chi_{13}\Delta^\chi_{24}\Big) \\
&+ \Big(e^{ip_1\cdot x_1}e^{ip_2\cdot x_1}e^{-ip_1\cdot x_3}e^{-ip_2\cdot x_3} + e^{ip_1\cdot x_3}e^{ip_2\cdot x_3}e^{-ip_1\cdot x_1}e^{-ip_2\cdot x_1}\Big)
\end{aligned}
$$

$$
\begin{aligned}
&\Big(2\Delta^{\phi}_{24}\,\langle\phi_2\phi_4\rangle - (\Delta^{\phi}_{24})^2\Big)\Big(2\,\Delta^{\chi}_{12}\,\langle\chi_3\,\chi_4\rangle - \Delta^{\chi}_{12}\Delta^{\chi}_{34}\Big)\\
&+\Big(e^{ip_1\cdot x_1}e^{ip_2\cdot x_1}e^{-ip_1\cdot x_3}e^{-ip_2\cdot x_3} + e^{ip_1\cdot x_3}e^{ip_2\cdot x_3}e^{-ip_1\cdot x_1}e^{-ip_2\cdot x_1}\Big)\\
&\Big(\langle\phi_2^2\phi_4^2\rangle - 2\Delta^{\phi}_{24}\,\langle\phi_2\phi_4\rangle + (\Delta^{\phi}_{24})^2\Big)\Big(\Delta^{\chi}_{12}\Delta^{\chi}_{34}\Big)\\
&+\Big(e^{ip_1\cdot x_1}e^{ip_2\cdot x_1}e^{-ip_1\cdot x_4}e^{-ip_2\cdot x_4} + e^{ip_1\cdot x_4}e^{ip_2\cdot x_4}e^{-ip_1\cdot x_1}e^{-ip_2\cdot x_1}\Big)\\
&\Big(2\Delta^{\phi}_{23}\,\langle\phi_2\phi_3\rangle - (\Delta^{\phi}_{23})^2\Big)\Big(2\Delta^{\chi}_{12}\,\langle\chi_3\,\chi_4\rangle - \Delta^{\chi}_{12}\Delta^{\chi}_{34}\Big)\\
&+\Big(e^{ip_1\cdot x_1}e^{ip_2\cdot x_1}e^{-ip_1\cdot x_4}e^{-ip_2\cdot x_4} - e^{ip_1\cdot x_4}e^{ip_2\cdot x_4}e^{-ip_1\cdot x_1}e^{-ip_2\cdot x_1}\Big)\\
&\Big(2\Delta^{\phi}_{23}\,\langle\phi_2\phi_3\rangle - (\Delta^{\phi}_{23})^2\Big)\Big(\Delta^{\chi}_{12}\Delta^{\chi}_{34}\Big)\Bigg)\,, \qquad (5.5.16)
\end{aligned}
$$

where $\langle\ldots\rangle$ is shorthand for $\langle 0^{\phi}|\ldots|0^{\phi}\rangle$ or $\langle 0^{\chi}|\ldots|0^{\chi}\rangle$. Due to the appearance of $\Theta_{1234}$ (from Eq. (4.1.9)), we can interpret the product of fields as the time-ordered product, and thus use Wick's theorem (as in Eq. (5.4.13)). This results in Feynman propagators. Eq. (5.5.16) is shown diagrammatically in Fig. 5.11, after simplification. Again, the diagrams are exactly those which would be generated using the rules in Section 5.2.

The disconnected diagrams can be ignored since they are not kinematically allowed for positive, non-zero initial momenta. All of the remaining diagrams have the topology shown in Fig. 5.12, differing by their time-orderings and combinations of Feynman and retarded propagators. Since our approach is inclusive over final states, these diagrams include two annihilation processes: the 2-to-2 process $\psi\psi \to \chi \to \phi\phi$ and the 2-to-1 process $\psi\psi \to \chi$ with a $\phi$ self-energy loop. The Feynman diagrams for these processes are shown in Fig. 5.13. Note that the diagrams in Fig. 5.11 do not neatly separate into 2-to-2 and 2-to-1 diagrams, highlighting the intrinsic inclusivity of the result.

The diagrams in Fig. 5.11 can be simplified by introducing the retarded self-energy [113, 114],

$$
\Pi^{\mathrm{R}}_{xy} = \frac{g_{\phi}^2}{2}\left[2\,F^{\phi}_{xy}\,R^{\phi}_{xy} - \left(R^{\phi}_{xy}\right)^2\right]\,, \qquad (5.5.17)
$$

reflecting the manifest causality of this approach. This simplification suggests the existence of more fundamental rules than those given in Section 5.2, i.e., rules involving manifestly causal objects such as the retarded self-energy.

The diagrams can be simplified further by realising that all possible time orderings are present, since the retarded propagators vanish when the arguments are not time-

$$\mathbb{P} = -\frac{4g_\psi^2 g_\phi^2}{(2\omega_{\mathbf{p}_1})(2\omega_{\mathbf{p}_2})} \int \mathrm{d}^4x_1\, \mathrm{d}^4x_2\, \mathrm{d}^4x_3\, \mathrm{d}^4x_4\, \Theta_{1234} \times$$

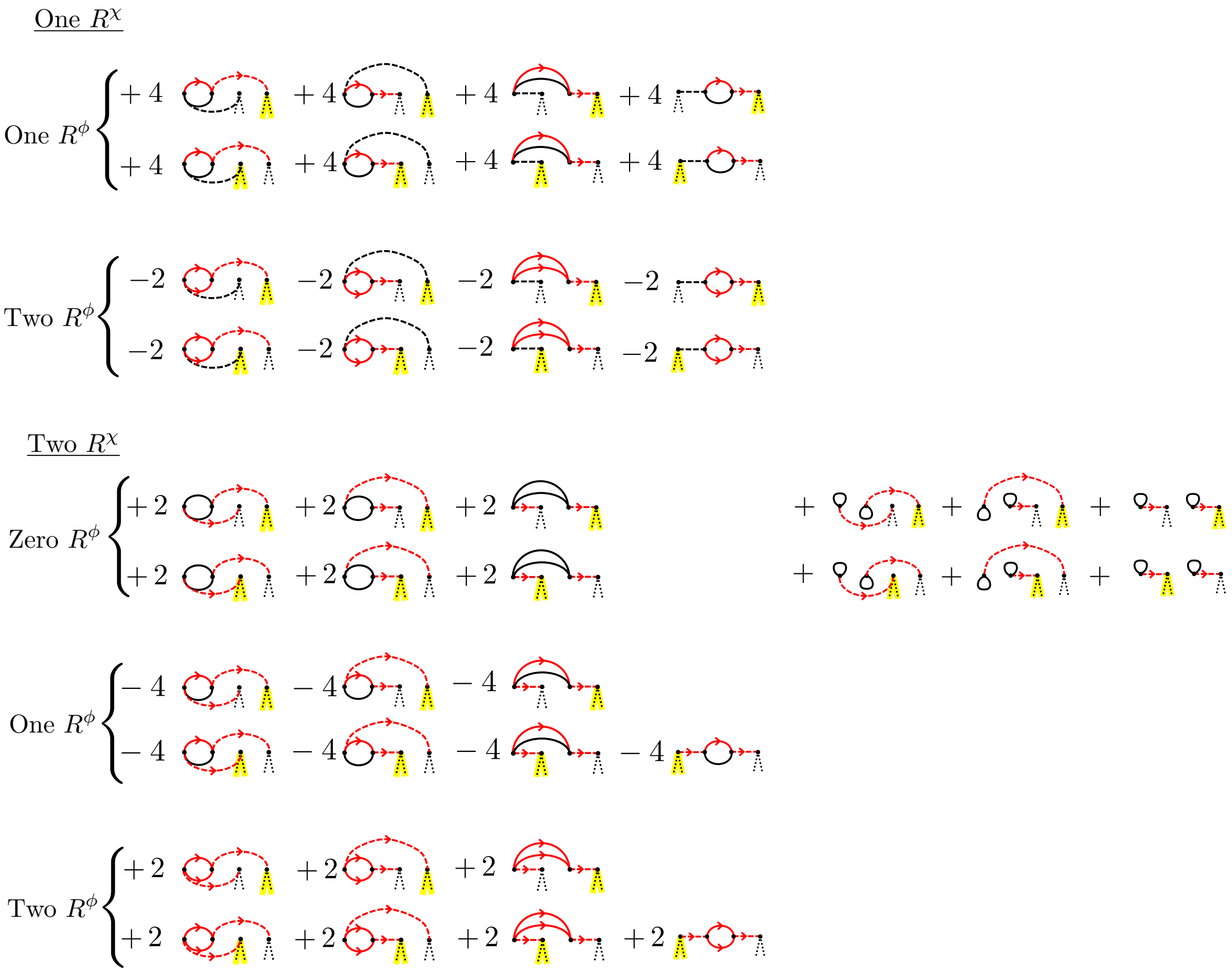

**Fig. 5.11.** Probability-level diagrammatic representation of the annihilation process expressed in Eq. (5.5.16). Terms are sorted by the number of retarded propagators for each field.

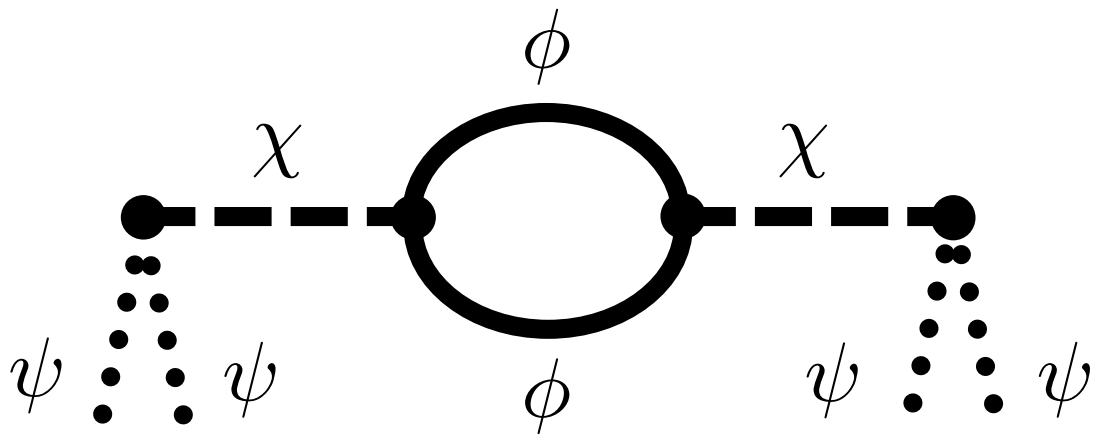

**Fig. 5.12.** The topology of all of the diagrams in Fig. 5.11.

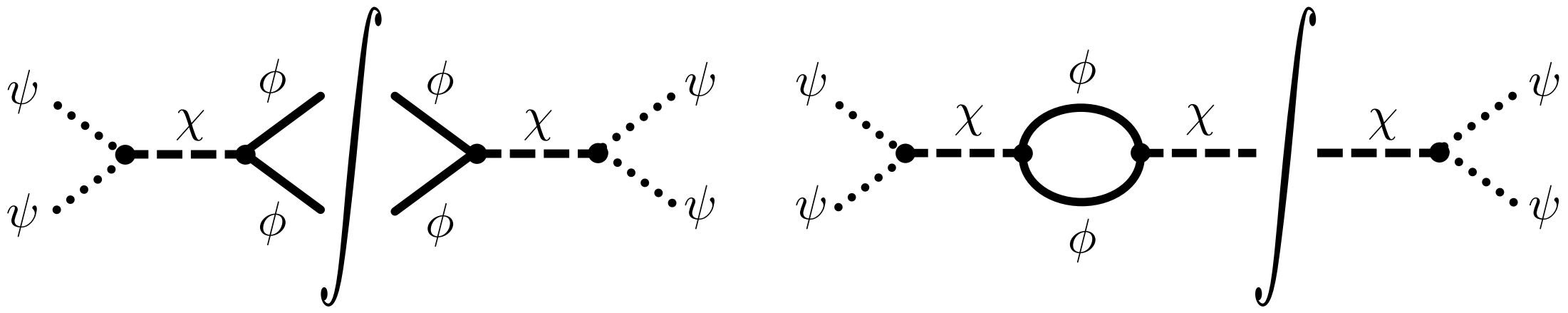

**Fig. 5.13.** Feynman diagrams (and conjugate diagrams) for the two annihilation processes. The curved line separates an amplitude from a conjugate amplitude. These are the annihilation processes which are encoded in the diagrams in Fig. 5.11. **Left:** $\psi\psi \to \chi \to \phi\phi$. **Right:** $\psi\psi \to \chi$ with a $\phi$ self-energy loop. The conjugate of this diagram also contributes.

ordered (see Eq. (2.3.6)). Consequently, the time-ordered integral can be replaced by an integral over all times, and we need only keep one of each unique diagram. This process is shown in Fig. 5.14, and is proven analytically in Appendix B for the diagrams with a loop of Feynman propagators (a similar procedure has been used for all diagrams).

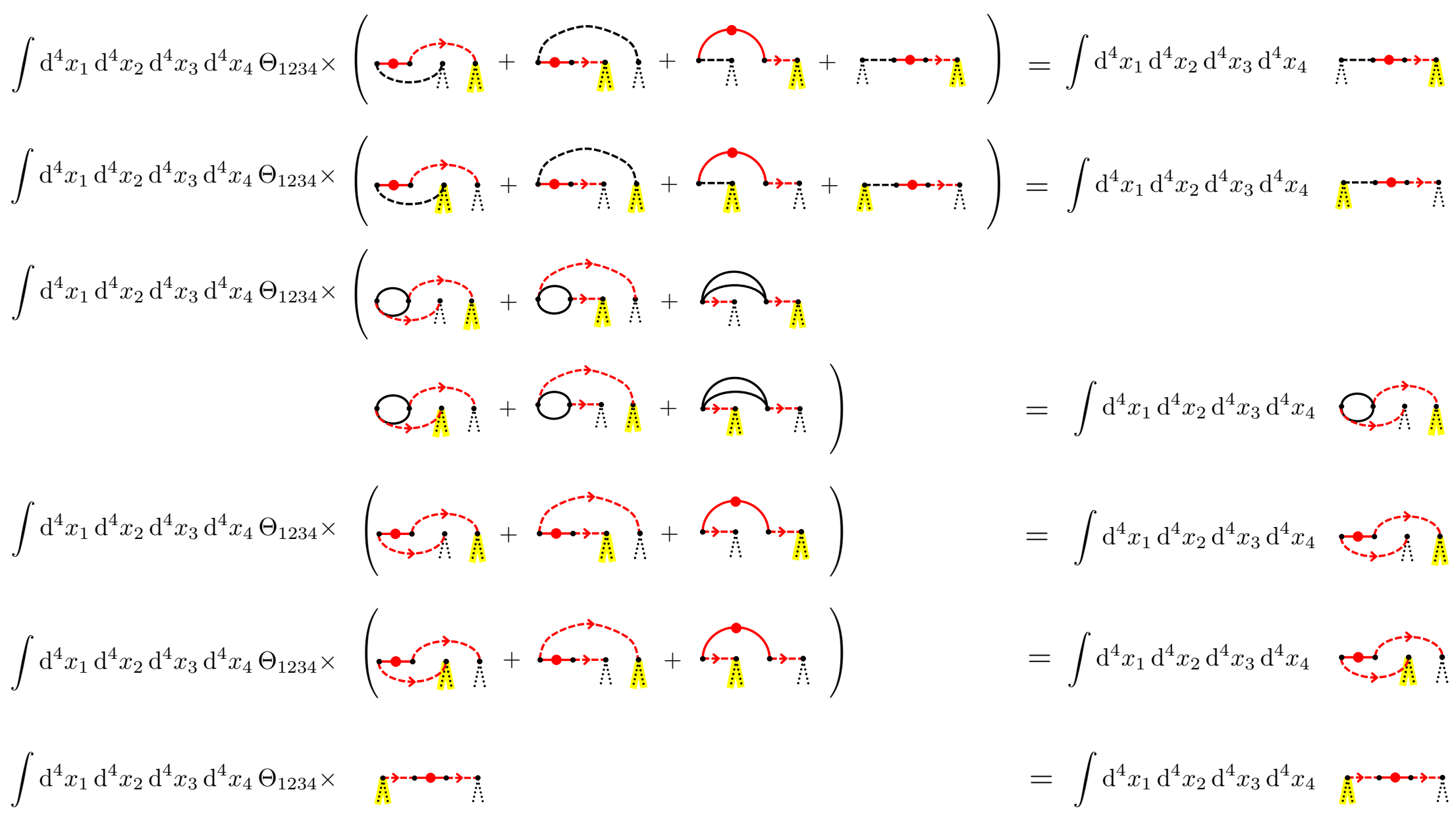

**Fig. 5.14.** Simplification of the annihilation diagrams. Note the absence of time-ordering on the right-hand side. The retarded self-energy, given by Eq. (5.5.17), is denoted by a red line with a solid red circle in the middle.

The diagrams resulting from these simplifications are shown in Fig. 5.15. Further simplification using propagator identities is possible, but we choose to keep the result solely in terms of Feynman and retarded propagators to highlight the causal structure.

$$\mathbb{P} = -\frac{4g_\psi^2 g_\phi^2}{(2\omega_{\mathbf{p}_1})(2\omega_{\mathbf{p}_2})} \int \mathrm{d}^4x_1\, \mathrm{d}^4x_2\, \mathrm{d}^4x_3\, \mathrm{d}^4x_4 \;\times$$

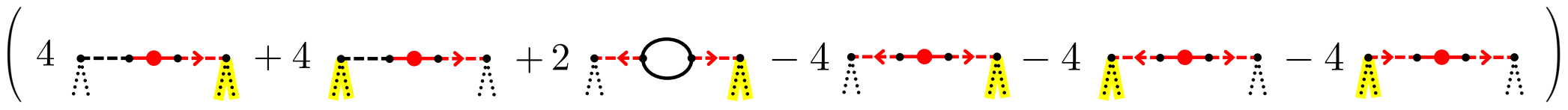

**Fig. 5.15.** The simplified result for the fully inclusive annihilation probability. The retarded self-energy, given by Eq. (5.5.17), is denoted by a red line with a solid red circle in the middle.

## 5.6 Summary

We have applied a novel, probability-level QFT formalism to scalar-field scattering processes in which causality is manifest. In scalar field theory, causality is encoded in the commutator of fields, which appear as a result of applying the Baker-Campbell-Hausdorff lemma to the transition probability. This formalism results in new probability-level diagrams, and we have presented the rules to generate the complete set of all diagrams for a scattering process which is fully inclusive over final states that do not contain initial-state particles. We have used the algebraic formalism to calculate the total probabilities for particle decay and the annihilation of two particles, both at fixed order. These results align with those expected from a traditional calculation and corroborate the general diagrammatic rules.

Since the diagrams correspond to the probability directly and involve retarded propagators, causality is manifest. The appearance of other causal structures, such as the retarded self-energy, suggests the existence of a more fundamental set of rules in terms of these causal objects. In particular, there may be a link between these rules and the Kobes-Semenoff unitary cutting rules [116, 117] in a similar fashion to the discussion in Ref. [32].

The diagrammatic method developed in this chapter will generalise to gauge theories such as QED and QCD. In these theories, individual contributions are IR-divergent but finite once regularised and combined, as shown in Chapter 3. Our probability-level method results in retarded propagators appearing in loops, and ‘real emission’ contributions are accounted for in the ‘self-energy’ and ‘vertex’ terms. We hope that these features may help avoid the IR divergences in gauge theories by satisfying the BN and KLN theorems implicitly.

Future work should include understanding how the rules are adapted for semi-inclusive effect operators, such as those discussed in Ref. [105]. Moreover, it remains to establish how manifestly causal probabilities can be extracted from causal $n$-point functions by means of an LSZ-like reduction procedure.

# Chapter 6

# The Unruh Effect

The trajectories of uniformly accelerating observers (*Rindler observers*) are restricted to a region of Minkowski spacetime (the *Rindler wedge*), and they are causally disconnected from another region of Minkowski spacetime (the opposite Rindler wedge). The mode expansion of a quantum field employed by a Rindler observer is different from that employed by a Minkowski observer. Thus, accelerated and inertial observers may disagree on the particle content of a field. Remarkably, a Rindler observer would associate a thermal bath of Rindler particles to the no-particle (vacuum) Minkowski state. This is the Unruh effect [18–21].

The Unruh effect is a direct mathematical consequence of quantum field theory. To probe the physics of the Unruh effect, localised particle detector models were developed and applied for a uniformly accelerating path [18, 21, 118]. The conclusion is that the detector's non-zero response rate per unit proper time along the detector's trajectory as measured by a Minkowski observer (who, using inertial measuring apparatus, otherwise experiences a vacuum) is identical to that measured by a Rindler observer (who experiences a thermal heat bath). The effect can be understood as a consequence of the presence of a horizon, which appears between the Rindler wedge and the rest of the universe. Therefore, similar methods to those used to study the Unruh effect can be used to study horizons in curved spacetimes [119, 120], reproducing the thermal properties of black holes [15, 16, 121, 122] and de Sitter space [123]. The mathematical relationship between Minkowski and Rindler coordinates (explained in Section 6.1) is very similar to that between Schwarzschild and Kruskal coordinates for black holes (defined in Chapter 7). A significant difference is that Hawking radiation from a black hole is detectable at infinity, since Schwarzschild coordinates become inertial at large distances. Close to the horizon, an observer at a fixed radial position would detect thermal effects that a free-falling observer would not [18, 124, 125], and this can be attributed to the acceleration required to maintain constant ra-dial position. There are also key similarities between the rotational Unruh effect and rotating black holes [126, 127]. Consequently, the Unruh effect offers an excellent avenue into understanding important features of quantum field theory (QFT) that are also relevant to Hawking radiation and black holes.

These key features of quantum field theory are explored further in the field of relativistic quantum information (RQI), which focuses on the relationship between relativistic quantum field theories and quantum information [109, 110]. This has resulted in important applications of the Unruh effect and the Unruh-DeWitt (UdW) detector, such as entanglement harvesting [128–135], entanglement degradation [136–140], corrections to quantum teleportation fidelity [141–143], quantum energy teleportation [144], curvature measurement [145, 146], and avoiding difficulties with field measurements [147–150].

This chapter starts with a review of the foundational theory of the Unruh effect in Section 6.1. In Section 6.2, the transition rate of a uniformly accelerated UdW monopole detector is calculated using the general, probabilistic method described in Chapter 4. This verifies that this causal method gives the correct results whilst also laying the foundations for its use in more complicated scenarios involving the Unruh effect. The transition rate is calculated for a measurement a finite time after preparing the initial state. Specifying the field to initially be in the Minkowski vacuum state causes transients, which decay as the measurement time increases, and these transients are investigated. In Section 6.3, the same transition rate is calculated from the perspective of an accelerating Rindler observer. The result is the same, including the finite-time transient effects. The corresponding transition rate for an inertial detector in a bath of Minkowski particles is calculated in Section 6.4, and it is shown that this is different, except for the massless case. Section 6.5 presents new numerical results, and Section 6.7 concludes. This chapter is adapted from Ref. [2].

## 6.1 Background

We review the concepts which are key to understanding the Unruh effect, including Rindler spacetime, the Minkowski vacuum state as viewed from a Rindler observer's perspective, the model of the UdW detector, and an accelerated UdW detector's transition rate. A thorough review of the Unruh effect is given in Ref. [22].

### 6.1.1 Rindler Spacetime

Consider four-dimensional Minkowski spacetime which has the line element [151–153],

$$\mathrm{d}s^2 = \mathrm{d}t^2 - \mathrm{d}x^2 - \mathrm{d}y^2 - \mathrm{d}z^2 = \mathrm{d}u\,\mathrm{d}v - \mathrm{d}y^2 - \mathrm{d}z^2\,, \tag{6.1.1}$$

where the null coordinates, $u, v$, are defined by

$$\begin{aligned} u &= t - x\,, \\ v &= t + x\,. \end{aligned} \tag{6.1.2}$$

To describe the physics as seen by a uniformly accelerating observer, we introduce Rindler coordinates $(\eta, \xi)$ [154] which are adapted to observers undergoing constant proper acceleration in Minkowski spacetime. The transformation between Minkowski coordinates $(t, x)$ and Rindler coordinates is given by:

$$t = \alpha^{-1}\, e^{\alpha\xi}\, \sinh \alpha\eta\,, \quad x = \alpha^{-1}\, e^{\alpha\xi}\, \cosh \alpha\eta\,, \quad y = y\,, \quad z = z\,. \tag{6.1.3}$$

where $\alpha$ is a positive constant and $-\infty < (\eta, \xi) < \infty$. In terms of the null coordinates, the equivalent coordinate transformation is

$$\begin{aligned} \overline{u} &= -\alpha^{-1}\, e^{-\alpha u}\,, \\ \overline{v} &= \alpha\, e^{\alpha v}\,. \end{aligned} \tag{6.1.4}$$

This coordinate transformation is very similar to the coordinate transform between Schwarzschild and Kruskal coordinates (Eq. (7.3.4)) commonly used when studying black holes. This highlights the intrinsic link between the Unruh effect and black hole physics, which is studied in Chapter 7.

In terms of these new coordinates, the line element is

$$\mathrm{d}s^2 = e^{2\alpha\xi}\left(\mathrm{d}\eta^2 - \mathrm{d}\xi^2\right) - \mathrm{d}y^2 - \mathrm{d}z^2 = e^{2\alpha\xi}\mathrm{d}\overline{u}\,\mathrm{d}\overline{v} - \mathrm{d}y^2 - \mathrm{d}z^2\,. \tag{6.1.5}$$

These coordinates only cover the $x > |t|$ quadrant of Minkowski space. This quadrant is called the *right Rindler wedge*, pictured in Fig. 6.1. On this spacetime diagram, lines of constant $\eta$ are straight and lines of constant $\xi$ are hyperbolae (since $x^2 - t^2 = \alpha^{-2}e^{2\alpha\xi}$ = constant). These hyperbolae thus represent the world lines

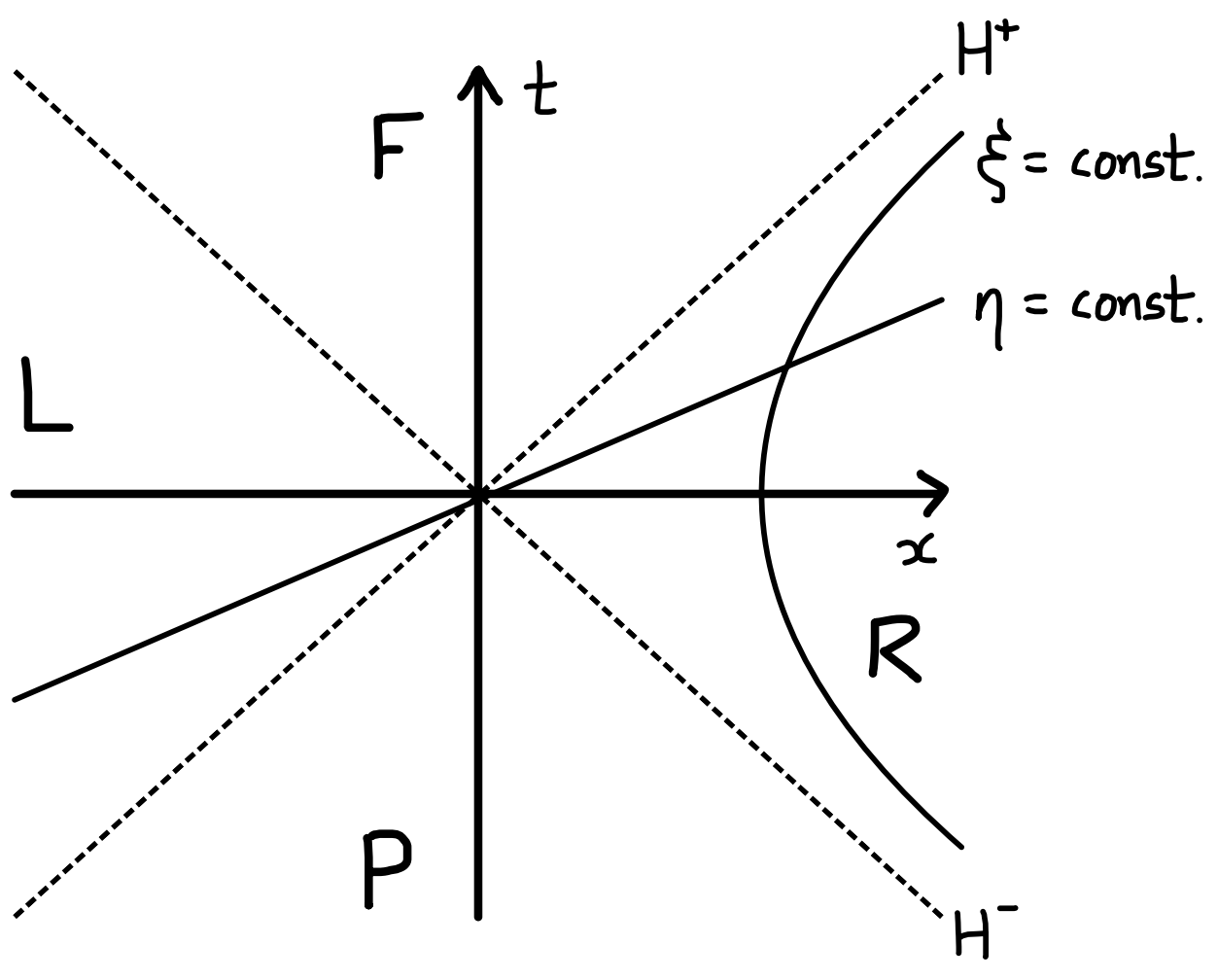

**Fig. 6.1.** The spacetime diagram of Rindler space. The future ($H^+$) and past ($H^-$) horizons separate Minkowski spacetime into four quadrants: the right Rindler wedge (R), the left Rindler wedge (L), the future wedge (F), and the past wedge (P).

for uniformly accelerated observers.

The proper acceleration of an observer at fixed $\xi$ is given by

$$\text{proper acceleration } = \alpha e^{-\alpha\xi} \tag{6.1.6}$$

and the observer's proper time is given by

$$\text{proper time } \equiv \tau = e^{\alpha\xi}\eta\,. \tag{6.1.7}$$

In this chapter, we consider a detector with Rindler coordinate $\xi = 0$ such that the proper acceleration is $\alpha$ and the proper time $\tau = \eta$, without loss of generality.

A second Rindler wedge can be constructed with similar coordinate transformations but with additional minus signs. These different Rindler coordinates cover a different quadrant of Minkowski spacetime known as the *left Rindler wedge*. The surfaces $x = |t|$ (or equivalently $\xi = -\infty$, or $u = v = 0$ in null coordinates) form a boundary that separates the right (R) and left (L) Rindler wedges from the rest of Minkowski space. These null surfaces, $H^+$ and $H^-$, constitute event horizons from the perspective of an observer in either wedge. This means that an observer in R cannot communicate with an observer in L, and vice versa. This causal structure is pictured in Fig. 6.1 and can be understood in terms of the light cones in Minkowski space. Any signal sent from within R travels at $45°$ on a spacetime diagram, and thus can never reach L. This implies that an accelerating observer in R is permanently causally dis-connected from L. This causal disconnection plays a crucial role in the emergence of the Unruh effect, as will be shown in the next section.

### 6.1.2 The Minkowski Vacuum as a Thermal State

To understand how the Minkowski vacuum appears to an accelerating observer, we decompose a scalar field $\phi(x)$ in terms of Rindler modes. The usual decomposition in Minkowski space is (see Section 2.1):

$$\phi(x) = \int \mathrm{d}^3\mathbf{k} \left( a_\mathbf{k} u_\mathbf{k} + a_\mathbf{k}^\dagger u_\mathbf{k}^* \right), \tag{6.1.8}$$

where $u_\mathbf{k}$ are the plane wave solutions to the Klein-Gordon equation and $a_\mathbf{k}$ and $a_\mathbf{k}^\dagger$ are annihilation and creation operators, respectively, which defined the Minkowski Fock basis and the Minkowski vacuum state, $a_\mathbf{k} |0_\mathrm{M}\rangle = 0$.

In Rindler coordinates, the field $\phi(x)$ can instead be decomposed as [22, 155]

$$\phi(x) = \int \mathrm{d}^3\mathbf{k} \left( b_\mathbf{k}^R v_\mathbf{k}^R + b_\mathbf{k}^{R\,\dagger} v_\mathbf{k}^{R*} + b_\mathbf{k}^L v_\mathbf{k}^L + b_\mathbf{k}^{L\,\dagger} v_\mathbf{k}^{L*} \right), \tag{6.1.9}$$

where $v_\mathbf{k}^R$ and $v_\mathbf{k}^L$ are solutions to the Klein-Gordon equation that are well-defined in the right and left Rindler wedges, respectively. These modes can be analytically continued into regions P and F such that Eq. (6.1.9) defines the scalar field across all of Minkowski spacetime [156, 157]. Further details of these modes can be found in Ref. [151]. The annihilation and creation operators, $b_\mathbf{k}^{R/L}$ and $b_\mathbf{k}^{R/L\,\dagger}$, define a new Fock basis and a different vacuum state, $b_\mathbf{k}^R |0_\mathrm{R}\rangle = b_\mathbf{k}^L |0_\mathrm{R}\rangle = 0$. This vacuum state is known as the *Rindler vacuum*. The definition of a different Fock basis means that the concept of particles is dependent on the basis modes which are used to decompose the field. Any given state will have a different number of *Minkowski particles* and *Rindler particles*. This even applies to the Minkowski vacuum state, in which an inertial (Minkowski) observer detects no particles. For a *Rindler observer* at constant acceleration, $\alpha$, in the right (left) Rindler wedge, the expectation value of the Rindler number operator in the Minkowski vacuum state is given by [22, 151]

$$\langle 0_\mathrm{M} | b_\mathbf{k}^{R(L)\,\dagger} b_\mathbf{k}^{R(L)} | 0_\mathrm{M} \rangle = \frac{1}{e^{2\pi\omega/\alpha} - 1}, \tag{6.1.10}$$

where $\omega = \sqrt{\mathbf{k}^2 + m^2}$. This is the number of ‘particles’ a Rindler observer would detect in mode **k**, if particles are defined using the Rindler basis modes. This result is exactly the Planck spectrum for a thermal bath of particles (bosons) at temperature

$$T_{\mathrm{U}} = \frac{\alpha}{2\pi}\,, \tag{6.1.11}$$

which is known as the *Unruh temperature*. Thus, the Minkowski vacuum state appears as a thermal bath of particles to an observer at constant acceleration.

This can also be understood by realising that the Minkowski vacuum state, $|0_{\mathrm{M}}\rangle$, can be expressed in terms of Rindler Fock states as [22]:

$$|0_M\rangle = \prod_i \left( \sqrt{1 - e^{-2\pi\omega_i/\alpha}} \sum_{n_i=0}^{\infty} e^{-\pi\omega n_i/\alpha} |n_{i,R}\rangle \otimes |n_{i,L}\rangle \right) . \tag{6.1.12}$$

A Rindler observer in the right Rindler wedge, for example, is causally disconnected from the left Rindler wedge due to the horizons, and thus can only probe the right Rindler wedge. The Minkowski vacuum state is therefore described by the density matrix which is obtained by tracing out the left Rindler states,

$$\rho_R = \prod_i \left( \left(1 - e^{-2\pi\omega_i/\alpha}\right) \sum_{n_i=0}^{\infty} e^{-2\pi\omega n_i/\alpha}\, |n_{i,R}\rangle\langle n_{i,R}| \right) . \tag{6.1.13}$$

This is the density matrix for a thermal bath of particles at $T = T_{\mathrm{U}}$.

### 6.1.3 Unruh-DeWitt Detectors

In order to place the mathematics of the Unruh effect on physical grounding, Unruh introduced two different detector models [18]. The first is a small box containing a non-relativistic particle which satisfies the Schrödinger equation with two energy states. The second is a fully relativistic detector model, involving a second scalar field for the field of interest to couple to, via a third scalar field. DeWitt then introduced the detector model [118] that is most commonly used today in literature regarding the Unruh effect, which consists of a two-level point-like monopole and is known as an UdW detector. Its two energy levels are denoted by $|1^D\rangle$ for the ground state and $|2^D\rangle$ for the excited state. This is a simple model which still captures the essential features of a detector's interaction with a field. The point-like nature of this detector leads to divergences, but these divergences can either be made to cancel by defining certain observables or explicitly regulated with integral cut-offs or spatial smearing functions [158–160].

The interaction Hamiltonian of an UdW detector coupled to a scalar field $\phi(x)$ is given by

$$H_{\text{int}} = \lambda\chi(\tau)M^D(\tau)\phi(\mathbf{x}^D(\tau),\tau), \tag{6.1.14}$$

where $\lambda$ is the coupling constant (which is set to 1 hereafter), $\chi(\tau)$ is a switching function controlling the interaction duration, $M^D(\tau)$ is the detector's monopole moment, and $\mathbf{x}^D$ is the detector trajectory. The detector's excitation rate from its ground state to its excited state is explicitly calculated in the next section (for an uniformly accelerated detector) using the manifestly causal formalism described in Chapter 4.

## 6.2 Excitation rate of an accelerated detector

### 6.2.1 Calculating the Transition Rate

We consider a point-like 'atom', $D$, which plays the role of a two-state, UdW detector. The atom interacts with a neutral scalar field $\phi(\mathbf{x},t)$, of mass $m$, where $\mathbf{x}$ and $t$ are coordinates in an inertial frame, and it is accelerated with a constant proper acceleration, $\alpha$, such that its position is given by

$$\mathbf{x}^D = \left(\frac{1}{\alpha}\cosh\alpha\tau\,,\,0\,,\,0\right) = \left(\frac{1}{\alpha}\sqrt{1+\alpha^2t^2}\,,\,0\,,\,0\right), \tag{6.2.1}$$

and the proper time of the atom is

$$\tau = \int_0^t \frac{\mathrm{d}t'}{\gamma(t')} = \int_0^t \frac{\mathrm{d}t'}{\sqrt{1+\alpha^2t'^2}} = \frac{1}{\alpha}\,\text{arcsinh}\,\alpha t\,. \tag{6.2.2}$$

The system is described by states living in a product of the Hilbert spaces of the atom and the field: $\mathscr{H} = \mathscr{H}^D\times\mathscr{H}^\phi$. For the Hamiltonian, we take $H(t) = H_0(t)+H_{\text{int}}(t)$, where $H_0(t) = H_0^D(t) + H_0^\phi(t)$. Under the free part of the Hamiltonian, $H_0$, the atom has a complete set of states $\{|1^D\rangle\,, |2^D\rangle\}$ (one ground state $|1^D\rangle$ and one excited state $|2^D\rangle$), with $H_0^D\,|n^D\rangle = \Omega_n\,|n^D\rangle$, $n = 1, 2$. In the inertial frame, we assume that the interaction-picture Hamiltonian is given by

$$H_0 \;=\; \sum_{n=1}^{2} \gamma^{-1}(t)\, \Omega_n \, |n^D\rangle \, \langle n^D| \;+\; \int \mathrm{d}^3\mathbf{x} \, \left( \tfrac{1}{2}(\partial_t \phi)^2 \;+\; \tfrac{1}{2}(\nabla\phi)^2 \;+\; \tfrac{1}{2} m^2 \phi^2 \right), \tag{6.2.3a}$$

$$H_{\text{int}} \;=\; M^D(t)\, \phi(\mathbf{x}^D, t)\,, \tag{6.2.3b}$$

where

$$M^D(t) \equiv \gamma^{-1}(t) \sum_{m,n=1}^{2} \mu_{mn}\, e^{i(\Omega_m - \Omega_n)\tau} \; |m^D\rangle \, \langle n^D| \tag{6.2.4}$$

represents a monopole interaction. Comparing to Eq. (4.4.6), we see that this detector is a two-level variant of the detector considered in Section 4.4. For future reference, we define $\mu \equiv \mu_{12} = \mu_{21}^*$ and $\Omega \equiv \Omega_2 - \Omega_1$. We will also assume that $\mu_{nn} = 0 \; \forall \, n$, so that the interaction always involves transitions between the states.

Suppose that the system is initially ($t \,=\, 0$) described by a density matrix $\rho_0$ and that the measurement outcome is described by an effect operator $E$. In general, $E$ is an element of a Positive Operator-Valued Measure, and it may be written as a sum over products of Hermitian operators:

$$E \;=\; \sum_{\kappa} E^D_{(\kappa)} \otimes E^{\phi}_{(\kappa)}\,. \tag{6.2.5}$$

The superscripts $D$ and $\phi$ denote the Hilbert space in which the operators act and $\kappa$ denotes different configurations of final states. The probability of the measurement outcome, $\mathbb{P}$, is then given by

$$\mathbb{P} \;=\; \mathrm{Tr}(E\rho_t)\,, \tag{6.2.6}$$

where

$$\rho_t \;\equiv\; U_{t,0}\, \rho_0 \, U_{t,0}^{\dagger} \tag{6.2.7}$$

is the density operator at time $t$ and

$$U_{t,0} \;=\; \mathrm{T} \exp\left( \frac{1}{i} \int_0^t \mathrm{d}t' \; H_{\text{int}}(t') \right) \tag{6.2.8}$$

is the unitary evolution operator (T indicates time ordering). Note that evolving the initial state from $t' \;=\; 0$ to $t' \;=\; t$ is mathematically equivalent to evolving it from

$t = -\infty$ to $t' = \infty$ but with a top-hat switching function (defined in Eq. (6.1.14)),

$$\chi(t') = \begin{cases} 1 & \text{for } 0 \leq t' \leq t\,, \\ 0 & \text{otherwise}\,. \end{cases} \tag{6.2.9}$$

We consider the case where the initial density operator is $\rho_0 \;=\; |1^D, 0^\phi_M\rangle\,\langle 1^D, 0^\phi_M|$, in which $|0^\phi_M\rangle$ denotes the Minkowski vacuum state, and the effect operator is $E \;=\; |2^D\rangle\,\langle 2^D| \otimes \mathbb{I}_\phi$. This effect operator describes a set of final states in which the atom is excited and the final state of the field is anything at all. Fixing the field in the Minkowski vacuum state, $|0^M_\phi\rangle$ at an instant in time $(t\,=\,0)$ is somewhat arbitrary and will result in transient effects.

We are interested in the excitation rate of the atom, $\varGamma(1\,\rightarrow\,2)$. The master equation for the probability of finding the detector in the excited state is given by

$$\frac{\mathrm{d}\mathbb{P}(2;t)}{\mathrm{d}t} \;=\; \varGamma(1\rightarrow 2)\mathbb{P}(1;t) \;-\; \varGamma(2\rightarrow 1)\mathbb{P}(2;t)\,, \tag{6.2.10}$$

where

$$\varGamma(1\rightarrow 2) \;=\; \frac{\mathrm{d}\mathbb{P}(2;t)}{\mathrm{d}t}\Big(1 \;+\; \mathcal{O}(|\mu|^2)\Big)\,, \tag{6.2.11}$$

and

$$\mathbb{P}(2;t) \equiv \langle 1^D, 0^\phi_M|\, U^\dagger_{t,0} E\, U_{t,0}\, |1^D, 0^\phi_M\rangle\;. \tag{6.2.12}$$

Following the formalism described in Chapter 4, we use a generalisation of the Baker-Campbell-Hausdorff lemma to commute the operator $E$ through the time-evolution operator, which gives

$$\mathbb{P}(2;t) \;=\; \sum_{j\,=\,0}^{\infty} \int_0^t \mathrm{d}t_1 \mathrm{d}t_2 \ldots \mathrm{d}t_j\; \Theta_{12\ldots j}\, \langle 1^D, 0^\phi_M|\, \mathcal{F}_j\, |1^D, 0^\phi_M\rangle\;, \tag{6.2.13}$$

where

$$\begin{aligned} \mathcal{F}_0 \;&=\; E\,, \\ \mathcal{F}_j \;&=\; \tfrac{1}{i}\Big[\mathcal{F}_{j-1}, H_{\text{int}}(t_j)\Big]\,, \end{aligned} \tag{6.2.14}$$

and $\Theta_{ijk\ldots} \equiv 1$ if $t_i > t_j > t_k \ldots$ and zero otherwise. Using the notation

$\phi_j^D \equiv \phi(\mathbf{x}_j^D, t_j)$, $M_j^D \equiv M^D(t_j)$ and $\mathbf{x}_j^D \equiv \mathbf{x}^D(t_j)$, we may write

$$\mathcal{F}_j \;=\; \tfrac{1}{i}\Big[\mathcal{F}_{j-1}\,,\; M_j^D \phi_j^D\Big]\,. \tag{6.2.15}$$

Eq. (6.2.13) includes contributions from all perturbative orders.

Thus far, we have introduced a general method for calculating observable probabilities that is based on effect operators and the evolution of the initial density matrix. This approach has the advantage that it can be used to treat both pure and mixed states, as well as exclusive, inclusive and semi-inclusive observables [105], all on equal footing. Moreover, it is an approach that has been shown to make physical principles, such as causality [32], manifest, as seen in Chapters 4 and 5. In what follows, we show how this probability-level approach can be used to treat the initial time-dependent response of the UdW detector and recover known results in the late-time limit. This is with a view to future applications to problems that may be less tractable at the amplitude level.

Proceeding to expand in $\mu$, the two lowest-order contributions are

$$\begin{aligned}
\mathcal{F}_1 \;&=\; i\gamma^{-1}(t_1)\phi_1^D\,\langle 2|2\rangle\,\Big(\mu\, e^{-i\Omega\tau_1}\,|1\rangle\langle 2| \;-\; \mu^* e^{i\Omega\tau_1}\,|2\rangle\langle 1|\Big)\,, &(6.2.16)\\
\mathcal{F}_2 \;&=\; \gamma^{-1}(t_1)\gamma^{-1}(t_2)\,\langle 2|2\rangle\,|\mu|^2\Big[\mathbb{I}_\phi\,\Delta_{12}\sin(\Omega\tau_{12})\Big(\,\langle 2|2\rangle\,|1\rangle\langle 1| \;+\; \langle 1|1\rangle\,|2\rangle\langle 2|\,\Big)\\
&\qquad +\; \{\phi_1^D, \phi_2^D\}\cos(\Omega\tau_{12})\Big(\,\langle 2|2\rangle\,|1\rangle\langle 1| \;-\; \langle 1|1\rangle\,|2\rangle\langle 2|\,\Big)\Big]\,, &(6.2.17)
\end{aligned}$$

where

$$\Delta_{12} \;\equiv\; \frac{1}{i}\;\langle 0_M^\phi|[\phi_1^D, \phi_2^D]|0_M^\phi\rangle \tag{6.2.18}$$

is the Pauli-Jordan function of the $\phi$ field (evaluated at points on the detector's path) and

$$\tau_{12} \equiv \tau_1 - \tau_2 > 0\,. \tag{6.2.19}$$

The lowest-order, non-vanishing contribution to $\mathbb{P}$ arises from $\mathcal{F}_2$ in Eq. (6.2.17), and is

$$\begin{aligned}
\mathbb{P}(2;t) \;&=\; |\mu|^2\int_0^t \frac{\mathrm{d}t_1}{\gamma(t_1)}\int_0^{t_1}\frac{\mathrm{d}t_2}{\gamma(t_2)}\;\Big[\Delta_{12}^R\sin(\Omega\tau_{12}) \;+\; \Delta_{12}^H\cos(\Omega\tau_{12})\Big]\\
&=\; |\mu|^2\int_0^\tau \mathrm{d}\tau_1\int_0^{\tau_1}\mathrm{d}\tau_2\;\Big[\Delta_{12}^R\sin(\Omega\tau_{12}) \;+\; \Delta_{12}^H\cos(\Omega\tau_{12})\Big]\,, &(6.2.20)
\end{aligned}$$

where, due to the time-ordering, $\Delta_{12}$ has become the retarded propagator

$$\Delta^R_{12} \equiv \Theta_{12}\,\Delta_{12}\,, \tag{6.2.21}$$

and

$$\Delta^H_{12} \equiv \langle 0^\phi_M|\{\phi^D_1, \phi^D_2\}|0^\phi_M\rangle \tag{6.2.22}$$

is the Hadamard propagator of the $\phi$ field (evaluated at points on the detector's path). Eq. (6.2.20) is equal to

$$\mathbb{P}(2;t) = |\mu|^2 \int_0^\tau \mathrm{d}\tau_1 \int_0^\tau \mathrm{d}\tau_2\, e^{-i\Omega(\tau_1-\tau_2)}\, \langle 0^\phi_M|\phi^D_1\,\phi^D_2|0^\phi_M\rangle \equiv |\mu|^2\, F(\Omega)\,, \tag{6.2.23}$$

as calculated by DeWitt [161] and seen in many studies thereafter (albeit usually with integration limits $-\infty < \tau_1, \tau_2 < \infty$). The detector response function, $F(\Omega)$, is defined such that it is the component of the transition probability which does not depend on the detector's internal properties, only its energy gap and trajectory [155, 159, 160, 162].

Since the commutator of interaction-picture fields is proportional to the identity operator, the free retarded propagator $\Delta^R$ does not depend on the initial state[1]. Therefore, writing the rate in terms of $\Delta^R$ and $\Delta^H$ separates it into a term which does not depend on the initial state and a term which does.

For time-like intervals, $\Delta^R_{12}$ and $\Delta^H_{12}$ are given by [33][2],

$$\Delta^R_{12} = \frac{m^2}{4\pi}\frac{J_1(ms^\alpha_{12})}{ms^\alpha_{12}} - \frac{\delta((s^\alpha_{12})^2)}{2\pi} \quad \text{and} \quad \Delta^H_{12} = \frac{m^2}{4\pi}\frac{Y_1(ms^\alpha_{12})}{ms^\alpha_{12}}\,, \tag{6.2.24}$$

where $J_1$ and $Y_1$ are Bessel functions of the first and second kind, and

$$\begin{aligned} s^\alpha_{12} &\equiv \sqrt{(x^\mu_1 - x^\mu_2)^2} = \sqrt{(t_1-t_2)^2 - \frac{1}{\alpha^2}\left(\sqrt{1+\alpha^2 t_1^2} - \sqrt{1+\alpha^2 t_2^2}\right)^2} \\ &= \frac{2}{\alpha}\sinh\frac{\alpha\tau_{12}}{2}\,, \end{aligned} \tag{6.2.25}$$

for the trajectory given in Eq. (6.2.1). The $\delta$-function in Eq. (6.2.24) only has support at $\tau_{12} = 0$ and will not contribute further (the coefficient of $\Delta^R_{12}$ will always vanish at

[1] This is no longer the case when the retarded propagator is dressed with self-energy corrections.

[2] Note that there is a sign error on the $Y_1(z_{ij})$ term in equation (A11) in Ref. [33] (see Ref. [74]).

$\tau_{12} = 0$). Eq. (6.2.20) thus becomes

$$\mathbb{P}(2;t) = |\mu|^2 \int_0^{\tau} \mathrm{d}\tau_1 \int_0^{\tau_1} \mathrm{d}\tau_2 \, \frac{m^2}{4\pi} \left[ \frac{J_1(m s_{12}^{\alpha})}{m s_{12}^{\alpha}} \sin(\Omega\tau_{12}) + \frac{Y_1(m s_{12}^{\alpha})}{m s_{12}^{\alpha}} \cos(\Omega\tau_{12}) \right] . \tag{6.2.26}$$

This expression is divergent for $\tau_{12} = 0$, since $Y_1(x) \to -\infty$ as $x \to 0$. This is because the two-point correlation function evaluated at a point is infinite, and the divergence is not present if one considers a detector of finite spatial extent [18, 155, 158, 161]. The expression can be regularised by considering a spatial profile [159, 162, 163], but it would remain the case that a measurement at $\tau = 0$ would be divergent. Changing variables and introducing a lower limit on the integral to cut off the divergent part, the transition probability becomes

$$\mathbb{P}(2;t) = |\mu|^2 \int_{1/\Lambda}^{\tau} \mathrm{d}\tau_1 \int_{1/\Lambda}^{\tau_1} \mathrm{d}\tau_{12} \, \frac{m^2}{4\pi} \left[ \frac{J_1(m s_{12}^{\alpha})}{m s_{12}^{\alpha}} \sin(\Omega\tau_{12}) + \frac{Y_1(m s_{12}^{\alpha})}{m s_{12}^{\alpha}} \cos(\Omega\tau_{12}) \right] . \tag{6.2.27}$$

The lowest-order transition rate is then

$$\frac{\partial \mathbb{P}}{\partial \tau} = |\mu|^2 \int_{1/\Lambda}^{\tau} \mathrm{d}\tau_{12} \left[ \Delta_{12}^{R} \sin(\Omega\tau_{12}) + \Delta_{12}^{H} \cos(\Omega\tau_{12}) \right] \tag{6.2.28}$$

$$= \frac{m^2|\mu|^2}{4\pi} \int_{1/\Lambda}^{\tau} \mathrm{d}\tau_{12} \left[ \frac{J_1(m s_{12}^{\alpha})}{m s_{12}^{\alpha}} \sin \Omega\tau_{12} + \frac{Y_1(m s_{12}^{\alpha})}{m s_{12}^{\alpha}} \cos \Omega\tau_{12} \right] . \tag{6.2.29}$$

Consider this transition rate with the inertial ($\alpha = 0$) case subtracted,

$$\frac{\partial \mathbb{P}}{\partial \tau} - \left.\frac{\partial \mathbb{P}}{\partial \tau}\right|_{\alpha=0} = \frac{m^2|\mu|^2}{4\pi} \int_{1/\Lambda}^{\tau} \mathrm{d}\tau_{12} \left[ \left( \frac{J_1(m s_{12}^{\alpha})}{m s_{12}^{\alpha}} - \frac{J_1(m\tau_{12})}{m\tau_{12}} \right) \sin \Omega\tau_{12} + \left( \frac{Y_1(m s_{12}^{\alpha})}{m s_{12}^{\alpha}} - \frac{Y_1(m\tau_{12})}{m\tau_{12}} \right) \cos \Omega\tau_{12} \right] . \tag{6.2.30}$$

The divergence as $\tau_{12} \to 0$ in Eq. (6.2.29) is independent of $\alpha$ and cancels. Thus, Eq. (6.2.30) is independent of $\Lambda$ as $\frac{1}{\Lambda} \to 0$. Subtracting the inertial rate also gives an intuitive interpretation of the expression: it is the transition rate *due to the detector's acceleration.*

If we are not fully inclusive over the final state of the radiation field and instead require that we remain in the Minkowski vacuum, then $E = |0^{\phi}\rangle\langle 0^{\phi}| \otimes |2^{D}\rangle\langle 2^{D}|$. In this case, we can quickly convince ourselves that the excitation probability—now a single

matrix element squared—is zero, i.e.

$$\mathbb{P}(2;t) \;=\; \left|\langle 0^\phi, 2^D|U_{t,0}|0^\phi, 1^D\rangle\right|^2 \;=\; 0\,, \tag{6.2.31}$$

since for this to be non-zero, $U_{t,0}$ would need an even number of field operators, $\phi(t, \mathbf{x}^D(t))$, and an odd number of monopole operators, $M^D(t)$. Since $H^{\text{int}}(t)$ is linear in both $\phi(t, \mathbf{x}^D(t))$ and $M^D(t)$, this matrix element must therefore equal zero.

### 6.2.2 Different Limits

Considering Eq. (6.2.29) in different limits can simplify the expression and act as a cross-check for numerical results, since all results derived in this section conform with the numerical results in Section 6.5. First, the case of a massless scalar field ($m = 0$) is considered. It is also shown that, for small acceleration $\alpha$, the subtracted rate scales as $\alpha^2$. At early times, the subtracted rate is independent of mass. At late times, the subtracted rate exhibits decaying oscillations as it tends to a constant, with the period and zeroes of the integrand of Eq. (6.2.43) agreeing with the period and extrema of the numerical results.

**Massless limit** $(m \to 0)$

When $m \to 0$, $\Delta^R_{12} \to 0$ and $\Delta^H_{12} \to -1/2\pi^2|s^\alpha_{12}|^2$, which leaves

$$\mathbb{P}|_{m=0} = \int_0^t \mathrm{d}t_1 \mathrm{d}t_2 \frac{-|\mu|^2}{\gamma(t_1)\gamma(t_2)}\,\Theta_{12}\frac{\cos(\Omega\tau_{12})}{2\pi^2(s^\alpha_{12})^2}\;; \tag{6.2.32}$$

$$\left.\frac{\partial\mathbb{P}}{\partial\tau}\right|_{m=0} = -\frac{|\mu|^2\alpha^2}{8\pi^2}\int_{1/\Lambda}^{\tau}\mathrm{d}\tau_{12}\,\frac{\cos(\Omega\tau_{12})}{\sinh^2\frac{1}{2}\alpha\tau_{12}}\;; \tag{6.2.33}$$

$$\left.\frac{\partial\mathbb{P}}{\partial\tau}\right|_{m=0} - \left.\frac{\partial\mathbb{P}}{\partial\tau}\right|_{m,\alpha=0} = -\frac{|\mu|^2}{8\pi^2}\int_{1/\Lambda}^{\tau}\mathrm{d}\tau_{12}\,\cos(\Omega\tau_{12})\left(\frac{\alpha^2}{\sinh^2\frac{1}{2}\alpha\tau_{12}} - \frac{4}{\tau_{12}^2}\right). \tag{6.2.34}$$

**Small acceleration** $(\alpha \ll 1/\tau)$

For $\alpha \ll 1/\tau$,

$$\frac{J_1(ms^\alpha_{12})}{ms^\alpha_{12}} - \frac{J_1(m\tau_{12})}{m\tau_{12}} \to -\frac{\alpha^2\tau^2}{24}J_2(m\tau) + \mathcal{O}((\alpha\tau)^4)\,, \tag{6.2.35}$$

$$\frac{Y_1(ms_{12}^{\alpha})}{ms_{12}^{\alpha}} - \frac{Y_1(m\tau_{12})}{m\tau_{12}} \to -\frac{\alpha^2\tau^2}{24}Y_2(m\tau) + \mathcal{O}((\alpha\tau)^4)\,, \tag{6.2.36}$$

and Eq. (6.2.30) becomes

$$\frac{\partial \mathbb{P}}{\partial \tau} - \frac{\partial \mathbb{P}}{\partial \tau}\bigg|_{\alpha=0} = \frac{m^2\alpha^2|\mu|^2}{96\pi}\int_{1/\Lambda}^{\tau} \mathrm{d}\tau_{12}\,\Big[J_2(m\tau_{12})\sin\Omega\tau_{12} + Y_2(m\tau_{12})\cos\Omega\tau_{12}\Big]\,. \tag{6.2.37}$$

Thus, for small $\alpha$, the rate scales as $\alpha^2$.

**Early times** $(\alpha\tau,\ m\tau \to 0)$

For $\alpha\tau$, $m\tau \to 0$, we use

$$\frac{J_1(ms_{12}^{\alpha})}{ms_{12}^{\alpha}} - \frac{J_1(m\tau_{12})}{m\tau_{12}} \to 0\,, \tag{6.2.38}$$

$$\frac{Y_1(ms_{12}^{\alpha})}{ms_{12}^{\alpha}} - \frac{Y_1(m\tau_{12})}{m\tau_{12}} \to \frac{\alpha^2}{6m^2\pi}\,, \tag{6.2.39}$$

and Eq. (6.2.30) becomes

$$\frac{\partial \mathbb{P}}{\partial \tau} - \frac{\partial \mathbb{P}}{\partial \tau}\bigg|_{\alpha=0} = \frac{\alpha^2|\mu|^2}{24\pi^2}\tau\,. \tag{6.2.40}$$

**Late times** $(\tau \gg 1/\alpha)$

As $\tau_{12} \to \infty$, the integrand of Eq. (6.2.29) goes to zero. This means that the rate tends to a constant as $\tau \to \infty$. This constant value is calculated in Section 6.2.3 (Eq. (6.2.49)). However, as the rate tends to a constant, it also oscillates about the constant value. This is because, for large arguments,

$$J_1(x) \to \sqrt{\frac{2}{\pi x}}\cos\left(x - \frac{3\pi}{4}\right), \tag{6.2.41}$$

$$Y_1(x) \to \sqrt{\frac{2}{\pi x}}\sin\left(x - \frac{3\pi}{4}\right), \tag{6.2.42}$$

such that (after taking $s_{12}^{\alpha} \gg \tau_{12}$),

$$\frac{\partial \mathbb{P}}{\partial \tau} - \frac{\partial \mathbb{P}}{\partial \tau}\bigg|_{\alpha=0} = \text{constant } + |\mu|^2\sqrt{\frac{m}{8\pi^3}}\int_{\tau_0}^{\tau} \mathrm{d}\tau_{12}\tau_{12}^{-3/2}\sin\left((m+\Omega)\tau_{12} + \frac{\pi}{4}\right), \tag{6.2.43}$$

where $\tau_0$ is a time large enough for the late-time limit to apply.

### 6.2.3 Momentum space

The propagators in Eq. (6.2.20) are Lorentz invariant and can be evaluated in any frame. Evaluating the momentum-space expressions for the propagators in the frame in which $x^0_{12} = s^\alpha_{12}$ and $\mathbf{x}_{12} = 0$ gives

$$\Delta_{12} = -\int \frac{\mathrm{d}^3\mathbf{p}}{(2\pi)^3} \frac{e^{i\mathbf{p}\cdot\mathbf{x}_{12}}}{\omega_\mathbf{p}} \sin\big(\omega_\mathbf{p} x^0_{12}\big) = -\int \frac{\mathrm{d}^3\mathbf{p}}{(2\pi)^3} \frac{\sin(\omega_\mathbf{p} s^\alpha_{12})}{\omega_\mathbf{p}} , \tag{6.2.44}$$

$$\Delta^H_{12} = \int \frac{\mathrm{d}^3\mathbf{p}}{(2\pi)^3} \frac{e^{i\mathbf{p}\cdot\mathbf{x}_{12}}}{\omega_\mathbf{p}} \cos\big(\omega_\mathbf{p} x^0_{12}\big) = \int \frac{\mathrm{d}^3\mathbf{p}}{(2\pi)^3} \frac{\cos(\omega_\mathbf{p} s^\alpha_{12})}{\omega_\mathbf{p}} , \tag{6.2.45}$$

where

$$\omega_\mathbf{p} = \sqrt{\mathbf{p}^2 + m^2} \,. \tag{6.2.46}$$

Inserting these expressions into Eq. (6.2.20) gives

$$\begin{aligned} \mathbb{P}(2;t) \;&=\; \frac{|\mu|^2}{8\pi^3} \int_0^\tau \mathrm{d}\tau_1 \int_0^{\tau_1} \mathrm{d}\tau_2 \int \mathrm{d}^2\mathbf{p}_\perp \int_{-\infty}^{\infty} \frac{\mathrm{d}p_x}{\omega_\mathbf{p}} \\ &\quad \Big[ -\sin(\omega_\mathbf{p} s^\alpha_{12}) \sin(\Omega\tau_{12}) \;+\; \cos(\omega_\mathbf{p} s^\alpha_{12}) \cos(\Omega\tau_{12}) \Big] \\ &=\; \frac{|\mu|^2}{8\pi^3} \int_0^\tau \mathrm{d}\tau_1 \int_0^{\tau_1} \mathrm{d}\tau_2 \int \mathrm{d}^2\mathbf{p}_\perp \int_{-\infty}^{\infty} \frac{\mathrm{d}p_x}{\omega_\mathbf{p}} \cos(\omega_\mathbf{p} s^\alpha_{12} + \Omega\tau_{12}) \,. \end{aligned} \tag{6.2.47}$$

Thus, the rate is

$$\begin{aligned} \frac{\partial\mathbb{P}}{\partial\tau} &= \frac{|\mu|^2}{8\pi^3} \int_0^\tau \mathrm{d}\tau_{12} \int \mathrm{d}^2\mathbf{p}_\perp \int_{-\infty}^{\infty} \frac{\mathrm{d}p_x}{\omega_\mathbf{p}} \cos\left[\frac{2\omega_\mathbf{p}}{\alpha} \sinh(\alpha\tau_{12}/2) + \Omega\tau_{12}\right] \\ &= \frac{|\mu|^2}{16\pi^3} \int_{-\tau}^\tau \mathrm{d}\tau_{12} \int \mathrm{d}^2\mathbf{p}_\perp \int_{-\infty}^{\infty} \frac{\mathrm{d}p_x}{\omega_\mathbf{p}} \cos\left[\frac{2\omega_\mathbf{p}}{\alpha} \sinh(\alpha\tau_{12}/2) + \Omega\tau_{12}\right] , \end{aligned} \tag{6.2.48}$$

agreeing with the real part of Eq. (3.12) of [22], which considers an initial state defined in the infinite past and a measurement taken in the infinite future (i.e., $\tau \to \infty$). In this limit, following [22],

$$\frac{\partial\mathbb{P}(\tau\to\infty)}{\partial\tau} = \frac{|\mu|^2}{2\pi^2\alpha} e^{-\frac{\pi\Omega}{\alpha}} \int_m^\infty \mathrm{d}\nu\, \nu \left| K_{i\Omega/\alpha}\left(\frac{\nu}{\alpha}\right)\right|^2 , \tag{6.2.49}$$

where $\nu = \sqrt{\mathbf{p}_\perp^2 + m^2}$. In the limit $\alpha \to 0$, the integrand vanishes as

$$\frac{8\pi\alpha\nu}{\Omega} e^{-\frac{\pi\Omega}{\alpha}} \sin^2\!\left(\frac{\nu^2}{4\Omega\alpha} + \ldots\right) ,$$

which means an inertial detector does not undergo excitation in a vacuum at $\tau \to \infty$. As a result, at $\tau \to \infty$, the subtracted rate given by Eq. (6.2.30) is equal to the unsubtracted rate given by Eq. (6.2.29).

Note that for an inertial path, $\mathbf{x}^D = (vt, 0, 0)$, the parameter $s_{12}^\alpha$ is replaced by

$$s_{12} = \sqrt{(t_1 - t_2)^2 - v^2\,(t_1 - t_2)^2} = \gamma^{-1}\, t_{12} = \tau_{12}\,, \tag{6.2.50}$$

such that the transition rate becomes

$$\frac{\partial \mathbb{P}}{\partial \tau} = \frac{|\mu|^2}{16\pi^3} \int_{-\tau}^{\tau} \mathrm{d}\tau_{12} \int \mathrm{d}^2\mathbf{p}_\perp \int_{-\infty}^{\infty} \frac{\mathrm{d}p_x}{\omega_\mathbf{p}} \cos\left[\omega_\mathbf{p}\tau_{12} + \Omega\tau_{12}\right]\,. \tag{6.2.51}$$

As we will now show, we can start from this expression and rederive Eq. (6.2.48). A derivation of the Unruh effect along these lines (for $m = 0$) appears in Ref. [164].

For the uniformly accelerated trajectory, the modes are subject to a characteristic, time-dependent Doppler shift, such that

$$\omega'_\mathbf{p}(\tau) = \omega_\mathbf{p} \cosh(\alpha\tau) - p_x \sinh(\alpha\tau)\,, \tag{6.2.52}$$

reducing to

$$\omega'_\mathbf{p}(\tau) = \omega_\mathbf{p} e^{\mp\alpha\tau}\,, \qquad p_x \gtrless 0\,, \tag{6.2.53}$$

in the massless limit and in one spatial dimension, as used in Ref. [164]. To account for this, we can proceed from Eq. (6.2.51) by replacing

$$\begin{aligned} \omega_\mathbf{p}\tau_{12} \longrightarrow \int_{\tau_2}^{\tau_1} \mathrm{d}\tau'\,\omega'_\mathbf{p}(\tau') &= \frac{\omega_\mathbf{p}}{\alpha}\left[\sinh(\alpha\tau_1) - \sinh(\alpha\tau_2)\right] - \frac{p_x}{\alpha}\left[\cosh(\alpha\tau_1) - \cosh(\alpha\tau_2)\right] \\ &= \frac{2}{\alpha} \sinh(\alpha\tau_{12}/2)\left[\omega_\mathbf{p}\cosh(\alpha\bar{\tau}) - p_x \sinh(\alpha\bar{\tau})\right]\,, \end{aligned} \tag{6.2.54}$$

where $\bar{\tau} = (\tau_1 + \tau_2)/2$. The transition rate should not depend on $\bar{\tau}$. To see this, we boost to the instantaneous rest frame of the modes via the transformations

$$\omega''_\mathbf{p} = \omega_\mathbf{p} \cosh(\alpha\bar{\tau}) - p_x \sinh(\alpha\bar{\tau})\,, \tag{6.2.55a}$$

$$p''_x = p_x \cosh(\alpha\bar{\tau}) - \omega_\mathbf{p} \sinh(\alpha\bar{\tau})\,. \tag{6.2.55b}$$

The measure transforms as $\mathrm{d}p_x/\omega_\mathbf{p} = \mathrm{d}p''_x/\omega''_\mathbf{p}$, and we recover Eq. (6.2.48) after relabelling the integration variables.

### 6.2.4 Transients

The integral over $\tau_{12} \in [-\tau, \tau]$ in Eq. (6.2.48) can be expressed as an integral over the whole real line by inserting a top-hat distribution of width $2\tau$, centred on the origin. Replacing the latter with its Fourier transform, we can write

$$\begin{aligned}\frac{\partial \mathbb{P}}{\partial \tau} &= \frac{|\mu|^2}{16\pi^3} \mathrm{Re} \int_{-\infty}^{\infty} \mathrm{d}\tau_{12} \int_{-\infty}^{\infty} \frac{\mathrm{d}k}{\pi} \frac{\sin(k\tau)}{k} \int \mathrm{d}^2\mathbf{p}_\perp \\ &\quad \times \int_{-\infty}^{\infty} \frac{\mathrm{d}p_x}{\omega_\mathbf{p}} \exp\left\{ i \left[ \frac{2\omega_\mathbf{p}}{\alpha} \sinh(\alpha\tau_{12}/2) + (\Omega - k)\tau_{12} \right] \right\} \\ &= \frac{|\mu|^2}{16\pi^3} \int_{-\infty}^{\infty} \mathrm{d}\tau_{12} \int_{-\infty}^{\infty} \frac{\mathrm{d}k}{\pi} \frac{\sin(k\tau)}{k} \int \mathrm{d}^2\mathbf{p}_\perp \\ &\quad \times \int_{-\infty}^{\infty} \frac{\mathrm{d}p_x}{\omega_\mathbf{p}} \cos\left[ \frac{2\omega_\mathbf{p}}{\alpha} \sinh(\alpha\tau_{12}/2) + (\Omega - k)\tau_{12} \right] . \end{aligned} \tag{6.2.56}$$

Swapping the order of the $\tau_{12}$ and $k$ integrals, we recognise the integral from Ref. [22] with $\Omega \to \Omega - k$, such that we have

$$\frac{\partial \mathbb{P}(\tau)}{\partial \tau} = \frac{|\mu|^2}{2\pi^2\alpha} \int_{-\infty}^{\infty} \frac{\mathrm{d}k}{\pi} \frac{\sin(k\tau)}{k} e^{-\frac{\pi(\Omega-k)}{\alpha}} \int_m^{\infty} \mathrm{d}\nu\, \nu \left| K_{i(\Omega-k)/\alpha}\left(\frac{\nu}{\alpha}\right)\right|^2 . \tag{6.2.57}$$

In the limit, $\tau \to \infty$, we have

$$\lim_{\tau\to\infty} \frac{1}{\pi} \frac{\sin(k\tau)}{k} = \delta(k) , \tag{6.2.58}$$

and we recover Eq. (6.2.49).

Thus, we see that the transients arise from a convolution with the Fourier transform of the top-hat distribution. It is as if we fixed the field configuration at $t = -\infty$ and discontinuously turned on the interaction at $t = 0$. If instead we turned the interaction on smoothly using some switching function then the transients would arise from a convolution with the Fourier transform of this switching function.

## 6.3 Excitation rate in a Rindler thermal bath

Before examining the results of the previous section, we shall compute the corresponding quantities from the perspective of a Rindler observer confined to the right Rindler wedge. The transformation from Minkowski to Rindler coordinates, $(\eta, \xi, y, z)$,

is given by Eq. (6.1.3)

$$t = \alpha^{-1} e^{\alpha\xi} \sinh \alpha\eta\,, \quad x = \alpha^{-1} e^{\alpha\xi} \cosh \alpha\eta\,, \quad y = y\,, \quad z = z\,. \tag{6.3.1}$$

In these coordinates, the atom is stationary at $\xi = 0$ (recovering Eqs. (6.2.1) and (6.2.2)), and the Minkowski vacuum state is exactly equivalent to a thermal state of Rindler particles (as explained in Section 6.1.2). Therefore, from a Rindler observer's perspective, the detector is stationary in a thermal bath of Rindler particles at temperature $T$. In other words, the thermal bath is in an effective gravitational field (in accordance with the equivalence principle), and hence is different to a free-falling (inertial) thermal bath, which is considered in Section 6.4.

The calculation proceeds in an identical fashion up to Eq. (6.2.28). We may expand the field using creation and annihilation operators for Rindler particles, i.e., in the right Rindler wedge, as in Eq. (6.1.9),

$$\phi(x) = \int \mathrm{d}\omega_{\mathbf{p}} \mathrm{d}^2\mathbf{p}_\perp [v^R_{\omega_{\mathbf{p}}\mathbf{p}_\perp} b^R_{\omega_{\mathbf{p}}\mathbf{p}_\perp} + \text{H.c.}]\,, \tag{6.3.2}$$

where 'H.c.' stands for Hermitian conjugate and [22, 155]

$$\begin{aligned} v^R_{\omega_{\mathbf{p}}\mathbf{p}_\perp} &= \left[\frac{\sinh(\pi\omega_{\mathbf{p}}/\alpha)}{4\pi^4\alpha}\right]^{1/2} K_{i\omega_{\mathbf{p}}/\alpha}\left[\frac{\sqrt{\mathbf{p}_\perp^2 + m^2}}{\alpha e^{-\alpha\xi}}\right] e^{-i\omega_{\mathbf{p}}\tau + i\mathbf{p}_\perp\cdot\mathbf{x}_\perp} \\ &= \left[\frac{\sinh(\omega_{\mathbf{p}}/2\,T)}{8\pi^5 T}\right]^{1/2} K_{i\omega_{\mathbf{p}}/2\pi T}\left[\frac{\sqrt{\mathbf{p}_\perp^2 + m^2}}{2\pi T}\right] e^{-i\omega_{\mathbf{p}}\tau + i\mathbf{p}_\perp\cdot\mathbf{x}_\perp}\,, \end{aligned} \tag{6.3.3}$$

where $T = T_{\mathrm{U}} \equiv \alpha/2\pi$ is the Unruh temperature measured by an observer at the Rindler coordinate $\xi = 0$, and $\mathbf{x}_\perp = (y, z)$. The Rindler operators $b^R_{\omega_{\mathbf{p}}\mathbf{p}_\perp}$ and $b^{R\dagger}_{\omega_{\mathbf{p}}\mathbf{p}_\perp}$ define the Fock space for Rindler particles, and give the Minkowski vacuum expectation values [22]

$$\begin{aligned} \langle 0^\phi_M | b^{R\dagger}_{\omega_{\mathbf{p}}\mathbf{p}_\perp} b^R_{E'_{\mathbf{p}}\mathbf{p}'_\perp} | 0^\phi_M\rangle &= (e^{\omega_{\mathbf{p}}/T} - 1)^{-1}\, \delta(\omega_{\mathbf{p}} - E'_{\mathbf{p}})\, \delta^2(\mathbf{p}_\perp - \mathbf{p}'_\perp) \\ &= n\, \delta(\omega_{\mathbf{p}} - E'_{\mathbf{p}})\, \delta^2(\mathbf{p}_\perp - \mathbf{p}'_\perp)\,, \end{aligned} \tag{6.3.4}$$

$$\begin{aligned} \langle 0^\phi_M | b^R_{\omega_{\mathbf{p}}\mathbf{p}_\perp} b^{R\dagger}_{\omega_{\mathbf{p}}\mathbf{p}_\perp} | 0^\phi_M\rangle &= (1 - e^{-\omega_{\mathbf{p}}/T})^{-1}\, \delta(\omega_{\mathbf{p}} - E'_{\mathbf{p}})\, \delta^2(\mathbf{p}_\perp - \mathbf{p}'_\perp) \\ &= (n+1)\, \delta(\omega_{\mathbf{p}} - E'_{\mathbf{p}})\, \delta^2(\mathbf{p}_\perp - \mathbf{p}'_\perp)\,, \end{aligned} \tag{6.3.5}$$

where

$$n \equiv n(\omega_{\mathbf{p}}) = (e^{\omega_{\mathbf{p}}/T} - 1)^{-1} \tag{6.3.6}$$

is the Bose-Einstein distribution for a thermal bath at temperature $T$. For a stationary trajectory in Rindler coordinates at $\xi = 0$, the Pauli-Jordan and Hadamard functions can be expressed as

$$\begin{aligned}
\Delta_{12} &\equiv \frac{1}{i}\langle 0_M^\phi|[\phi_1,\phi_2]|0_M^\phi\rangle \\
&= -\int \mathrm{d}\omega_{\mathbf{p}}\mathrm{d}^2\mathbf{p}_\perp \frac{\sinh(\omega_{\mathbf{p}}/2T)}{4\pi^5 T}\left|K_{i\omega_{\mathbf{p}}/2\pi T}\left[\frac{\sqrt{\mathbf{p}_\perp^2+m^2}}{2\pi T}\right]\right|^2 \sin(\omega_{\mathbf{p}}\tau_{12})\,, \qquad (6.3.7)\\
\Delta_{12}^H &\equiv \langle 0_M^\phi|\{\phi_1,\phi_2\}|0_M^\phi\rangle \\
&= \int \mathrm{d}\omega_{\mathbf{p}}\mathrm{d}^2\mathbf{p}_\perp \frac{\sinh(\omega_{\mathbf{p}}/2T)}{4\pi^5 T}\left|K_{i\omega_{\mathbf{p}}/2\pi T}\left[\frac{\sqrt{\mathbf{p}_\perp^2+m^2}}{2\pi T}\right]\right|^2 \cos(\omega_{\mathbf{p}}\tau_{12})\big(2n+1\big)\,.
\end{aligned}$$

(6.3.8)

The transition rate (Eq. (6.2.28)) then becomes

$$\begin{aligned}
\frac{\partial \mathbb{P}}{\partial \tau} &= \frac{|\mu|^2}{4\pi^5 T}\int_0^\tau \mathrm{d}\tau_{12}\int_0^\infty \mathrm{d}\omega_{\mathbf{p}}\mathrm{d}^2\mathbf{p}_\perp \sinh\left(\frac{\omega_{\mathbf{p}}}{2T}\right)\left|K_{i\omega_{\mathbf{p}}/2\pi T}\left(\frac{\sqrt{\mathbf{p}_\perp^2+m^2}}{2\pi T}\right)\right|^2 \times \\
&\quad \Big(-\sin(\omega_{\mathbf{p}}\tau_{12})\sin(\Omega\tau_{12}) + (2n+1)\cos(\omega_{\mathbf{p}}\tau_{12})\cos(\Omega\tau_{12})\Big) \\
&= \frac{|\mu|^2}{2\pi^4 T}\int_0^\infty \mathrm{d}\omega_{\mathbf{p}}\int_m^\infty \mathrm{d}\nu\,\nu\sinh\left(\frac{E}{2T}\right)\left|K_{i\omega_{\mathbf{p}}/2\pi T}\left(\frac{\nu}{2\pi T}\right)\right|^2 \times \\
&\quad \left((n+1)\frac{\sin\big((\omega_{\mathbf{p}}+\Omega)\,\tau\big)}{\omega_{\mathbf{p}}+\Omega} + n\,\frac{\sin\big((\omega_{\mathbf{p}}-\Omega)\,\tau\big)}{\omega_{\mathbf{p}}-\Omega}\right) \\
&= \frac{|\mu|^2}{4\pi^4 T}\int_0^\infty \mathrm{d}\omega_{\mathbf{p}}\int_m^\infty \mathrm{d}\nu\,\nu\left|K_{i\omega_{\mathbf{p}}/2\pi T}\left(\frac{\nu}{2\pi T}\right)\right|^2 \times \\
&\quad \left(e^{\omega_{\mathbf{p}}/2T}\frac{\sin\big((\omega_{\mathbf{p}}+\Omega)\,\tau\big)}{\omega_{\mathbf{p}}+\Omega} + e^{-\omega_{\mathbf{p}}/2T}\,\frac{\sin\big((\omega_{\mathbf{p}}-\Omega)\,\tau\big)}{\omega_{\mathbf{p}}-\Omega}\right). 
\end{aligned} \tag{6.3.9}$$

The term proportional to $(n+1)$ in Eq. (6.3.9) corresponds to the *emission* rate from the detector to the Rindler thermal bath, and the term proportional to $n$ corresponds to the *absorption* rate of the detector from the Rindler thermal bath. The $\nu$ integral can be performed when $m = 0$ [165]:

$$\left.\frac{\partial \mathbb{P}}{\partial \tau}\right|_{m=0} = \frac{|\mu|^2 T}{\pi^2}\int_0^\infty \mathrm{d}\omega_{\mathbf{p}}\sinh\left(\frac{\omega_{\mathbf{p}}}{2T}\right)\left|\Gamma\left(1+\frac{i\omega_{\mathbf{p}}}{2\pi T}\right)\right|^2$$

$$\times\left((n+1)\frac{\sin\left((\omega_{\mathbf{p}}+\Omega)\,\tau\right)}{\omega_{\mathbf{p}}+\Omega}+n\,\frac{\sin\left((\omega_{\mathbf{p}}-\Omega)\,\tau\right)}{\omega_{\mathbf{p}}-\Omega}\right)$$

$$=\frac{|\mu|^2}{2\pi^2}\int_0^\infty \mathrm{d}\omega_{\mathbf{p}}\,\omega_{\mathbf{p}}\left((n+1)\frac{\sin\left((\omega_{\mathbf{p}}+\Omega)\,\tau\right)}{\omega_{\mathbf{p}}+\Omega}+n\,\frac{\sin\left((\omega_{\mathbf{p}}-\Omega)\,\tau\right)}{\omega_{\mathbf{p}}-\Omega}\right), \tag{6.3.10}$$

where $\Gamma$ is the Gamma function defined by Eq.(2.6.15) and we have used [22]

$$\left|\Gamma\left(1+\frac{i\omega_{\mathbf{p}}}{2\pi T}\right)\right|^2=\frac{\omega_{\mathbf{p}}}{2T\sinh\left(\omega_{\mathbf{p}}/2T\right)}\,. \tag{6.3.11}$$

In order to match the calculation in Section 6.2, the inertial transition rate must be subtracted from Eq. (6.3.9). From the Rindler observer's perspective, this corresponds to the $T=0$ limit of the transition rate. This is simply the transition rate for an inertial observer in the Minkowski vacuum, which is calculated and expressed as an integral over energy in Appendix C. Note that the inertial rate is *not* the same as simply taking the part of the Rindler thermal rate which is not proportional to $n$; doing this would give the transition rate for an accelerating detector in the Rindler vacuum. Subtracting Eq. (C.0.4), the Rindler rate becomes

$$\begin{aligned}\frac{\partial\mathbb{P}}{\partial\tau}-\left.\frac{\partial\mathbb{P}}{\partial\tau}\right|_{T=0}=&\frac{|\mu|^2}{4\pi^4T}\int_0^\infty \mathrm{d}\omega_{\mathbf{p}}\int_m^\infty \mathrm{d}\nu\,\nu\left|K_{i\omega_{\mathbf{p}}/2\pi T}\left(\frac{\nu}{2\pi T}\right)\right|^2\times\\&\left(e^{\omega_{\mathbf{p}}/2T}\frac{\sin\left((\omega_{\mathbf{p}}+\Omega)\,\tau\right)}{\omega_{\mathbf{p}}+\Omega}+e^{-\omega_{\mathbf{p}}/2T}\,\frac{\sin\left((\omega_{\mathbf{p}}-\Omega)\,\tau\right)}{\omega_{\mathbf{p}}-\Omega}\right)\\&-\frac{|\mu|^2}{2\pi^2}\int_m^\infty \mathrm{d}\omega_{\mathbf{p}}\,\sqrt{\omega_{\mathbf{p}}^2-m^2}\,\frac{\sin[(\omega_{\mathbf{p}}+\Omega)\tau]}{\omega_{\mathbf{p}}+\Omega}\,.\end{aligned} \tag{6.3.12}$$

It can be confirmed numerically that this is the same result as Eq. (6.2.30) for all times. Eq. (6.3.12) can be shown to vanish when $T\to 0$, since

$$\lim_{T\to 0}\int_m^\infty \mathrm{d}\nu\,\nu\left|K_{i\omega_{\mathbf{p}}/2\pi T}\left(\frac{\nu}{2\pi T}\right)\right|^2=e^{-\omega_{\mathbf{p}}/2T}\,2\pi^2T\,\Theta(\omega_{\mathbf{p}}-m)\sqrt{\omega_{\mathbf{p}}^2-m^2}\,, \tag{6.3.13}$$

where $\Theta(x)$ is the Heaviside function [85]. Taking the $\tau\to\infty$ limit of Eq. (6.3.12) gives the transition rate of a detector coupled to a massive scalar field after the transient effects have subsided. In this limit, Eq. (6.3.12) becomes,

$$\frac{\partial\mathbb{P}(\tau\to\infty)}{\partial\tau}-\left.\frac{\partial\mathbb{P}(\tau\to\infty)}{\partial\tau}\right|_{T=0}=\frac{|\mu|^2}{4\pi^3T}e^{-\frac{\Omega}{2T}}\int_m^\infty \mathrm{d}\nu\,\nu\left|K_{i\Omega/2\pi T}\left(\frac{\nu}{2\pi T}\right)\right|^2, \tag{6.3.14}$$

since

$$\lim_{\tau\to\infty}\frac{\sin[(\omega_{\mathbf{p}}\pm\Omega)\tau]}{\omega_{\mathbf{p}}\pm\Omega}=\pi\delta(\omega_{\mathbf{p}}\pm\Omega)\,, \tag{6.3.15}$$

and only the delta function $\delta(\omega_{\mathbf{p}}-\Omega)$ has support over the domain of the energy integrals. This corresponds to the emission rate vanishing at late times (once the transients have decayed), so the detector only absorbs Rindler particles and does not emit them. This is the same result as Eq. (6.2.49).

For a massless scalar field

$$\left.\frac{\partial\mathbb{P}}{\partial\tau}\right|_{m=0}-\left.\frac{\partial\mathbb{P}}{\partial\tau}\right|_{T,m=0}=\frac{|\mu|^2}{2\pi^2}\int_0^\infty \mathrm{d}\omega_{\mathbf{p}}\,\omega_{\mathbf{p}}\,n\left(\frac{\sin\big((\omega_{\mathbf{p}}+\Omega)\,\tau\big)}{\omega_{\mathbf{p}}+\Omega}+\frac{\sin\big((\omega_{\mathbf{p}}-\Omega)\,\tau\big)}{\omega_{\mathbf{p}}-\Omega}\right), \tag{6.3.16}$$

which, in the $\tau\to\infty$ limit, reduces to

$$\left.\frac{\partial\mathbb{P}(\tau\to\infty)}{\partial\tau}\right|_{m=0}-\left.\frac{\partial\mathbb{P}(\tau\to\infty)}{\partial\tau}\right|_{m,T=0}=\frac{|\mu|^2}{2\pi}\frac{\Omega}{e^{\beta\Omega}-1}\,. \tag{6.3.17}$$

Coincidentally, this is the the expected transition rate for a detector in an inertial Bose-Einstein thermal bath of massless Minkowski particles, as will be shown in the next section.

## 6.4 Excitation rate in a Minkowski thermal bath

We shall now calculate the response of a detector in a thermal bath of Minkowski particles and show that this is different to the Rindler case for $m\neq 0$. The zero-temperature, Minkowski Hadamard function is given by

$$\Delta^H_{12}\big|_{T=0}=\left.\Delta^>_{12}\right|_{T=0}+\left.\Delta^<_{12}\right|_{T=0}=\int\frac{\mathrm{d}^4p}{(2\pi)^3}\,e^{-ip_\mu x^\mu_{12}}\,\delta(p^2-m^2)\,. \tag{6.4.1}$$

In equilibrium with a thermal bath at temperature, $T$, the Wightman functions are [155, 166]

$$\Delta^>_{12}(T)=\int\frac{\mathrm{d}^4p}{(2\pi)^3}\Big[\Theta(p^0)\,(1+n)+\Theta(-p^0)\,n\Big]e^{-ip_\mu x^\mu_{12}}\,\delta(p^2-m^2)\,, \tag{6.4.2}$$

$$\Delta^<_{12}(T)=\int\frac{\mathrm{d}^4p}{(2\pi)^3}\Big[\Theta(-p^0)\,(1+n)+\Theta(p^0)\,n\Big]e^{-ip_\mu x^\mu_{12}}\,\delta(p^2-m^2)\,. \tag{6.4.3}$$

where $n \equiv n(|p^0|) = (\exp(|p^0|/T) - 1)^{-1}$. The thermal Minkowski Hadamard function is therefore

$$\Delta^H_{12}(T) = \int \frac{\mathrm{d}^4p}{(2\pi)^3}\Big[\Theta(p^0) + \Theta(-p^0)\Big]\big(1+2n\big)e^{-ip_\mu x^\mu_{12}}\,\delta(p^2-m^2) \tag{6.4.4}$$

$$= \int \frac{\mathrm{d}^4p}{(2\pi)^3}\big(1+2n\big)e^{-ip_\mu x^\mu_{12}}\,\delta(p^2-m^2)\,. \tag{6.4.5}$$

Comparing with Eq. (6.4.1), we see that the thermal piece of the Hadamard function is simply the zero-temperature piece multiplied by $2n$. Since the detector is static, $\mathbf{x}_{12} = 0$ (as long as the detector is inertial, we can boost to this frame due to its time-like trajectory). Therefore, the thermal part of $\Delta^H_{12}$ takes the form,

$$\Delta^H_{12} \supset \int \frac{\mathrm{d}^4p}{(2\pi)^3}\, e^{-ip^0t_{12}}\,(2n)\,\delta(p^2-m^2) = \frac{1}{\pi^2}\int_m^\infty \mathrm{d}\omega_\mathbf{p}\,\sqrt{\omega_\mathbf{p}^2-m^2}\;n\,\cos(\omega_\mathbf{p}t_{12})\,. \tag{6.4.6}$$

The retarded propagator does not pick up a thermal part (i.e., it does not have any temperature dependence, since terms proportional to $n$ cancel when taking the difference of positive and negative Wightman functions). The thermal contribution to the excitation rate is thus

$$\begin{aligned}\frac{\partial\mathbb{P}}{\partial t} - \frac{\partial\mathbb{P}}{\partial t}\bigg|_{T=0} &= \frac{|\mu|^2}{\pi^2}\int_0^t \mathrm{d}t'\int_m^\infty \mathrm{d}\omega_\mathbf{p}\sqrt{\omega_\mathbf{p}^2-m^2}\;n\,\cos\Omega t'\cos\omega_\mathbf{p}t' \\ &= \frac{|\mu|^2}{2\pi^2}\int_m^\infty \mathrm{d}\omega_\mathbf{p}\,\sqrt{\omega_\mathbf{p}^2-m^2}\;n\left\{\frac{\sin[(\omega_\mathbf{p}-\Omega)t]}{\omega_\mathbf{p}-\Omega} + \frac{\sin[(\omega_\mathbf{p}+\Omega)t]}{\omega_\mathbf{p}+\Omega}\right\}.\end{aligned} \tag{6.4.7}$$

This expression is not equal to Eq. (6.3.9). However, with $m = 0$, it is exactly the same as Eq. (6.3.16). This means that the response of a *monopole detector* coupled to a *massless* scalar field is insensitive to the difference between inertial and accelerating thermal distributions. This misleading example may lead one to the conclusion that an accelerated detector responds identically to an inertial detector in an ordinary (Minkowski) Bose gas at finite temperature. This is not the statement of the Unruh effect. It is clearly not true for a massive scalar field, nor is it true for vector fields or other detector models [22, 167, 168].

In the limit $t \to \infty$, we obtain

$$\frac{\partial\mathbb{P}}{\partial t} - \frac{\partial\mathbb{P}}{\partial t}\bigg|_{T=0} = \frac{\partial\mathbb{P}}{\partial t}$$

$$= \frac{|\mu|^2}{2\pi} \int_m^\infty \mathrm{d}\omega_{\mathbf{p}} \sqrt{\omega_{\mathbf{p}}^2 - m^2} \, \frac{1}{e^{\beta\omega_{\mathbf{p}}} - 1} \left[ \delta(\omega_{\mathbf{p}} - \Omega) + \delta(\omega_{\mathbf{p}} + \Omega) \right] . \tag{6.4.8}$$

Only the first delta function has support, and only when $\Omega \geq m$, so we arrive at the result

$$\frac{\partial \mathbb{P}}{\partial t} = \frac{|\mu|^2}{2\pi} \frac{\sqrt{\Omega^2 - m^2}}{e^{\beta\Omega} - 1} \Theta(\Omega - m) \, , \tag{6.4.9}$$

agreeing with Eq. (3.73) of Ref. [151]. Thus, a detector in a thermal bath of Minkowski particles requires $\Omega > m$. This is not the case for the Rindler thermal bath, and this highlights a crucial physical difference. For the Rindler thermal bath, as $\tau \to \infty$, it remains true that the energy, $\omega_{\mathbf{p}}$, of the absorbed Rindler particle must equal the detector's energy gap, $\Omega$, but the transition rate is non-zero even if the energy gap is less than the mass of the field ($\Omega < m$). This is because $\omega_{\mathbf{p}} \geq m$ is a flat spacetime constraint, and a general field quantisation does not lead to a simple dispersion relation relating a particle's energy to its mass. This difference can be seen when comparing Figs 6.3 and 6.4. Further discussion of Rindler particles with energy $\omega_{\mathbf{p}} < m$ is given in Section III.A.3 of Ref. [22].

## 6.5 Numerical Results

The transition rate for a uniformly accelerating detector, with the inertial rate subtracted, is given by the identical Eqs. (6.2.30) and (6.3.12). The explicit dependence of the transition rate on the detector's proper time is shown in Fig. 6.2. Specifying the state at $\tau = 0$ causes transients and the frequency of these transients is dependent on $m/\Omega$ and independent of $\alpha/\Omega$. As $\Omega\tau \to \infty$, the transients decay and the rate tends to a constant value. We can also observe that the transients subside more rapidly for larger accelerations.

The constant, late-time value is given by Eq. (6.2.49) (and Eq. (6.3.14)). The dependence of this 'equilibrium' rate on temperature (and hence acceleration via $T = \alpha/2\pi$) and mass is shown in Fig. 6.3. The values of $\alpha/\Omega$ and $m/\Omega$ are chosen so as to scan a wide range of dimensionless ratios. Fig. 6.3a shows that the larger the detector's acceleration, the larger the transition rate at $\tau \to \infty$. It also shows that for a scalar field with larger mass, a larger acceleration is required for the detector to 'switch on', with

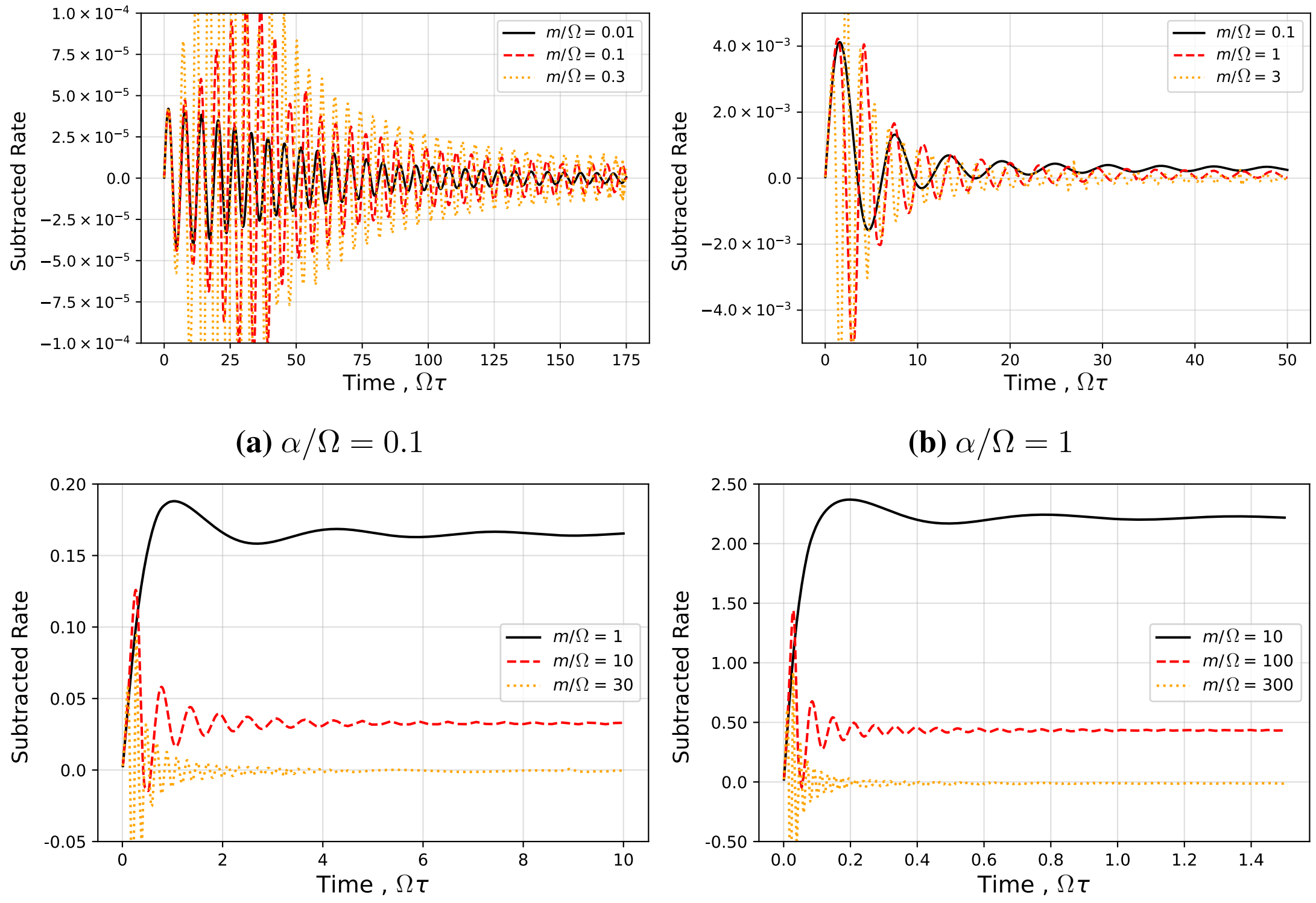

**(a)** $\alpha/\Omega = 0.1$

**(b)** $\alpha/\Omega = 1$

**(c)** $\alpha/\Omega = 10$

**(d)** $\alpha/\Omega = 100$

**Fig. 6.2.** The transition rate for an accelerated detector against time, with the inertial transition rate subtracted (given by Eq. (6.2.30)), i.e., $\frac{1}{\Omega|\mu|^2}\left(\frac{\mathrm{d}\mathbb{P}}{\mathrm{d}\tau} - \frac{\mathrm{d}\mathbb{P}}{\mathrm{d}\tau}\Big|_{\alpha=0}\right)$. Each plot is for a given acceleration, with three different values of the mass of the scalar field. The transition rate exhibits transient effects, but at late times tends to a constant value.

the transition rate becoming non-negligible at $T \sim m/4\pi$ ($\alpha \sim m/2$). This is superseded by another requirement: the detector 'switches on' at $T/\Omega \sim 1/2\pi$ ($\alpha \sim \Omega$). This leads to the sensible conclusion that the detector's transition rate begins to increase when its acceleration is above its energy gap. At large accelerations, the gradient becomes independent of the mass, meaning that the 'sensitivity' of the transition rate to acceleration (defined as $\mathrm{d}^2\mathbb{P}/\mathrm{d}\tau\mathrm{d}\alpha$) is independent of mass. Fig. 6.3b shows how the transition rate at $\tau \to \infty$ depends on the mass of the scalar field. As the mass increases, the transition rate tends to zero. However, it remains non-zero above $m/\Omega = 1$, which reflects the fact that an accelerating detector can absorb quanta of larger mass than its energy gap.

To highlight the difference between a Rindler thermal bath and a Minkowski thermal bath, the transition rate at $\tau \to \infty$ for a detector in a Minkowski thermal bath is plotted in Fig. 6.4. Fig. 6.4a shows that, like the Rindler bath case, the transition rate for the Minkowski thermal bath also 'switches on' at $T/\Omega \sim 1/2\pi$ ($\alpha \sim \Omega$), but there is no longer any requirement that $T \gtrsim m/4\pi$ ($\alpha \gtrsim m/2$) and the gradient of the rate (the sensitivity) is dependent on the mass even at large accelerations. Also, the transi-tion rate at $\tau \to \infty$ is zero for $m \geq \Omega$, regardless of the temperature. This is due to the flat-spacetime constraint $E \geq m$, which means the detector cannot absorb a particle of mass larger than its energy gap. Figs. 6.3b and 6.4b illustrate that, if $m = 0$, the $\tau \to \infty$ transition rate of a detector in a Rindler thermal bath is identical to that of a detector in a Minkowski thermal bath. The transition rate then differs as $m/\Omega$ is increased. This is true for all times, not just $\tau \to \infty$, as explained in Section 6.4.

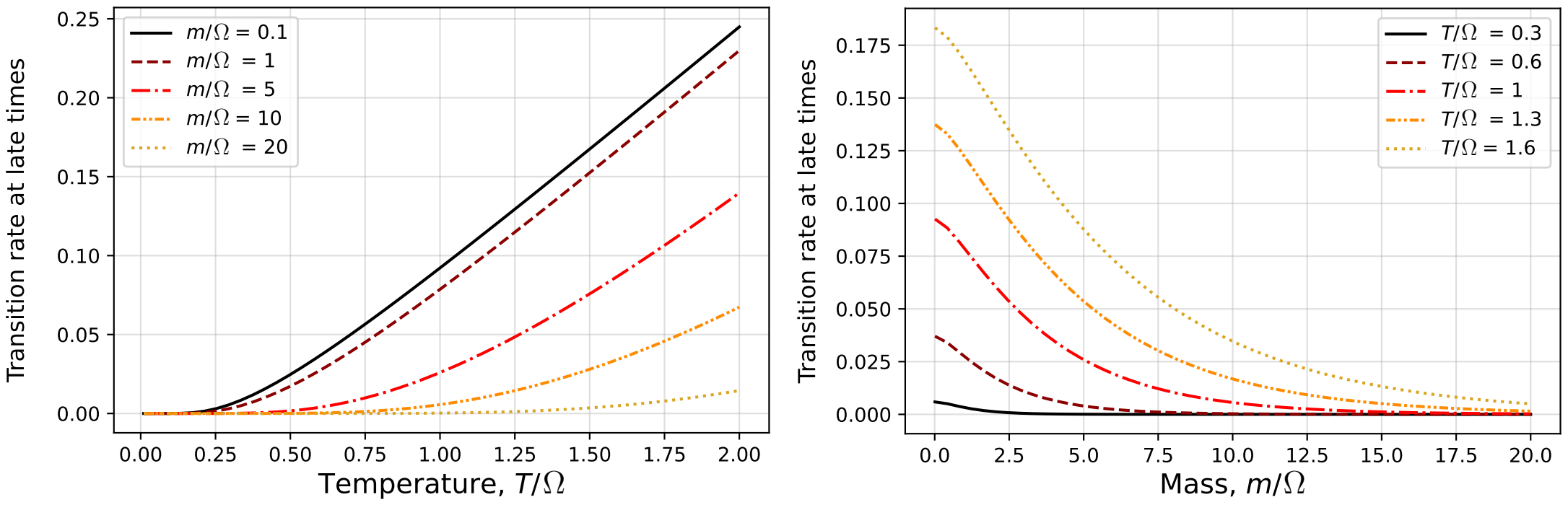

**(a)** **(b)**

**Fig. 6.3.** The transition rate for an *accelerated* detector tends to a constant at late times, given by the identical Eqs. (6.2.49) and (6.3.14). This constant depends on the acceleration (left) and the mass of the scalar field (right) as plotted here.

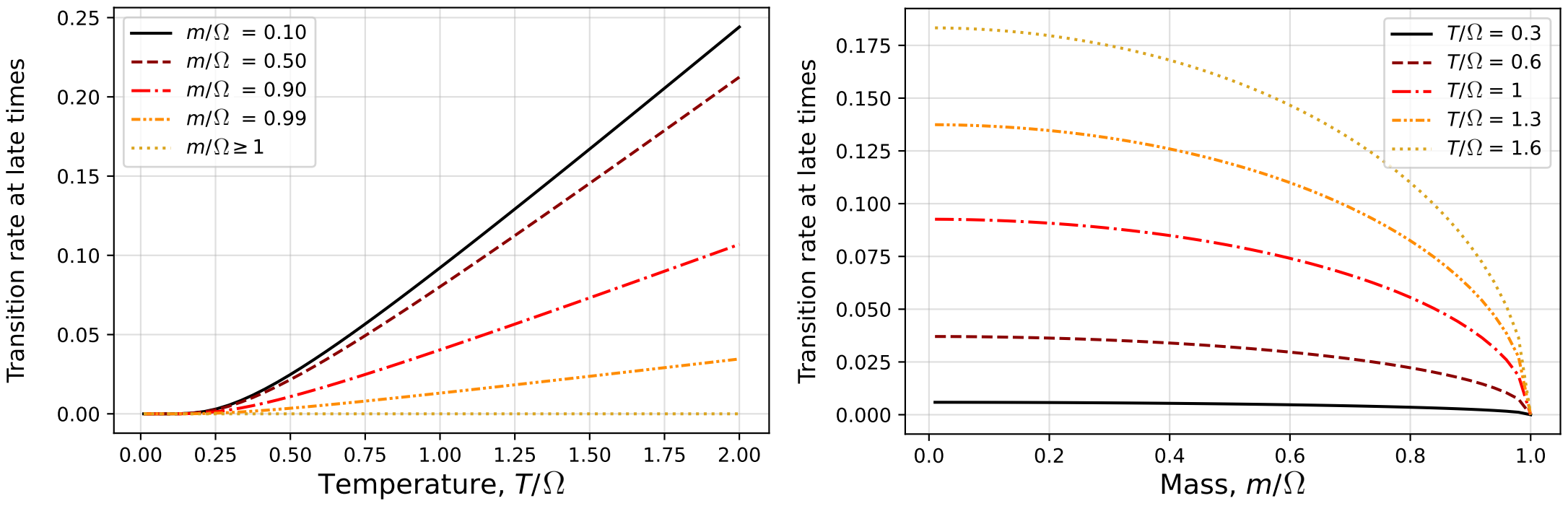

**(a)** **(b)**

**Fig. 6.4.** The transition rate for an *inertial* detector in a Minkowski thermal bath tends to a constant at late times, given by Eq. (6.4.9). This constant depends on the acceleration (left) and the mass of the scalar field (right) as plotted here.

## 6.6 Dipole detector

The manifestly causal, probabilistic formalism can also be applied to other detector models. In this section, we derive the response of a uniformly accelerated dipole detector [169]. Adapting the interaction Hamiltonian for a monopole detector (Eq. (6.2.3b)) to a general dipole detector is

$$H_{\text{int}}(t) \;=\; D^{\mu}(t)\,\partial_{\mu}\phi(\mathbf{x}^{D},t)\;, \tag{6.6.1}$$

where $D^{\mu}(t) \;\equiv\; \gamma^{-1}(t)\Big(\xi^{\mu}e^{-i\Omega\tau}\,|1\rangle\,\langle 2| \,+\, (\xi^{\mu})^{*}e^{i\Omega\tau}\,|2\rangle\,\langle 1|\,\Big)$ and $\xi^{\mu}$ is the 4-vector equivalent of the scalar $\mu$ in the monopole operator. With initial state $\rho_0 = |1, 0^{\phi}\rangle\,\langle 1, 0^{\phi}|$ and measurement operator $E = \mathbb{I}^{\phi} \otimes |2^{D}\rangle\,\langle 2^{D}|$ as before, we have

$$\mathcal{F}_1 \;=\; \frac{1}{i\gamma(t_1)}\Bigg(-\,\xi^{\mu_1}\partial_{\mu_1}\phi_1 e^{-i\Omega\tau}\,|1\rangle\,\langle 2| + (\xi^{\mu_1})^{*}\partial_{\mu_1}\phi_1 e^{i\Omega\tau}\,|2\rangle\,\langle 1|\Bigg)\,, \tag{6.6.2}$$

$$\begin{aligned}
\mathcal{F}_2 \;=&\; \frac{1}{i^2\gamma(t_1)\gamma(t_2)}\Bigg(-\,\xi^{\mu_1}(\xi^{\mu_2})^{*}\partial_{\mu_1}\phi_1\partial_{\mu_2}\phi_2 e^{-i\Omega\tau_{12}}\,|1\rangle\,\langle 1| \\
&\quad + (\xi^{\mu_1})^{*}\xi^{\mu_2}\partial_{\mu_1}\phi_1\partial_{\mu_2}\phi_2 e^{i\Omega\tau_{12}}\,|2\rangle\,\langle 2| \\
&\quad - \xi^{\mu_2}(\xi^{\mu_1})^{*}\partial_{\mu_2}\phi_2\partial_{\mu_1}\phi_1 e^{i\Omega\tau_{12}}\,|1\rangle\,\langle 1| \\
&\quad + (\xi^{\mu_2})^{*}\xi^{\mu_1}\partial_{\mu_2}\phi_2\partial_{\mu_1}\phi_1 e^{-i\Omega\tau_{12}}\,|2\rangle\,\langle 2|\Bigg) \\
=&\; \frac{|1\rangle\,\langle 1|}{\gamma(t_1)\gamma(t_2)}\Big(\tfrac{1}{2}\,\mathrm{Re}\,\big(\xi^{\mu_1}\xi^{\mu_2 *}e^{-i\Omega\tau}\big)\partial_{\mu_1}\partial_{\mu_2}\{\phi_1,\phi_2\} \\
&\quad + \tfrac{i}{2}\,\mathrm{Im}\,\big(\xi^{\mu_1}\xi^{\mu_2 *}e^{-i\Omega\tau}\big)\partial_{\mu_1}\partial_{\mu_2}[\phi_1,\phi_2]\Big) \\
&+ \frac{|2\rangle\,\langle 2|}{\gamma(t_1)\gamma(t_2)}\Big(-\,\tfrac{1}{2}\,\mathrm{Re}\,\big(\xi^{\mu_1}\xi^{\mu_2 *}e^{-i\Omega\tau}\big)\partial_{\mu_1}\partial_{\mu_2}\{\phi_1,\phi_2\} \\
&\quad + \tfrac{i}{2}\,\mathrm{Im}\,\big(\xi^{\mu_1}\xi^{\mu_2 *}e^{-i\Omega\tau}\big)\partial_{\mu_1}\partial_{\mu_2}[\phi_1,\phi_2]\Big)\,.
\end{aligned} \tag{6.6.3}$$

Once we take the expectation value of $\mathcal{F}_j$ in the initial state as in Eq. (6.2.13), we see that the lowest order contribution comes from the term in $\mathcal{F}_2$ which is proportional to $|1\rangle\,\langle 1|$, giving

$$\begin{aligned}
\mathbb{P}(\alpha,t) \;=&\; \int_0^t \mathrm{d}t_1\mathrm{d}t_2\frac{\Theta_{12}}{\gamma(t_1)\gamma(t_2)}\Bigg[\Big(\mathrm{Re}\,\big(\xi^{\mu_1}\xi^{\mu_2 *}\big)\cos\Omega\tau_{12} \\
&\quad + \mathrm{Im}\,\big(\xi^{\mu_1}\xi^{\mu_2 *}\big)\sin\Omega\tau_{12}\Big)\partial_{\mu_1}\partial_{\mu_2}\Delta^{H}_{12} \\
&\quad + \Big(-\,\mathrm{Im}\,\big(\xi^{\mu_1}\xi^{\mu_2 *}\big)\cos\Omega\tau_{12} \\
&\quad + \mathrm{Re}\,\big(\xi^{\mu_1}\xi^{\mu_2 *}\big)\sin\Omega\tau_{12}\Big)\partial_{\mu_1}\partial_{\mu_2}\Delta^{R}_{12}\Bigg]\,.
\end{aligned} \tag{6.6.4}$$

Writing $x_{12}^\mu \equiv x_1^\mu - x_2^\mu$, the relevant propagator derivatives are as follows:

$$\begin{aligned}
\frac{\partial}{\partial x^{\mu_1}}\frac{\partial}{\partial x^{\mu_2}}\Delta_{12}^H &= \frac{m^2}{4\pi}\frac{\partial}{\partial x^{\mu_1}}\frac{\partial}{\partial x^{\mu_2}}\frac{Y_1(ms_{12})}{ms_{12}} \\
&= \frac{m^2}{4\pi}\frac{\partial}{\partial x^{\mu_1}}\left[-\frac{Y_2(ms_{12})}{ms_{12}}\frac{\partial}{\partial x^{\mu_2}}(ms_{12})\right] \\
&= \frac{m^4}{4\pi}\frac{\partial}{\partial x^{\mu_1}}\left[\frac{Y_2(ms_{12})}{ms_{12}}g_{\mu_2\nu_2}\frac{x_{12}^{\nu_2}}{ms_{12}}\right] \\
&= \frac{m^4}{4\pi}\left[\frac{Y_2(ms_{12})}{(ms_{12})^2}g_{\mu_2\mu_1} - m^2\frac{Y_3(ms_{12})}{(ms_{12})^2}g_{\mu_1\nu_1}\frac{x_{12}^{\nu_1}}{ms_{12}}g_{\mu_2\nu_2}x_{12}^{\nu_2}\right] \\
&= \frac{m^4}{4\pi}\left[\frac{Y_2(ms_{12})}{(ms_{12})^2}g_{\mu_2\mu_1} - \frac{Y_3(ms_{12})}{(ms_{12})^3}m^2 x_{12}^{\mu_1}x_{12}^{\mu_2}\right], \qquad (6.6.5)\\
\frac{\partial}{\partial x^{\mu_1}}\frac{\partial}{\partial x^{\mu_2}}\Delta_{12}^R &= \frac{m^2}{4\pi}\frac{\partial}{\partial x^{\mu_1}}\frac{\partial}{\partial x^{\mu_2}}\frac{J_1(ms_{12})}{ms_{12}} \\
&= \frac{m^4}{4\pi}\left[\frac{J_2(ms_{12})}{(ms_{12})^2}g_{\mu_2\mu_1} - \frac{J_3(ms_{12})}{(ms_{12})^3}m^2 x_{12}^{\mu_1}x_{12}^{\mu_2}\right], \qquad (6.6.6)
\end{aligned}$$

where $J_i(x)$ and $Y_i(x)$ are Bessel functions of the first and second kind of order $i$, respectively, $g_{\mu\nu}$ is the Minkowski spacetime metric, and we have used $s_{12} \equiv \sqrt{g_{\mu\nu}x_{12}^\mu x_{12}^\nu}$. Let the dipole moment be $\xi^\mu = (\chi, \xi\sin\theta\cos\phi, \xi\sin\theta\sin\phi, \xi\cos\theta)$ in the frame of reference of the detector. The contractions appearing in the probability are then

$$\xi^{\mu_1}\xi^{\mu_2 *}g_{\mu_2\mu_1} = |\chi|^2 - |\xi|^2 \qquad (6.6.7)$$

$$\xi^{\mu_1}\xi^{\mu_2 *}x_{12}^{\mu_1}x_{12}^{\mu_2} = |\chi|^2\tau_{12}^2 \qquad (6.6.8)$$

confirming that the transition probability is independent of the orientation of the detector. This is consistent with (but does not imply) an isotropic thermal bath, which is interesting since the acceleration defines a privileged direction. However, it should be noted that this calculation is not directly sensitive to the direction of propagation of the radiation, since the field is traced over and we do not attempt to compare modes of the scalar field with different directions of momentum, **k**. This would make for an interesting future calculation to contribute to the discourse on whether the Unruh thermal bath is entirely isotropic [22, 170–177]. It may also be illuminating to consider a dipole detector coupled to a vector field instead of a scalar field.

The transition rate in the accelerated frame may now be expressed as

$$\frac{\partial \mathbb{P}}{\partial \tau} = \frac{m^4}{4\pi}\int_0^\tau \mathrm{d}\tau' \left[\left(\frac{Y_2(ms')}{(ms')^2}\left(|\chi|^2 - |\xi|^2\right) - \frac{Y_3(ms')}{(ms')^3}m^2|\chi|^2\tau'^2\right)\cos\Omega\tau'\right.$$

$$+\left(\frac{J_2(ms')}{(ms')^2}\left(|\chi|^2-|\xi|^2\right)-\frac{J_3(ms')}{(ms')^3}m^2|\chi|^2\tau'^2\right)\sin\Omega\tau'\Bigg], \quad (6.6.9)$$

where $s'=\frac{2}{\alpha}\sinh\frac{\alpha\tau'}{2}$.

## 6.7 Summary

We have employed the manifestly causal, probabilistic method explained in Chapter 4 to calculate the first-order transition rate of a uniformly accelerated UdW monopole detector from the ground state to the excited state, with the inertial rate subtracted (Eq. (6.2.30)). The transition rate has been expressed as a sum of two terms; one is proportional to the retarded propagator and independent of the initial state, and one is proportional to the Hadamard function and encapsulates all initial state dependence. Eq. (6.3.9) is the same transition rate, calculated from the perspective of a Rindler (accelerating) observer, who describes the detector as stationary in Rindler coordinates in a thermal bath of Rindler particles. The two expressions are equal at all times, including the transient effects which arise due to specifying the field to initially be in the Minkowski vacuum state. This is due to the Unruh effect: an observer accelerating through the Minkowski vacuum experiences a thermal bath of Rindler particles. Eq. (6.4.7) is the corresponding transition rate for an inertial detector in a 'real' (Minkowski) thermal bath. This rate is different and is unrelated to the Unruh effect. It is only coincidentally equal for a massless scalar field. The Unruh effect has also been presented as the result of a time-dependent Doppler shift of the field modes. The numerical results are new and highlight the dependence of the transients on the mass of the scalar field, the acceleration and the energy gap of the detector. The late-time behaviour of the transition rate has been explored numerically and compared to the transition rate for an inertial detector in the Minkowski thermal bath. Finally, we used our formalism to calculate the transition rate of a uniformly accelerated dipole detector.

The probability-level framework presented here can be utilised to study the response of an accelerated detector with a different model of the detector, a smooth switching function for the interaction (resulting in different transients), or different background spacetimes. It can also be used to study two accelerating UdW detectors, where the appearance of retarded propagators will then indicate clear causal connections be-tween the two detectors and which contributions are acausal, as in the Fermi two-atom system (Section 4.4). It also has the advantage of being able to treat mixed states and (semi-)inclusive observables, potentially simplifying calculations that are more complex at the amplitude level. This approach may also prove useful in studies of RQI, where working with mixed states arises naturally.

# Chapter 7

# Black Holes

Following our exploration of the Unruh effect in the previous chapter, we now turn our attention to a closely related topic in quantum field theory (QFT): black holes. The Unruh effect itself emerged from Unruh's study [18] of black holes and Hawking radiation [15, 16]. Unruh recognised that an observer undergoing constant proper acceleration would perceive a thermal bath of particles, akin to the radiation observed near the event horizon of a black hole. This insight suggested a connection between acceleration and gravitational effects in the context of QFT, as one may expect through the strong equivalence principle of general relativity [152, 178, 179].

There are further interesting parallels between the Unruh effect and Hawking radiation which we encounter in this chapter: both can be thought of as the result of a horizon (a Rindler horizon or a black hole event horizon); the Unruh temperature, $T_{\mathrm{U}} = \alpha/2\pi$ is of a similar form to the Hawking temperature, $T_{\mathrm{H}} = \kappa/2\pi$; and the coordinate transformation between inertial and accelerated coordinates is mathematically similar to the transformation between Schwarzschild and Kruskal-Szekeres coordinates. As such, the study of the Unruh effect naturally leads to the investigation of black holes, with acceleration serving as a simpler analogue to the gravitational field encountered near a black hole.

The primary goal of this chapter is to serve as a literature review and a pedagogical introduction to the subject of the response of Unruh-DeWitt (UdW) detectors in $(3+1)$-dimensional Schwarzschild spacetime, which describes a spherically symmetric black hole. The material covered here is central not only to the study of black hole thermodynamics and Hawking radiation, but also to our broader understanding of quantum gravity. By engaging with these ideas and investigating how quantum fields behave under extreme conditions, we begin to uncover the nature of spacetime itself, via the thermodynamic properties of gravity [121, 180–184], the idea that spacetime is emergent from entanglement [185–188], and the holographic principle [71, 189–191]. The response of detectors in black hole spacetimes provides valuable insights into the interplay between quantum field theory and general relativity, and lays the groundwork for future research in the quest to understand the quantum nature of gravity.

Aside from $(3+1)$-dimensional Schwarzschild black holes, progress has been made studying other black hole spacetimes such as BTZ black holes [192–196], where the pullback of the Wightman function to an infalling trajectory can be computed as a sum over images instead of the unavoidable mode sum in the Schwarzschild case. Interesting related research extends to the more tractable $(1{+}1)$-dimensional Schwarzschild spacetime [132, 197–199], de Sitter space [200–203], and analogue gravity models [204–213]. Some studies define quantities such as an effective temperature [214–220], which give physical insight without considering the response of an UdW detector.

In this chapter, we first review the foundational concepts of Penrose diagrams (Section 7.1), Killing vectors (Section 7.2), the Schwarzschild metric (Section 7.3), and the quantisation of a scalar field in the Schwarzschild geometry, examining the various vacuum states that can arise in this context (Sections 7.4 and 7.5). Section 7.6 then summarises the results in the literature regarding how UdW detectors respond for different trajectories in these different vacua. Section 7.7 concludes and suggests interesting future work.

## 7.1 Penrose Diagrams

The structure of spacetime is encapsulated by the spacetime metric, $g_{\mu\nu}$, which relates the spacetime coordinates, $x^\mu$, to the line element, $\mathrm{d}s^2$, by [151–153, 178]

$$\mathrm{d}s^2 = g_{\mu\nu}(x)\mathrm{d}x^\mu\mathrm{d}x^\nu\,, \tag{7.1.1}$$

where $\mu, \nu = 1, 2, \ldots, n-1$ for an $n$-dimensional spacetime. In order to visualise the causal structure of spacetime, we introduce Penrose diagrams [221], which provide a compact representation of infinite spacetimes using *conformal transformations*. These transformations rescale the spacetime metric, $g_{\mu\nu}$, by a position-dependent fac-tor, effectively shrinking or stretching the geometry. Mathematically, this is expressed as:

$$g_{\mu\nu}(x) \to \overline{g}_{\mu\nu}(x) = \Omega^2(x) g_{\mu\nu}(x)\,, \tag{7.1.2}$$

where $\Omega(x)$ is a real, finite, continuous, and non-vanishing function. Conformal transformations change the metric tensor and hence the geometry of the spacetime itself, and so must be distinguished from *coordinate transformations*, $x^\mu \to x'^\mu$, which simply relabel the coordinates without altering the underlying geometry.

A key feature of Penrose diagrams is that rays of light (null geodesics) always travel at $45°$, as conformal transformations preserve angles. This property highlights the causal structure of spacetime by clearly delineating which regions are causally connected to a given event. The boundaries of these regions are defined by the event's past and future *light cones*. Consequently, Penrose diagrams are an important tool for analysing causal relationships.

To better understand Penrose diagrams, consider the simple example of the infinite, flat, two-dimensional Minkowski spacetime described by the metric,

$$\mathrm{d}s^2 = \mathrm{d}t^2 - \mathrm{d}x^2 = \mathrm{d}u\,\mathrm{d}v\,, \tag{7.1.3}$$

where we have introduced the null coordinates, $u$ and $v$, defined by

$$\begin{aligned} u &= t - x\,, \\ v &= t + x\,. \end{aligned} \tag{7.1.4}$$

The coordinate $u$ is known as the outgoing null coordinate because radial outgoing null geodesics are lines of constant $u$. Similarly, the coordinate $v$ is known as the ingoing null coordinate. In the null coordinate system, the Minkowski metric is thus

$$g_{\mu\nu} = \frac{1}{2}\begin{pmatrix} 0 & 1 \\ 1 & 0 \end{pmatrix}\,. \tag{7.1.5}$$

We now perform a coordinate transformation,

$$\begin{aligned} u' &= 2\tan^{-1} u \quad , \quad -\pi \le u' \le \pi\,, \\ v' &= 2\tan^{-1} v \quad , \quad -\pi \le v' \le \pi\,, \end{aligned} \tag{7.1.6}$$

such that

$$\mathrm{d}s^2 = \frac{1}{4}\sec^2\left(\frac{u'}{2}\right)\sec^2\left(\frac{v'}{2}\right)\mathrm{d}u'\,\mathrm{d}v'\,, \tag{7.1.7}$$

therefore

$$g_{\mu\nu}(u',v') = \frac{1}{8}\sec^2\left(\frac{u'}{2}\right)\sec^2\left(\frac{v'}{2}\right)\begin{pmatrix}0 & 1\\ 1 & 0\end{pmatrix}\,. \tag{7.1.8}$$

This is still the usual Minkowski metric, expressed in the $u', v'$ coordinate system. We now perform a conformal transformation with

$$\Omega^2(x) = \left(\frac{1}{4}\sec^2\left(\frac{u'}{2}\right)\sec^2\left(\frac{v'}{2}\right)\right)^{-1}\,. \tag{7.1.9}$$

This rescales the Minkowski metric as

$$g_{\mu\nu}(u',v') \to \overline{g}_{\mu\nu}(u',v') = \frac{1}{2}\begin{pmatrix}0 & 1\\ 1 & 0\end{pmatrix}\,, \tag{7.1.10}$$

which results in a line element,

$$\mathrm{d}\overline{s}^2 = \mathrm{d}u'\mathrm{d}v'\,. \tag{7.1.11}$$

This is identical to Eq. (7.1.3), except this line element spans the finite, compact region $-\pi \leq \{u', v'\} \leq \pi$. As a result, we can draw the entirety of this spacetime as a Penrose diagram, as shown in Fig. 7.1. The conformal transformation has rescaled the infinities of Minkowski spacetime to the boundary lines of the Penrose diagram. Null rays will travel (at $45°$) to the diagonal boundaries labelled $\mathscr{I}^+$ (future null infinity) and $\mathscr{I}^-$ (past null infinity). Timelike lines travel from the point labelled $i^-$ (past timelike infinity) to the point labelled $i^+$ (future timelike infinity). Spacelike lines extend to the points labelled $i^0$ (spacelike infinity).

In four dimensions, the calculation is similar and the Penrose diagram is identical, where each point of the diagram represents a 2-sphere except for points along the vertical axis and $i^0$, which are spacetime points. The null infinities are thus 3-surfaces.

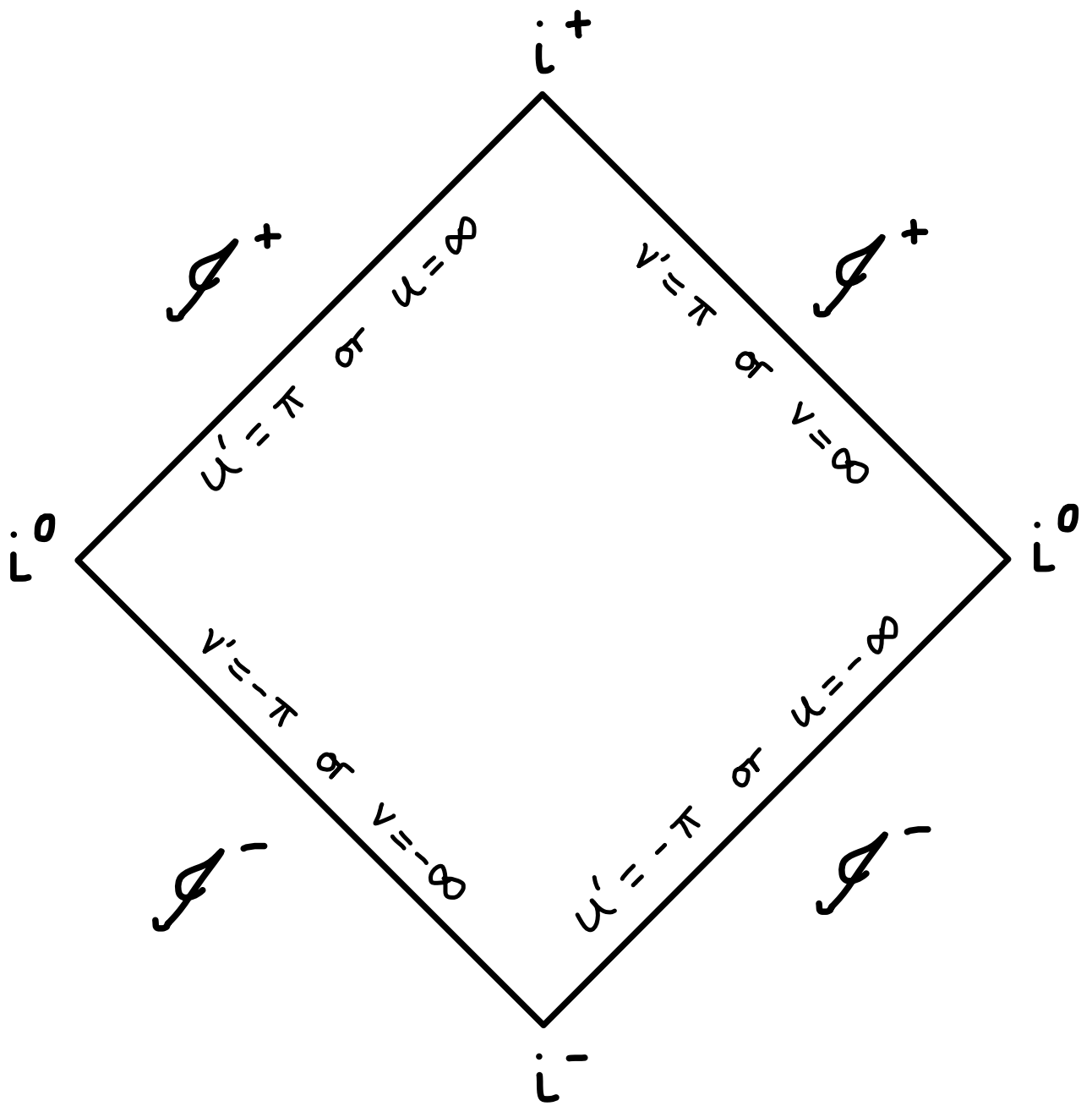

**Fig. 7.1.** The Penrose diagram for Minkowski space. The compact region $-\pi \leq \{u', v'\} \leq \pi$ is conformal to the entirety of Minkowski space, $-\infty \leq \{u, v\} \leq \infty$. The labels show future null infinity ($\mathscr{I}^+$), past null infinity ($\mathscr{I}^-$), past timelike infinity ($i^-$), future timelike infinity ($i^+$), and spacelike infinity ($i^0$).

## 7.2 Killing Vectors

Spacetimes with special geometrical symmetries can be described by *Killing vectors*. A Killing vector, $\xi^\mu$, is a solution to Killing's equation [151, 152, 179, 222],

$$\mathscr{L}_\xi g_{\mu\nu}(x) = 0\,, \tag{7.2.1}$$

where $\mathscr{L}_\xi$ is the *Lie derivative* along the vector field $\xi^\mu$. Physically, a Killing vector field is the vector field that generates an isometry of spacetime, meaning it corresponds to a symmetry under which the metric remains invariant. If a spacetime admits a timelike Killing vector it provides a natural way to define time translations and, consequently, a preferred notion of positive frequency modes. Specifically, a mode function $\phi_\omega$ is said to have positive frequency if it satisfies

$$\mathscr{L}_\xi \phi_\omega = -i\omega\phi_\omega, \quad \omega > 0\,. \tag{7.2.2}$$

This choice ensures a well-defined vacuum state. However, in spacetimes without a globally timelike Killing vector, such as the Schwarzschild spacetime considered in this chapter, the notion of positive frequency modes becomes observer-dependent, and thus there is no obvious choice of vacuum state.

## 7.3 The Schwarzschild Metric

Schwarzschild spacetime [223] is a unique solution to the Einstein field equations of general relativity which describes the spacetime surrounding a spherically symmetric, non-rotating, uncharged body. In this chapter, we will use it to describe a black hole with such features. The line element for four-dimensional Schwarzschild spacetime is

$$\mathrm{d}s^2 = \left(1 - \frac{2M}{r}\right)\mathrm{d}t^2 - \left(1 - \frac{2M}{r}\right)^{-1}\mathrm{d}r^2 - r^2(\mathrm{d}\theta^2 + \sin^2\theta\,\mathrm{d}\phi^2)\,, \tag{7.3.1}$$

where $M$ is the mass of the black hole. There are a number of useful coordinate systems used when working with the Schwarzschild metric. First, the *tortoise coordinate*, $r_*$, is given by,

$$r_* = r + 2M\ln\left|\frac{r}{2M} - 1\right|. \tag{7.3.2}$$

The outgoing and ingoing null coordinates are then defined by

$$\begin{aligned} u &= t - r_*\,, \\ v &= t + r_*\,. \end{aligned} \tag{7.3.3}$$

We now define the *Kruskal-Szekeres* coordinates [224],

$$\begin{aligned} \overline{u} &= -4Me^{-u/4M}\,, \\ \overline{v} &= 4Me^{v/4M}\,, \end{aligned} \tag{7.3.4}$$

which are very similar to the coordinates defined in Eq. (6.1.4), reinforcing the relevance of studying the Unruh effect as a flat-spacetime analogue of black hole physics.

The coordinates in Eq. (7.3.4) are defined such that the Schwarzschild line element, given by Eq. (7.3.1), can be written in a familiar form,

$$\mathrm{d}s^2 = \frac{2M}{r}\,e^{-r/2M}\,\mathrm{d}\overline{u}\,\mathrm{d}\overline{v} - r^2(\mathrm{d}\theta^2 + \sin^2\theta\,\mathrm{d}\phi^2)\,. \tag{7.3.5}$$

The first term is conformal to the two-dimensional Minkowski line element in Eq. (7.1.3). Applying the same coordinate and conformal transformations described by Eqs. (7.1.6) and (7.1.10), the resulting Penrose diagram is almost identical to that for Minkowski spacetime (shown in Fig. 7.1). The key difference is that neither of the two left boundaries of the diagram are null infinity, $\mathscr{I}$, since the Kruskal-Szekeres coordinates are only defined for $-\infty < \overline{u} \leq 0$ and $0 \leq \overline{v} < \infty$, as follows from Eq. (7.3.4). The bottom-left boundary ($\overline{v} = 0$) and the top-left boundary ($\overline{u} = 0$) instead both correspond to $r = 2M$ (and/or $t = \mp\infty$). By inspecting Eq. (7.3.1), one can see that $r = 2M$ results in a singularity. However, this is not a *physical singularity*, but merely a *coordinate singularity* which results from our choice of coordinate system. Upon inspection of Eq. (7.3.5), it is clear that there is no singularity at $r = 2M$ in the $\{\overline{u}, \overline{v}\}$ coordinate system. As a result, we can analytically extend the spacetime beyond the boundaries $\overline{v} = 0$ and $\overline{u} = 0$. The result is known as the *maximally extended Kruskal manifold* (or the *maximally extended Schwarzschild spacetime*), and its Penrose diagram is shown in Fig. 7.2.

This spacetime is also a solution to the vacuum Einstein equation at all spacetime points, except for at the physical singularity at $r = 0$, which is denoted by a horizontal zig-zag line on the Penrose diagram (pictured in the future and the past). Region I ($\overline{u} < 0$, $\overline{v} > 0$, or $r > 2M$, $-\infty < t < \infty$) is the usual Schwarzschild spacetime. Region II ($\overline{u} > 0$, $\overline{v} < 0$) is identical to Region I, except the direction of time is reversed (i.e., $t \rightarrow -t$). Region I and Region II are causally disconnected. Physically, Region II is sometimes interpreted as a separate, causally disconnected universe.

The null ray $\overline{u} = 0$ is the latest null ray to reach future null infinity, $\mathscr{I}^+$. All null rays for which $\overline{u} > 0$ terminate on the singularity at $r = 0$. Therefore, since $\overline{u} = 0$ is a 2-surface in four dimensions, it represents the surface which separates regions in which null geodesics either travel to the singularity or to future infinity. This is the *event horizon*, which separates the exterior Region I from the interior of a black hole. Similarly, $\overline{v} = 0$ represents the boundary of Region I from which anything not from past null infinity, $\mathscr{I}^-$, must have originated. Regions from which all world lines leave are known as *white holes*. Therefore, Region III ($\overline{u} > 0$, $\overline{v} > 0, \overline{u}\,\overline{v} < (4M)^2$) is the black hole interior and Region IV ($\overline{u} < 0$, $\overline{v} < 0, \overline{u}\,\overline{v} < (4M)^2$)) is the white hole interior. The world lines of all observers in Region I and II start at $i^-$ and end at $i^+$, as long as they do not originate from the white hole and avoid the black hole. Spacetimes of astrophysical black holes (formed by the collapse of matter) do not include Regions II and IV, so in this sense they are not physical regions [151].

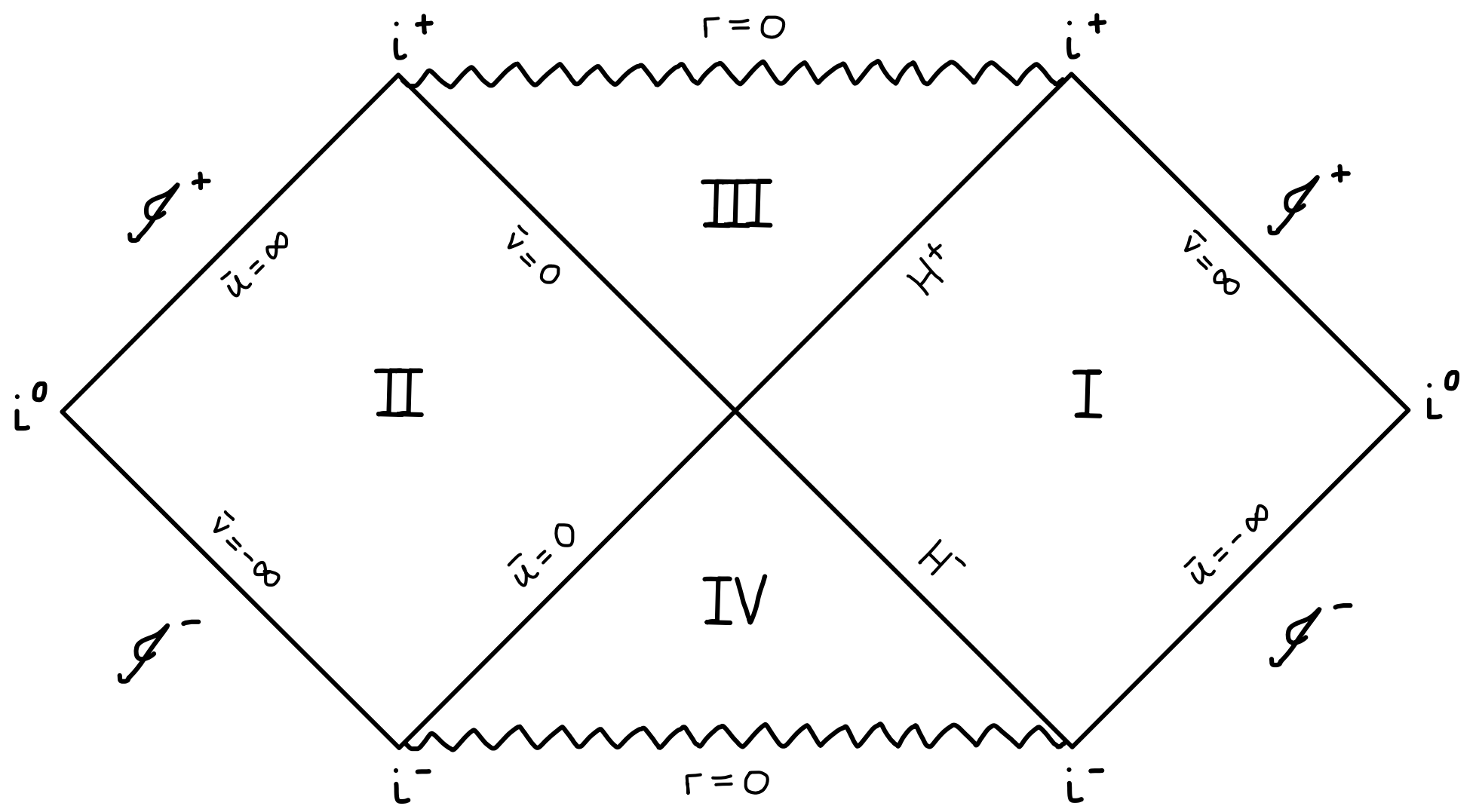

**Fig. 7.2.** The Penrose diagram for the maximally extended Schwarzschild spacetime. Region I is the exterior to the black hole, Region II is universe causally disconnected from Region I, Region III is the black hole interior, and Region IV is the white hole interior. The future and past horizons are labelled $H^+$ and $H^-$ respectively, and the physical singularity at $r = 0$ is marked by a zigzag line.

## 7.4 Classical Scalar Field in Schwarzschild Spacetime

In this section, we will review the method and results for solving the Klein-Gordon equation,

$$\nabla_\mu \nabla^\mu \phi = 0\,, \tag{7.4.1}$$

for a massless scalar field, $\phi(x)$, in Schwarzschild spacetime. For more details of this process, see Refs. [151, 214, 225].

For now, let's only consider Region I. In Schwarzschild coordinates $(t, r, \theta, \varphi)$, the Klein-Gordon equation is separable and its general solution is a linear combination of the basis modes [118]

$$\Phi_{\omega\ell}(r) Y_{\ell m}(\theta, \varphi) e^{\pm i\omega t}\,, \tag{7.4.2}$$

where $\omega > 0$ and $Y_{\ell m}$ is a spherical harmonic [85], with the usual definitions of $\ell \in \{0, 1, ...\}$, $m \in \{-\ell, ..., \ell\}$. Since we will only consider spherically symmetric situations, we can set

$$Y_{\ell m}(0, \phi) = \begin{cases} \left(\dfrac{2\ell + 1}{4\pi}\right)^{1/2}, & m = 0\,, \\ 0, & |m| = 1, 2, 3, \ldots\,, \end{cases} \tag{7.4.3}$$

thus imposing $m = 0$ (instead of summing over $m$) without loss of generality. For now, however, we shall keep it general. The radial function $\Phi_{\omega\ell}(r)$ is a solution to

$$\frac{\mathrm{d}^2\Phi}{\mathrm{d}r^2}+\frac{2\,(r-M)}{r\,(r-2M)}\frac{\mathrm{d}\Phi}{\mathrm{d}r}+\left[\frac{\omega^2 r^2}{(r-2M)^2}-\frac{\ell(\ell+1)}{r\,(r-2M)}\right]\Phi=0. \tag{7.4.4}$$

This equation is known as the generalised spheroidal wave equation [226]. With respect to the timelike Killing vector, $\partial_t$, the positive frequency basis modes are proportional to $e^{-i\omega t}$ and the negative frequency basis modes are proportional to $e^{+i\omega t}$.

If we define the dimensionless variable $\rho(r)\equiv r\,\Phi(r)$, Eq. (7.4.4) takes the form of the time-independent Schrödinger equation in terms of the tortoise coordinate $r_*$,

$$\frac{\mathrm{d}^2\rho}{\mathrm{d}r_*^2}=\left[V_\ell(r)-\omega^2\right]\rho, \tag{7.4.5}$$

with the effective potential

$$V_\ell(r)\equiv\left(1-\frac{2M}{r}\right)\left(\frac{\ell(\ell+1)}{r^2}+\frac{2M}{r^3}\right). \tag{7.4.6}$$

In the asymptotic regions $r\to 2M$ (close to the event horizon) and $r\to\infty$ (far from the black hole), $V_\ell(r)\to 0$. From Eq. (7.4.5), we conclude that $\rho_{\omega\ell}(r)$ is a linear combination of $e^{\pm i\omega r_*}$ in these asymptotic regions. We now introduce two independent solutions, characterised by their asymptotic behaviour,

$$\rho^{\text{in}}_{\omega\ell}(r)\to\begin{cases}B^{\text{in}}_{\omega\ell}\,e^{-i\omega r_*}, & r\to 2M,\\ e^{-i\omega r_*}+A^{\text{in}}_{\omega\ell}\,e^{+i\omega r_*}, & r\to\infty,\end{cases} \tag{7.4.7}$$

and

$$\rho^{\text{up}}_{\omega\ell}(r)\to\begin{cases}A^{\text{up}}_{\omega\ell}\,e^{-i\omega r_*}+e^{+i\omega r_*}, & r\to 2M,\\ B^{\text{up}}_{\omega\ell}\,e^{+i\omega r_*}, & r\to\infty,\end{cases} \tag{7.4.8}$$

where the $A$ and $B$ terms are reflection and transmission coefficients, respectively. The values of these coefficients are calculated via normalisation conditions in Ref. [225]. We now have two independent solutions to Eq. (7.4.4),

$$\Phi^{\text{in/up}}_{\omega\ell}(r)\equiv\rho^{\text{in/up}}_{\omega\ell}(r)/r, \tag{7.4.9}$$

and can thus use Eq. (7.4.2) to define the normalised positive-frequency modes

$$u^{\text{in/up}}_{\omega\ell m}(t, r, \theta, \varphi) \equiv \frac{Y_{\ell m}(\theta, \varphi)}{\sqrt{4\pi\omega}} \Phi^{\text{in/up}}_{\omega\ell}(r) e^{-i\omega t}. \tag{7.4.10}$$

The set $\{u^{\text{in/up}}_{\omega\ell m}, u^{\text{in/up}*}_{\omega\ell m}\}$ for all $\omega, \ell, m$ is a complete set of solutions to the Klein-Gordon equation in Region I.

The modes $u^{\text{in}}_{\omega\ell m}$ are called *in modes* because *near the horizon* they have the asymptotic behaviour

$$u^{\text{in}}_{\omega\ell m} \propto e^{-i\omega(t+r_*)}, \quad r \to 2M, \tag{7.4.11}$$

which represents waves travelling *into* the horizon with no outgoing component. At early times, these waves originate infinitely far from the black hole and travel towards the horizon, scattering off the gravitational potential of the black hole. This results in part of each wave travelling into the horizon and part travelling back out to infinity. Consequently, these modes are a superposition of ingoing and outgoing components at $r \to \infty$, as seen in Eq. (7.4.7). 'Out modes' would be defined as having no incoming component near the horizon.

Conversely, the modes $u^{\text{up}}_{\omega\ell m}$ are called *up modes* because *far from the black hole* they have the asymptotic behaviour

$$u^{\text{up}}_{\omega\ell m} \propto e^{-i\omega(t-r_*)}, \quad r \to \infty, \tag{7.4.12}$$

which represents waves travelling *up* from the black hole (at $r \to \infty$) with no ingoing component. These waves originate on the past horizon of the white hole, travelling away from the horizon and scattering off the potential. This causes part of each wave to travel back towards the black hole horizon and part to travel out to infinity. Consequently, these modes are a superposition of ingoing and outgoing components at $r \to 2M$, as seen in Eq. (7.4.8). 'Down modes' would be defined as having no outgoing component at $r \to \infty$.

Fig. 7.3 visualises these modes on a Penrose diagram for the exterior region in Schwarzschild spacetime.

Since Region II is identical to Region I except for time-reversal, the mode solutions for Region II take the same form as Eq. (7.4.10). However, due to the time-reversal, the future-directed Killing vector with which we should define the positive frequency modes is $\partial_{-t}$ instead of $\partial_t$. The result of this is that the mode solutions in Region II, which we denote as $v(x)$ instead of $u(x)$, are the complex conjugate of Eq. (7.4.10),

$$v^{\text{in/up}}_{\omega\ell m}(t,r,\theta,\varphi) \equiv \frac{Y^*_{\ell m}(\theta,\varphi)}{\sqrt{4\pi\omega}}\Phi^{\text{in/up}*}_{\omega\ell}(r)e^{i\omega t}. \tag{7.4.13}$$

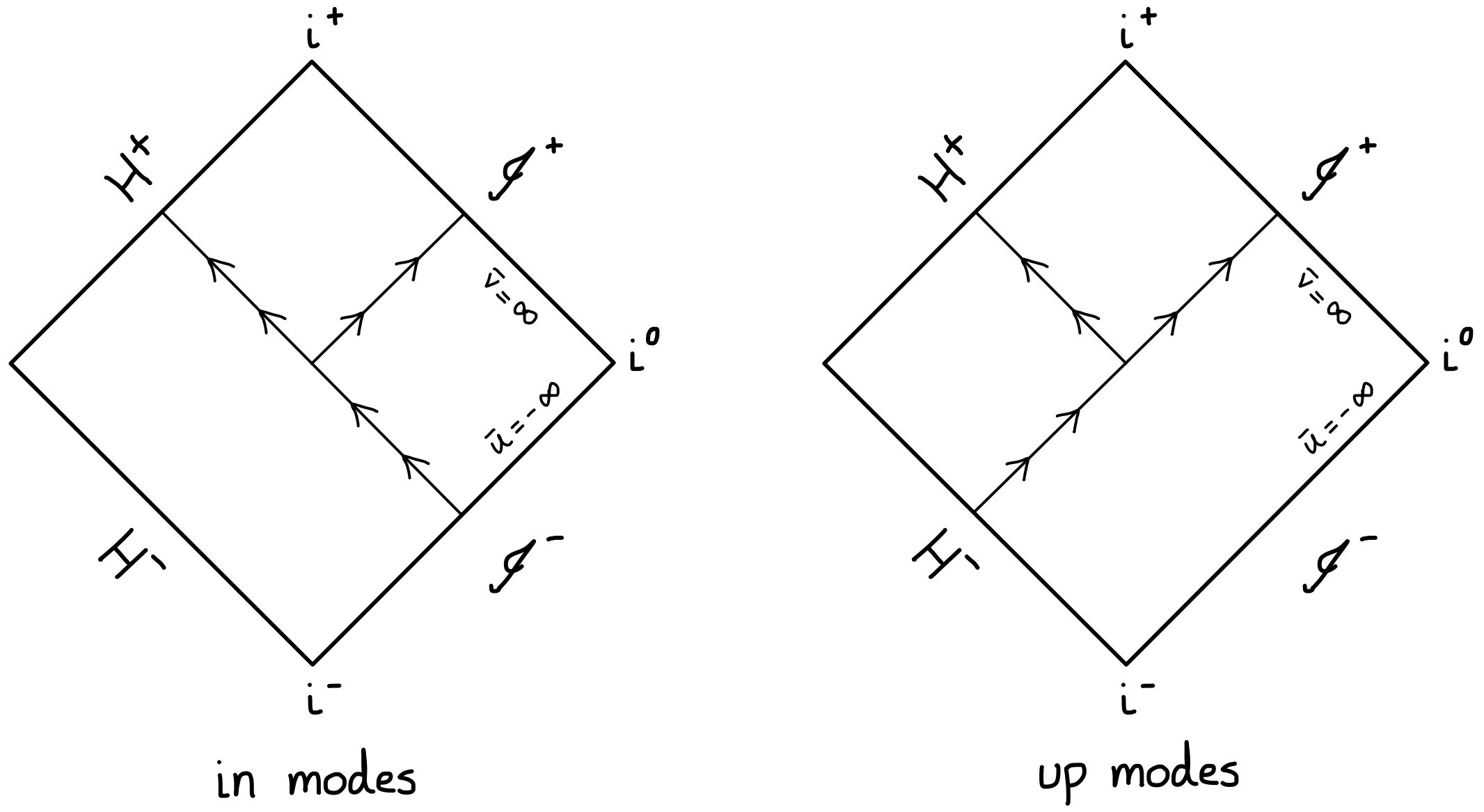

**Fig. 7.3.** Illustrations of the 'in' (shown left) and 'up' (shown right) modes, defined in Eq.(7.4.10). The modes are shown in Region I of the Schwarzschild spacetime, i.e., the exterior to the black hole.

For Regions III and IV, the in modes can be analytically continued from Regions I or II, and thus are defined in the same way (such that they are continuous across the horizon) [227]. Due to the presence of the singularity, and the absence of spatial infinity, there is no analogue of the up modes in Regions III and IV. The in modes and their complex conjugates suffice as an orthogonal basis set to span these regions. We now have basis modes which are solutions to the Klein-Gordon equation in each region of the maximally extended Schwarzschild spacetime.

## 7.5 Quantising the Scalar Field

Now that we understand the classical solutions for the scalar field modes in Schwarzschild spacetime, we must quantise the theory. In Minkowski spacetime, the standard way to do this is to take our complete set of orthonormal basis modes and define creation/annihilation operators to couple them with. Our choice of basis modes and creation/annihilation operators will then define a vacuum state and corresponding Fock states.

This is relatively straightforward due to the presence of a global timelike Killing vector, $\partial/\partial t$, which allows a natural definition of positive and negative frequency modes, $(2\omega(2\pi)^3)^{-1/2}e^{\mp ik\cdot x}$. However, in curved spacetime, choosing basis modes and defining a vacuum state is more subtle due to the lack of a unique global timelike Killing vector and the presence of event horizons. In Schwarzschild spacetime, we typically choose different vacuum states associated with different basis modes. If an eternal black hole were to exist in the universe, we could observe which quantum state is actually realised. Nevertheless, having a choice of vacua is advantageous in the sense that we can choose states to represent different processes of interest. The standard choices of vacuum states in Schwarzschild spacetime are the Boulware, Hartle-Hawking, and Unruh vacua, each of which is motivated by different physical considerations.

### 7.5.1 The Boulware Vacuum

The Boulware vacuum [156] is defined by requiring that the mode functions be positive frequency with respect to the Schwarzschild Killing vector $\partial_t$ (where $t$ is Schwarzschild time) at spatial infinity in Region I, and the Killing vector $\partial_{-t}$ in Region II. In other words, we choose basis modes for the scalar wave equation proportional to $e^{-i\omega u}$, $e^{-i\omega v}$, where $\{u, v\}$ are the Schwarzschild null coordinates. These basis modes oscillate infinitely quickly at the event horizon. The mode decomposition of the scalar field is thus

$$\phi(x) = \sum_{lm} \int_0^\infty \mathrm{d}\omega \left( b^{\text{in}}_{\omega lm} u^{\text{in}}_{\omega lm} + b^{\text{up}}_{\omega lm} u^{\text{up}}_{\omega lm} + b'^{\text{in}}_{\omega lm} v^{\text{in}}_{\omega lm} + b'^{\text{up}}_{\omega lm} v^{\text{up}}_{\omega lm} + \text{H.c.} \right), \quad (7.5.1)$$

where 'H.c' stands for the Hermitian conjugate of the preceding terms and the annihilation operators $b_{\omega lm}$ and $b'_{\omega lm}$ define the Boulware vacuum $|0_{\text{B}}\rangle$ by $b^{\text{in/up}}_{\omega lm} |0_{\text{B}}\rangle = b'^{\text{in/up}}_{\omega lm} |0_{\text{B}}\rangle = 0$.

Physically, the Boulware vacuum resembles the Minkowski vacuum at large distances from the black hole but leads to divergent stress-energy tensor components near the event horizon. Specifically, the expectation value of the renormalised stress-energy tensor $\langle T_{\mu\nu}\rangle$ diverges at the horizon, signalling that this vacuum is unphysical for describing a black hole with any thermal radiation [17]. This behaviour is analogous to the Rindler vacuum, $|0_{\text{R}}\rangle$, of an accelerating observer. An observer at $R \to 2M$ in $|0_{\text{B}}\rangle$ in Schwarzschild spacetime is similar to an accelerating observer at $\alpha \to \infty$ in $|0_\mathrm{R}\rangle$ in Minkowski spacetime. Neither detect particles due to the definition of the vacuum, but the vacuum itself diverges. Likewise, an observer at $R \to \infty$ in $|0_\mathrm{B}\rangle$ in Schwarzschild spacetime is similar to an accelerating observer at $\alpha \to 0$ in $|0_\mathrm{R}\rangle$ in Minkowski spacetime (which reduces to the Minkowski vacuum, $|0_\mathrm{M}\rangle$, for $\alpha \to 0$).

The Boulware vacuum is a suitable choice for Schwarzschild spacetime when there is no physical event horizon, such as in the case of the spacetime exterior to the Earth's surface [228]. Since the Earth's event horizon would lie deep within its interior, a different metric is required there. The absence of an event horizon implies that no Hawking radiation is present in the spacetime surrounding the Earth, even though this region can still be described by the Schwarzschild metric.

### 7.5.2 The Hartle-Hawking Vacuum

The Hartle-Hawking vacuum [229] is defined by requiring that the modes be positive frequency with respect to the Kruskal Killing vector $\partial_t$ in all regions (i.e., they are proportional to $e^{-i\omega\overline{u}}, e^{-i\omega\overline{v}}$, where $\{\overline{u}, \overline{v}\}$ are the Kruskal null coordinates). In terms of in and up modes, the Hartle-Hawking vacuum is defined by the normalised superpositions of $u(x)$ and $v(x)$,

$$w^{\text{in/up}}_{\omega\ell m} \equiv \frac{u^{\text{in/up}}_{\omega\ell m} + e^{-4\pi M\omega} v^{\text{in/up}*}_{\omega\ell m}}{\sqrt{1 - e^{-8\pi M\omega}}}, \tag{7.5.2a}$$

$$\bar{w}^{\text{in/up}}_{\omega\ell m} \equiv \frac{e^{-4\pi M\omega} u^{\text{in/up}*}_{\omega\ell m} + v^{\text{in/up}}_{\omega\ell m}}{\sqrt{1 - e^{-8\pi M\omega}}}. \tag{7.5.2b}$$

The set of basis modes $\{w^{\text{in/up}}_{\omega\ell m}, \bar{w}^{\text{in/up}}_{\omega\ell m}, w^{\text{in/up*}}_{\omega\ell m}, \bar{w}^{\text{in/up*}}_{\omega\ell m}\}$ form a complete orthonormal set of solutions to the Klein-Gordon equation. We can hence quantise the scalar field in terms of these modes,

$$\phi(x) = \sum_{lm} \int_0^\infty \mathrm{d}\omega \left( d^{\text{in}}_{\omega lm} w^{\text{in}}_{\omega lm} + d^{\text{up}}_{\omega lm} w^{\text{up}}_{\omega lm} + \bar{d}^{\text{in}}_{\omega lm} \bar{w}^{\text{in}}_{\omega lm} + \bar{d}^{\text{up}}_{\omega lm} \bar{w}^{\text{up}}_{\omega lm} + \text{H.c.} \right), \tag{7.5.3}$$

where the Hartle-Hawking vacuum $|0_\mathrm{H}\rangle$ is defined by $d^{\text{in/up}}_{\omega lm}|0_\mathrm{H}\rangle = \bar{d}^{\text{in/up}}_{\omega lm}|0_\mathrm{H}\rangle = 0$.

The Hartle-Hawking vacuum state is regular across the entire extended Schwarzschild spacetime, except for at the physical singularities. Observers at $r \to \infty$ observe a thermal bath of radiation with a temperature equal to the Hawking temperature [179],

$$T_{\mathrm{H}} = \frac{\kappa}{2\pi} = \frac{1}{8\pi M}, \tag{7.5.4}$$

where $\kappa = 1/4M$ is the surface gravity of the black hole.

The Hartle-Hawking state describes a black hole in thermal equilibrium with its surroundings and its own Hawking radiation, which models a black hole enclosed in a reflecting cavity. The appearance of $|0_{\mathrm{H}}\rangle$, which is defined on the entire manifold, as a thermal bath to an observer at infinity, confined to Region I, is a consequence of the horizons causally disconnecting parts of the spacetime. In doing so, information about the modes in Region II is lost, which is naturally associated with a non-zero entropy and thus a thermal state, instead of a pure state. This is similar to the appearance of the Minkowski vacuum as a thermal bath to an accelerating observer, due to the horizon in Rindler space.

### 7.5.3 The Unruh Vacuum

The Unruh vacuum [18] uses a mix of the previously defined basis modes, using the in modes of the Boulware vacuum state and the up modes of the Hartle-Hawking vacuum state. The associated mode decomposition of the scalar field is thus,

$$\phi(x) = \sum_{\ell,m} \int_0^\infty d\omega \left( b^{\mathrm{in}}_{\omega\ell m} u^{\mathrm{in}}_{\omega\ell m} + d^{\mathrm{up}}_{\omega\ell m} w^{\mathrm{up}}_{\omega\ell m} + b'^{\mathrm{in}}_{\omega\ell m} v^{\mathrm{in}}_{\omega\ell m} + \bar{d}^{\mathrm{up}}_{\omega\ell m} \bar{w}^{\mathrm{up}}_{\omega\ell m} + \text{h.c.} \right), \tag{7.5.5}$$

where the Unruh vacuum $|0_{\mathrm{U}}\rangle$ is defined by $b^{\mathrm{in}}_{\omega\ell m} |0_{\mathrm{U}}\rangle = d^{\mathrm{up}}_{\omega\ell m} |0_{\mathrm{U}}\rangle = b'^{\mathrm{in}}_{\omega\ell m} |0_{\mathrm{U}}\rangle = \bar{d}^{\mathrm{up}}_{\omega\ell m} |0_{\mathrm{U}}\rangle = 0$.

The Unruh vacuum is defined to model a black hole that forms from gravitational collapse and emits Hawking radiation. The vacuum condition is imposed such that ingoing modes are defined as positive frequency with respect to the Schwarzschild time at past null infinity, ensuring no incoming radiation from $\mathscr{I}^-$, while outgoing modes are positive frequency with respect to Kruskal coordinates. Outgoing modes appear as thermal flux at future null infinity, at the Hawking temperature, $T_{\mathrm{H}}$, capturing the effect of Hawking radiation. This vacuum state arises in models of the formation of black holes via gravitational collapse in the late-time, near-horizon limit [230]. The choice of basis modes is a way of recreating the physical effects of such a col-lapse [151]. Due to this, $|0_{\rm U}\rangle$ is a physically realistic vacuum state for observers falling into astrophysical black holes sufficiently long after their formation.

## 7.6 Unruh-DeWitt Detectors in Schwarzschild Spacetime

Just like in Minkowski space, the interaction Hamiltonian for an UdW detector coupled to a real scalar field, $\phi(x)$, in curved spacetime is

$$H_{\rm int} = \lambda\,\chi(t)\,M^D(t)\,\phi(\mathbf{x}^D, t) \tag{7.6.1}$$

where $\lambda$ is a coupling constant, $\chi(t)$ is the switching function, $M^D(t)$ is the monopole operator representing the detector, and $\mathbf{x}^D$ is the trajectory of the detector.

The detector response function, $F(\Omega)$, as defined in Eq. (6.2.23), is still given by

$$F(\Omega) = \int_{-\infty}^{\infty} \mathrm{d}\tau_1 \int_{-\infty}^{\infty} \mathrm{d}\tau_2 e^{-i\Omega(\tau_1-\tau_2)}\chi(\tau_1)\chi(\tau_2)W(\tau_1,\tau_2)\,, \tag{7.6.2}$$

where $W(\tau_1, \tau_2) \equiv W(\mathbf{x}^D(\tau_1), \mathbf{x}^D(\tau_2))$ is the pull-back of the positive Wightman function,

$$W(x_1, x_2) \equiv \langle\Psi|\phi(x_1)\phi(x_2)|\Psi\rangle \tag{7.6.3}$$

to the trajectory of the detector, for a given state of the scalar field, $|\Psi\rangle$. The integration is now over $-\infty < \tau_1, \tau_2 < \infty$ and the interaction duration is thus controlled solely by the switching function. All of the information about the trajectory of the detector and the state of the field in Eq. (7.6.2) is contained in the Wightman function, so it is this we must alter when investigating different detector trajectories in curved spacetimes for different vacuum states. Given a vacuum state, $|0\rangle$, defined by the set of orthonormal basis states $\{u_k(x), u_k^*(x)\}$ for frequency mode $k$, the positive Wightman function can be expressed as

$$W(x_1, x_2) = \sum_k u_k(x_1)u_k^*(x_2)\,. \tag{7.6.4}$$

Eq. (7.6.2) then becomes

$$F(\Omega) = \sum_k \left| \int_{-\infty}^{\infty} \mathrm{d}\tau_1 e^{-i\Omega\tau_1}\chi(\tau_1)u_k(\mathbf{x}(\tau_1)) \right|^2 . \tag{7.6.5}$$

We can now simply substitute in the basis modes for each vacuum of interest.

A stationary trajectory is defined by the time-translational invariance of the Wightman function, such that $W(\tau_1, \tau_2) = W(\tau_1 - \tau_2)$. For these cases, the algebraic expressions can simplify, and integrals may become analytically solvable. Furthermore, stationarity allows for the derivative of $F(\Omega)$ with respect to $\tau$ to be approximated by $F(\Omega)/\tau$ as $\tau \to \infty$, where $\tau$ is the total duration of the measurement in proper time, since the transition rate is time-independent. This defines the response rate, $\dot{F}$, which can be interpreted as the average rate of particles detected along the trajectory[1]. In these stationary cases, the response rate is given by

$$\dot{F}(\Omega) = \int_{-\infty}^{\infty} \mathrm{d}\tau_{12}\, e^{-i\Omega\tau_{12}}\, W(\tau_{12})\,, \tag{7.6.6}$$

where $\tau_{12} \equiv \tau_1 - \tau_2$ and the time-translation invariance means we are free to push the switch-on and switch-off time of the detector to the asymptotic past and future, respectively [227], effectively setting the switching function to $\chi(t) = 1$.

Defining $\dot{F}$ is more subtle for non-stationary trajectories, since defining an instantaneous transition rate requires carefully taking the limit of the switching function becoming arbitrarily sharp at the time of measurement, leading to divergences due to the structure of the Wightman function [160, 162, 163, 214]. Due to its non-trivial time dependence, the response rate is also typically more computationally demanding.

## 7.6.1 Fixed Radial Distance

A static detector at fixed radius in Schwarzschild spacetime is studied in Ref. [227], including numerical plots of the response rate for different vacuum states, detector energy gaps, and radial distances. Only Region I, the exterior of the black hole, is studied. Here, we summarise the results.

For a static detector, $r = R$ at all times. The difference in Schwarzschild time, $t_{12} \equiv t_1 - t_2$, is then $t_{12} = \tau_{12}/\sqrt{1 - 2M/R}$. The angular coordinates can be set to $\theta = \phi = 0$ without loss of generality. Due to stationarity, we can use Eq. (7.6.6) to compute the response rate for different vacuum states (different Wightman functions). Stationarity also allows us to set the measurement time to $\tau \to \infty$ such that transient effects have subsided and we are measuring the 'steady-state' response rate. As such, the choice of switching function is arbitrary, and it is effectively taken to be a top-hat function.

[1] $F(\Omega)$ gives the fraction of detectors in a given ensemble which have undergone a transition, but a measurement alters the dynamics of the system, meaning this interpretation no longer applies after a measurement. Thus, to measure $\dot{F}(\Omega)$, one requires a set of identical ensembles, with each ensemble used to measure $F(\Omega)$ at a single value of $\tau$.

Throughout this section, the results are given in terms of the normalised radial solutions

$$\tilde{\Phi}^{\text{in}}_{w\ell} = \Phi^{\text{in}}_{w\ell}\,, \quad \tilde{\Phi}^{\text{up}}_{w\ell} = \frac{\Phi^{\text{up}}_{w\ell}}{2M}\,, \tag{7.6.7}$$

where $\Phi^{\text{up/in}}_{w\ell}$ are solutions to Eq. (7.4.4) with specific boundary conditions given by Eqs. (7.4.7), (7.4.8), and (7.4.9). There are no known exact analytic expressions for these radial solutions, so they must be solved computationally. These solutions a described further in Refs. [225, 227] (but note the opposite tilde notation).

### Boulware Vacuum

In the Boulware vacuum state, the Wightman function evaluated on a static trajectory is given by

$$W_{\text{B}}(x, x') = \sum_{\ell=0}^{\infty} \int_0^{\infty} \mathrm{d}\omega\, \frac{(2\ell+1)}{16\pi^2\omega} e^{-i\omega\tau_{12}/\sqrt{1-2M/R}} \left( |\tilde{\Phi}^{\text{up}}_{\omega\ell}(R)|^2 + |\tilde{\Phi}^{\text{in}}_{\omega\ell}(R)|^2 \right) . \tag{7.6.8}$$

Substituting this into Eq. (7.6.6), the response rate is

$$\dot{F}_{\text{B}}(\Omega) = \frac{\Theta(-\Omega)}{8\pi|\Omega|} \sum_{\ell=0}^{\infty} (2\ell+1) \left( |\tilde{\Phi}^{\text{up}}_{\tilde{\omega}\ell}(R)|^2 + |\tilde{\Phi}^{\text{in}}_{\tilde{\omega}\ell}(R)|^2 \right) , \tag{7.6.9}$$

where $\tilde{\omega} \equiv \Omega\sqrt{1 - 2M/R}$. This response rate is zero for positive detector energy gaps, $\Omega > 0$. In other words, the excitation rate is zero, since $\Omega < 0$ corresponds to de-excitations. This is similar to the transition rate of an inertial detector in Minkowski spacetime, $-\Omega\Theta(-\Omega)/2\pi$ [227], which is consistent with Eq. (6.4.9) in that it vanishes for positive energy gap (excitations) and massless fields. A vanishing excitation rate makes sense for the Boulware vacuum state, which is defined as detecting no incoming or outgoing particles. The flat-spacetime analogy is an accelerating detector's excitation rate is zero in the Rindler vacuum state.

The modifications to this rate are due to the curvature of spacetime, and can be thought of as the local density of states [231]. For $R \to \infty$, the response rate becomes identical to that of an inertial detector in Minkowski spacetime in the Minkowski vacuum state. This is analogous to the Rindler vacuum state becoming equal to the Minkowski vacuum state when $\alpha \to 0$, for an accelerating detector in Minkowski spacetime, as studied in Chapter 6.

### Hartle-Hawking Vacuum

In the Hartle-Hawking vacuum state, the Wightman function evaluated on a static trajectory is given by

$$\begin{aligned} W_{\text{H}}(x,x') = \sum_{\ell=0}^{\infty} \int_0^{\infty} \mathrm{d}\omega \, \frac{(2\ell+1)}{16\pi^2 \omega \sinh{(4\pi M \omega)}} \left( |\tilde{\Phi}^{\text{up}}_{\omega\ell}(R)|^2 + |\tilde{\Phi}^{\text{in}}_{\omega\ell}(R)|^2 \right) \\ \times \cosh \left[ 4\pi M \omega - \frac{i\omega\tau_{12}}{\sqrt{1-2M/R}} \right] , \end{aligned} \tag{7.6.10}$$

The response rate is then

$$\dot{F}_{\text{H}}(\Omega) = \frac{1}{8\pi\Omega} \frac{1}{e^{\Omega/T_{\text{loc}}} - 1} \sum_{\ell=0}^{\infty} (2\ell+1) \left( |\tilde{\Phi}^{\text{up}}_{\tilde{\omega}\ell}(R)|^2 + |\tilde{\Phi}^{\text{in}}_{\tilde{\omega}\ell}(R)|^2 \right) , \tag{7.6.11}$$

where $\tilde{\omega} \equiv \Omega\sqrt{1-2M/R}$ and $T_{\text{loc}}$ is the local Hawking temperature, defined by

$$T_{\text{loc}} \equiv \frac{1}{8\pi M \sqrt{1-2M/R}} . \tag{7.6.12}$$

Since the mode functions, $\tilde{\Phi}^{\text{up/in}}_{\tilde{\omega}\ell}$, only depend on the absolute value of $\tilde{\omega}$, this response rate obeys the principle of detailed balance [232–236],

$$\dot{F}_{\text{H}}(\Omega) = e^{-\Omega/T_{\text{loc}}} \dot{F}(-\Omega) , \tag{7.6.13}$$

which means that excitation rate out of the ground state equals the de-excitation rate into the ground state, corrected with the Boltzmann factor expressing the relative population of the excited state with respect to the ground state, provided the detector is in thermal equilibrium at temperature $T$. Thus, this condition, which follows from the *Kubo-Martin-Schwinger (KMS) condition* [237–239], is often used to define thermal equilibrium [155]. This confirms the thermality of the Hartle-Hawking state in the KMS sense.

The physical interpretation is that the detector, fixed at radial distance $R$, is in equilibrium with a thermal bath at temperature $T_{\text{loc}}$ (which indeed tends to $T_{\text{H}} = 1/8\pi M$ at $r \to \infty$). This situation is analogous to an accelerating UdW detector being in equilibrium with a thermal bath of Rindler particles (once the transients have subsided) in the Minkowski vacuum state in Minkowski spacetime, as studied in Chapter 6.

**Unruh Vacuum**

In the Unruh vacuum state, the Wightman function evaluated on a static trajectory is given by

$$W_{\text{U}}(x,x') = \sum_{\ell=0}^{\infty} \int_0^{\infty} \mathrm{d}\omega \, \frac{(2\ell+1)}{16\pi^2\omega} \times \left[ \frac{|\tilde{\Phi}^{\text{up}}_{\omega\ell}(R)|^2}{2\sinh(4\pi M\omega)} \left( e^{4\pi\omega - i\omega\tau_{12}/\sqrt{1-2M/R}} + e^{-4\pi\omega + i\omega\tau_{12}/\sqrt{1-2M/R}} \right) + |\tilde{\Phi}^{\text{in}}_{\omega\ell}(R)|^2 e^{-i\omega\tau_{12}/\sqrt{1-2M/R}} \right] . \tag{7.6.14}$$

Substituting this into Eq. (7.6.6), we get

$$\dot{F}_{\text{U}}(\Omega) = \sum_{l=0}^{\infty} \frac{(2\ell+1)}{4\pi} \left[ \frac{|\tilde{\Phi}^{\text{up}}_{\tilde{\omega}\ell}(R)|^2}{2\Omega \left(e^{\Omega/T_{\text{loc}}} - 1\right)} - \frac{|\tilde{\Phi}^{\text{in}}_{\tilde{\omega}\ell}(R)|^2}{2\Omega} \Theta(-\Omega) \right] , \tag{7.6.15}$$

where $\tilde{\omega} \equiv \Omega\sqrt{1-2M/R}$ and $T_{\text{loc}}$ is given by (7.6.12). The first term (involving up modes) is identical to the term in involving up modes in the Hartle-Hawking response rate (Eq. (7.6.11)). The second term (involving in modes) is identical to the term involving in modes in the Boulware response rate (Eq. (7.6.9)). This reflects how the Unruh vacuum, $|0_{\text{U}}\rangle$, is constructed as a combination of the field modes relevant to $|0_{\text{H}}\rangle$ and $|0_{\text{B}}\rangle$. The contribution from the in modes vanishes for $\Omega > 0$ (detector excitations). The contribution from the up modes represents a thermal flux of outgoing radiation, which vanishes as $R^{-2}$ as $R \to \infty$ [227]. Consequently, the Unruh response rate mimics the Boulware response rate at infinity, in that it tends to zero.

## 7.6.2 Circular Orbit

In this section, we summarise the results in Ref. [227] for the transition rate of a detector orbiting the Schwarzschild black hole on a circular geodesic. We once again consider the Boulware, Hartle-Hawking, and Unruh vacuum states. The detector trajectory is

$$t = a\tau\,, \quad r = R\,, \quad \theta = \pi/2\,, \quad \phi = b\tau\,, \tag{7.6.16}$$

where $R > 3M$ and

$$\begin{aligned} a &\equiv \sqrt{R/(R-3M)}\,, \\ b &\equiv \frac{1}{a}\frac{d\phi}{dt} = \frac{1}{a}\sqrt{M/R^3}\,. \end{aligned} \tag{7.6.17}$$

There are no circular orbits for $R < 3M$, trajectories with $3M < R \leq 6M$ are unstable, and trajectories with $R > 6M$ are stable [152, 240–242]—although the stability of the orbit has no qualitative effect on the transition rate [227]. The normalised solutions $\tilde{\Phi}^{\text{up/in}}_{w\ell}$ are the same as those in Section 7.6.1.

### Boulware Vacuum

In the Boulware vacuum state, the Wightman function evaluated on a circular-orbit trajectory is given by

$$\begin{aligned} W_{\text{B}}(x,x') = \sum_{\ell=0}^{\infty}\sum_{m=-\ell}^{\ell}\int_0^{\infty} \mathrm{d}\omega\, &\frac{(\ell-m)!(2\ell+1)|P_\ell^m(0)|^2}{16\pi^2\omega(\ell+m)!} e^{imb\tau_{12}-ia\omega\tau_{12}} \\ &\times \left(|\tilde{\Phi}^{\text{up}}_{\omega\ell}(R)|^2 + |\tilde{\Phi}^{\text{in}}_{\omega\ell}(R)|^2\right)\,, \end{aligned} \tag{7.6.18}$$

where $P_\ell^m(0)$ are the associated Legendre polynomials [243]. The response rate is then

$$\begin{aligned} \dot{F}_{\text{B}}(\Omega) = \frac{1}{a}\sum_{\ell=0}^{\infty}\sum_{m=-\ell}^{\ell} &\frac{(\ell-m)!(2\ell+1)|P_\ell^m(0)|^2}{8\pi\omega_-(\ell+m)!} \\ &\times \left(|\tilde{\Phi}^{\text{up}}_{\omega_-\ell}(R)|^2 + |\tilde{\Phi}^{\text{in}}_{\omega_-\ell}(R)|^2\right)\Theta\left(mb-\Omega\right)\,, \end{aligned} \tag{7.6.19}$$

with

$$\omega_- \equiv (mb-\Omega)/a\,. \tag{7.6.20}$$

Unlike the case of a static detector in the Boulware vacuum (Eq. (7.6.9)), the response rate in Eq. (7.6.19) has a non-vanishing excitation component that decreases as the radius increases. This is consistent with the Boulware vacuum state approximating to the Minkowski vacuum state as $R \to \infty$, as well as the circular-orbit detector behaving like a a static detector in the same limit.

**Hartle-Hawking Vacuum**

In the Hartle-Hawking vacuum state, the Wightman function evaluated on a circular-orbit trajectory is given by

$$W_{\rm H}(x,x') = \sum_{\ell=0}^{\infty}\sum_{m=-\ell}^{+\ell}\int_0^{\infty} {\rm d}\omega\, \frac{(\ell-m)!(2\ell+1)|P_\ell^m(0)|^2}{32\pi^2\omega(l+m)!\sinh(4\pi M\omega)} \times \left(|\tilde{\Phi}^{\rm up}_{\omega\ell}(R)|^2 + |\tilde{\Phi}^{\rm in}_{\omega\ell}(R)|^2\right) \times \left[e^{4\pi M\omega - ia\omega\tau_{12} + imb\tau_{12}} + e^{-4\pi M\omega + ia\omega\tau_{12} - imb\tau_{12}}\right]. \tag{7.6.21}$$

The response rate is then

$$\dot{F}_{\rm H}(\Omega) = \sum_{\ell=0}^{\infty}\sum_{m=-\ell}^{+\ell} \frac{(\ell-m)!(2\ell+1)|P_\ell^m(0)|^2}{16\pi(l+m)!} \times \Bigg[\frac{\left(|\tilde{\Phi}^{\rm up}_{\omega_-\ell}(R)|^2 + |\tilde{\Phi}^{\rm in}_{\omega_-\ell}(R)|^2\right) e^{4\pi M\omega_-}}{a\omega_-\sinh(4\pi M\omega_-)}\Theta(mb-\Omega) + \frac{\left(|\tilde{\Phi}^{\rm up}_{\omega_+\ell}(R)|^2 + |\tilde{\Phi}^{\rm in}_{\omega_+\ell}(R)|^2\right) e^{-4\pi M\omega_-}}{a\omega_+\sinh(4\pi M\omega_+)}\Theta(mb+\Omega)\Bigg], \tag{7.6.22}$$

with

$$\omega_\pm \equiv (mb \pm \Omega)/a\,. \tag{7.6.23}$$

In the limit of a large energy gap, $\Omega$, Eq. (7.6.22) becomes thermal (in the KMS sense, defined by Eq. (7.6.13)) at a temperature that is higher than the local Hawking temperature. This discrepancy is larger than the Dopper blueshift factor due to the velocity of the circular geodesic with respect to the static detectors, as found in Ref. [244] for a model that suppressed the angular dependence of the field. Thus, the physical explanation seems to be that the transition rate at large excitation energies is dominated by the most energetic field quanta, and these are seen by the detector from a head-on direction and are hence blueshifted more than by the Doppler shift factor that accounts for just the time dilation.

Due to the analogy between the static UdW detector in the Hartle-Hawking vacuum state in Schwarzschild spacetime and the accelerated UdW detector in the Minkowski vacuum state in Minkowski spacetime, it seems natural to ask what the analogy of circular motion is in the flat-spacetime case [245]. Ref. [227] suggests that the answer is a uniformly accelerating UdW detector in Minkowski space with a constant veloc-ity drift in the transverse direction. They show the transition rate in such a scenario aligns with Schwarzschild circular-orbit detector rate for large orbital radius.

#### Unruh Vacuum

In the Unruh vacuum state, the Wightman function evaluated on a circular-orbit trajectory is given by

$$W_{\mathrm{U}}(x,x') = \sum_{\ell=0}^{\infty}\sum_{m=-\ell}^{\ell}\int_0^{\infty} \mathrm{d}\omega\, \frac{(\ell-m)!(2\ell+1)|P_\ell^m(0)|^2}{16\pi^2(\ell+m)!}\times$$
$$\times\left[\frac{|\tilde{\Phi}^{\mathrm{up}}_{\omega\ell}(R)|^2\left(e^{4\pi M\omega - ia\omega\tau_{12}+imb\tau_{12}} + e^{-4\pi M\omega + ia\omega\tau_{12}-imb\tau_{12}}\right)}{2\omega\sinh\left(4\pi M\omega\right)}\right.$$
$$\left. + \frac{|\tilde{\Phi}^{\mathrm{in}}_{\omega\ell}(R)|^2 e^{-ia\omega\tau_{12}+imb\tau_{12}}}{\omega}\right]. \quad (7.6.24)$$

The response rate is then

$$\dot{F}_{\mathrm{U}}(\Omega) = \frac{1}{a}\sum_{\ell=0}^{\infty}\sum_{m=-\ell}^{\ell}\frac{(\ell-m)!(2\ell+1)|P_\ell^m(0)|^2}{8\pi(\ell+m)!}$$
$$\times\left[\left(\frac{|\tilde{\Phi}^{\mathrm{up}}_{\omega_-\ell}(R)|^2}{2\omega_-\sinh\left(4\pi M\omega_-\right)}e^{4\pi M\omega_-} + \frac{|\tilde{\Phi}^{\mathrm{in}}_{\omega_-\ell}(R)|^2}{\omega_-}\right)\Theta(mb-\Omega)\right.$$
$$\left. + \frac{|\tilde{\Phi}^{\mathrm{up}}_{\omega_+\ell}(R)|^2}{2\omega_+\sinh\left(4\pi M\omega_+\right)}e^{-4\pi M\omega_+}\Theta(mb+\Omega)\right], \quad (7.6.25)$$

with

$$\omega_\pm \equiv (mb\pm\Omega)/a\,. \quad (7.6.26)$$

Similarly to the static case, Ref. [227] shows that, as the radius increases, the circular-geodesic rates in the Boulware and Unruh states (Eq. (7.6.19) and (7.6.25)) become equal, whereas the ratio of the Hartle-Hawking rate (Eq. (7.6.22)) to Unruh rate becomes large. As is the case for the circular-orbit detector in the Hartle-Hawking state, Eq. (7.6.25) becomes thermal in the KMS sense in the limit of large energy gap.

### 7.6.3 Radial Free Fall

An UdW detector falling radially into a black hole is a non-stationary situation. One result of this is that defining the instantaneous transition *rate* is more subtle, since we cannot use time-translation invariance to move the switching on and off into the asymptotic past and future, respectively (which previously allowed us differentiate out the formally divergent integral over proper time). Furthermore, the sharp-switching limit (i.e., the switching function becoming a top-hat function) can introduce divergences that must be regularised by parameterising the Wightman function, which is formally a distribution, and carefully taking the limit of the switching function becoming arbitrarily sharp [160, 162, 163, 214]. The non-trivial time dependence must therefore be accounted for when attempting to define an instantaneous transition rate.

In addition, we are interested in the case of a $(3+1)$-dimensional Schwarzschild black hole, in which the expressions relevant for an infalling UdW detector (such as the radial solutions to Eq. (7.4.4)) are not analytically solvable. Numerically, the equations for the transition rate are much more computationally demanding and have not yet been made to converge [214, 225, 246]. Specifically, integrating the Wightman functions over time, $\tau_{12}$, for a non-stationary trajectory no longer results in $\delta$-functions which collapse the integral over $\omega$ (which was performed implicitly in the results of Sections 7.6.1 and 7.6.2), meaning the numerical evaluation of the transition rate becomes significantly more involved. For example, the Wightman function is divergent at short distances, and while it is known how the divergent parts are subtracted in the expressions for the transition probability and transition rate [162, 225], the challenge in numerical calculations is to implement these subtractions term-by-term in a mode sum. As a result, this section will focus on the simpler case of the transition probability instead of the transition rate.

Since the detector is destined to reach the singularity, we *must* cut off the measurement at a finite time. Furthermore, in order to probe the local structure of the spacetime as we fall into the black hole, we should only switch on the detector for a short, finite time. Otherwise, we may be able to conclude that the detector was excited, but we would not be able to tell where on its trajectory this occurred. However, if the detector is switched on for too short a time, the transition probability will be dominated by transient switching effects.

Ref. [214] is the latest paper to study this situation for a $(3+1)$-dimensional Schwarzschild black hole, and the authors conclude that the time spent near the event horizon is shorter than the time taken for transient effects to subside (for all choices of vacuum state), meaning that an infalling UdW detector could never be used to unambiguously detect physical radiation (such as Hawking radiation) near the event horizon of a black hole. Nevertheless, they derive the relevant equations to calculate the (albeit transients-dominated) transition probability of a radially infalling UdW detector in the Boulware, Hartle-Hawking, and Unruh vacuum states. In this section, we will review the results.

Ref. [246] also studies the transition probability of an UdW detector falling radially into a $(3+1)$-dimensional Schwarzschild black hole (for the same vacuum states), and they arrive at slightly different conclusions. The main difference to Ref. [214] is that Ref. [246] reports a small local extremum in transition probability as the detector crosses the event horizon. This result would contradict the generally-accepted prediction motivated from the equivalence principle: An observer crossing the event horizon of a black hole is locally in free fall, and hence experiences nothing special as they cross the horizon (for a black hole massive enough such that tidal effects are negligible near the horizon). Ref. [214] disagrees with this result, and the original authors have admitted there is a lack of analytic corroboration [192]. However, more evidence for its existence would be incredibly interesting.

In Kruskal coordinates, the trajectory of a radially infalling detector is given by

$$\begin{aligned} \overline{u} &= -4M(z-1)e^{z+z^2/2+z^3/3}\,, \\ \overline{v} &= 4M(z+1)e^{-z+z^2/2-z^3/3}\,, \end{aligned} \tag{7.6.27}$$

where $z = (\tau/\tau_{\mathrm{H}})^{1/3}$. $\tau \in (-\infty, 0)$ is the proper time, and $\tau_{\mathrm{H}} = -4M/3$. In these coordinates, the detector starts its journey in the asymptotic past, $\tau \to -\infty$, crosses the horizon at $\tau = \tau_{\mathrm{H}}$, and ends at the singularity as $\tau \to 0^-$. Equivalently, the trajectory is described by,

$$\begin{aligned} \frac{\mathrm{d}r}{\mathrm{d}\tau} &= \eta\sqrt{2M/r - 2M/R}\,, \\ \frac{\mathrm{d}v}{\mathrm{d}\tau} &= \frac{\sqrt{1-2M/R} + \eta\sqrt{2M/r-2M/R}}{1-2M/r}\,, \end{aligned} \tag{7.6.28}$$

where $v$ is the ingoing null coordinate defined in Eq. (7.3.3).

### Boulware Vacuum

Substituting the mode functions in Eq. (7.5.1) into Eq. (7.6.5),

$$F_{\mathrm{B}} = F_{\mathrm{B}}^{\mathrm{in}} + F_{\mathrm{B}}^{\mathrm{up}}\,, \tag{7.6.29}$$

with

$$F_{\mathrm{B}}^{\mathrm{in/up}} = \sum_{\ell=0}^{\infty} \int_0^{\infty} \mathrm{d}\omega \frac{2\ell+1}{16\pi^2\omega} \left| \int_{-\infty}^{\infty} \mathrm{d}\tau\, \chi(\tau) e^{-i\Omega\tau} I_{\omega\ell}^{\mathrm{in/up}}(\tau) \right|^2 , \tag{7.6.30}$$

where

$$I_{\omega\ell}^{\mathrm{in}}(\tau) = e^{i\omega r_*} \Phi_{\omega\ell}^{\mathrm{in}}(r) V^{-i4M\omega}, \tag{7.6.31a}$$

$$I_{\omega\ell}^{\mathrm{up}}(\tau) = \begin{cases} e^{-i\omega r_*} \Phi_{\omega\ell}^{\mathrm{up}}(r)(-U)^{i4M\omega}, & r > 2M, \\ \dfrac{A_{\omega\ell}^{\mathrm{up}}}{B_{\omega\ell}^{\mathrm{in}}} e^{i\omega r_*} \Phi_{\omega\ell}^{\mathrm{in}}(r) V^{-i4M\omega} + e^{-4\pi M\omega} \dfrac{1}{B_{\omega\ell}^{\mathrm{in}*}} e^{-i\omega r_*} \Phi_{\omega\ell}^{\mathrm{in}*}(r) U^{i4M\omega}, & r < 2M, \end{cases} \tag{7.6.31b}$$

$U = \overline{u}/4M$, $V = \overline{v}/4M$, $r_*$ is defined by Eq. (7.3.2), $\chi(\tau)$ is the switching function which controls the measurement time and duration, and $\Phi_{\omega\ell}^{\mathrm{up/in}}(r)$ are the radial solutions described in Section 7.4. The spacetime coordinates $U$, $V$, and $r$ are evaluated on the detector trajectory.

The Boulware vacuum diverges at the event horizon, making it a poor candidate of vacuum choice to study the response of an infalling detector. The Hartle-Hawking and Unruh vacua give more physically realistic descriptions of near-horizon physics. Ref. [246] plots the transition probability for an infalling detector in the Boulware vacuum state up to $r \sim 1.3(2M)$ and shows that the probability decreases as the detector approaches the horizon. Note that this is not a description of what happens at the event horizon, since the entire integration region must fall outside of the horizon, meaning the measurement must be averaged over a region large enough for transients to subside.

### Hartle-Hawking Vacuum

Substituting the mode functions in Eq. (7.5.2) into Eq. (7.6.5),

$$F_{\mathrm{H}} = F_{\mathrm{H}}^{\mathrm{in}} + F_{\mathrm{H}}^{\mathrm{up}} + F_{\mathrm{H}}^{\overline{\mathrm{in}}} + F_{\mathrm{H}}^{\overline{\mathrm{up}}}\,, \tag{7.6.32}$$

where

$$F_{\mathrm{H}}^{\mathrm{in/up},\overline{\mathrm{in/up}}} = \sum_{\ell=0}^{\infty} \int_0^{\infty} \frac{\mathrm{d}\omega}{1-e^{-8\pi M\omega}} \frac{2\ell+1}{16\pi^2\omega} \left| \int_{-\infty}^{\infty} \mathrm{d}\tau\, \chi(\tau) e^{-i\Omega\tau} I_{\omega\ell}^{\mathrm{in/up},\overline{\mathrm{in/up}}}(\tau) \right|^2 \tag{7.6.33}$$

and

$$I_{\omega\ell}^{\overline{\mathrm{in}}}(\tau) = e^{-4\pi M\omega} e^{-i\omega r_*} \Phi_{\omega\ell}^{\mathrm{in}*}(r) V^{i4M\omega}, \tag{7.6.34a}$$

$$I_{\omega\ell}^{\overline{\mathrm{up}}}(\tau) = \begin{cases} e^{-4\pi M\omega} e^{i\omega r_*} \Phi_{\omega\ell}^{\mathrm{up}*}(r)(-U)^{-i4M\omega}, & r > 2M, \\ e^{-4\pi M\omega} \dfrac{A_{\omega\ell}^{\mathrm{up}*}}{B_{\omega\ell}^{\mathrm{in}*}} e^{-i\omega r_*} \Phi_{\omega\ell}^{\mathrm{in}*}(r) V^{i4M\omega} + \dfrac{1}{B_{\omega\ell}^{\mathrm{in}}} e^{i\omega r_*} \Phi_{\omega\ell}^{\mathrm{in}}(r) U^{-i4M\omega}, & r < 2M. \end{cases} \tag{7.6.34b}$$

The functions and variables $I_{\omega\ell}^{\mathrm{in}}(\tau)$, $I_{\omega\ell}^{\mathrm{up}}(\tau)$, $U$, $V$, $r_*$, $\chi(\tau)$, and $\Phi_{\omega\ell}^{\mathrm{up/in}}(r)$ are defined as in Eq. (7.6.31).

Ref. [214] shows that the magnitude of the response function, $F_{\mathrm{H}}$, decreases for larger detector energy gaps, $\Omega$. For all detector energy gaps, $F_{\mathrm{H}}$ increases smoothly and monotonically as the detector approaches the black hole, with no distinctive features at the event horizon. This aligns with the equivalence principle conclusion that there is no way to discern whether you are crossing the event horizon or not, for a black hole of sufficient mass. This conclusion contrasts with the conclusions of Ref. [246], which reports a local extremum in $F_{\mathrm{H}}$ near the horizon.

The Hartle-Hawking vacuum state describes a thermal bath of particles, but Ref. [214] argues that, near the horizon, the detector does not spend enough time ‘switched on’ to reach thermal equilibrium with this thermal state. As a result, the behaviour of $F_{\mathrm{H}}$ is a reflection of local interactions with the field due to switching, and not a physical measurement of the thermal state. Allowing enough time for transient effects to subside requires switching the detector on over a large portion of its trajectory, resulting in a cumulative measurement and preventing local measurements.

### Unruh Vacuum

Substituting the mode functions in Eq. (7.5.5) into Eq. (7.6.5) gives

$$F_{\mathrm{U}} = F_{\mathrm{B}}^{\mathrm{in}} + F_{\mathrm{H}}^{\mathrm{up}} + F_{\mathrm{H}}^{\overline{\mathrm{up}}}, \tag{7.6.35}$$

where $F_{\mathrm{B}}^{\mathrm{in}}$ is defined by Eq. (7.6.30) and $F_{\mathrm{H}}^{\mathrm{up}/\overline{\mathrm{up}}}$ is defined by Eq. (7.6.33).

The detector response in the Unruh vacuum, $F_{\mathrm{U}}$, is similar to the response in the Hartle-Hawking vacuum, $F_{\mathrm{H}}$, except it is slightly smaller in magnitude. Larger detector energy gaps, $\Omega$, result in $F_{\mathrm{U}}$ being of a smaller magnitude, and $F_{\mathrm{U}}$ increases smoothly and monotonically as the detector approaches the black hole. Again, this encapsulates the detector's response due to switching, and not a measurement of Hawking radiation (as one might expect from the definition of the Unruh vacuum state).

## 7.7 Summary

Four-dimensional Schwarzschild black holes are a complicated environment to study the response of UdW detectors. The response functions are not analytically solvable, and solving them numerically is computationally demanding. In the case of a detector falling radially into the black hole, defining the instantaneous transition rate is challenging and we are typically restricted to studying the transition probability. Moreover, Ref. [214] claims it is impossible for an infalling observer to unambiguously measure Hawking radiation near the horizon, since the detector's response is dominated by transients. Nevertheless, it is possible to calculate expressions for the response function, even for radial free fall, and study the results. Results are presented in Ref. [227] for stationary scenarios and Ref. [214] for free fall.

Finding new techniques to compute the local instantaneous transition rate of an infalling detector in $(3 + 1)$-dimensional Schwarzschild spacetime would be an exciting direction for future work. If it is possible to isolate the contribution of transient effects, in a similar fashion to Section 6.2.4 and Ref. [247], then we may be able to study an UdW detector's response to Hawking radiation. Studying the interaction between multiple UdW detectors in various black hole spacetimes could also give new insights into the quantum mechanical properties of black holes [128–140, 202, 248–250]. These multiple-detector situations may benefit from the manifestly causal method outlined in Chapter 4, as they represent a more complex extension of the Fermi two-atom problem discussed in Section 4.4.

# Chapter 8

# Conclusions

In this thesis, we have developed a new, manifestly causal formalism of quantum field theory (QFT), in which calculations are performed directly at the probability level. Chapter 4 introduces this formalism and provides a versatile toolbox for its application to a variety of quantum systems. By generalising the Baker-Campbell-Hausdorff (BCH) lemma, commutators and anticommutators naturally arise in the definition of a probability. Since these structures encode causality in QFT, every intermediate step and result in our approach is manifestly causal. This is not the case for traditional calculations in QFT, where the Feynman propagator is ubiquitous.

The formalism employs a broad definition of measurement (represented by an element of a Positive Operator-Valued Measure) that can be rendered inclusive from the outset. In line with the Bloch-Nordsieck (BN) [27] and Kinoshita-Lee-Nauenberg (KLN) [28, 29] theorems, infrared (IR) divergences always cancel in the computation of physical observables, since the observables must be sufficiently inclusive. Thus, our causal formalism may offer a novel perspective on the cancellation of IR divergences by incorporating inclusiveness at the probability (observable) level.

Motivated by the investigation of divergences, Chapter 3 provided a rigorous calculation of a scattering cross section with first-order quantum chromodynamics (QCD) corrections, using conventional QFT methods. We demonstrated that exclusive observables, such as the cross section for the process $e^-e^+ \to q\bar{q}$, were IR-divergent, and required the use of dimensional regularisation to cancel these infinite divergences. Chapter 5 then investigated similar calculations using the causal approach. We used scalar field theory as a simplified model of particle scattering, avoiding the subtleties and analytic complexities of gauge theories. We presented a new set of rules to generate probability-level diagrams, whose sum yields the total transition probability for a given process. Using the causal formalism, we explicitly calculated the tran-sition probabilities for inclusive particle decay and pair-annihilation, obtaining results that conformed with the rules. Notably, the appearance of retarded propagators and related causal structures, such as the retarded self-energy, underscores the manifest causality of our formalism. We also observed that terms corresponding to 'real emissions' (as defined in the traditional calculation) were not separable from other contributions, suggesting an intrinsic summation over degenerate processes. This is a promising avenue for further exploration of the cancellation of IR divergences, or even their potential absence, in this framework.

Chapter 6 applied the causal formalism to investigate the Unruh effect. We derived the excitation rate of a uniformly accelerating Unruh-DeWitt (UdW) detector in terms of commutators and anticommutators. We carefully calculated this transition rate from both the perspective of an inertial observer and an accelerating observer, showing that the rate was equal despite their differing physical interpretations. By considering an inertial observer in a Minkowski thermal bath, we emphasised that the Rindler thermal bath perceived by an accelerated observer is distinct from the inertial definition of a thermal bath. This is the essence of the Unruh effect—the concept of a particle, and hence the vacuum state, is inherently observer-dependent.

This is a natural prerequisite for the study of a detector's response in curved spacetimes, for which Chapter 7 served as an instructive introduction and literature review. Black holes offer an extreme environment in which to explore quantum fields in curved spacetime and emergent phenomena such as Hawking radiation. Although the detector response to Hawking radiation remains an open question, with transient effects potentially obscuring the underlying physics, we have summarised results from Refs. [227] and [214] for the response of an UdW detector in $(3+1)$-dimensional Schwarzschild spacetime along static, circular, and radially infalling trajectories.

Looking ahead, natural questions arise regarding the extension of this causal formalism to gauge theories and semi-inclusive observables. In particular, how can the results and rules established in Chapter 5 be generalised and applied to the calculations presented in Chapter 3? Furthermore, instead of considering observables that are fully inclusive over gluon final states, we would like to explore how the outcomes change when one adopts a semi-inclusive final state that accounts for the real emission of gluons only in the soft and collinear momentum regions. Ref. [105] discusses further semi-inclusive observables that can be analysed using the causal formalism.

Additionally, our formalism allows for a straightforward generalisation of initial and final states to mixed states. This feature makes it well-suited for studies in relativistic quantum information (RQI), where mixed states naturally arise.

Further investigation of the quantum properties of black holes may provide crucial insights into a fundamental theory that successfully unites QFT and gravity. In this context, a deeper understanding of the response of an UdW detector in black hole spacetimes represents a promising avenue for future research. The causal formalism presented here may contribute to this endeavour, as it addresses the often counterintuitive role of causality in the presence of an event horizon and the complex behaviour of quantum entanglement between multiple detectors in such spacetimes.

# References

[1] Robert Dickinson et al. “Towards a Manifestly Causal Approach to Particle Scattering”. In: (Feb. 2025). arXiv: 2502.18551 [hep-ph] (cited on pp. 10, 20, 82, 94).

[2] Robert Dickinson et al. “A new study of the Unruh effect”. In: *Class. Quant. Grav.* 42.2 (2025), p. 025014. DOI: 10.1088/1361-6382/ad9c12. arXiv: 2409.12697 [hep-th] (cited on pp. 10, 20, 82, 121).

[3] D. Hanneke, S. Fogwell, and G. Gabrielse. “New Measurement of the Electron Magnetic Moment and the Fine Structure Constant”. In: *Phys. Rev. Lett.* 100 (12 Mar. 2008), p. 120801. DOI: 10.1103/PhysRevLett.100.120801. URL: https://link.aps.org/doi/10.1103/PhysRevLett.100.120801 (cited on p. 18).

[4] T. Aoyama et al. “Revised value of the eighth-order QED contribution to the anomalous magnetic moment of the electron”. In: *Phys. Rev. D* 77 (2008), p. 053012. DOI: 10.1103/PhysRevD.77.053012. arXiv: 0712.2607 [hep-ph] (cited on p. 18).

[5] Tatsumi Aoyama, Toichiro Kinoshita, and Makiko Nio. “Theory of the Anomalous Magnetic Moment of the Electron”. In: *Atoms* 7.1 (2019), p. 28. DOI: 10.3390/atoms7010028 (cited on p. 18).

[6] Sheldon L. Glashow. “Partial-symmetries of weak interactions”. In: *Nuclear Physics* 22.4 (1961), pp. 579–588. ISSN: 0029-5582. DOI: https://doi.org/10.1016/0029-5582(61)90469-2. URL: https://www.sciencedirect.com/science/article/pii/0029558261904692 (cited on p. 18).

[7] Steven Weinberg. “A Model of Leptons”. In: *Phys. Rev. Lett.* 19 (21 Nov. 1967), pp. 1264–1266. DOI: 10.1103/PhysRevLett.19.1264. URL: https://link.aps.org/doi/10.1103/PhysRevLett.19.1264 (cited on pp. 18, 28).

[8] Abdus Salam. "Weak and Electromagnetic Interactions". In: *Conf. Proc. C* 680519 (1968), pp. 367–377. doi: 10.1142/9789812795915_0034 (cited on p. 18).

[9] Alexander Altland and Ben Simons. *Condensed Matter Field Theory*. Cambridge University Press, Aug. 2023. isbn: 978-1-108-78124-4. doi: 10.1017/9781108781244 (cited on p. 18).

[10] Eduardo H. Fradkin. *Field Theories of Condensed Matter Physics*. Vol. 82. Cambridge, UK: Cambridge Univ. Press, Feb. 2013. isbn: 978-0-521-76444-5 (cited on p. 18).

[11] John L. Cardy. *Scaling and renormalization in statistical physics*. 1996 (cited on p. 18).

[12] K. G. Wilson and John B. Kogut. "The Renormalization group and the epsilon expansion". In: *Phys. Rept.* 12 (1974), pp. 75–199. doi: 10.1016/0370-1573(74)90023-4 (cited on p. 18).

[13] John F. Donoghue. "General relativity as an effective field theory: The leading quantum corrections". In: *Phys. Rev. D* 50 (1994), pp. 3874–3888. doi: 10.1103/PhysRevD.50.3874. arXiv: gr-qc/9405057 (cited on p. 18).

[14] C. P. Burgess. "Quantum gravity in everyday life: General relativity as an effective field theory". In: *Living Rev. Rel.* 7 (2004), pp. 5–56. doi: 10.12942/lrr-2004-5. arXiv: gr-qc/0311082 (cited on p. 18).

[15] S. W. Hawking. "Particle creation by black holes". In: *Commun. Math. Phys.* 43 (1975), pp. 199–220. doi: 10.1007/BF02345020 (cited on pp. 18, 120, 149).

[16] S. W. Hawking. "Black hole explosions". In: *Nature* 248 (1974), pp. 30–31. doi: 10.1038/248030a0 (cited on pp. 18, 120, 149).

[17] S.M. Christensen and S.A. Fulling. "Trace anomalies and the Hawking effect". In: *Phys. Rev. D* 15 (1977), p. 2088. doi: 10.1103/PhysRevD.15.2088 (cited on pp. 18, 160).

[18] W. Unruh. "Notes on black hole evaporation". In: *Phys. Rev. D* 14.4 (1976), p. 870. doi: 10.1103/PhysRevD.14.870 (cited on pp. 18, 120, 125, 131, 149, 162).

[19] P. C. W. Davies. “Scalar particle production in Schwarzschild and Rindler metrics”. In: *J. Phys. A* 8 (1975), pp. 609–616. doi: 10.1088/0305-4470/8/4/022 (cited on pp. 18, 120).

[20] S. A. Fulling. “Nonuniqueness of canonical field quantization in Riemannian space-time”. In: *Phys. Rev. D* 7 (1973), pp. 2850–2862. doi: 10.1103/PhysRevD.7.2850 (cited on pp. 18, 120).

[21] W. Unruh. “Particle detectors and black holes”. In: *1st Marcel Grossmann Meeting On General Relativity*. 1977, pp. 527–536 (cited on pp. 18, 120).

[22] L. C. B. Crispino, A. Higuchi, and G. E. Matsas. “The Unruh effect and its applications”. In: *Rev. Mod. Phys.* 80 (2008), pp. 787–838. doi: 10.1103/RevModPhys.80.787. eprint: arXiv:0710.5373[gr-qc] (cited on pp. 18, 121, 124, 125, 134, 136, 137, 139, 141, 142, 146).

[23] Michael E. Peskin and Daniel V. Schroeder. *An Introduction to quantum field theory*. Reading, USA: Addison-Wesley, 1995. isbn: 978-0-201-50397-5 (cited on pp. 19, 21–23, 25, 26, 29, 30, 34–37, 39, 43, 45, 46, 55, 61–64, 70, 72, 73, 80).

[24] Steven Weinberg. *The Quantum Theory of Fields*. Cambridge University Press, 1995 (cited on pp. 19, 23, 39, 44).

[25] Matthew D. Schwartz. *Quantum Field Theory and the Standard Model*. Cambridge University Press, Mar. 2014. isbn: 978-1-107-03473-0 (cited on pp. 19, 29, 30, 35–37, 39, 46, 55, 60–62, 64, 65, 70–73).

[26] John C. Collins. *Renormalization : An Introduction to Renormalization, the Renormalization Group and the Operator-Product Expansion*. Vol. 26. Cambridge Monographs on Mathematical Physics. Cambridge: Cambridge University Press, 1984. isbn: 978-0-521-31177-9. doi: 10.1017/9781009401807 (cited on pp. 19, 44).

[27] F. Bloch and A. Nordsieck. “Note on the Radiation Field of the electron”. In: *Phys. Rev.* 52 (1937), pp. 54–59. doi: 10.1103/PhysRev.52.54 (cited on pp. 19, 43, 94, 176).

[28] Toichiro Kinoshita. “Mass Singularities of Feynman Amplitudes”. In: *Journal of Mathematical Physics* 3.4 (July 1962), pp. 650–677. ISSN: 0022-2488. DOI: 10.1063/1.1724268. eprint: https://pubs.aip.org/aip/jmp/article-pdf/3/4/650/19167464/650_1_online.pdf. URL: https://doi.org/10.1063/1.1724268 (cited on pp. 19, 43, 58, 94, 176).

[29] T. D. Lee and M. Nauenberg. “Degenerate Systems and Mass Singularities”. In: *Phys. Rev.* 133 (6B Mar. 1964), B1549–B1562. DOI: 10.1103/PhysRev.133.B1549. URL: https://link.aps.org/doi/10.1103/PhysRev.133.B1549 (cited on pp. 19, 43, 58, 94, 176).

[30] Rudolf Haag. *Local Quantum Physics*. Theoretical and Mathematical Physics. Berlin: Springer, 1996. ISBN: 978-3-540-61049-6. DOI: 10.1007/978-3-642-61458-3 (cited on p. 19).

[31] Kasia Rejzner. *Perturbative Algebraic Quantum Field Theory: An Introduction for Mathematicians*. Mathematical Physics Studies. New York: Springer, 2016. ISBN: 978-3-319-25899-7. DOI: 10.1007/978-3-319-25901-7 (cited on p. 19).

[32] R. Dickinson et al. “Manifest causality in quantum field theory with sources and detectors”. In: *J. High Energy Phys.* 2014.6 (2014), p. 49. DOI: 10.1007/JHEP06(2014)049. arXiv: 1312.3871 [hep-th] (cited on pp. 19, 82, 119, 129).

[33] R. Dickinson, J. Forshaw, and P. Millington. “Probabilities and signalling in quantum field theory”. In: *Phys. Rev. D* 93.6 (2016), p. 065054. DOI: 10.1103/PhysRevD.93.065054. arXiv: 1601.07784 [hep-th] (cited on pp. 19, 82, 87, 89, 92, 93, 130).

[34] R. Dickinson, J. Forshaw, and P. Millington. “Working directly with probabilities in quantum field theory”. In: *J. Phys. Conf. Ser.* 880.1 (2017), p. 012041. DOI: 10.1088/1742-6596/880/1/012041. arXiv: 1702.04602 [hep-th] (cited on pp. 19, 82).

[35] Emma Albertini et al. “In-in correlators and scattering amplitudes on a causal set”. In: *Phys. Rev. D* 109.10 (2024), p. 106014. DOI: 10.1103/PhysRevD.109.106014. arXiv: 2402.08555 [hep-th] (cited on pp. 19, 82).

[36] Selomit Ramírez-Uribe et al. "Rewording Theoretical Predictions at Colliders with Vacuum Amplitudes". In: *Phys. Rev. Lett.* 133.21 (2024), p. 211901. DOI: 10.1103/PhysRevLett.133.211901. arXiv: 2404.05491 [hep-ph] (cited on p. 19).

[37] J. Jesus Aguilera-Verdugo et al. "Open Loop Amplitudes and Causality to All Orders and Powers from the Loop-Tree Duality". In: *Phys. Rev. Lett.* 124.21 (2020), p. 211602. DOI: 10.1103/PhysRevLett.124.211602. arXiv: 2001.03564 [hep-ph] (cited on p. 19).

[38] E. T. Tomboulis. "Causality and Unitarity via the Tree-Loop Duality Relation". In: *JHEP* 05 (2017), p. 148. DOI: 10.1007/JHEP05(2017)148. arXiv: 1701.07052 [hep-th] (cited on p. 19).

[39] Ian Jubb. "Interacting Quantum Scalar Field Theory on a Causal Set". In: 2024. DOI: 10.1007/978-981-19-3079-9_76-1. arXiv: 2306.12484 [hep-th] (cited on p. 19).

[40] Pierre Ramond. *Field theory: a modern primer*. Vol. 51. 1981 (cited on p. 21).

[41] Leonhard Euler. *Methodus Inveniendi Lineas Curvas*. Leipzig: Impensis A. Haude et J. D. Kummer, 1755 (cited on p. 21).

[42] Joseph-Louis Lagrange. *Mécanique Analytique*. Chez la Veuve Courcier, 1788 (cited on p. 21).

[43] Herbert Goldstein. *Classical Mechanics*. 2nd. Addison-Wesley, 1980 (cited on p. 21).

[44] V. I. Arnold. *Mathematical Methods of Classical Mechanics*. 2nd. Springer, 1989 (cited on p. 21).

[45] Oskar Klein. "Quantentheorie und fünfdimensionale Relativitätstheorie". In: *Zeitschrift für Physik* 37 (1926), pp. 895–906. DOI: 10.1007/BF01397481 (cited on p. 22).

[46] Walter Gordon. "Der Comptoneffekt nach der Schrödingerschen Theorie". In: *Zeitschrift für Physik* 40 (1926), pp. 117–133. DOI: 10.1007/BF01400290 (cited on p. 22).

[47] Claude Itzykson and Jean-Bernard Zuber. *Quantum Field Theory*. McGraw-Hill, 1980 (cited on pp. 22, 61).

[48] P. A. M. Dirac. *The Principles of Quantum Mechanics*. Oxford University Press, 1930 (cited on p. 22).

[49] Richard P. Feynman and Albert R. Hibbs. *Quantum Mechanics and Path Integrals*. McGraw-Hill, 1965 (cited on p. 22).

[50] L. D. Landau and E. M. Lifshitz. *Quantum Mechanics: Non-Relativistic Theory*. Pergamon Press, 1977 (cited on p. 22).

[51] David J. Griffiths. *Introduction to Quantum Mechanics*. Pearson Education, 2005 (cited on pp. 22, 27).

[52] W. Pauli and V. Weisskopf. "The Quantization of the Scalar Relativistic Wave Equation". In: *Helvetica Physica Acta* 7 (1934). English translation: A. I. Miller, *Early Quantum Electrodynamics*, Cambridge University Press, Cambridge (1994), pp. 188–205, pp. 709–731 (cited on p. 22).

[53] A. Zee. *Quantum Field Theory in a Nutshell*. Princeton University Press, 2003 (cited on p. 23).

[54] M. Srednicki. *Quantum field theory*. Cambridge University Press, Jan. 2007. ISBN: 978-0-521-86449-7. DOI: `10.1017/CBO9780511813917` (cited on pp. 24, 29, 35, 36, 39, 55, 64, 73, 104).

[55] George F. Sterman. *An Introduction to quantum field theory*. Cambridge University Press, Aug. 1993. ISBN: 978-0-521-31132-8 (cited on pp. 24, 112).

[56] G. C. Wick. "The Evaluation of the Collision Matrix". In: *Phys. Rev.* 80 (2 Oct. 1950), pp. 268–272. DOI: `10.1103/PhysRev.80.268`. URL: `https://link.aps.org/doi/10.1103/PhysRev.80.268` (cited on p. 25).

[57] Paul A. M. Dirac. "The quantum theory of the electron". In: *Proc. Roy. Soc. Lond. A* 117 (1928), pp. 610–624. DOI: `10.1098/rspa.1928.0023` (cited on p. 26).

[58] W. K. Clifford. "Applications of Grassmann's Extensive Algebra". In: *American Journal of Mathematics* 1 (1878), pp. 350–358. DOI: `10.2307/2369379` (cited on p. 26).

[59] W. K. Clifford. *Mathematical Papers*. Macmillan, 1878 (cited on p. 26).

[60] W. Pauli. “Über den Zusammenhang des Abschlusses der Elektronengruppen im Atom mit der Komplexstruktur der Spektren”. In: *Zeitschrift für Physik* 31.1 (Feb. 1925), pp. 765–783. ISSN: 0044-3328. DOI: 10.1007/BF02980631. URL: https://doi.org/10.1007/BF02980631 (cited on p. 27).

[61] Enrico Fermi. “Sulla quantizzazione del gas perfetto monoatomico”. Italian. In: *Rendiconti Lincei* 3 (1926). Translated as Zannoni, Alberto (1999-12-14). ”On the Quantization of the Monoatomic Ideal Gas”, pp. 145–149 (cited on p. 28).

[62] Paul A. M. Dirac. “On the Theory of Quantum Mechanics”. In: *Proceedings of the Royal Society A* 112.762 (1926), pp. 661–677. DOI: 10.1098/rspa.1926.0133 (cited on p. 28).

[63] R.K. Pathria and P.D. Beale. *Statistical Mechanics*. 3rd. Elsevier, 2011 (cited on p. 28).

[64] Stephen J. Blundell and Katherine M. Blundell. *Concepts in Thermal Physics*. Oxford University Press, Oct. 2009. ISBN: 9780199562091. DOI: 10.1093/acprof:oso/9780199562091.001.0001. URL: https://doi.org/10.1093/acprof:oso/9780199562091.001.0001 (cited on p. 28).

[65] Satyendranath Bose. “Planck’s law and the light quantum hypothesis”. In: *Journal of Astrophysics and Astronomy* 15.1 (Mar. 1994), pp. 3–7. ISSN: 0973-7758. DOI: 10.1007/BF03010400. URL: https://doi.org/10.1007/BF03010400 (cited on p. 28).

[66] Albert Einstein. “Quantentheorie des einatomigen idealen Gases”. German. In: *Königliche Preußische Akademie der Wissenschaften. Sitzungsberichte* (1924), pp. 261–267 (cited on p. 28).

[67] Ian Duck, E. C. G. Sudarshan, and Arthur S. Wightman. “Pauli and the Spin-Statistics Theorem”. In: *American Journal of Physics* 67.8 (Aug. 1999), pp. 742–746. ISSN: 0002-9505. DOI: 10.1119/1.19365. eprint: https://pubs.aip.org/aapt/ajp/article-pdf/67/8/742/7528272/742_1_online.pdf. URL: https://doi.org/10.1119/1.19365 (cited on p. 28).

[68] J. C. Ward. “An Identity in Quantum Electrodynamics”. In: *Phys. Rev.* 78 (2 Apr. 1950), pp. 182–182. DOI: 10.1103/PhysRev.78.182. URL: https://link.aps.org/doi/10.1103/PhysRev.78.182 (cited on pp. 29, 74).

[69] Y. Takahashi. “On the generalized Ward identity”. In: *Il Nuovo Cimento (1955-1965)* 6.2 (Aug. 1957), pp. 371–375. ISSN: 1827-6121. DOI: 10.1007/BF02832514. URL: https://doi.org/10.1007/BF02832514 (cited on pp. 29, 74).

[70] L.D. Faddeev and V.N. Popov. “Feynman diagrams for the Yang-Mills field”. In: *Physics Letters B* 25.1 (1967), pp. 29–30. ISSN: 0370-2693. DOI: https://doi.org/10.1016/0370-2693(67)90067-6. URL: https://www.sciencedirect.com/science/article/pii/0370269367900676 (cited on p. 31).

[71] Gerard ’t Hooft. “Dimensional Reduction in Quantum Gravity”. In: *arXiv preprint gr-qc/9310026* (1993). arXiv: gr-qc/9310026 (cited on pp. 31, 150).

[72] J.C. Taylor. “Ward identities and charge renormalization of the Yang-Mills field”. In: *Nuclear Physics B* 33.2 (1971), pp. 436–444. ISSN: 0550-3213. DOI: https://doi.org/10.1016/0550-3213(71)90297-5. URL: https://www.sciencedirect.com/science/article/pii/0550321371902975 (cited on pp. 31, 73).

[73] A. A. Slavnov. “Ward identities in gauge theories”. In: *Theoretical and Mathematical Physics* 10.2 (Feb. 1972), pp. 99–104. ISSN: 1573-9333. DOI: 10.1007/BF01090719. URL: https://doi.org/10.1007/BF01090719 (cited on pp. 31, 73).

[74] Walter Greiner and Joachim Reinhardt. *Field Quantization*. Springer, 1996. ISBN: 978-3-642-61485-9. DOI: 10.1007/978-3-642-61485-9 (cited on pp. 32, 82, 84, 130).

[75] W. Heisenberg. “Die „beobachtbaren Größen“ in der Theorie der Elementarteilchen”. In: *Zeitschrift für Physik* 120.7 (July 1943), pp. 513–538. ISSN: 0044-3328. DOI: 10.1007/BF01329800. URL: https://doi.org/10.1007/BF01329800 (cited on p. 35).

[76] F. J. Dyson. “The $S$ Matrix in Quantum Electrodynamics”. In: *Phys. Rev.* 75 (11 June 1949), pp. 1736–1755. DOI: 10.1103/PhysRev.75.1736. URL: https://link.aps.org/doi/10.1103/PhysRev.75.1736 (cited on pp. 35, 39).

[77] John F. Donoghue and Lorenzo Sorbo. *A Prelude to Quantum Field Theory*. Princeton University Press, Mar. 2022. ISBN: 978-0-691-22349-0 (cited on pp. 36, 44).

[78] Johannes Skaar, University of Oslo. *The S Matrix and the LSZ Reduction Formula*. Accessed: 21 March 2025. 2023. URL: `https://www.uio.no/studier/emner/matnat/fys/FYS4170/h23/lsz.pdf` (cited on p. 37).

[79] H. Lehmann, K. Symanzik, and W. Zimmermann. "Zur Formulierung quantisierter Feldtheorien". In: *Il Nuovo Cimento (1955-1965)* 1 (1955), pp. 205–225. DOI: `10.1007/BF02731765` (cited on p. 37).

[80] R. P. Feynman. "Space-Time Approach to Quantum Electrodynamics". In: *Phys. Rev.* 76 (6 Sept. 1949), pp. 769–789. DOI: `10.1103/PhysRev.76.769`. URL: `https://link.aps.org/doi/10.1103/PhysRev.76.769` (cited on pp. 39, 46).

[81] G. C. Wick. "Properties of Bethe-Salpeter Wave Functions". In: *Phys. Rev.* 96 (4 Nov. 1954), pp. 1124–1134. DOI: `10.1103/PhysRev.96.1124`. URL: `https://link.aps.org/doi/10.1103/PhysRev.96.1124` (cited on p. 47).

[82] Leonhard Euler. "De progressionibus transcendentibus seu quarum termini generales algebraice dari nequeunt". In: *Commentarii Academiae Scientiarum Petropolitanae* 5 (1738). Originally presented in 1729, pp. 36–57 (cited on pp. 48, 49).

[83] Adrien-Marie Legendre. "Exercices de Calcul Intégral". In: *Courcier, Paris* 1 (1809) (cited on p. 48).

[84] Milton Abramowitz and Irene A. Stegun. *Handbook of Mathematical Functions with Formulas, Graphs, and Mathematical Tables*. Dover Publications, 1972 (cited on pp. 48, 49).

[85] Frank W. J. Olver et al. *NIST Handbook of Mathematical Functions*. Cambridge University Press, 2010 (cited on pp. 48, 49, 139, 156).

[86] Pierre Laurent. "Mémoire sur les séries entières". In: *Journals of the École Polytechnique* 1 (1843), pp. 3–20 (cited on p. 49).

[87] George B. Arfken and Hans J. Weber. *Mathematical Methods for Physicists*. 7th. Elsevier, 2013 (cited on p. 49).

[88] Leonhard Euler. “De progressionibus harmonicis observationes”. In: *Commentarii Academiae Scientiarum Petropolitanae* 7 (1734), pp. 150–161 (cited on p. 49).

[89] Victor Ilisie. “One-Loop Two and Three-Point Functions”. In: *Concepts in Quantum Field Theory: A Practitioner’s Toolkit*. Cham: Springer International Publishing, 2016, pp. 113–140. ISBN: 978-3-319-22966-9. DOI: 10.1007/978-3-319-22966-9_8. URL: https://doi.org/10.1007/978-3-319-22966-9_8 (cited on pp. 49, 53, 65, 70, 71).

[90] Yulian V. Sokhotski. “On definite integrals and functions used in series expansions”. Russian. In: *St. Petersburg University Proceedings* 1 (1873), pp. 1–34 (cited on p. 52).

[91] Jakob Plemelj. *Riemannsche Funktionenscharen mit gegebener Monodromiegruppe*. Leipzig, Germany: B. G. Teubner, 1908 (cited on p. 52).

[92] S. Mandelstam. “Determination of the Pion-Nucleon Scattering Amplitude from Dispersion Relations and Unitarity. General Theory”. In: *Phys. Rev.* 112 (4 Nov. 1958), pp. 1344–1360. DOI: 10.1103/PhysRev.112.1344. URL: https://link.aps.org/doi/10.1103/PhysRev.112.1344 (cited on p. 55).

[93] David J. Griffiths. *Introduction to Elementary Particles*. 2nd. Wiley-VCH, 2008 (cited on p. 57).

[94] Roger A. Horn and Charles R. Johnson. *Matrix Analysis*. Cambridge University Press, 1990 (cited on p. 61).

[95] R. D. Field. *Applications of Perturbative QCD*. Vol. 77. 1989 (cited on pp. 64, 78).

[96] G. ’tHooft. “Renormalization of massless Yang-Mills fields”. In: *Nuclear Physics B* 33.1 (1971), pp. 173–199. ISSN: 0550-3213. DOI: https://doi.org/10.1016/0550-3213(71)90395-6. URL: https://www.sciencedirect.com/science/article/pii/0550321371903956 (cited on p. 73).

[97] N. N. Bogolyubov and D. V. Shirkov. *Introduction to the Theory of Quantized Fields*. Vol. 3. 1959 (cited on p. 82).

[98] J. D. Franson and Michelle M. Donegan. "Perturbation theory for quantum-mechanical observables". In: *Physical Review A* 65.5 (Apr. 2002). doi: 10.1103/physreva.65.052107. url: https://doi.org/10.1103%2Fphysreva.65.052107 (cited on pp. 82, 84).

[99] Henry Frederick Baker. In: *Proceedings of the London Mathematical Society*. 1st ser. 34 (1902), pp. 347–360 (cited on pp. 82, 84).

[100] H. Baker. In: *Proceedings of the London Mathematical Society*. 1st ser. 35 (1903), pp. 333–374 (cited on pp. 82, 84).

[101] H. Baker. In: *Proceedings of the London Mathematical Society*. 2nd ser. 3 (1905), pp. 24–47 (cited on pp. 82, 84).

[102] John Edward Campbell. In: *Proceedings of the London Mathematical Society* 28 (1897). See pp. 386–387, pp. 381–390 (cited on pp. 82, 84).

[103] J. Campbell. In: *Proceedings of the London Mathematical Society* 29 (1898), pp. 14–32 (cited on pp. 82, 84).

[104] Felix Hausdorff. "Die symbolische Exponentialformel in der Gruppentheorie". In: *Berichte über die Verhandlungen der Sächsischen Akademie der Wissenschaften zu Leipzig* 58 (1906), pp. 19–48 (cited on pp. 82, 84).

[105] R. Dickinson, J. Forshaw, and P. Millington. "Fock-space projection operators for semi-inclusive final states". In: *Phys. Lett. B* 774 (2017), pp. 706–709. doi: 10.1016/j.physletb.2017.10.037. arXiv: 1702.04131 [hep-th] (cited on pp. 82, 119, 129, 177).

[106] M. Cliche and A. Kempf. "Relativistic quantum channel of communication through field quanta". In: *Physical Review A* 81.1 (Jan. 2010). doi: 10.1103/physreva.81.012330. url: https://doi.org/10.1103%2Fphysreva.81.012330 (cited on p. 82).

[107] M. A. Nielsen and I. L. Chuang. *Quantum Computation and Quantum Information*. Cambridge University Press, 2000. isbn: 978-1-107-00217-3 (cited on p. 83).

[108] Asher Peres and Daniel R. Terno. "Quantum information and relativity theory". In: *Rev. Mod. Phys.* 76 (1 Jan. 2004), pp. 93–123. doi: 10.1103/RevModPhys.76.93. url: https://link.aps.org/doi/10.1103/RevModPhys.76.93 (cited on p. 83).

[109] R B Mann and T C Ralph. “Relativistic quantum information”. In: *Classical and Quantum Gravity* 29.22 (Nov. 2012), p. 220301. DOI: 10.1088/0264-9381/29/22/220301. URL: https://dx.doi.org/10.1088/0264-9381/29/22/220301 (cited on pp. 85, 121).

[110] Erickson Tjoa and Finnian Gray. “The Unruh–DeWitt model and its joint interacting Hilbert space”. In: *Journal of Physics A: Mathematical and Theoretical* 57.32 (Oct. 2024), p. 325301. DOI: 10.1088/1751-8121/ad6365. URL: https://dx.doi.org/10.1088/1751-8121/ad6365 (cited on pp. 85, 121).

[111] Enrico Fermi. “Quantum Theory of Radiation”. In: *Rev. Mod. Phys.* 4 (1 Jan. 1932), pp. 87–132. DOI: 10.1103/RevModPhys.4.87. URL: https://link.aps.org/doi/10.1103/RevModPhys.4.87 (cited on p. 90).

[112] M. I. Shirokov. “The Velocity of Electromagnetic Retardation in Quantum Electrodynamics”. In: *Soviet Journal of Nuclear Physics* 4 (1967), p. 774 (cited on p. 90).

[113] M.A. van Eijck and Ch.G. van Weert. “Finite-temperature retarded and advanced Green functions”. In: *Physics Letters B* 278.3 (1992), pp. 305–310. ISSN: 0370-2693. DOI: https://doi.org/10.1016/0370-2693(92)90198-D. URL: https://www.sciencedirect.com/science/article/pii/037026939290198D (cited on pp. 95, 116).

[114] R. Kobes. “Retarded functions, dispersion relations, and Cutkosky rules at zero and finite temperature”. In: *Phys. Rev. D* 43 (4 Feb. 1991), pp. 1269–1282. DOI: 10.1103/PhysRevD.43.1269. URL: https://link.aps.org/doi/10.1103/PhysRevD.43.1269 (cited on pp. 95, 116).

[115] Mark K. Transtrum and Jean-François S. Van Huele. “Commutation relations for functions of operators”. In: *Journal of Mathematical Physics* 46.6 (June 2005), p. 063510. ISSN: 0022-2488. DOI: 10.1063/1.1924703. eprint: https://pubs.aip.org/aip/jmp/article-pdf/doi/10.1063/1.1924703/14869459/063510_1_online.pdf. URL: https://doi.org/10.1063/1.1924703 (cited on p. 101).

[116] Randal L. Kobes and Gordon W. Semenoff. “Discontinuities of Green functions in field theory at finite temperature and density”. In: *Nuclear Physics B* 260.3 (1985), pp. 714–746. ISSN: 0550-3213. DOI: https://doi.org/10.

1016/0550-3213(85)90056-2. URL: https://www.sciencedirect.com/science/article/pii/0550321385900562 (cited on p. 119).

[117] Randal L. Kobes and Gordon W. Semenoff. "Discontinuities of Green functions in field theory at finite temperature and density (II)". In: *Nuclear Physics B* 272.2 (1986), pp. 329–364. ISSN: 0550-3213. DOI: https://doi.org/10.1016/0550-3213(86)90006-4. URL: https://www.sciencedirect.com/science/article/pii/0550321386900064 (cited on p. 119).

[118] B. S. DeWitt. "Quantum Field Theory in Curved Space-Time". In: *Phys. Rept.* 19 (1975), pp. 295–357. DOI: 10.1016/0370-1573(75)90051-4 (cited on pp. 120, 125, 156).

[119] G. Sewell. "Quantum fields on manifolds: PCT and gravitationally induced thermal states". In: *Annals Of Physics* 141 (1982), pp. 201–224 (cited on p. 120).

[120] B. Kay and R. Wald. "Theorems on the uniqueness and thermal properties of stationary, nonsingular, quasifree states on spacetimes with a bifurcate Killing horizon". In: *Physics Reports* 207 (1991), pp. 49–136 (cited on p. 120).

[121] J. Bekenstein. "Black Holes and Entropy". In: *Phys. Rev. D* 7 (Apr. 1973), pp. 2333–2346. DOI: 10.1103/PhysRevD.7.2333 (cited on pp. 120, 149).

[122] G. W. Gibbons and M. J. Perry. "Black Holes and Thermal Green's Functions". In: *Proc. Roy. Soc. Lond. A* 358 (1978), pp. 467–494. DOI: 10.1098/rspa.1978.0022 (cited on p. 120).

[123] G. W. Gibbons and S. W. Hawking. "Cosmological event horizons, thermodynamics, and particle creation". In: *Phys. Rev. D* 15 (1977), pp. 2738–2751. DOI: 10.1103/PhysRevD.15.2738 (cited on p. 120).

[124] W. G. Unruh. "Origin of the Particles in Black Hole Evaporation". In: *Phys. Rev. D* 15 (1977), pp. 365–369. DOI: 10.1103/PhysRevD.15.365 (cited on p. 120).

[125] S. A. Fulling. "Alternative Vacuum States in Static Space-Times with Horizons". In: *J. Phys. A* 10 (1977), pp. 917–951. DOI: 10.1088/0305-4470/10/6/014 (cited on p. 120).

[126] A. A. Starobinsky. "Amplification of waves reflected from a rotating "black hole"". In: *Sov. Phys. JETP* 37.1 (1973), pp. 28–32 (cited on p. 121).

[127] W. G. Unruh. "Second quantization in the Kerr metric". In: *Phys. Rev. D* 10 (1974), pp. 3194–3205. DOI: 10.1103/PhysRevD.10.3194 (cited on p. 121).

[128] I. Fuentes-Schuller and R. B. Mann. "Alice Falls into a Black Hole: Entanglement in Noninertial Frames". In: *Phys. Rev. Lett.* 95 (12 Sept. 2005), p. 120404. DOI: 10.1103/PhysRevLett.95.120404. URL: https://link.aps.org/doi/10.1103/PhysRevLett.95.120404 (cited on pp. 121, 175).

[129] Antony Valentini. "Non-local correlations in quantum electrodynamics". In: *Physics Letters A* 153.6 (1991), pp. 321–325. ISSN: 0375-9601. DOI: https://doi.org/10.1016/0375-9601(91)90952-5. URL: https://www.sciencedirect.com/science/article/pii/0375960191909525 (cited on pp. 121, 175).

[130] Grant Salton, Robert B Mann, and Nicolas C Menicucci. "Acceleration-assisted entanglement harvesting and rangefinding". In: *New Journal of Physics* 17.3 (Mar. 2015), p. 035001. DOI: 10.1088/1367-2630/17/3/035001. URL: https://dx.doi.org/10.1088/1367-2630/17/3/035001 (cited on pp. 121, 175).

[131] Zhihong Liu et al. "Does acceleration assist entanglement harvesting?" In: *Phys. Rev. D* 105 (8 Apr. 2022), p. 085012. DOI: 10.1103/PhysRevD.105.085012. URL: https://link.aps.org/doi/10.1103/PhysRevD.105.085012 (cited on pp. 121, 175).

[132] Kensuke Gallock-Yoshimura, Erickson Tjoa, and Robert B. Mann. "Harvesting entanglement with detectors freely falling into a black hole". In: *Phys. Rev. D* 104 (2 July 2021), p. 025001. DOI: 10.1103/PhysRevD.104.025001. URL: https://link.aps.org/doi/10.1103/PhysRevD.104.025001 (cited on pp. 121, 150, 175).

[133] Erickson Tjoa and Robert B. Mann. "Harvesting correlations in Schwarzschild and collapsing shell spacetimes". In: *Journal of High Energy Physics* 2020.8 (Aug. 2020), p. 155. ISSN: 1029-8479. DOI: 10.1007/JHEP08(2020)155. URL: https://doi.org/10.1007/JHEP08(2020)155 (cited on pp. 121, 175).

[134] C. Suryaatmadja, R. B. Mann, and W. Cong. "Entanglement harvesting of inertially moving Unruh-DeWitt detectors in Minkowski spacetime". In: *Phys.*

*Rev. D* 106 (7 Oct. 2022), p. 076002. DOI: 10.1103/PhysRevD.106.076002. URL: https://link.aps.org/doi/10.1103/PhysRevD.106.076002 (cited on pp. 121, 175).

[135] David E. Bruschi et al. "Unruh effect in quantum information beyond the single-mode approximation". In: *Phys. Rev. A* 82 (4 Oct. 2010), p. 042332. DOI: 10.1103/PhysRevA.82.042332. URL: https://link.aps.org/doi/10.1103/PhysRevA.82.042332 (cited on pp. 121, 175).

[136] Jürgen Audretsch, Michael Mensky, and Rainer Müller. "Continuous measurement and localization in the Unruh effect". In: *Phys. Rev. D* 51 (4 Feb. 1995), pp. 1716–1727. DOI: 10.1103/PhysRevD.51.1716. URL: https://link.aps.org/doi/10.1103/PhysRevD.51.1716 (cited on pp. 121, 175).

[137] Pieter Kok and Ulvi Yurtsever. "Gravitational decoherence". In: *Phys. Rev. D* 68 (8 Oct. 2003), p. 085006. DOI: 10.1103/PhysRevD.68.085006. URL: https://link.aps.org/doi/10.1103/PhysRevD.68.085006 (cited on pp. 121, 175).

[138] Eduardo Martín-Martínez, Luis J. Garay, and Juan León. "Unveiling quantum entanglement degradation near a Schwarzschild black hole". In: *Phys. Rev. D* 82 (6 Sept. 2010), p. 064006. DOI: 10.1103/PhysRevD.82.064006. URL: https://link.aps.org/doi/10.1103/PhysRevD.82.064006 (cited on pp. 121, 175).

[139] Eduardo Martín-Martínez and Juan León. "Fermionic entanglement that survives a black hole". In: *Phys. Rev. A* 80 (4 Oct. 2009), p. 042318. DOI: 10.1103/PhysRevA.80.042318. URL: https://link.aps.org/doi/10.1103/PhysRevA.80.042318 (cited on pp. 121, 175).

[140] Eduardo Martín-Martínez and Juan León. "Quantum correlations through event horizons: Fermionic versus bosonic entanglement". In: *Phys. Rev. A* 81 (3 Mar. 2010), p. 032320. DOI: 10.1103/PhysRevA.81.032320. URL: https://link.aps.org/doi/10.1103/PhysRevA.81.032320 (cited on pp. 121, 175).

[141] Paul M. Alsing, David McMahon, and G. J. Milburn. "Teleportation in a non-inertial frame". In: *J. Opt. B* 6 (2004), p. 834. DOI: 10.1088/1464-4266/6/8/033. arXiv: quant-ph/0311096 (cited on p. 121).

[142] P. M. Alsing et al. “Entanglement of Dirac fields in noninertial frames”. In: *Phys. Rev. A* 74 (3 Sept. 2006), p. 032326. DOI: 10.1103/PhysRevA.74.032326. URL: https://link.aps.org/doi/10.1103/PhysRevA.74.032326 (cited on p. 121).

[143] Paul M. Alsing and G. J. Milburn. “Teleportation with a Uniformly Accelerated Partner”. In: *Phys. Rev. Lett.* 91 (18 Oct. 2003), p. 180404. DOI: 10.1103/PhysRevLett.91.180404. URL: https://link.aps.org/doi/10.1103/PhysRevLett.91.180404 (cited on p. 121).

[144] Masahiro Hotta, Jiro Matsumoto, and Go Yusa. “Quantum energy teleportation without a limit of distance”. In: *Phys. Rev. A* 89 (1 Jan. 2014), p. 012311. DOI: 10.1103/PhysRevA.89.012311. URL: https://link.aps.org/doi/10.1103/PhysRevA.89.012311 (cited on p. 121).

[145] Eduardo Martín-Martínez and Nicolas C Menicucci. “Cosmological quantum entanglement”. In: *Classical and Quantum Gravity* 29.22 (Oct. 2012), p. 224003. DOI: 10.1088/0264-9381/29/22/224003. URL: https://dx.doi.org/10.1088/0264-9381/29/22/224003 (cited on p. 121).

[146] Aida Ahmadzadegan, Eduardo Martín-Martínez, and Robert B. Mann. “Cavities in curved spacetimes: The response of particle detectors”. In: *Phys. Rev. D* 89 (2 Jan. 2014), p. 024013. DOI: 10.1103/PhysRevD.89.024013. URL: https://link.aps.org/doi/10.1103/PhysRevD.89.024013 (cited on p. 121).

[147] Rafael D. Sorkin. “Impossible measurements on quantum fields”. In: *Directions in General Relativity: An International Symposium in Honor of the 60th Birthdays of Dieter Brill and Charles Misner*. Feb. 1993. arXiv: gr-qc/9302018 (cited on p. 121).

[148] Dionigi M T Benincasa et al. “Quantum information processing and relativistic quantum fields”. In: *Classical and Quantum Gravity* 31.7 (Mar. 2014), p. 075007. DOI: 10.1088/0264-9381/31/7/075007. URL: https://dx.doi.org/10.1088/0264-9381/31/7/075007 (cited on p. 121).

[149] Henning Bostelmann, Christopher J. Fewster, and Maximilian H. Ruep. “Impossible measurements require impossible apparatus”. In: *Phys. Rev. D* 103 (2

Jan. 2021), p. 025017. DOI: 10.1103/PhysRevD.103.025017. URL: https://link.aps.org/doi/10.1103/PhysRevD.103.025017 (cited on p. 121).

[150] Andrzej Dragan et al. "Localized projective measurement of a quantum field in non-inertial frames". In: *Classical and Quantum Gravity* 30.23 (Oct. 2013), p. 235006. DOI: 10.1088/0264-9381/30/23/235006. URL: https://dx.doi.org/10.1088/0264-9381/30/23/235006 (cited on p. 121).

[151] N. D. Birrell and P. C. W. Davies. *Quantum Fields in Curved Space*. Cambridge University Press, 1984. ISBN: 978-0-521-27858-4 (cited on pp. 122, 124, 142, 150, 153, 155, 156, 163).

[152] Charles W. Misner, Kip S. Thorne, and John A. Wheeler. *Gravitation*. W. H. Freeman and Company, 1973. ISBN: 0-7167-0344-0 (cited on pp. 122, 149, 150, 153, 168).

[153] Robert M. Wald. *General Relativity*. Chicago, USA: Chicago Univ. Pr., 1984. DOI: 10.7208/chicago/9780226870373.001.0001 (cited on pp. 122, 150).

[154] W. Rindler. "Hyperbolic Motion in Curved Space Time". In: *Phys. Rev.* 119 (6 Sept. 1960), pp. 2082–2089. DOI: 10.1103/PhysRev.119.2082. URL: https://link.aps.org/doi/10.1103/PhysRev.119.2082 (cited on p. 122).

[155] S. Takagi. "Vacuum Noise and Stress Induced by Uniform Acceleration: Hawking-Unruh Effect in Rindler Manifold of Arbitrary Dimension". In: *Prog. Theor. Phys. Suppl.* 88 (1986), pp. 1–142. DOI: 10.1143/PTP.88.1 (cited on pp. 124, 130, 131, 137, 140, 166).

[156] D.G. Boulware. "Quantum field theory in Schwarzschild and Rindler spaces". In: *Phys. Rev. D* 11 (1975), p. 1404. DOI: 10.1103/PhysRevD.11.1404 (cited on pp. 124, 160).

[157] David G. Boulware. "Spin-$\frac{1}{2}$ quantum field theory in Schwarzschild space". In: *Phys. Rev. D* 12 (2 July 1975), pp. 350–367. DOI: 10.1103/PhysRevD.12.350. URL: https://link.aps.org/doi/10.1103/PhysRevD.12.350 (cited on p. 124).

[158] P. G. Grove and A. C. Ottewill. "Notes on Particle Detectors". In: *J. Phys. A* 16 (1983), pp. 3905–3920. DOI: 10.1088/0305-4470/16/16/029 (cited on pp. 126, 131).

[159] J. Louko and A. Satz. “How often does the Unruh-DeWitt detector click? Regularisation by a spatial profile”. In: *Class. Quant. Grav.* 23 (2006), pp. 6321–6344. DOI: 10.1088/0264-9381/23/22/015. arXiv: gr-qc/0606067 [gr-qc] (cited on pp. 126, 130, 131).

[160] Alejandro Satz. “Then again, how often does the Unruh–DeWitt detector click if we switch it carefully?” In: *Classical and Quantum Gravity* 24.7 (Mar. 2007), p. 1719. DOI: 10.1088/0264-9381/24/7/003. URL: https://dx.doi.org/10.1088/0264-9381/24/7/003 (cited on pp. 126, 130, 164, 171).

[161] B. S. DeWitt. “Quantum Gravity: The New Synthesis”. In: *General Relativity*. Ed. by S. W. Hawking and W. Israel. Cambridge: Cambridge University Press, 1979, pp. 680–743 (cited on pp. 130, 131).

[162] J. Louko and A. Satz. “Transition rate of the Unruh-DeWitt detector in curved spacetime”. In: *Class. Quant. Grav.* 25 (2008), p. 055012. DOI: 10.1088/0264-9381/25/5/055012. arXiv: 0710.5671 [gr-qc] (cited on pp. 130, 131, 164, 171).

[163] S. Schlicht. “Considerations on the Unruh effect: Causality and regularization”. In: *Class. Quant. Grav.* 21 (2004), pp. 4647–4660. DOI: 10.1088/0264-9381/21/19/011. arXiv: gr-qc/0306022 [gr-qc] (cited on pp. 131, 164, 171).

[164] P. M. Alsing and P. W. Milonni. “Simplified derivation of the Hawking-Unruh temperature for an accelerated observer in vacuum”. In: *Am. J. Phys.* 72 (2004), pp. 1524–1529. DOI: 10.1119/1.1761064. arXiv: quant-ph/0401170 [quant-ph] (cited on p. 135).

[165] I. S. Gradshteyn and I. M. Ryzhik. *Table of Integrals, Series, and Products (Eighth Edition)*. Academic Press, 2014. Chap. 6-7: Definite Integrals of Special Functions, p. 692. ISBN: 9780123849335 (cited on p. 138).

[166] Peter Millington. “Thermal quantum field theory and perturbative non-equilibrium dynamics”. PhD thesis. New York: Manchester U., 2012. DOI: 10.1007/978-3-319-01186-8 (cited on p. 140).

[167] W. Unruh and R. Wald. “What happens when an accelerating observer detects a Rindler particle”. In: *Phys. Rev. D* 29 (1984), pp. 1047–1056. DOI: 10.1103/PhysRevD.29.1047 (cited on p. 141).

[168] S. A. Fulling and G. E. A. Matsas. “Unruh effect”. In: *Scholarpedia* 9.10 (2014). revision #143950, p. 31789. DOI: 10.4249/scholarpedia.31789 (cited on p. 141).

[169] Robert Dickinson. *Private communication*. Private communication, February 2025. 2025 (cited on p. 144).

[170] K. Hinton, P. C. W. Davies, and J. Pfautsch. “ACCELERATED OBSERVERS DO NOT DETECT ISOTROPIC THERMAL RADIATION”. In: *Phys. Lett. B* 120 (1983), pp. 88–90. DOI: 10.1016/0370-2693(83)90629-9 (cited on p. 146).

[171] E. E. Kholupenko. “On the Possible Anisotropy of the Unruh Radiation. Part I: Massless Scalar Field in (1+1)D Space-Time”. In: *Grav. Cosmol.* 25.3 (2019), pp. 213–225. DOI: 10.1134/S0202289319030071 (cited on p. 146).

[172] E. E. Kholupenko. “On the Possible Anisotropy of the Unruh Radiation. Part II: Massive Scalar Field in (3+1) D Space-Time”. In: *Grav. Cosmol.* 28.2 (2022), pp. 139–152. DOI: 10.1134/S0202289322020074 (cited on p. 146).

[173] S. Takagi. “A DIRECTIONAL RINDLER PARTICLE DETECTOR”. In: *Phys. Lett. B* 148 (1984), pp. 116–118. DOI: 10.1016/0370-2693(84)91621-6 (cited on p. 146).

[174] W. Israel and J.M. Nester. “Is acceleration radiation isotropic?” In: *Physics Letters A* 98.7 (1983), pp. 329–331. ISSN: 0375-9601. DOI: https://doi.org/10.1016/0375-9601(83)90228-1. URL: https://www.sciencedirect.com/science/article/pii/0375960183902281 (cited on p. 146).

[175] Ulrich H. Gerlach. “Absolute nature of the thermal ambience of accelerated observers”. In: *Phys. Rev. D* 27 (10 May 1983), pp. 2310–2315. DOI: 10.1103/PhysRevD.27.2310. URL: https://link.aps.org/doi/10.1103/PhysRevD.27.2310 (cited on p. 146).

[176] P G Grove and A C Ottewill. “Is acceleration radiation isotropic?” In: *Classical and Quantum Gravity* 2.3 (May 1985), p. 373. DOI: 10.1088/0264-9381/2/3/013. URL: https://dx.doi.org/10.1088/0264-9381/2/3/013 (cited on p. 146).

[177] Sanved Kolekar. "Directional dependence of the Unruh effect for spatially extended detectors". In: *Phys. Rev. D* 101 (2 Jan. 2020), p. 025002. DOI: 10.1103/PhysRevD.101.025002. URL: https://link.aps.org/doi/10.1103/PhysRevD.101.025002 (cited on p. 146).

[178] J. B. Hartle. *Gravity: An introduction to Einstein's general relativity*. 2003. ISBN: 978-0-8053-8662-2 (cited on pp. 149, 150).

[179] Sean M. Carroll. *Spacetime and Geometry: An Introduction to General Relativity*. Cambridge University Press, July 2019. ISBN: 978-0-8053-8732-2. DOI: 10.1017/9781108770385 (cited on pp. 149, 153, 162).

[180] Jacob D. Bekenstein. "Generalized second law of thermodynamics in black-hole physics". In: *Phys. Rev. D* 9 (12 June 1974), pp. 3292–3300. DOI: 10.1103/PhysRevD.9.3292. URL: https://link.aps.org/doi/10.1103/PhysRevD.9.3292 (cited on p. 149).

[181] Ted Jacobson. "Thermodynamics of Spacetime: The Einstein Equation of State". In: *Physical Review Letters* 75 (1995), pp. 1260–1263. DOI: 10.1103/PhysRevLett.75.1260 (cited on p. 149).

[182] Erik Verlinde. "On the Origin of Gravity and the Laws of Newton". In: *Journal of High Energy Physics* 2011.4 (2011), p. 29. DOI: 10.1007/JHEP04(2011)029 (cited on p. 149).

[183] Thanu Padmanabhan. "Thermodynamical Aspects of Gravity: New Insights". In: *Reports on Progress in Physics* 73.4 (2010), p. 046901. DOI: 10.1088/0034-4885/73/4/046901 (cited on p. 149).

[184] Thanu Padmanabhan. "Emergent Gravity Paradigm: Recent Progress". In: *Modern Physics Letters A* 29.37 (2014), p. 1430010. DOI: 10.1142/S0217732314300101 (cited on p. 149).

[185] Mark Van Raamsdonk. "Building Up Spacetime with Quantum Entanglement". In: *General Relativity and Gravitation* 42 (2010), pp. 2323–2329. DOI: 10.1142/S0218271810018529 (cited on p. 150).

[186] Shinsei Ryu and Tadashi Takayanagi. "Holographic derivation of entanglement entropy from AdS/CFT". In: *Physical Review Letters* 96.18 (2006), p. 181602. DOI: 10.1103/PhysRevLett.96.181602 (cited on p. 150).

[187] Juan Maldacena and Leonard Susskind. "Cool horizons for entangled black holes". In: *Fortschritte der Physik* 61 (2013), pp. 781–811. DOI: 10.1002/prop.201300020 (cited on p. 150).

[188] Brian Swingle. "Entanglement Renormalization and Holography". In: *Physical Review D* 86 (2012), p. 065007. DOI: 10.1103/PhysRevD.86.065007 (cited on p. 150).

[189] Leonard Susskind. "The World as a Hologram". In: *Journal of Mathematical Physics* 36 (1995), pp. 6377–6396. DOI: 10.1063/1.531249 (cited on p. 150).

[190] Juan Maldacena. "The Large N Limit of Superconformal Field Theories and Supergravity". In: *International Journal of Theoretical Physics* 38 (1999), pp. 1113–1133. DOI: 10.1023/A:1026654312961 (cited on p. 150).

[191] Edward Witten. "Anti-de Sitter Space and Holography". In: *Advances in Theoretical and Mathematical Physics* 2 (1998), pp. 253–291. arXiv: hep-th/9802150 (cited on p. 150).

[192] Massimiliano Spadafora et al. "Deep in the knotted black hole". In: *arXiv preprint* (2025). Citing arXiv version 1 (v1) specifically. arXiv: 2412.02755v1 [gr-qc]. URL: https://arxiv.org/abs/2412.02755v1 (cited on pp. 150, 172).

[193] María R. Preciado-Rivas et al. "More excitement across the horizon". In: *Phys. Rev. D* 110 (2 July 2024), p. 025002. DOI: 10.1103/PhysRevD.110.025002. URL: https://link.aps.org/doi/10.1103/PhysRevD.110.025002 (cited on p. 150).

[194] Sijia Wang et al. "Singular excitement beyond the horizon of a rotating black hole". In: *Phys. Rev. D* 110.6 (2024), p. 065013. DOI: 10.1103/PhysRevD.110.065013. arXiv: 2407.01673 [gr-qc] (cited on p. 150).

[195] Lee Hodgkinson and Jorma Louko. "Static, stationary, and inertial Unruh-DeWitt detectors on the BTZ black hole". In: *Phys. Rev. D* 86 (6 Sept. 2012), p. 064031. DOI: 10.1103/PhysRevD.86.064031. URL: https://link.aps.org/doi/10.1103/PhysRevD.86.064031 (cited on p. 150).

[196] Ireneo James Membrere et al. "Tripartite Entanglement Extraction from the Black Hole Vacuum". In: *Advanced Quantum Technologies* 6.9 (2023), p. 2300125. DOI: https://doi.org/10.1002/qute.202300125. eprint: https://advanced.onlinelibrary.wiley.com/doi/pdf/10.1002/qute.202300125. URL: https://advanced.onlinelibrary.wiley.com/doi/abs/10.1002/qute.202300125 (cited on p. 150).

[197] Benito A Juárez-Aubry and Jorma Louko. "Onset and decay of the 1 + 1 Hawking–Unruh effect: what the derivative-coupling detector saw". In: *Classical and Quantum Gravity* 31.24 (Nov. 2014), p. 245007. DOI: 10.1088/0264-9381/31/24/245007. URL: https://dx.doi.org/10.1088/0264-9381/31/24/245007 (cited on p. 150).

[198] Benito A. Juárez-Aubry. "Asymptotics in the time-dependent Hawking and Unruh effects". PhD thesis. Nottingham U., 2016. arXiv: 1708.09430 [gr-qc] (cited on p. 150).

[199] Benito A. Juárez-Aubry and Jorma Louko. "Quantum kicks near a Cauchy horizon". In: *AVS Quantum Science* 4.1 (Feb. 2022), p. 013201. ISSN: 2639-0213. DOI: 10.1116/5.0073373. eprint: https://pubs.aip.org/avs/aqs/article-pdf/doi/10.1116/5.0073373/16493576/013201_1_online.pdf. URL: https://doi.org/10.1116/5.0073373 (cited on p. 150).

[200] G. W. Gibbons and S. W. Hawking. "Cosmological event horizons, thermodynamics, and particle creation". In: *Phys. Rev. D* 15 (10 May 1977), pp. 2738–2751. DOI: 10.1103/PhysRevD.15.2738. URL: https://link.aps.org/doi/10.1103/PhysRevD.15.2738 (cited on p. 150).

[201] Laura Niermann and Luis C. Barbado. "Particle detectors in superposition in de Sitter spacetime". In: (Mar. 2024). arXiv: 2403.02087 [gr-qc] (cited on p. 150).

[202] Sourav Bhattacharya and Shagun Kaushal. "Entanglement generation between two comoving Unruh-DeWitt detectors in the cosmological de Sitter spacetime". In: (Apr. 2024). arXiv: 2404.11931 [gr-qc] (cited on pp. 150, 175).

[203] Shahnewaz Ahmed, Mir Mehedi Faruk, and Muktadir Rahman. "Accelerated paths and Unruh effect: finite time detector response in (anti) de Sitter space-

time and Huygen’s principle”. In: *The European Physical Journal C* 83.11 (Nov. 2023), p. 1087. ISSN: 1434-6052. DOI: 10.1140/epjc/s10052-023-12245-9. URL: https://doi.org/10.1140/epjc/s10052-023-12245-9 (cited on p. 150).

[204] Germain Rousseaux et al. “Horizon effects with surface waves on moving water”. In: *New J. Phys.* 12 (2010), p. 095018. DOI: 10.1088/1367-2630/12/9/095018. arXiv: 1004.5546 [gr-qc] (cited on p. 150).

[205] Chia-Shun Yih. “Surface waves in flowing water”. In: *Journal of Fluid Mechanics* 51.2 (1972), pp. 209–220. DOI: 10.1017/S002211207200117X (cited on p. 150).

[206] Andrew J. S. Hamilton and Jason P. Lisle. “The River model of black holes”. In: *Am. J. Phys.* 76 (2008), pp. 519–532. DOI: 10.1119/1.2830526. arXiv: gr-qc/0411060 (cited on p. 150).

[207] T. G. Philbin. “An exact solution for the Hawking effect in a dispersive fluid”. In: *Phys. Rev. D* 94.6 (2016), p. 064053. DOI: 10.1103/PhysRevD.94.064053. arXiv: 1607.03743 [gr-qc] (cited on p. 150).

[208] Silke Weinfurtner et al. “Measurement of Stimulated Hawking Emission in an Analogue System”. In: *Phys. Rev. Lett.* 106 (2 Jan. 2011), p. 021302. DOI: 10.1103/PhysRevLett.106.021302. URL: https://link.aps.org/doi/10.1103/PhysRevLett.106.021302 (cited on p. 150).

[209] Jeff Steinhauer. “Observation of quantum Hawking radiation and its entanglement in an analogue black hole”. In: *Nature Physics* 12.10 (Oct. 2016), pp. 959–965. ISSN: 1745-2481. DOI: 10.1038/nphys3863. URL: https://doi.org/10.1038/nphys3863 (cited on p. 150).

[210] Carlos Barceló, Stefano Liberati, and Matt Visser. “Analogue Gravity”. In: *Living Reviews in Relativity* 14.1 (May 2011), p. 3. ISSN: 1433-8351. DOI: 10.12942/lrr-2011-3. URL: https://doi.org/10.12942/lrr-2011-3 (cited on p. 150).

[211] Matt Visser. “Acoustic black holes: horizons, ergospheres and Hawking radiation”. In: *Classical and Quantum Gravity* 15.6 (June 1998), p. 1767. DOI: 10.1088/0264-9381/15/6/024. URL: https://dx.doi.org/10.1088/0264-9381/15/6/024 (cited on p. 150).

[212] W. G. Unruh. “Experimental Black-Hole Evaporation?” In: *Phys. Rev. Lett.* 46 (21 May 1981), pp. 1351–1353. DOI: 10.1103/PhysRevLett.46.1351. URL: https://link.aps.org/doi/10.1103/PhysRevLett.46.1351 (cited on p. 150).

[213] W. G. Unruh. “Dumb holes: Analogues for black holes”. In: *Phil. Trans. Roy. Soc. Lond. A* 366 (2008), pp. 2905–2913. DOI: 10.1098/rsta.2008.0062 (cited on p. 150).

[214] Christopher J. Shallue and Sean M. Carroll. *What Hawking Radiation Looks Like as You Fall into a Black Hole*. 2025. arXiv: 2501.06609 [gr-qc]. URL: https://arxiv.org/abs/2501.06609 (cited on pp. 150, 156, 164, 171, 172, 174, 175, 177).

[215] S Massar, R Parentani, and R Brout. “Energy-momentum tensor of the evaporating black hole and local Bogoljubov transformations”. In: *Classical and Quantum Gravity* 10.11 (Nov. 1993), p. 2431. DOI: 10.1088/0264-9381/10/11/025. URL: https://dx.doi.org/10.1088/0264-9381/10/11/025 (cited on p. 150).

[216] Yongwan Gim and Wontae Kim. “A quantal Tolman temperature”. In: *The European Physical Journal C* 75.11 (Nov. 2015), p. 549. ISSN: 1434-6052. DOI: 10.1140/epjc/s10052-015-3765-2. URL: https://doi.org/10.1140/epjc/s10052-015-3765-2 (cited on p. 150).

[217] Myungseok Eune, Yongwan Gim, and Wontae Kim. “Effective Tolman temperature induced by trace anomaly”. In: *The European Physical Journal C* 77.4 (Apr. 2017), p. 244. ISSN: 1434-6052. DOI: 10.1140/epjc/s10052-017-4812-y. URL: https://doi.org/10.1140/epjc/s10052-017-4812-y (cited on p. 150).

[218] Wontae Kim. “The effective Tolman temperature in curved spacetimes”. In: *International Journal of Modern Physics D* 26.14 (2017), p. 1730025. DOI: 10.1142/S0218271817300257. eprint: https://doi.org/10.1142/S0218271817300257. URL: https://doi.org/10.1142/S0218271817300257 (cited on p. 150).

[219] Jessica Santiago and Matt Visser. “Tolman-like temperature gradients in stationary spacetimes”. In: *Phys. Rev. D* 98 (6 Sept. 2018), p. 064001. DOI: 10.

1103/PhysRevD.98.064001. URL: https://link.aps.org/doi/10.1103/PhysRevD.98.064001 (cited on p. 150).

[220] Erling J. Brynjolfsson and Larus Thorlacius. “Taking the temperature of a black hole”. In: *Journal of High Energy Physics* 2008.09 (Sept. 2008), p. 066. DOI: 10.1088/1126-6708/2008/09/066. URL: https://dx.doi.org/10.1088/1126-6708/2008/09/066 (cited on p. 150).

[221] R. Penrose. “Conformal treatment of infinity”. In: (1964). Ed. by C. DeWitt and B. DeWitt, pp. 565–586. DOI: 10.1007/s10714-010-1110-5 (cited on p. 150).

[222] Eric Poisson. *A Relativist’s Toolkit: The Mathematics of Black-Hole Mechanics*. Cambridge University Press, Dec. 2009. DOI: 10.1017/CBO9780511606601 (cited on p. 153).

[223] Karl Schwarzschild. “Über das Gravitationsfeld eines Massenpunktes nach der Einsteinschen Theorie”. In: *Sitzungsberichte der Königlich Preußischen Akademie der Wissenschaften (Berlin)* (1916), pp. 189–196 (cited on p. 154).

[224] M. D. Kruskal. “Maximal Extension of Schwarzschild Metric”. In: *Phys. Rev.* 119 (5 Sept. 1960), pp. 1743–1745. DOI: 10.1103/PhysRev.119.1743. URL: https://link.aps.org/doi/10.1103/PhysRev.119.1743 (cited on p. 154).

[225] Lee Hodgkinson. “Particle detectors in curved spacetime quantum field theory”. PhD thesis. Nottingham U., 2013. arXiv: 1309.7281 [gr-qc] (cited on pp. 156, 157, 165, 171).

[226] E. W. Leaver. “Solutions to a generalized spheroidal wave equation: Teukolsky’s equations in general relativity, and the two-center problem in molecular quantum mechanics”. In: *Journal of Mathematical Physics* 27.5 (May 1986), pp. 1238–1265. ISSN: 0022-2488. DOI: 10.1063/1.527130. eprint: https://pubs.aip.org/aip/jmp/article-pdf/27/5/1238/19245696/1238_1_online.pdf. URL: https://doi.org/10.1063/1.527130 (cited on p. 157).

[227] Lee Hodgkinson, Jorma Louko, and Adrian C. Ottewill. “Static detectors and circular-geodesic detectors on the Schwarzschild black hole”. In: *Phys. Rev. D* 89 (10 May 2014), p. 104002. DOI: 10.1103/PhysRevD.89.104002. URL:

https://link.aps.org/doi/10.1103/PhysRevD.89.104002 (cited on pp. 159, 164, 165, 167–170, 175, 177).

[228] Douglas Singleton and Steve Wilburn. "Hawking Radiation, Unruh Radiation, and the Equivalence Principle". In: *Phys. Rev. Lett.* 107 (8 Aug. 2011), p. 081102. DOI: 10.1103/PhysRevLett.107.081102. URL: https://link.aps.org/doi/10.1103/PhysRevLett.107.081102 (cited on p. 161).

[229] J.B. Hartle and S.W. Hawking. "Path integral derivation of black hole radiance". In: *Phys. Rev. D* 13 (1976), p. 2188. DOI: 10.1103/PhysRevD.13.2188 (cited on p. 161).

[230] Alessandro Fabbri and Jose Navarro-Salas. *Modeling Black Hole Evaporation*. Jan. 2005, p. 335. ISBN: 978-1-86094-527-4. DOI: 10.1142/P378 (cited on p. 162).

[231] Adrian Ottewill and Shin Takagi. "Particle Detector Response for Thermal States in Static Space-Times". In: *Progress of Theoretical Physics* 77.2 (Feb. 1987), pp. 310–321. ISSN: 0033-068X. DOI: 10.1143/PTP.77.310. eprint: https://academic.oup.com/ptp/article-pdf/77/2/310/5359234/77-2-310.pdf. URL: https://doi.org/10.1143/PTP.77.310 (cited on p. 166).

[232] Ludwig Boltzmann. *Lectures on Gas Theory*. Berkeley, CA, USA: University of California Press, 1964 (cited on p. 166).

[233] Richard C. Tolman. *The Principles of Statistical Mechanics*. London, UK: Oxford University Press, 1938 (cited on p. 166).

[234] James Clerk Maxwell. "On the Dynamical Theory of Gases". In: *Philosophical Transactions of the Royal Society of London* 157 (1867), pp. 49–88 (cited on p. 166).

[235] Lars Onsager. "Reciprocal Relations in Irreversible Processes. I". In: *Physical Review* 37 (1931), pp. 405–426 (cited on p. 166).

[236] Lars Onsager. "Reciprocal Relations in Irreversible Processes. II". In: *Physical Review* 38 (1931), pp. 2265–2279 (cited on p. 166).

[237] Ryogo Kubo. "Statistical-Mechanical Theory of Irreversible Processes. I. General Theory and Simple Applications to Magnetic and Conduction Problems". In: *Journal of the Physical Society of Japan* 12.6 (1957), pp. 570–586. DOI: 10.1143/JPSJ.12.570. eprint: https://doi.org/10.1143/JPSJ.12.570. URL: https://doi.org/10.1143/JPSJ.12.570 (cited on p. 166).

[238] Paul C. Martin and Julian Schwinger. "Theory of Many-Particle Systems. I". In: *Phys. Rev.* 115 (6 Sept. 1959), pp. 1342–1373. DOI: 10.1103/PhysRev.115.1342. URL: https://link.aps.org/doi/10.1103/PhysRev.115.1342 (cited on p. 166).

[239] R. Haag, N. M. Hugenholtz, and M. Winnink. "On the equilibrium states in quantum statistical mechanics". In: *Communications in Mathematical Physics* 5.3 (June 1967), pp. 215–236. ISSN: 1432-0916. DOI: 10.1007/BF01646342. URL: https://doi.org/10.1007/BF01646342 (cited on p. 166).

[240] Sean M. Carroll. "Lecture notes on general relativity". In: (Dec. 1997). arXiv: gr-qc/9712019 (cited on p. 168).

[241] Zeyu Fan and Edward Teo. "Fundamental photon orbits in the double Schwarzschild space-time". In: *Eur. Phys. J. C* 82.11 (2022), p. 1016. DOI: 10.1140/epjc/s10052-022-10996-5 (cited on p. 168).

[242] David D. Nolte. "Inner-Most Stable Circular Orbit". In: *Galileo Unbound* (Aug. 2021). URL: https://galileo-unbound.blog/2021/08/28/inner-most-stable-circular-orbit/ (cited on p. 168).

[243] H. Kautzleben. "S. L. Belousov: Tables of Normalized Associated Legendre Polynomials. Mathematical Tables Series, Vol. 18. Pergamon Press, Oxford/London/New York/Paris, 1962. 379 S. Preis £ 7. - net". In: *Astronomische Nachrichten* 287.5-6 (1963), pp. 277–278. DOI: https://doi.org/10.1002/asna.19632870512. eprint: https://onlinelibrary.wiley.com/doi/pdf/10.1002/asna.19632870512. URL: https://onlinelibrary.wiley.com/doi/abs/10.1002/asna.19632870512 (cited on p. 168).

[244] Matteo Smerlak and Suprit Singh. "New perspectives on Hawking radiation". In: *Phys. Rev. D* 88 (10 Nov. 2013), p. 104023. DOI: 10.1103/PhysRevD.88.104023. URL: https://link.aps.org/doi/10.1103/PhysRevD.88.104023 (cited on p. 169).

[245] Shohreh Abdolrahimi. “Velocity effects on an accelerated Unruh–DeWitt detector”. In: *Classical and Quantum Gravity* 31.13 (June 2014), p. 135009. DOI: 10.1088/0264-9381/31/13/135009. URL: https://dx.doi.org/10.1088/0264-9381/31/13/135009 (cited on p. 169).

[246] Keith K. Ng et al. “A little excitement across the horizon”. In: *New J. Phys.* 24.10 (2022), p. 103018. DOI: 10.1088/1367-2630/ac9547. arXiv: 2109.13260 [gr-qc] (cited on pp. 171–174).

[247] P. C. W. Davies and Adrian C. Ottewill. “Detection of negative energy: 4-dimensional examples”. In: *Phys. Rev. D* 65 (10 May 2002), p. 104014. DOI: 10.1103/PhysRevD.65.104014. URL: https://link.aps.org/doi/10.1103/PhysRevD.65.104014 (cited on p. 175).

[248] Subhajit Barman, Dipankar Barman, and Bibhas Ranjan Majhi. “Entanglement harvesting from conformal vacuums between two Unruh-DeWitt detectors moving along null paths”. In: *JHEP* 09 (2022), p. 106. DOI: 10.1007/JHEP09(2022)106. arXiv: 2112.01308 [gr-qc] (cited on p. 175).

[249] Dipankar Barman, Subhajit Barman, and Bibhas Ranjan Majhi. “Role of thermal field in entanglement harvesting between two accelerated Unruh-DeWitt detectors”. In: *Journal of High Energy Physics* 2021.7 (July 2021), p. 124. ISSN: 1029-8479. DOI: 10.1007/JHEP07(2021)124. URL: https://doi.org/10.1007/JHEP07(2021)124 (cited on p. 175).

[250] Kensuke Gallock-Yoshimura and Robert B. Mann. “Entangled detectors nonperturbatively harvest mutual information”. In: *Phys. Rev. D* 104.12 (2021), p. 125017. DOI: 10.1103/PhysRevD.104.125017. arXiv: 2109.07495 [quant-ph] (cited on p. 175).

# Appendices

# Appendix A

# First-Order Decay Expressions

Eqs. (A.0.1)–(A.0.12) display the full expressions of each term in Eq. (5.4.6).

$$
\begin{aligned}
\langle E^{h}_{\underline{2}3}\, E^{\chi}_{14}\, \mathcal{E}^{\chi hh\chi}_{\underline{1}2\underline{3}\underline{4}} \rangle = {} & 2\,\Delta^{h}_{23}\,\frac{1}{2\omega_p}\big(-e^{ip\cdot x_1}\,e^{-ip\cdot x_4} - \,e^{ip\cdot x_4}\,e^{-ip\cdot x_1}\big) \times 16\, g_h^2 g_\chi^2\, \Delta^{\phi}_{12} \\
& \langle 0^{\phi}| \Bigg( 2\ \phi_1\,\phi_2\,\phi_3^2\,\phi_4^2 - \,\Delta^{\phi}_{12}\,\phi_3^2\,\phi_4^2 \\
& - 2\,\Delta^{\phi}_{13}\,\phi_2\,\phi_3\,\phi_4^2 - 2\,\Delta^{\phi}_{23}\,\phi_1\,\phi_3\,\phi_4^2 + 2\,\Delta^{\phi}_{13}\,\Delta^{\phi}_{23}\,\phi_4^2 \\
& - 2\,\Delta^{\phi}_{14}\Big[\,\phi_2\,\phi_3^2\,\phi_4 - \Delta^{\phi}_{23}\,\phi_3\,\phi_4\Big] - 2\,\Delta^{\phi}_{24}\Big[\,\phi_1\,\phi_3^2\,\phi_4 - \Delta^{\phi}_{13}\ \phi_3\,\phi_4\Big] \\
& - 2\,\Delta^{\phi}_{34}\Big[2\ \phi_1\,\phi_2\,\phi_3\,\phi_4 - \Delta^{\phi}_{12}\,\phi_3\,\phi_4 - \Delta^{\phi}_{13}\,\phi_2\,\phi_4 - \Delta^{\phi}_{23}\,\phi_1\,\phi_4\Big] \\
& + 2\,\Delta^{\phi}_{14}\,\Delta^{\phi}_{24}\,\phi_3^2 + 2\,\Delta^{\phi}_{14}\,\Delta^{\phi}_{34}\Big[2\ \phi_2\,\phi_3 - \,\Delta^{\phi}_{23}\Big] \\
& + 2\,\Delta^{\phi}_{24}\,\Delta^{\phi}_{34}\Big[2\ \phi_1\,\phi_3 - \,\Delta^{\phi}_{13}\Big] + (\Delta^{\phi}_{34})^2\Big[2\ \phi_1\,\phi_2 - \Delta^{\phi}_{12}\Big]\Bigg)\,|0^{\phi}\rangle
\end{aligned}
\tag{A.0.1}
$$

$$
\begin{aligned}
\langle E^{h}_{\underline{2}3}\, E^{\chi}_{1\underline{4}}\, \mathcal{E}^{\chi hh\chi}_{\underline{1}2\underline{3}4} \rangle = {} & 2\,\Delta^{h}_{23}\,\frac{1}{2\omega_p}\big(-e^{ip\cdot x_1}\,e^{-ip\cdot x_4} + \,e^{ip\cdot x_4}\,e^{-ip\cdot x_1}\big) \times 16\, g_h^2 g_\chi^2\, \Delta^{\phi}_{12} \\
& \langle 0^{\phi}| \Bigg( 2\,\Delta^{\phi}_{14}\Big[\,\phi_2\,\phi_3^2\,\phi_4 - \Delta^{\phi}_{23}\ \phi_3\,\phi_4\Big] + 2\,\Delta^{\phi}_{24}\Big[\,\phi_1\,\phi_3^2\,\phi_4 - \Delta^{\phi}_{13}\ \phi_3\,\phi_4\Big] \\
& + 2\,\Delta^{\phi}_{34}\Big[2\ \phi_1\,\phi_2\,\phi_3\,\phi_4 - \Delta^{\phi}_{12}\,\phi_3\,\phi_4 - \Delta^{\phi}_{13}\,\phi_2\,\phi_4 - \Delta^{\phi}_{23}\,\phi_1\,\phi_4\Big] \\
& - 2\,\Delta^{\phi}_{14}\,\Delta^{\phi}_{24}\ \phi_3^2 - 2\,\Delta^{\phi}_{14}\,\Delta^{\phi}_{34}\Big[2\ \phi_2\,\phi_3 - \Delta^{\phi}_{23}\ \Big] \\
& - 2\,\Delta^{\phi}_{24}\,\Delta^{\phi}_{34}\Big[2\ \phi_1\,\phi_3 - \Delta^{\phi}_{13}\ \Big] - (\Delta^{\phi}_{34})^2\Big[2\ \phi_1\,\phi_2 - \Delta^{\phi}_{12}\Big]\Bigg)\,|0^{\phi}\rangle
\end{aligned}
\tag{A.0.2}
$$

$$
\begin{aligned}
\langle E^{h}_{\underline{2}\underline{3}}\, E^{\chi}_{14}\, \mathcal{E}^{\chi hh\chi}_{\underline{1}23\underline{4}} \rangle = {}& 2\, \Delta^{h(H)}_{23}\, \frac{1}{2\omega_p} \big( -e^{ip\cdot x_1}\, e^{-ip\cdot x_4} -\, e^{ip\cdot x_4}\, e^{-ip\cdot x_1} \big) \times 32\, g_h^2 g_\chi^2\, \Delta^{\phi}_{12} \\
& \langle 0^{\phi} | \Bigg( \Delta^{\phi}_{13} \phi_2\, \phi_3\, \phi_4^2 + \Delta^{\phi}_{23}\, \phi_1\, \phi_3\, \phi_4^2 - \Delta^{\phi}_{13}\, \Delta^{\phi}_{23}\, \phi_4^2 \\
& \quad - \Delta^{\phi}_{14} \Delta^{\phi}_{23}\, \phi_3\, \phi_4 - \Delta^{\phi}_{24} \Delta^{\phi}_{13}\, \phi_3\, \phi_4 \\
& \quad - \Delta^{\phi}_{34} \Big[ \Delta^{\phi}_{13}\, \phi_2\, \phi_4 + \Delta^{\phi}_{23}\, \phi_1\, \phi_4 - \Delta^{\phi}_{14}\, \Delta^{\phi}_{23} - \Delta^{\phi}_{24}\, \Delta^{\phi}_{13} \Big] \Bigg) |0^{\phi}\rangle
\end{aligned}
\tag{A.0.3}
$$

$$
\begin{aligned}
\langle E^{h}_{\underline{2}\underline{3}}\, E^{\chi}_{1\underline{4}}\, \mathcal{E}^{\chi hh\chi}_{\underline{1}234} \rangle = {}& 2\, \Delta^{h(H)}_{23}\, \frac{1}{2\omega_p} \big( -e^{ip\cdot x_1}\, e^{-ip\cdot x_4} +\, e^{ip\cdot x_4}\, e^{-ip\cdot x_1} \big) \times 32\, g_h^2 g_\chi^2\, \Delta^{\phi}_{12} \\
& \langle 0^{\phi} | \Bigg( \Delta^{\phi}_{14} \Delta^{\phi}_{23}\, \phi_3\, \phi_4 + \Delta^{\phi}_{24} \Delta^{\phi}_{13}\, \phi_3\, \phi_4 \\
& \quad + \Delta^{\phi}_{34} \Big[ \Delta^{\phi}_{13}\, \phi_2\, \phi_4 + \Delta^{\phi}_{23}\, \phi_1\, \phi_4 - \Delta^{\phi}_{14}\, \Delta^{\phi}_{23} - \Delta^{\phi}_{24}\, \Delta^{\phi}_{13} \Big] \Bigg) |0^{\phi}\rangle
\end{aligned}
\tag{A.0.4}
$$

$$
\begin{aligned}
\langle E^{h}_{\underline{2}4}\, E^{\chi}_{13}\, \mathcal{E}^{\chi h\chi h}_{\underline{1}2\underline{3}\underline{4}} \rangle = {}& 2\, \Delta^{h}_{24}\, \frac{1}{2\omega_p} \big( -e^{ip\cdot x_1}\, e^{-ip\cdot x_3} -\, e^{ip\cdot x_3}\, e^{-ip\cdot x_1} \big) \times 16\, g_h^2 g_\chi^2\, \Delta^{\phi}_{12} \\
& \langle 0^{\phi} | \Bigg( 2\ \phi_1\, \phi_2\, \phi_3^2\, \phi_4^2 -\ \Delta^{\phi}_{12}\, \phi_3^2\, \phi_4^2 \\
& - 2\, \Delta^{\phi}_{13}\, \phi_2\, \phi_3\, \phi_4^2 - 2\, \Delta^{\phi}_{23}\, \phi_1\, \phi_3\, \phi_4^2 + 2\, \Delta^{\phi}_{13}\, \Delta^{\phi}_{23}\, \phi_4^2 \\
& - 2\, \Delta^{\phi}_{14} \Big[ \phi_2\, \phi_3^2\, \phi_4 - \Delta^{\phi}_{23}\, \phi_3\, \phi_4 \Big] - 2\, \Delta^{\phi}_{24} \Big[ \phi_1\, \phi_3^2\, \phi_4 - \Delta^{\phi}_{13}\ \phi_3\, \phi_4 \Big] \\
& - 2\, \Delta^{\phi}_{34} \Big[ 2\ \phi_1\, \phi_2\, \phi_3\, \phi_4 - \Delta^{\phi}_{12}\, \phi_3\, \phi_4 - \Delta^{\phi}_{13}\, \phi_2\, \phi_4 - \Delta^{\phi}_{23}\, \phi_1\, \phi_4 \Big] \\
& + 2\, \Delta^{\phi}_{14}\, \Delta^{\phi}_{24}\, \phi_3^2 + 2\, \Delta^{\phi}_{14}\, \Delta^{\phi}_{34} \Big[ 2\ \phi_2\, \phi_3 -\ \Delta^{\phi}_{23} \Big] \\
& + 2\, \Delta^{\phi}_{24}\, \Delta^{\phi}_{34} \Big[ 2\ \phi_1\, \phi_3 -\ \Delta^{\phi}_{13} \Big] + (\Delta^{\phi}_{34})^2 \Big[ 2\ \phi_1\, \phi_2 - \Delta^{\phi}_{12} \Big] \Bigg) |0^{\phi}\rangle
\end{aligned}
\tag{A.0.5}
$$

$$
\begin{aligned}
\langle E^{h}_{\underline{2}4}\, E^{\chi}_{1\underline{3}}\, \mathcal{E}^{\chi h\chi h}_{\underline{1}23\underline{4}} \rangle = {}& 2\, \Delta^{h}_{24}\, \frac{1}{2\omega_p} \big( -e^{ip\cdot x_1}\, e^{-ip\cdot x_3} +\, e^{ip\cdot x_3}\, e^{-ip\cdot x_1} \big) \times 32\, g_h^2 g_\chi^2\, \Delta^{\phi}_{12} \\
& \langle 0^{\phi} | \Bigg( \Delta^{\phi}_{13} \phi_2\, \phi_3\, \phi_4^2 + \Delta^{\phi}_{23}\, \phi_1\, \phi_3\, \phi_4^2 - \Delta^{\phi}_{13}\, \Delta^{\phi}_{23}\, \phi_4^2
\end{aligned}
$$

$$
\begin{aligned}
&- \Delta^{\phi}_{14}\Delta^{\phi}_{23}\,\phi_3\,\phi_4 - \Delta^{\phi}_{24}\Delta^{\phi}_{13}\,\phi_3\,\phi_4 \\
&- \Delta^{\phi}_{34}\Big[\Delta^{\phi}_{13}\,\phi_2\,\phi_4 + \Delta^{\phi}_{23}\,\phi_1\,\phi_4 - \Delta^{\phi}_{14}\,\Delta^{\phi}_{23} - \Delta^{\phi}_{24}\,\Delta^{\phi}_{13}\Big]\Bigg)\,|0^{\phi}\rangle
\end{aligned}
\tag{A.0.6}
$$

$$
\begin{aligned}
\langle E^{h}_{\underline{2}\underline{4}}\,E^{\chi}_{13}\,\mathcal{E}^{\chi h\chi h}_{\underline{1}2\underline{3}4}\rangle = {}& 2\,\Delta^{h(H)}_{24}\,\frac{1}{2\omega_p}\big(-e^{ip\cdot x_1}\,e^{-ip\cdot x_3} - \,e^{ip\cdot x_3}\,e^{-ip\cdot x_1}\big)\times 16\,g_h^2 g_\chi^2\,\Delta^{\phi}_{12} \\
&\langle 0^{\phi}|\Bigg(2\,\Delta^{\phi}_{14}\Big[\phi_2\,\phi_3^2\,\phi_4 - \Delta^{\phi}_{23}\;\phi_3\,\phi_4\Big] + 2\,\Delta^{\phi}_{24}\Big[\phi_1\,\phi_3^2\,\phi_4 - \Delta^{\phi}_{13}\;\phi_3\,\phi_4\Big] \\
&+ 2\,\Delta^{\phi}_{34}\Big[2\;\phi_1\,\phi_2\,\phi_3\,\phi_4 - \Delta^{\phi}_{12}\,\phi_3\,\phi_4 - \Delta^{\phi}_{13}\,\phi_2\,\phi_4 - \Delta^{\phi}_{23}\,\phi_1\,\phi_4\Big] \\
&- 2\,\Delta^{\phi}_{14}\,\Delta^{\phi}_{24}\;\phi_3^2 - 2\,\Delta^{\phi}_{14}\,\Delta^{\phi}_{34}\Big[2\;\phi_2\,\phi_3 - \Delta^{\phi}_{23}\Big] \\
&- 2\,\Delta^{\phi}_{24}\,\Delta^{\phi}_{34}\Big[2\;\phi_1\,\phi_3 - \Delta^{\phi}_{13}\Big] - (\Delta^{\phi}_{34})^2\Big[2\;\phi_1\,\phi_2 - \Delta^{\phi}_{12}\Big]\Bigg)\,|0^{\phi}\rangle
\end{aligned}
\tag{A.0.7}
$$

$$
\begin{aligned}
\langle E^{h}_{\underline{2}\underline{4}}\,E^{\chi}_{1\underline{3}}\,\mathcal{E}^{\chi hh\chi}_{\underline{1}234}\rangle = {}& 2\,\Delta^{h(H)}_{24}\,\frac{1}{2\omega_p}\big(-e^{ip\cdot x_1}\,e^{-ip\cdot x_3} + \,e^{ip\cdot x_3}\,e^{-ip\cdot x_1}\big)\times 32\,g_h^2 g_\chi^2\,\Delta^{\phi}_{12} \\
&\langle 0^{\phi}|\Bigg(\Delta^{\phi}_{14}\Delta^{\phi}_{23}\,\phi_3\,\phi_4 + \Delta^{\phi}_{24}\Delta^{\phi}_{13}\,\phi_3\,\phi_4 \\
&+ \Delta^{\phi}_{34}\Big[\Delta^{\phi}_{13}\,\phi_2\,\phi_4 + \Delta^{\phi}_{23}\,\phi_1\,\phi_4 - \Delta^{\phi}_{14}\,\Delta^{\phi}_{23} - \Delta^{\phi}_{24}\,\Delta^{\phi}_{13}\Big]\Bigg)\,|0^{\phi}\rangle
\end{aligned}
\tag{A.0.8}
$$

$$
\begin{aligned}
\langle E^{h}_{\underline{3}4}\,E^{\chi}_{12}\,\mathcal{E}^{\chi\chi hh}_{\underline{1}\underline{2}3\underline{4}}\rangle = {}& 2\,\Delta^{h}_{34}\,\frac{1}{2\omega_p}\big(-e^{ip\cdot x_1}\,e^{-ip\cdot x_2} - \,e^{ip\cdot x_2}\,e^{-ip\cdot x_1}\big)\times 16\,g_h^2 g_\chi^2 \\
&\langle 0^{\phi}|\Bigg(2\,\Delta^{\phi}_{13}\Big(\phi_1\,\phi_2^2\,\phi_3\,\phi_4^2 - \,\Delta^{\phi}_{12}\,\phi_2\,\phi_3\,\phi_4^2\Big) \\
&+ 2\,\Delta^{\phi}_{23}\Big(\phi_1^2\,\phi_2\,\phi_3\,\phi_4^2 - \,\Delta^{\phi}_{12}\,\phi_1\,\phi_3\,\phi_4^2\Big) \\
&- 2\,\Delta^{\phi}_{13}\,\Delta^{\phi}_{23}\Big(2\,\phi_1\,\phi_2\,\phi_4^2 - \,\Delta^{\phi}_{12}\,\phi_4^2\Big) \\
&- (\Delta^{\phi}_{13})^2\,\phi_2^2\,\phi_4^2 - (\Delta^{\phi}_{23})^2\phi_1^2\,\phi_4^2 \\
&+ 2\,\Delta^{\phi}_{14}\Big[-\Delta^{\phi}_{13}\,\phi_2^2\,\phi_3\,\phi_4 - \Delta^{\phi}_{23}\big(2\,\phi_1\,\phi_2\,\phi_3\,\phi_4 - \,\Delta^{\phi}_{12}\,\phi_3\,\phi_4\big) \\
&\qquad + 2\,\Delta^{\phi}_{13}\,\Delta^{\phi}_{23}\,\phi_2\,\phi_4 + (\Delta^{\phi}_{23})^2\,\phi_1\,\phi_4\Big]
\end{aligned}
$$

$$
\begin{aligned}
&+ 2\,\Delta^{\phi}_{24}\Big[-\Delta^{\phi}_{23}\,\phi_1^2\,\phi_3\,\phi_4 - \Delta^{\phi}_{13}\big(2\,\phi_1\,\phi_2\,\phi_3\,\phi_4 - \Delta^{\phi}_{12}\,\phi_3\,\phi_4\big) \\
&\qquad + 2\,\Delta^{\phi}_{13}\,\Delta^{\phi}_{23}\,\phi_1\,\phi_4 + (\Delta^{\phi}_{13})^2\,\phi_2\,\phi_4\Big] \\
&+ 2\,\Delta^{\phi}_{34}\Big[-\Delta^{\phi}_{13}\big(\phi_1\,\phi_2^2\,\phi_4 - \Delta^{\phi}_{12}\,\phi_2\,\phi_4\big) \\
&\qquad - \Delta^{\phi}_{23}\big(\phi_1^2\,\phi_2\,\phi_4 - \Delta^{\phi}_{12}\,\phi_1\,\phi_4\big)\Big] \\
&+ 4\,\Delta^{\phi}_{14}\,\Delta^{\phi}_{24}\Big[\Delta^{\phi}_{13}\,\phi_2\,\phi_3 + \Delta^{\phi}_{23}\,\phi_1\,\phi_3 - \Delta^{\phi}_{13}\,\Delta^{\phi}_{23}\Big] \\
&+ 2\,\Delta^{\phi}_{14}\,\Delta^{\phi}_{34}\Big[\Delta^{\phi}_{13}\,\phi_2^2 + \Delta^{\phi}_{23}\big(2\,\phi_1\,\phi_2 - \Delta^{\phi}_{12}\big)\Big] \\
&+ 2\,\Delta^{\phi}_{24}\,\Delta^{\phi}_{34}\Big[\Delta^{\phi}_{23}\,\phi_1^2 + \Delta^{\phi}_{13}\big(2\,\phi_1\,\phi_2 - \Delta^{\phi}_{12}\big)\Big] \\
&+ (\Delta^{\phi}_{14})^2\Big[2\,\Delta^{\phi}_{23}\,\phi_2\,\phi_3 - (\Delta^{\phi}_{23})^2\Big] \\
&+ (\Delta^{\phi}_{24})^2\Big[2\,\Delta^{\phi}_{13}\,\phi_1\,\phi_3 - (\Delta^{\phi}_{13})^2\Big]\Bigg)\,|0^{\phi}\rangle
\end{aligned}
\tag{A.0.9}
$$

$$
\begin{aligned}
\langle E^{h}_{\underline{3}4}\,E^{\chi}_{1\underline{2}}\,\mathcal{E}^{\chi\chi hh}_{\underline{1}23\underline{4}}\rangle = {} & 2\,\Delta^{h}_{34}\,\frac{1}{2\omega_p}\big(-e^{ip\cdot x_1}\,e^{-ip\cdot x_2} + e^{ip\cdot x_2}\,e^{-ip\cdot x_1}\big) \times 32\,g_h^2 g_\chi^2\,\Delta^{\phi}_{12} \\
& \langle 0^{\phi}|\Bigg(\Delta^{\phi}_{13}\phi_2\,\phi_3\,\phi_4^2 + \Delta^{\phi}_{23}\,\phi_1\,\phi_3\,\phi_4^2 - \Delta^{\phi}_{13}\,\Delta^{\phi}_{23}\,\phi_4^2 \\
& - \Delta^{\phi}_{14}\Delta^{\phi}_{23}\,\phi_3\,\phi_4 - \Delta^{\phi}_{24}\Delta^{\phi}_{13}\,\phi_3\,\phi_4 \\
& - \Delta^{\phi}_{34}\Big[\Delta^{\phi}_{13}\,\phi_2\,\phi_4 + \Delta^{\phi}_{23}\,\phi_1\,\phi_4 - \Delta^{\phi}_{14}\,\Delta^{\phi}_{23} - \Delta^{\phi}_{24}\,\Delta^{\phi}_{13}\Big]\Bigg)\,|0^{\phi}\rangle
\end{aligned}
\tag{A.0.10}
$$

$$
\begin{aligned}
\langle E^{h}_{\underline{3}\underline{4}}\,E^{\chi}_{12}\,\mathcal{E}^{\chi\chi hh}_{\underline{1}\underline{2}34}\rangle = {} & 2\,\Delta^{h(H)}_{34}\,\frac{1}{2\omega_p}\big(-e^{ip\cdot x_1}\,e^{-ip\cdot x_2} - e^{ip\cdot x_2}\,e^{-ip\cdot x_1}\big) \times 16\,g_h^2 g_\chi^2 \\
& \langle 0^{\phi}|\Bigg(2\,\Delta^{\phi}_{14}\Big[\Delta^{\phi}_{13}\,\phi_2^2\,\phi_3\,\phi_4 + \Delta^{\phi}_{23}\big(2\,\phi_1\,\phi_2\,\phi_3\,\phi_4 - \Delta^{\phi}_{12}\,\phi_3\,\phi_4\big) \\
& \qquad - 2\,\Delta^{\phi}_{13}\,\Delta^{\phi}_{23}\,\phi_2\,\phi_4 - (\Delta^{\phi}_{23})^2\,\phi_1\,\phi_4\Big] \\
& 2\,\Delta^{\phi}_{24}\Big[\Delta^{\phi}_{23}\,\phi_1^2\,\phi_3\,\phi_4 + \Delta^{\phi}_{13}\big(2\,\phi_1\,\phi_2\,\phi_3\,\phi_4 - \Delta^{\phi}_{12}\,\phi_3\,\phi_4\big) \\
& \qquad - 2\,\Delta^{\phi}_{13}\,\Delta^{\phi}_{23}\,\phi_1\,\phi_4 - (\Delta^{\phi}_{13})^2\,\phi_2\,\phi_4\big)\Big] \\
& 2\,\Delta^{\phi}_{34}\Big[\Delta^{\phi}_{13}\big(\phi_1\,\phi_2^2\,\phi_4 - \Delta^{\phi}_{12}\,\phi_2\,\phi_4\big)
\end{aligned}
$$

$$
\begin{aligned}
&+ \Delta^{\phi}_{23}\big(\phi_1^2\,\phi_2\,\phi_4 - \Delta^{\phi}_{12}\,\phi_1\,\phi_4\big)\Big] \\
&- 4\,\Delta^{\phi}_{14}\,\Delta^{\phi}_{24}\Big[\Delta^{\phi}_{13}\,\phi_2\,\phi_3 + \Delta^{\phi}_{23}\,\phi_1\,\phi_3 - \Delta^{\phi}_{13}\,\Delta^{\phi}_{23}\Big] \\
&- 2\,\Delta^{\phi}_{14}\,\Delta^{\phi}_{34}\Big[\Delta^{\phi}_{13}\,\phi_2^2 + \Delta^{\phi}_{23}\big(2\,\phi_1\,\phi_2 - \Delta^{\phi}_{12}\big)\Big] \\
&- 2\,\Delta^{\phi}_{24}\,\Delta^{\phi}_{34}\Big[\Delta^{\phi}_{23}\,\phi_1^2 + \Delta^{\phi}_{13}\big(2\,\phi_1\,\phi_2 - \Delta^{\phi}_{12}\big)\Big] \\
&- (\Delta^{\phi}_{14})^2\Big[2\,\Delta^{\phi}_{23}\,\phi_2\,\phi_3 - (\Delta^{\phi}_{23})^2\Big] \\
&- (\Delta^{\phi}_{24})^2\Big[2\,\Delta^{\phi}_{13}\,\phi_1\,\phi_3 - (\Delta^{\phi}_{13})^2\Big]\Bigg)\,|0^{\phi}\rangle \qquad \text{(A.0.11)}
\end{aligned}
$$

$$
\begin{aligned}
\langle E^{h}_{\underline{3}\underline{4}}\,E^{\chi}_{1\underline{2}}\,\mathcal{E}^{\chi\chi hh}_{\underline{1}\underline{2}34}\rangle = {}& 2\,\Delta^{h(H)}_{34}\,\frac{1}{2\omega_p}\big(-e^{ip\cdot x_1}\,e^{-ip\cdot x_2} + e^{ip\cdot x_2}\,e^{-ip\cdot x_1}\big) \times 32\,g_h^2 g_\chi^2\,\Delta^{\phi}_{12} \\
&\langle 0^{\phi}|\Bigg(\Delta^{\phi}_{14}\Delta^{\phi}_{23}\,\phi_3\,\phi_4 + \Delta^{\phi}_{24}\Delta^{\phi}_{13}\,\phi_3\,\phi_4 \\
&+ \Delta^{\phi}_{34}\Big[\Delta^{\phi}_{13}\,\phi_2\,\phi_4 + \Delta^{\phi}_{23}\,\phi_1\,\phi_4 - \Delta^{\phi}_{14}\,\Delta^{\phi}_{23} - \Delta^{\phi}_{24}\,\Delta^{\phi}_{13}\Big]\Bigg)\,|0^{\phi}\rangle
\end{aligned}
\qquad \text{(A.0.12)}
$$

# Appendix B

# Symmetrising the Time-Ordering in the Annihilation Probability

Here, we show analytically how the time-ordered integral in Eq. (5.5.16) can be symmetrised and written as an integral over all times, as is shown in Fig. 5.14. We only show this for the diagrams containing Feynman loops, but the same procedure can be applied to each line in Fig. 5.14.

The terms containing Feynman loops in the annihilation probability are

$$
\begin{aligned}
\mathbb{P} \supset -\frac{4\, g_\psi^2 g_\phi^2}{(2\omega_{p_1})(2\omega_{p_2})} \int & \mathrm{d}^4x_1\, \mathrm{d}^4x_2\, \mathrm{d}^4x_3\, \mathrm{d}^4x_4 \\
& \Bigg( e^{ip_1\cdot x_1} e^{ip_2\cdot x_1} e^{-ip_1\cdot x_2} e^{-ip_2\cdot x_2} \left(F^\phi_{34}\right)^2 R^\chi_{13} R^\chi_{24}\, \Theta_{12}\Theta_{23}\Theta_{34} \\
& + e^{ip_1\cdot x_1} e^{ip_2\cdot x_1} e^{-ip_1\cdot x_2} e^{-ip_2\cdot x_2} \left(F^\phi_{34}\right)^2 R^\chi_{14} R^\chi_{23}\, \Theta_{12}\Theta_{23}\Theta_{34} \\
& + e^{ip_1\cdot x_1} e^{ip_2\cdot x_1} e^{-ip_1\cdot x_3} e^{-ip_2\cdot x_3} \left(F^\phi_{24}\right)^2 R^\chi_{12} R^\chi_{34}\, \Theta_{12}\Theta_{23}\Theta_{34} \\
& + e^{-ip_1\cdot x_1} e^{-ip_2\cdot x_1} e^{ip_1\cdot x_2} e^{ip_2\cdot x_2} \left(F^\phi_{34}\right)^2 R^\chi_{13} R^\chi_{24}\, \Theta_{12}\Theta_{23}\Theta_{34} \\
& + e^{-ip_1\cdot x_1} e^{-ip_2\cdot x_1} e^{ip_1\cdot x_2} e^{ip_2\cdot x_2} \left(F^\phi_{34}\right)^2 R^\chi_{14} R^\chi_{23}\, \Theta_{12}\Theta_{23}\Theta_{34} \\
& + e^{-ip_1\cdot x_1} e^{-ip_2\cdot x_1} e^{ip_1\cdot x_3} e^{ip_2\cdot x_3} \left(F^\phi_{24}\right)^2 R^\chi_{12} R^\chi_{34}\, \Theta_{12}\Theta_{23}\Theta_{34} \Bigg) ,
\end{aligned}
\tag{B.0.1}
$$

where the $\Theta$-function has been rewritten as $\Theta_{1234} = \Theta_{12}\Theta_{23}\Theta_{34}$. We are free to relabel the integration variables in each term as

$$
\begin{aligned}
\mathbb{P} \supset -\frac{4\, g_\psi^2 g_\phi^2}{(2\omega_{p_1})(2\omega_{p_2})} \int & \mathrm{d}^4x_1\, \mathrm{d}^4x_2\, \mathrm{d}^4x_3\, \mathrm{d}^4x_4 \\
& \Bigg( e^{ip_1\cdot x_1} e^{ip_2\cdot x_1} e^{-ip_1\cdot x_2} e^{-ip_2\cdot x_2} \left(F^\phi_{34}\right)^2 R^\chi_{13} R^\chi_{24}\, \Theta_{12}\Theta_{23}\Theta_{34}
\end{aligned}
$$

$$+ e^{ip_1\cdot x_1} e^{ip_2\cdot x_1} e^{-ip_1\cdot x_2} e^{-ip_2\cdot x_2} \left(F^{\phi}_{34}\right)^2 R^{\chi}_{13} R^{\chi}_{24}\, \Theta_{12}\Theta_{24}\Theta_{43}$$

$$+ e^{ip_1\cdot x_1} e^{ip_2\cdot x_1} e^{-ip_1\cdot x_2} e^{-ip_2\cdot x_2} \left(F^{\phi}_{34}\right)^2 R^{\chi}_{13} R^{\chi}_{24}\, \Theta_{13}\Theta_{32}\Theta_{24}$$

$$+ e^{-ip_1\cdot x_1} e^{-ip_2\cdot x_1} e^{ip_1\cdot x_2} e^{ip_2\cdot x_2} \left(F^{\phi}_{34}\right)^2 R^{\chi}_{14} R^{\chi}_{23}\, \Theta_{12}\Theta_{24}\Theta_{43}$$

$$+ e^{-ip_1\cdot x_1} e^{-ip_2\cdot x_1} e^{ip_1\cdot x_2} e^{ip_2\cdot x_2} \left(F^{\phi}_{34}\right)^2 R^{\chi}_{14} R^{\chi}_{23}\, \Theta_{12}\Theta_{23}\Theta_{34}$$

$$+ e^{-ip_1\cdot x_1} e^{-ip_2\cdot x_1} e^{ip_1\cdot x_2} e^{ip_2\cdot x_2} \left(F^{\phi}_{34}\right)^2 R^{\chi}_{14} R^{\chi}_{23}\, \Theta_{14}\Theta_{42}\Theta_{23} \Bigg)\,, \tag{B.0.2}$$

where we have used $F^{\phi}_{xy} = F^{\phi}_{yx}$. Using the relations

$$\Theta_{xy}\Theta_{yz} = \Theta_{xy}\Theta_{yz}\Theta_{xz} \tag{B.0.3}$$

$$\Theta_{xy} + \Theta_{yx} = 1 \tag{B.0.4}$$

$$\Theta_{xy} R^{\chi}_{xy} = R^{\chi}_{xy}\,, \tag{B.0.5}$$

we obtain

$$\mathbb{P} \supset -\frac{4\, g_{\psi}^2 g_{\phi}^2}{(2\omega_{p_1})(2\omega_{p_2})} \int \mathrm{d}^4x_1\, \mathrm{d}^4x_2\, \mathrm{d}^4x_3\, \mathrm{d}^4x_4 \Bigg( e^{ip_1\cdot x_1} e^{ip_2\cdot x_1} e^{-ip_1\cdot x_2} e^{-ip_2\cdot x_2} \left(F^{\phi}_{34}\right)^2 R^{\chi}_{13} R^{\chi}_{24}\, \Theta_{12}$$
$$+ e^{-ip_1\cdot x_1} e^{-ip_2\cdot x_1} e^{ip_1\cdot x_2} e^{ip_2\cdot x_2} \left(F^{\phi}_{34}\right)^2 R^{\chi}_{14} R^{\chi}_{23}\, \Theta_{12} \Bigg)\,. \tag{B.0.6}$$

Re-labelling $x_1 \leftrightarrow x_2$ in the second term yields

$$\mathbb{P} \supset -\frac{4\, g_{\psi}^2 g_{\phi}^2}{(2\omega_{p_1})(2\omega_{p_2})} \int \mathrm{d}^4x_1\, \mathrm{d}^4x_2\, \mathrm{d}^4x_3\, \mathrm{d}^4x_4\; e^{ip_1\cdot x_1} e^{ip_2\cdot x_1} e^{-ip_1\cdot x_2} e^{-ip_2\cdot x_2} \left(F^{\phi}_{34}\right)^2 R^{\chi}_{13} R^{\chi}_{24}\, (\Theta_{12} + \Theta_{21})$$
$$\supset -\frac{4\, g_{\psi}^2 g_{\phi}^2}{(2\omega_{p_1})(2\omega_{p_2})} \int \mathrm{d}^4x_1\, \mathrm{d}^4x_2\, \mathrm{d}^4x_3\, \mathrm{d}^4x_4 e^{ip_1\cdot x_1} e^{ip_2\cdot x_1} e^{-ip_1\cdot x_2} e^{-ip_2\cdot x_2} \left(F^{\phi}_{34}\right)^2 R^{\chi}_{13} R^{\chi}_{24}\,. \tag{B.0.7}$$

This result is equal to the first term in Eq. (B.0.1), but integrated over all times since there is no time-ordering $\Theta$-function.

# Appendix C

# Excitation Rate of an Inertial Detector in the Minkowski Vacuum

Here we consider the excitation probability of an inertial detector in the Minkowski vacuum. For a time-like interval, one can always boost to a frame in which $\mathbf{x}_1 - \mathbf{x}_2 = 0$, such that,

$$\Delta_{12}^{R}\big|_{\alpha=0} = \int \frac{\mathrm{d}^4 p}{(2\pi)^4} \frac{e^{-ip^0 t_{12}}}{(p^0 + i\epsilon)^2 - \omega_{\mathbf{p}}^2} = -\frac{1}{2\pi^2}\int_m^\infty \mathrm{d}\omega_{\mathbf{p}}\, \sqrt{\omega_{\mathbf{p}}^2 - m^2}\, \sin(\omega_{\mathbf{p}} t_{12}), \tag{C.0.1a}$$

$$\Delta_{12}^{H}\big|_{\alpha=0} = \int \frac{\mathrm{d}^4 p}{(2\pi)^4}\, e^{-ip^0 t_{12}} 2\pi\delta(p^2 - m^2) = \frac{1}{2\pi^2}\int_m^\infty \mathrm{d}\omega_{\mathbf{p}}\, \sqrt{\omega_{\mathbf{p}}^2 - m^2}\, \cos(\omega_{\mathbf{p}} t_{12}). \tag{C.0.1b}$$

Using equation (6.2.20), we have

$$\begin{aligned}\mathbb{P}(2;t)\big|_{\alpha=0} &= \frac{|\mu|^2}{2\pi^2}\int_0^t \mathrm{d}t_1 \int_0^{t_1} \mathrm{d}t_2 \int_m^\infty \mathrm{d}\omega_{\mathbf{p}}\, \sqrt{\omega_{\mathbf{p}}^2 - m^2}\big[\cos(\Omega t_{12})\cos(\omega_{\mathbf{p}} t_{12}) \\ &\qquad - \sin(\Omega t_{12})\sin(\omega_{\mathbf{p}} t_{12})\big] \\ &= \frac{|\mu|^2}{2\pi^2}\int_0^t \mathrm{d}t_1 \int_0^{t_1} \mathrm{d}t_2 \int_m^\infty \mathrm{d}\omega_{\mathbf{p}}\, \sqrt{\omega_{\mathbf{p}}^2 - m^2}\, \cos[(\omega_{\mathbf{p}} + \Omega)t_{12}].\end{aligned} \tag{C.0.2}$$

Performing the time integrals, we have

$$\mathbb{P}(2;t)\big|_{\alpha=0} = \frac{|\mu|^2}{2\pi^2}\int_m^\infty \mathrm{d}\omega_{\mathbf{p}}\, \sqrt{\omega_{\mathbf{p}}^2 - m^2}\, \frac{1 - \cos[(\omega_{\mathbf{p}} + \Omega)t]}{(\omega_{\mathbf{p}} + \Omega)^2}, \tag{C.0.3}$$

such that the transition rate is

$$\frac{\partial \mathbb{P}(2;t)}{\partial t}\bigg|_{\alpha=0} = \frac{|\mu|^2}{2\pi^2}\int_m^\infty \mathrm{d}\omega_{\mathbf{p}}\, \sqrt{\omega_{\mathbf{p}}^2 - m^2}\, \frac{\sin[(\omega_{\mathbf{p}} + \Omega)t]}{\omega_{\mathbf{p}} + \Omega}. \tag{C.0.4}$$

This transition rate exhibits an ultraviolet (UV) divergence due to the treatment of the detector as point-like. By considering the difference of two transition rates in Eq. (6.2.30) (e.g., accelerated rate minus inertial rate), we subtract this divergence.